\documentclass[smallextended]{svjour3}       
\smartqed  
\usepackage{graphicx}
\usepackage{multirow}
\usepackage{threeparttable}
\usepackage{mathrsfs}
 \journalname{Bulletin of Mathematical Biology}

\usepackage{hyperref}

\usepackage{amsmath} \usepackage{amsfonts,euscript,amssymb}
\usepackage{graphicx} \usepackage{subcaption} \usepackage[font=scriptsize]{caption}
\usepackage{epsf}
\usepackage{subcaption}
\usepackage{tabularx}

\usepackage{mathtools}
\usepackage{calc}   

\usepackage{color}

\makeatletter
\newcommand{\vast}{\bBigg@{4}}
\makeatother

\definecolor{Darkgreen}{rgb}{0,0.45,0}
\newcommand{\rs}[1]{\textcolor{black}{#1}}
\newcommand{\dt}[1]{\textcolor{black}{#1}}

\newcommand{\eequ}{\end{equation}}
\newcommand{\bequ}{\begin{equation}}
\newcommand{\eequd}{\end{eqnarray*}}
\newcommand{\bequd}{\begin{eqnarray*}}

\def\K{\mathcal{K}}

\def\P{\mathcal{P}}

\def\Bila{\mathbf{B}}

\def\P{\mathcal{P}}

\def\K{\mathcal{K}}

\def\R{\mathbb{R}}

\def\Bila{\mathbf{B}}

\newcommand {\nor} [1]{\parallel #1 \parallel}

\begin{document}

\title{Cell-scale degradation of peritumoural extracellular matrix fibre network and its role within tissue-scale cancer invasion}


\titlerunning{Multiscale Modelling of Cancer Invasion}        

\author{Robyn Shuttleworth
            \and
            Dumitru Trucu}


\institute{Robyn Shuttleworth\at
            Division of Mathematics, University of Dundee, Dundee, DD1 4HN, Scotland, UK\\
            \email{r.shuttleworth@dundee.ac.uk}
	\and
	Dumitru Trucu \at
	Division of Mathematics, University of Dundee, Dundee, DD1 4HN, Scotland, UK\\
	\email{trucu@maths.dundee.ac.uk}
}

\date{Received: date/ Accepted: date}

\maketitle
\begin{abstract}
Local cancer invasion of tissue is a complex, multiscale process which plays an essential role in tumour progression. Occurring over many different temporal and spatial scales, the first stage of invasion is the secretion of matrix degrading enzymes (MDEs) by the cancer cells that consequently degrade the surrounding extracellular matrix (ECM). This process is vital for creating space in which the cancer cells can progress and it is driven by the activities of specific matrix metalloproteinases (MMPs). In this paper, we consider the key role of two MMPs by developing \dt{further} the novel two-part multiscale model introduced in \cite{Shutt_2018} \dt{to better relate at micro-scale the two micro-scale activities that were considered there}, namely, \dt{the micro-dynamics concerning the continuous rearrangement of the naturally oriented ECM fibres within the bulk of the tumour and MDEs proteolytic micro-dynamics that take place in an appropriate cell-scale neighbourhood of the tumour boundary.} 

Focussing primarily on the activities of the membrane-tethered MT1-MMP and the soluble MMP-2 with the fibrous ECM phase, \dt{in this work we} investigate the MT1-MMP/MMP-2 cascade and its overall effect on tumour progression. To that end, we will \dt{propose a} new multiscale \dt{modelling framework} by considering the degradation of the ECM fibres not only \dt{to take place at macroscale in the bulk of the tumour}  but also \dt{explicitly in the micro-scale neighbourhood of the tumour interface as a consequence of the interactions with molecular fluxes of MDEs that exercise their spatial dynamics at the invasive edge of the tumour.}

%

\end{abstract}
%
%

\section{Introduction}
Cancer cell invasion of tissue is a complex, multiscale process in which gene mutations in healthy cells promote enhanced proliferation and the production of proteolytic enzymes. One of the first steps of the local invasion of tissue is the secretion of matrix degrading enzymes (MDEs) and the consequential degradation of the extracellular matrix (ECM). The ECM is a non-cellular structure that provides not only support to surrounding cells and tissues, but also acts as a platform for cellular communication. This feature is of particular use to cancer cells, which take advantage of the molecular interactions mediated by various ECM components and favourably utilise these as means for achieving invasion of the surrounding tissue. 

The ECM is comprised of a variety of secreted proteins, and the main constituent of the ECM is the structural cross-linked collagen type I, a dense network of fibres that give the ECM its rigidity. In order for a tumour to progress, these strong fibres must be broken down and degraded to free space for the cancer cells. One of the first MDEs to interact with the ECM is the membrane tethered MMP, MT1-MMP, or MMP-14. This MMP exhibits strong collagenolytic capabilities in which they are able to cleave the cross-linked collagen type I fibres and break them into shorter, soluble fibres. These soluble fibres are then degraded by the freely diffusible MMP-2, activated through the cleavage of proMMP-2 molecules by MT1-MMP. This MT1-MMP/MMP-2 cascade is highly effective in promoting tumour invasion through ECM degradation, however, there are some disadvantages to both types of MMPs. MT1-MMP molecules can overcome high density regions of collagen type I, particularly the cross-linked fibres, however they do not degrade the collagen, rather cleaving the fibres into smaller fibrils \cite{Tam_2004}. On the other hand, MMP-2 cannot degrade the dense cross-linked fibres, but can degrade the smaller, soluble fibrils within the peritumoural region \cite{Doren_2015}. Consequently, these two MMPs work in harmony with one another for successful invasion of tissue. 

Besides the abnormal rate of MDE secretion, the invasive capabilities of a tumour are strengthened by many other processes, including increased proliferation rates and the ability to adapt cellular adhesion properties. Cell adhesion is an essential process in which cells interact and attach to neighbouring cells through calcium dependent cell-adhesion molecules, known as CAMs, on the cell surface \cite{Humphries2006}. Cell-cell adhesion is dependent on specific cell-signalling pathways that are formed between $\text{Ca}^{2+}$ ions and calcium sensing receptors in the ECM \cite{Ko_2001}. This calcium dependent cell-cell signalling is regulated by a subfamily of glycoproteins known as E-cadherins that bind with intra-cellular proteins known as catenins, typically $\beta$-catenin, forming the E-cadherin/catenin complex. Any alteration to the function of $\beta$-catenin will result in a loss of the ability of E-cadherin to initiate cell-cell adhesion \cite{Wijnhoven2000}. This connection between E-cadherin and $\beta$-catenin and the calcium cell-signalling mechanism was first recorded in colon carcinoma \cite{Bhagavathula2007}. Additionally, cells can also bind to the ECM through cell-matrix adhesion \cite{Lodish2000}. Mediated by calcium independent CAMs, known as integrins, cell-matrix adhesion enables binding of cells to various components of the ECM, such as collagen and fibronectin, with this contributing to cancer cell migration within the tissue. Moreover, enhanced cell-matrix adhesion paired with a loss in cell-cell adhesion facilitates a quicker spread of the cancer cells into the surrounding tissue \cite{CAVALLARO200139}.

Over the past $25$ years there has been an increasing interest in the mathematical modelling of cancer invasion, see \dt{for example} \cite{Andasari2011,Anderson2005,Anderson2000,Chaplain2011,Chaplain2006a,Gerisch2008,Peng2016,Ramis-Conde_et_al_2000,szymanska_08,Dumitru_et_al_2013}. Many models of cancer invasion have focussed on the interactions between cancer cells and the extracellular matrix through a variety of different approaches, ranging from continuous and discrete individual cell-based models to the more complex multiscale modelling approach. Numerous processes of cancer invasion have been addressed, including the effects of proliferation and cellular adhesion \cite{Bitsouni_2017,Chauviere_2007,Domschke_et_al_2014,Gerisch2008,Painter2008,Ramis-Conde_et_al_2000}, as well as models developed to describe the migration strategies of cancer cells in tissue networks \cite{Chauviere_2007,Hillen2006,Perumpanani_et_al_1998}. A model describing both the mesenchymal and amoeboid motion of cells through a fibre network \cite{Painter2008} concluded that a structured ECM can induce cell aggregation in amoeboid type cells, whereas the actions of both contact-guidance and ECM remodelling are sufficient processes for mesenchymal type cell invasion to occur. 

Numerous processes of cancer invasion have been investigated, including the effects of proliferation and cellular adhesion \cite{Bitsouni_2017,Chauviere_2007,Domschke_et_al_2014,Gerisch2008,Painter2008,Ramis-Conde_et_al_2000} as well as the secretion and interaction of proteolytic enzymes, specifically MMPs and uPAs with the ECM \cite{Andasari2011,Chaplain2005,Chaplain2006,Peng2016,shutt_chapter,Dumitru_et_al_2013} during tumour invasion. In particular, a \dt{single-scale} model developed by \cite{Deakin_Chaplain_2013} focussed on the roles of two MMP molecules, namely the membrane-bound MT1-MMP and the soluble MMP-2, where the MT1-MMP/MMP-2 cascade was \dt{considered}, \dt{highlighting the importance of } MT1-MMP matrix remodelling \dt{within} collagen-rich environments. These model dynamics are of great interest to us for the current investigation of the peritumoural MMP processes.

Finally, as the invasion process is naturally multiscale, with its dynamics ranging from sub-cellular-, cellular- to tissue-scale, major advances have been witnessed in the multiscale modelling of cancer invasion \cite{anderson_07,Peng2016,Ramis-Conde_et_al_2000,shutt_chapter,Dumitru_et_al_2013}. In particular, advancements have been made in towards two-scale approaches, modelling and appropriately linking the spatio-temporal dynamics occurring at different scales, as first proposed in \cite{Dumitru_et_al_2013} and extended upon in \cite{Peng2016,Shutt_twopop,Shutt_2018}.

In this paper, we aim to advance the novel two-part multiscale model developed in \cite{Shutt_2018} \dt{by considering the multiscale contribution of ECM fibres within the invasive edge proteolytic dynamics of MDEs. This will pave the way for the establishment of an additional cell-scale link between the peritumoural micro-scale fibres rearrangements and the micro-scale proteolytic MDEs activities at the tumour boundary, which ultimately extends and complements the two-parts multiscale framework introduced in \cite{Shutt_2018} where the linking of its two constituent multiscale systems was so far mediated only through the shared macro-dynamics.}

%

\section{The mathematical modelling approach}
\dt{As this work} build\dt{s} upon the two-part multiscale model introduced and developed in \cite{Shutt_2018} that investigates cancer invasion within a heterogeneous ECM, \dt{let us start by} revisiting the multi-component structure of the ECM and macroscopic tumour dynamics for one cancer cell population considered in \cite{Shutt_2018}, while adapting the model to include further integration between the leading edge MDE and fibres rearrangement micro-scale processes.  


Since this paper builds on the modelling developed in \cite{Shutt_2018} and \cite{Dumitru_et_al_2013}, let us start this section by recalling the key details of the framework terminology that we introduced there. Let us denote the support of the invading tumour region as $\Omega(t_{0})$, and assume this evolves in the maximal reference tissue cube $Y \in \mathbb{R}^N$ for $N=2,3$, centred at the origin of the space.  At any spatio-temporal node $(x,t) \in \Omega(t_{0}) \times [0,T]$, we consider the tumour to be comprised of a cancer cell population $c(x,t)$, integrated within a multiphase heterogeneous ECM density denoted by $v(x,t)$.  Specifically, the heterogeneous ECM is regarded as comprising of a \emph{fibres} component and \emph{non-fibres} soluble component. \dt{We denote the tissue-scale (macro-scale) density of }\emph{non-fibres} ECM phase by $l(x,t)$ and we consider to include all non-fibre components of the ECM, i.e., elastin, laminins, fibroblasts, etc. \dt{On the other hand, the macro-scale mass density of the} \emph{fibres} ECM phase \dt{is denoted by $F(x,t)$ and accounts for} all fibrous proteins such as collagen and fibronectin within the matrix.

For completion, \dt{in the following the following subsections, we will briefly revisit the key points about the ECM fibres multiscale structure and functionality alongside} the \dt{tumour} macro-dynamics as well as the associated terminology and notations \dt{that were introduced and derived in \cite{Shutt_2018}. }

%

\subsection{The multiscale ECM fibre structure and its contribution to the tissue dynamics}
\begin{figure}[h]
\centering
\includegraphics[scale=0.4]{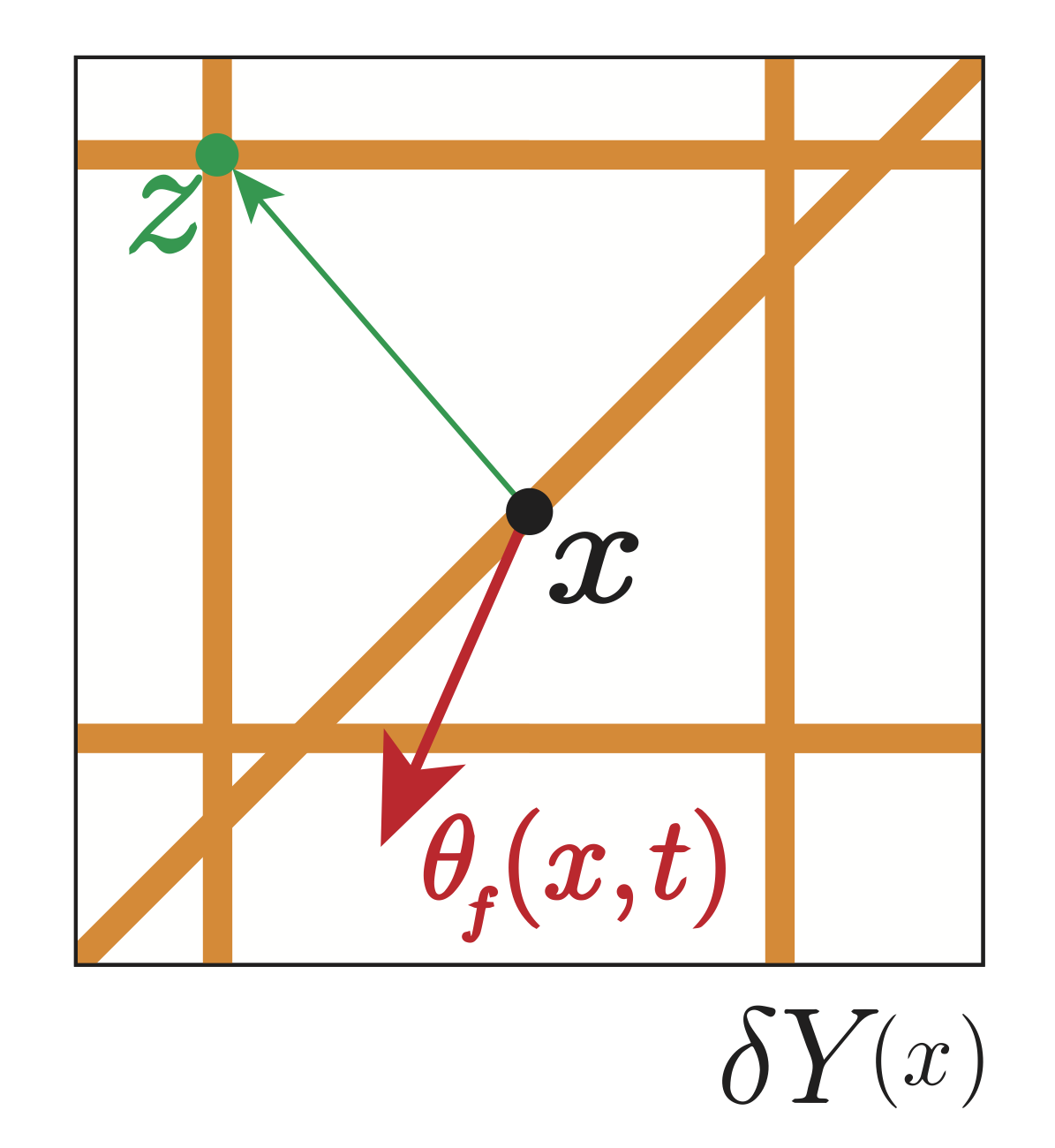}
\caption[Schematic of the micro-fibres distribution on the micro-domain $\delta Y(x)$, centred at $x$.]{\emph{Schematic of the micro-fibres distribution on the micro-domain $\delta Y(x)$, centred at $x$, with the barycentral position vector $\protect\overrightarrow{x \,z}:=z-x $ pointing towards an arbitrary micro-location $z\in \delta Y(x)$ illustrated by the green arrow.}}
\label{fig:fibre_cubes}
\end{figure}

\dt{As derived in \cite{Shutt_2018},} at any macro-position $x \in \Omega(t_{0})$, the ECM-fibre phase can be represented through a macro-scale vector field, \dt{denoted} $\theta_{f}(x,t)$, that \dt{captures} not only the macroscopic distribution of fibres $F(x,t)$, but also their naturally arising macroscopic orientation \dt{that is induced by their mass distribution of micro-fibres, \dt{denoted} $f(z,t)$, over a \dt{given} micro-domain $\delta Y(x)$ centred at the given macroscopic point $x$ of cell-scale $\delta>0$ (i.e., $\delta Y(x):=x+\delta Y$)}. An example of micro-fibres $f(z,t)$ patterns over a micro-domain  $\delta Y(x)$ (defined in Appendix \ref{microfibres}), alongside the naturally emerging macroscopic fibre orientation $\theta_{f}(x,t)$ that was derived in \cite{Shutt_2018}, are shown schematically in Figure \ref{fig:fibre_cubes}. 

In brief, while referring to its full derivation presented in \cite{Shutt_2018}, the naturally generated revolving barycentral orientation $\theta_{_{f,\delta Y(x)}}(x,t)$ associated with $\delta Y(x)$ is given by the \emph{Bochner-mean-value} of the position vectors function 
\[
\delta Y(x)\ni z\longmapsto z-x\in\mathbb{R}^{N}
\]
with respect to the density measure $f(z,t)\lambda(\cdot)$, where $\lambda(\cdot)$ is the usual Lebesgue measure (see \cite{yosida1980}), and so this is expressed mathematically as:
\bequ
\theta_{_{f,\delta Y(x)}}(x,t)=\frac{\int\limits_{\delta Y(x)}f(z,t)(z-x)dz}{\int\limits_{\delta Y(x)} f(z,t)  dz}. 
\eequ
Following on, at any spatio-temporal node $(x,t)$, \dt{this revolving barycentral orientation $\theta_{_{f,\delta Y(x)}}(x,t)$ induces the naturally arising macroscopic fibre orientation vector field representation that is defined as} 
\bequ\label{fiborien}
\theta_{_{f}}(x,t)=\frac{1}{\lambda (\delta Y(x))} \int_{\delta Y(x)} f(z,t) \ dz \cdot \frac{\theta_{_{f,\delta Y(x)}} (x,t)}{||\theta_{_{f,\delta Y(x)}} (x,t)||}.
\eequ
in which macroscopic mean-value fibre representation at any $(x,t)$ is given by the Euclidean magnitude of $\theta_{_{f}}(x,t)$, namely,
\bequ\label{fibmean}
F(x,t):=||\theta_{_{f}}(x,t)||_{2}.
\eequ
Thus, the total ECM distributed at any spatio-temporal node $(x,t)$ is therefore given by $v(x,t)=l(x,t)+F(x,t)$.

\subsection{Macro-scale dynamics}\label{macrodynamics}
For the tissue dynamics of a tumour consisting of a single cell population, again respecting the terminology and notations introduced in \cite{Shutt_2018}, we denote the macro-scale tumour population, including the surrounding two-phase ECM, by the vector
\[
\textbf{u}(x,t)=(c(x,t),F(x,t),l(x,t))
\]
with the tumour's \emph{volume of occupied space} given as
\[
\rho(\textbf{u}(x,t)) = \vartheta_{v}(F(x,t)+l(x,t)) + \vartheta_{c}c(x,t),
\]
where $\vartheta_{v}$ and $\vartheta_{c}$ represent the fractions of physical space occupied by the entire ECM and the cancer cells respectively. Focussing first on the cancer cell population, the spatial movement of the cancer cells is governed by random motility (approximated here by diffusion) and a cell-adhesion process that includes both cell-cell and cell-matrix adhesion, with cell-matrix adhesion accounting for both cell-fibre and cell-non-fibre adhesion. Assuming the cells are subject to a logistic proliferation law, the dynamics of the cancer cell population can be mathematically represented as
\bequ \label{eq:cancermacro}
\frac{\partial c}{\partial t} = \nabla \cdot [D_{1} \nabla c - c \mathcal{A}(x,t,\textbf{u}(\cdot,t),\theta_{f}(\cdot,t))] + \mu_{1}c(1 - \rho(\textbf{u}))
\eequ
where: $D_{1}$ and $\mu_{1}$ are the non-negative diffusion and proliferation rates, respectively. The non-local flux $\mathcal{A}(x,t,\textbf{u}(\cdot,t),\theta_{f}(\cdot,t))$ describes the spatial movement of the cancer cells due to cellular adhesion properties between each other (cell-cell adhesion) and the surrounding environment (cell-matrix adhesion accounting for both cell-fibre and cell-non-fibre adhesion). The adhesive flux considers the interactions of cancer cells within a \emph{sensing radius} $R$ with the other cancer cells and two-phase ECM distributed on the \emph{sensing region} $\textbf{B}(x,R):=x+\{\xi\in \R^{N}\,|\, \nor{\xi}_{_{2}}<R\}$, described by the following adhesive flux term
\bequ
\begin{split}
\mathcal{A}(t,x,\textbf{u}(\cdot, t), \theta_{_{f}}(\cdot, t))=\frac{1}{R} \int_{B(0,R)} \mathcal{K}(\nor{\!y\!}_{2}) & \big(n(y) (\textbf{S}_{_{cc}} c(x+y,t) + \textbf{S}_{_{cl}} l(x+y,t)) \\
&+ \hat{n}(y) \ \textbf{S}_{_{cF}} F(x+y,t) \big)(1-\rho(\textbf{u}))^{+}
\end{split}
\label{adhesionterm}
\eequ
Whilst the influence of the \dt{spatially distributed} adhesive interactions \dt{on the sensing region $\textbf{B}(x,R)$} is \dt{accounted} through \dt{a} radial kernel $\mathcal{K}(\cdot)$ detailed in Appendix \ref{kernelAppendix}, we explore the strength of the bonds created between the cells distributed at $x$ and the other cells or non-fibre ECM phase distributed at $y$ in the direction of the unit normal
\bequ
n(y):=
\left\{
\begin{array}{l}
y/||y||_2 \quad\text{if} \  y \in \Bila(0,R)\setminus\{(0,0)\}, \\[0.2cm]
(0,0) \quad \quad \text{if} \ y=(0,0).
\end{array}
\right.
\eequ
through the cell-cell and cell-non-fibre ECM adhesion strength coefficients $\textbf{S}_{cc}$ and $\textbf{S}_{cl}$, respectively. We consider $\textbf{S}_{cl}$ to be constant, whilst the coefficient representing cell-cell adhesion, $\textbf{S}_{cc}$, is monotonically dependent on the levels of extracellular $\text{Ca}^{2+}$ ions which enable \dt{robust} adhesive bonds between neighbouring cells \cite{Gu2014,Hofer2000}. Therefore we assume $\textbf{S}_{cc}$  is dependent on the underlying non-fibre ECM phase, smoothly ranging from $0$ to a $\text{Ca}^{2+}$-saturation level denoted $\textbf{S}_{_{max}}$ and is taken as 
\[
\textbf{S}_{_{cc}}(x,t):=\textbf{S}_{_{max}}e^{\big({1-\frac{1}{1-(1-l(x,t))^2}}\big)}.
\] 
The last term in \eqref{adhesionterm} considers the interactions between the cancer cells and the fibre ECM phase distributed within the region $\textbf{B}(x,R)$. The strength of this interaction is proportional to the distribution of macro-fibres at $F(x+y,t)$ and it is the orientation of these fibres, $\theta_{f}(x+y,t)$, that biases the direction in which adhesive bonds are formed in the direction $\hat{n}(\cdot)$

\bequ
\hat{n}(y):=
\left\{
\begin{array}{lll}
\frac{y+\theta_f(x+y)}{||y+\theta_f(x+y)||_{_{2}}},  \ & \text{if}& \  (y+\theta_f(x+y))  \neq (0,0), \\[0.2cm]
(0,0) \in \mathbb{R}^2, & &  \text{otherwise}.
\end{array}
\right.
\eequ

Finally, we describe the macroscopic dynamics of the ECM involving both the fibre and non-fibres phase on the invading tumour domain $\Omega(t)$, where the fibre and non-fibre ECM phase are subject to an overall degradation by the cancer cells and can be mathematically represented as 
\begin{align}
\frac{d F}{d t} &= -\gamma_{1} cF, \label{eq:fib} \\ 
\frac{d l}{d t} &= -\gamma_{2} cl, \label{eq:nonfib}
\end{align}
where $\gamma_{1}, \gamma_{2}$ are the degradation rates for fibres and non-fibres ECM phases respectively.

\subsection{Microscopic fibre rearrangement instigated by the macro-scale cell-flux}
As derived in \cite{Shutt_2018}, during the macro-dynamics, as the cancer cells invade the surrounding tissue they have the ability to push the fibres in the direction of \dt{their spatial flux} and rearrange the micro-fibre mass distributions, thereby reorienting the macro-fibres direction. Thus, at time $t$ and at any spatial location $x \in \Omega(t_{0})$, the micro-fibres $f(z,t), \ \forall z \in \delta Y(x)$ undergo a microscopic rearrangement process fuelled by the cancer cell-flux direction \dt{given mathematically as}
\[
\mathcal{F}(x,t) := D_{1} \nabla c(x,t) - c(x,t) \mathcal{A}(x,t,\textbf{u}(\cdot,t),\theta_{f}(\cdot,t)).
\]   
\dt{Hence, the micro-fibres $f(z,t), \ \forall z \in \delta Y(x)$ are acted upon uniformly by the resultant force of the \emph{rearrangement flux} vector-valued function
\[
r(\delta Y(x),t):=\omega(x,t)\mathcal{F}(x,t)+(1-\omega(x,t))\theta_{f}(x,t),
\]
where the spatial flux of the cancer cells $\mathcal{F}(x,t)$ is balanced in a weighted manner by the orientation $\theta_{f}(x,t)$ of the distribution of fibres at $(x,t)$ and the amount of cells exercising spatial transport at $(x,t)$, with a naturally emerging weight given by}
\[
\omega(x,t)=\frac{c(x,t)}{c(x,t)+F(x,t)}.
\]
As detailed in \cite{Shutt_2018}, under the influence of the force $r(\delta Y(x),t)$, an appropriate level of micro-fibres mass $f(z,t)$ will undergo a spatial movement towards a new position 
\[
z^{*}:=z + \nu_{_{\delta Y(x)}}(z,t)
\]
where the relocation direction and magnitude is given by
\begin{equation}
\nu_{_{\delta Y(x)}}(z,t)=\left(x_\text{dir}(z) + r(\delta Y(x), t)\right) \cdot \frac{f(z,t)(f_{\max}-f(z,t))}{f^{*}+||r(\delta Y(x)) - x_\text{dir}(z)||_{2}} \cdot \chi_{_{\{f(\cdot,t)>0\}}}
\label{eq:fibnu}
\end{equation}
\dt{with $f_{\max}$ representing the maximum amount of micro-fibres at $z\in\delta Y(x)$, and the relative degree of fibre's occupancy at $z$ being denoted by $f^{*}(z,t):=\frac{f(z,t)}{f_\text{max}}$.} Finally, the spatial transport of micro-fibres at $z$ will be exercised in accordance to the physical space available at the new position $z^{*}$, which is explored through a movement probability $p_{move}:=\max (0,1-f^{*}(z^{*},t))$, where an amount of micro-fibres $p_{move}f(z,t)$ will be moved to the new position $z^{*}$ while the rest of the micro-fibres will remain at $z$.

\section{Novel modelling for the MDE proteolytic micro-scale dynamics at the tumour interface}\label{microdynamics}

To incorporate the dynamics of peritumoural micro-scale fibres with the MDE micro-dynamics at the tumour invasive edge, let us start by exploring the influence of the macro-fibre distributions on the emergence of a cell-scale molecular source of MDE at the tumour interface. \dt{This will effectively enhance the \emph{top-down} macro-micro link derived and introduced in \cite{Dumitru_et_al_2013} (and which was considered in \cite{Shutt_2018}) for the boundary micro-dynamics and in return will influence the \emph{bottom-up} feedback link to macro-dynamics. Finally, MDEs micro-dynamics occurring within a cell-scale neighbourhood of the tumour boundary will be explored on an appropriately selected covering bundle of boundary micro-domains $\{\epsilon Y\}_{\epsilon Y \in \P}$, which was introduced and constructed with complete details in \cite{Dumitru_et_al_2013}.}

\paragraph{The top-down link} As discussed previously, the expansion of the tumour boundary is dependent on the peritumoural degradation of ECM by the matrix-degrading enzymes (MDEs). \dt{However}, the secretion of MDEs induced by the distribution of cancer cells \dt{in the outer proliferating rim is dependent upon the structure of the ECM and in particular upon the} fibre distribution. \dt{Indeed,} in the presence of a high distribution of \dt{ECM} fibres, the \dt{cancer} cells exhibit a strong rate of MT1-MMP secretion, \dt{which} in turn \dt{leads to} an increase in the activation of proMMP-2 molecules \cite{zigrino_2001}. \dt{Thus, since in} this \dt{emerging} MT1-MMP/MMP-2 cascade, the secretion rate of MT1-MMP \dt{correlates directly} to the \dt{amount} of MMP-2 molecules, \dt{we obtain therefore that the presence of the ECM fibres enhances the production of the MMP-2 molecules that are then} released at the tumour boundary. 

\dt{Thus,} the MMPs \dt{source} at each $y \in \epsilon Y \cap \Omega(t_{0})$ arises as a collective contribution of the cancer cells within the outer proliferating rim which are further enhanced by the presence of \dt{ECM} fibres, \dt{and therefore} this can be mathematically expressed as
\begin{align}
\begin{split}
1. \quad &g_{\epsilon Y}(y,\tau) = \frac{\int\limits_{\textbf{B}(y,\gamma)\cap\Omega(t_0)} \alpha c (x,t_0 + \tau) \ \widetilde{F}(x,t_0 + \tau) \ dx}{\widetilde{F} \cdot \lambda (\textbf{B}(y,\gamma)\cap\Omega(t_0))}, \quad y \in \epsilon Y \cap \Omega(t_0), \\[0.7cm]
2. \quad &g_{\epsilon Y}(y,\tau) = 0,\quad  y \in \epsilon Y \setminus \big( \Omega(t_0)+\{ y \in Y|  \ ||y||_2 < \gamma\}), 
\end{split}
\label{eq:sourceMDEs}
\end{align}
where $\gamma$ represents the maximum thickness of the outer proliferating rim, $\Bila(y,\gamma):=\{\xi\in \R^{N}\,|\, \nor{y-\xi}_{_{\infty}}\leq \gamma\}$ and $\alpha$ is an MMP secretion rate for the cancer cell population. \dt{Furthermore, } the function $\widetilde{F}(x,t + \tau):=1+F(x,t)$\dt{, $\forall \, t>0$} explores the \dt{spatially distributed} enhancement of the source of MMPs produced by the cancer cells \dt{that is enabled} through the presence of fibres \cite{zigrino_2001} \dt{(i.e., } a high\dt{er} distribution of fibres inducing a greater number of MMP-2 molecules\dt{)}. \dt{Finally}, $\widetilde{F} \cdot \lambda$ represents the underlying fibre density measure that is defined by 
\[
\widetilde{F} \cdot \lambda(G):=\int \limits_{G} \widetilde{F}(x, \dt{t_{0}+\tau}) \ dx, \quad G \in \Sigma(Y),
\]
where $\lambda$ is the standard Lebesgue measure on $\mathbb{R}^{2}$, and \dt{$G$ is a nonempty Borel subset set of $Y$, i.e.,  $G\in\Sigma(Y)$, with $\Sigma(Y)$ representing the Borel $\sigma$-algebra on $Y$}. 

In the presence of this source, \dt{on any micro-domain $\epsilon Y$, a cross-interface MMP-2 diffusive transport takes place, and as the MMP-2 find it easier to exercise their random movement in regions of lower micro-fibres density, the diffusion rate is micro-fibre density dependent, and therefore this spatio-temporal micro-dynamics can be mathematically formulated as}
\bequ \label{eq:mde}
\frac{\partial m}{\partial \tau} = \underbrace{D_m\dt{(f)} \Delta m}_{\text{diffusion}} \ \ +\underbrace{g_{\epsilon Y}(y,\tau)}_{\text{source term}},  \quad  y \in \epsilon Y, \quad \tau \in [0,\Delta t]
\eequ
\dt{with the fibre-dependent diffusion coefficient} 
\[
D_m\dt{(f)}=\frac{D}{1+\alpha_{m} \tilde{f}(y,t_{0}+\tau)},
\] 
where $D$ is the baseline diffusion rate \dt{and} $\alpha_{m}$ \dt{being} a \emph{``slowing down" constant factor} \dt{induced by the presence of the micro-fibres density $\tilde{f}(y,t_{0}+\tau)$, which is defined as follows. Considering first the micro-scale set-valued mapping that surveys the number of points $z$ corresponding to up to four fibres micro-domains $\delta Y(x)$ that represent the same micro-position $y$, namely}
\dt{\bequ
\begin{array}{cl}
&\epsilon Y\ni y\longmapsto z(y)\in \bigcup\limits_{x\in Y} \delta Y(x)\\
\textrm{is given by:}\\
z(y) \quad\textrm{ is the set of} & \textrm{the micro-spatial points within up to four $\delta Y(x)(y)$}\\
& \textrm{where}\\
& \textrm{each selected $\delta Y(x)(y)$ is a micro-cube within $\bigcup\limits_{x\in Y} \delta Y(x)$}\\
& \textrm{with the property that this contains $y\in \epsilon Y$}\\
& \textrm{(please see Figure \ref{fig:shapefunction} for an illustration of this situation)}. 
\end{array}
\eequ}
\dt{we have that
\bequ
\tilde{f}(y,t_{0}+\tau):=
\left\{
\begin{array}{lll}
f(z(y),t_{0}+\tau) & if & card(z(y))=1;\\
\frac{1}{card(z(y))}\sum\limits_{\zeta\in z(y)}f(\zeta, t_{0}+\tau) & if & card(z(y))>1. 
\end{array}
\right.
\eequ}

\paragraph{The bottom-up link} During their micro-dynamics, the MMPs diffuse into the surrounding ECM and it is the pattern of their advancing spatial distribution that control the degradation of the peritumoural ECM captured within each micro-domain $\epsilon Y$. This degradation ultimately leads to the movement of the tumour boundary, whereby a movement direction $\eta_{\epsilon Y}$ and displacement magnitude $\xi_{\epsilon Y}$ are derived from the pattern of ECM degradation in each micro-cube $\epsilon Y$ \dt{(and for full derivation we refer the reader to \cite{Dumitru_et_al_2013})}. The microscopic movement of the boundary is represented at the macro-scale through the movement from the boundary midpoint $x^{*}_{\epsilon Y}$ to a new spatial position $\widetilde{x^{*}_{\epsilon Y}}$. Thus, \dt{although} the \emph{bottom-up} link of the model is much akin to previous works \cite{shutt_chapter,Shutt_2018}, \dt{the specific context in which the presence of micro-fibres distribution causes on one enables on one hand an enhanced source of MMP and on the other hand acts as an impediment for their random motility}, the \dt{MMP micro-dynamics} within the peritumoural region of the tumour domain $\Omega(t_{0})$  \dt{incorporates now these important aspects, ultimately resulting in an improved estimate for the macroscopic boundary movement characteristics (i.e., encapsulated by the movement direction $\eta_{\epsilon Y}$ and displacement magnitude $\xi_{\epsilon Y}$ )}. \dt{T}his \dt{crucial micro-scale-induced boundary relocation} is then translated \dt{back} to the macro-scale, resulting in an expanded tumour domain $\Omega(t_{0}+\Delta t)$ on which the multiscale dynamics is continued \dt{(as detailed in \cite{Dumitru_et_al_2013})}.

\subsection{Microscopic fibre degradation}
\dt{Given the proteolytic properties of the MMPs, the cell-scale cross-interface MMP-2 micro-dynamics will result in a direct peritumoural ECM degradation, whose pattern not only depends on the amount of MMP-2 transported at a given location $y$ within a given micro-domain $\epsilon Y$, but also on the existing ECM micro-fibres spatial distribution at $y$.}
To \dt{address this} micro-scales \dt{interaction that takes place between the MMP-2 cross-interface transport and micro-fibres within any given micro-domain $\epsilon Y$, we consider that the spatial patterns of micro-fibres distributions that are aligned with the MMP-2 flux $\nabla m(\cdot, t_{0}+\tau)$ suffer less degradation than those that are positioned orthogonal to it. Thus, for any fibre micro-domain $\delta Y(x)$ that has non-empty intersection with $\epsilon Y$, by denoting $\Phi_{fm}(\cdot,\cdot):\delta Y(x)\times [0,\Delta t]\to [0,\pi]$ the function that explores these emerging angles, given by
\bequ
\begin{array}{l}
\Phi_{fm}(\!z,\tau\!)\!=\!\!
\left\{\!\!
\begin{array}{lll}
\arccos\!\!\left(\!\frac{<\nabla m(z,\tau\!) , \nabla f(z,t_{0}+\tau\!)>}{||\nabla m(z,\tau\!)||_{_{2}} || \nabla\! f(z,t_{0}+\tau\!)||_{_{2}}}\!\right)\chi_{_{\delta Y\!(x)\!\cap \!\epsilon Y}}(z), &  if & \nor{\!\!\nabla \!f(z, t_{0}\!+\!\tau\!)\!\!}_{_{2}}>\!0, \\[0.2cm]
\frac{\pi}{2} \chi_{_{\delta Y\!(x)\!\cap \!\epsilon Y}}(z),  & if & \nor{\!\!\nabla f(z, t_{0}\!+\!\tau\!)\!\!}_{_{2}}=\!0.
\end{array}
\right.
\end{array}
\eequ} 
where $\chi_{_{\delta Y\!(x)\!\cap \!\epsilon Y}}(\cdot)$ is the usual characteristic function of $\delta Y\!(x)\!\cap \!\epsilon Y$.

The strength of the micro-fibre degradation \dt{rate} is \dt{influenced by their degree of alignment with the flux of the MMPs, which is explored by the angle $\Phi_{fm}(\!z,\tau\!)$ }at which the flux of the MMP\dt{-2} molecules \dt{acts upon} \dt{the mass distribution of micro-fibres $f(z, t_{0}+\tau)$} at each micro-position $z$, \dt{and so this can therefore be mathematically formalised as}
\bequ
D_{g}(z,\tau):= d_{f} \cdot \exp\left(1-\frac{1}{1 - (1 - \Phi(z,\tau))^2}\right).
\eequ
\dt{where} $d_{f}$ is a non-negative degradation constant\dt{,} and the function $\Phi(\cdot,\cdot)$
\[
\Phi(z,\tau) = \left |\frac{\Phi_{fm}(z,\tau)}{\Phi_{\text{max}}}\right |,
\]
represents the influence of the collision angle relative to the maximum angle $\Phi_{\text{max}}\!:=\!\!\frac{\pi}{2}$ where the highest rate of degradation will occur (i.e., when the flux of MMPs \dt{is} perpendicular \dt{upon the micro-fibres mass}). Thus\dt{,} the micro-fibre dynamics at location $z$ in the $\delta$-sized micro-scale can be mathematically represented as
\bequ
\frac{\partial f}{\partial \tau} = - D_{g}(z,\tau) \cdot f(z,\dt{t_{0}+}\tau), \qquad z \in \delta Y, \ \tau\in[0,\Delta t]. 
\eequ

\section{Summary of model}
The \dt{tissue- and cell--scale dynamics of the new multiscale moving-boundary} model proposed here, \dt{can therefore be summarised as: } 
\begin{subequations}
\begin{align} 
\begin{split}
\text{\dt{Macro-scale}}&\,\,\text{\dt{ dynamics:}} \\
\frac{\partial c}{\partial t} &= \nabla \cdot [D_{1} \nabla c - c \mathcal{A}(t,x,\textbf{u}(\cdot,t), \theta_{_{f}}(\cdot, t))] +\mu_{1}c(1-\rho(\textbf{u})),  \\[2mm]
\frac{d F}{d t} &= -\gamma_{1} c F ,    \\[2mm]
\frac{d l}{d t} &= -\gamma_{2} c l + \omega(1-\rho(\textbf{u})),\\ 
\end{split}\label{eq:fullsystem-Macro}\\[2mm]
\begin{split}
\text{\dt{Micro-scale}}&\,\,\text{\dt{ boundary MMP-2-dynamics interacting with peritumoural micro-fibres:}}   \\
\frac{\partial m}{\partial \tau} &= D_{m} \dt{(f)}\Delta m +g_{\epsilon Y}(y,\tau),  \quad y \in \epsilon Y, \ \tau \in [0,\Delta t],   \\[2mm]
\frac{\partial f}{\partial \tau} &= - D_{g} \cdot f(z,\dt{t_{0}+}\tau),  \quad  z \in \delta Y, \ \tau \in [0,\Delta t].\\
\end{split}\label{eq:fullsystem-BMicro}
\end{align}\label{eq:fullsystem}
\end{subequations}
\dt{The macro- and micro-dynamics summarised in} \eqref{eq:fullsystem} \dt{aggregates} two interconnected multiscale frameworks that describe the interactions between two independent micro-scale systems \dt{and} share the same macro-scale tissue level dynamics (summarised in \eqref{eq:fullsystem-Macro}) \dt{that is} linked \dt{to the two micro-dynamics} through two double feedback loops, \dt{as established and detailed} in \cite{Shutt_2018}. 

Whi\dt{le} the macro-scale tissue dynamics describe the evolution of the spatial distribution of cancer cells and both the non-fibres and macro-fibres ECM phase, the micro-scale part of the first multiscale system governs the dynamic rearrangement of fibres\dt{. Specifically, the rearrangement of the fibres emerges as a consequence of the acting macro-scale flux of } the cancer cell spatial flux \dt{upon the mass distribution of micro-fibres distributed within micro-domains $\delta Y(x)$ which are centred at any given macroscopic point $x\in Y$. The redistributed micro-fibres naturally yield a new macroscopic vector field representation of the newly oriented fibres that will lave a cascade influence upon adhesion processes the cancer cell population tissue scale dynamics biasing their migration, as detailed in \cite{Shutt_2018}}. 

\dt{The second multiscale system involved in this modelling brings into the picture contribution of the micro-scale proteolytic dynamics occurring in a cell-scale neighbourhood of the tumour interface. Here we take forward the multiscale moving boundary approach initially introduced in \cite{Dumitru_et_al_2013} and recapitulated in \cite{Shutt_2018}, by considering the influence that the oriented macro-scale ECM fibres has upon the emergence of the micro-scale MMP-2 source and at the same time exploring the explicit cell-sale interaction that takes place between the MMP-2 spatial flux and the mass distribution of micro-fibres at micro-scale which results in fibre degradation, summarised in \eqref{eq:fullsystem-BMicro}, notably triggering changes in the macroscopic fibres orientation. Consequently, these leading edge MMPs dynamics instigate a change in the position of the tissue-scale tumour boundary that corresponds to the pattern of peritumoural ECM degradation, this way allowing the macro-dynamics to continue on the newly enlarged tumour region and thus the invasion process continues.}

\section{Numerical approaches and initial conditions for computations}
Expanding on the multiscale moving boundary framework developed in \cite{Shutt_2018} building on the model initially introduced in \cite{Dumitru_et_al_2013}, we developed a new \dt{modelling and} computational approach to address specifically the \dt{cell-scale peritumoural interaction} between the MDE and \dt{the micro-fibres mass distributions at} micro-scale. \dt{Specifically}, we explore the link between fibre distribution and MDE density at the tumour interface, in addition to the MDE induced \dt{micro-}fibre degradation \dt{at the cell-scale}. 

\subsection{\dt{Brief description of the numerical approach}}
To \dt{address} the tumour macro-dynamics, we use the novel predictor-corrector method developed and fully defined in \cite{Shutt_2018}, that accounts for the complexity of the cancer dynamics. \dt{For this, we consider a uniform spatial mesh of size $h=0.03125$, and we use a combination of central differences and mid-point methods to discretise the local spatial operators, while involving an off-grid approach (introduced and detailed in \cite{Shutt_2018}) for the calculation of the non-local adhesion terms (that enable the adhesive flux) at each spatio-temporal node.} 

\begin{figure}[h!]
\centering
\includegraphics[scale=0.45]{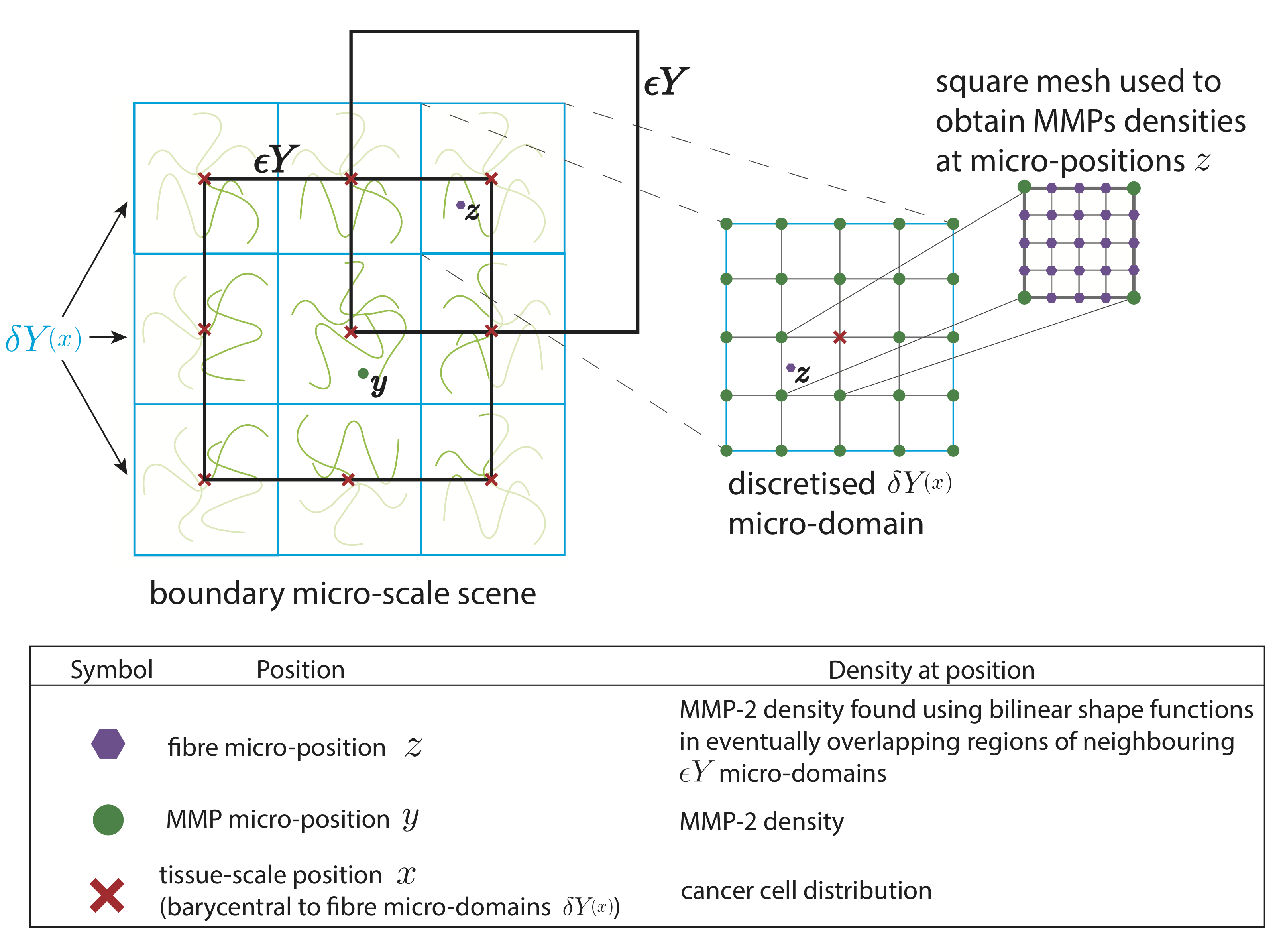}
\caption{Schematic illustrating the \dt{boundary} micro-scales \dt{computational setting where the micro-dynamics is explored}.}
\label{fig:shapefunction}
\end{figure}

Furthermore, to obtain the microscopic boundary relocation described in \cite{Shutt_chap}, we explore the \emph{top-down} and \emph{bottom-up} link, using a finite difference approach for computing the MDE micro-dynamics \dt{occurring on the bundle of boundary micro-domains $\{\epsilon Y\}_{\epsilon Y\in \P}$ over the time interval $[\tau_{0}+\Delta t]$. Hence, while considering a time discretisation into $p$ uniformly distributed time steps, i.e., $\delta t = \frac{\Delta t}{p}$,  each} micro-cube $\epsilon Y$ \dt{is also discretised uniformly, using a} spatial mesh of size $h_{\epsilon}$, i.e., $\Delta x_{\epsilon} = \Delta y_{\epsilon} = h_{\epsilon}$. \dt{Therefore,}  to discretise the reaction-diffusion equation \eqref{eq:mde}, \dt{we start by addressing the spatially discretised source term induced by the cancer cell and fibre distribution \eqref{eq:sourceMDEs} in the way it was described in \cite{Dumitru_et_al_2013}. Then, to solve the spatio-temporal dynamics in \eqref{eq:mde}, we develop a similar predictor-corrector method using trapezoidal corrector for the time marching, while for the spatial operators involved we use again a combination of mid-points and central differences, which in this context are given by} 
\begin{align}
\begin{split}
\nabla \cdot [\nabla m]_{i,j}^{n} &= \text{div}[\nabla m]^{n}_{i,j} \\[0.3cm] 
&\simeq \frac{[m_x]^n_{i+\frac{1}{2},j} - [m_x]^n_{i-\frac{1}{2},j}}{\Delta x_{\epsilon}} + \frac{[m_y]^n_{i,j+\frac{1}{2}} - [m_y]^n_{i,j-\frac{1}{2}}}{\Delta y_{\epsilon}}
\end{split}
\end{align}
where
\[\begin{cases}
[m_y]^n_{i,j+\frac{1}{2}} := \frac{m^n_{i,j+1} - m^n_{i,j}}{\Delta y_{\epsilon}} \\
[m_y]^n_{i,j-\frac{1}{2}} := \frac{m^n_{i,j} - m^n_{i,j-1}}{\Delta y_{\epsilon}} \\
[m_x]^n_{i+\frac{1}{2},j} := \frac{m^n_{i+1,j} - m^n_{i,j}}{\Delta x_{\epsilon}} \\
[m_x]^n_{i-\frac{1}{2},j} := \frac{m^n_{i,j} - m^n_{i-1,j}}{\Delta x_{\epsilon}} 
\end{cases}\] 
\dt{with $n = 0, \ldots , p$ and $i = 1, \ldots, q$, $j = 1,\ldots,q$ representing the indices for the $y$- and $x$-directions, respectively.}
\dt{Finally, important for capturing the degradation of the mass distribution of peritumoural micro-fibres at micro-scale, we use bilinear shape functions on square elements \cite{hughes_book} to appropriately interpolate the solutions and the associated fluxes of the MMP-2 micro-dynamics in the eventually overlapping regions of their $\epsilon Y$s boundary micro-domains. This way the appropriate MMP-2 flux information is obtained at any micro-position $z$ in any given intersecting $\delta Y(x)$ fibres micro-domains, i.e., within those fibres micro-domains $\delta Y (x)$ in the peritumoural region for which 
\[
\textrm{there exists a $\epsilon Y$ micro-domain such that $\delta Y (x)\cap \epsilon Y\neq \emptyset$}. 
\]}
\dt{as illustrated in Figure \ref{fig:shapefunction}}.

 All of the following simulations of the model have been performed on MATLAB.

\subsection{\dt{Initial conditions used in computational simulations}}

\begin{figure}[h!]
\centering
\includegraphics[scale=0.25]{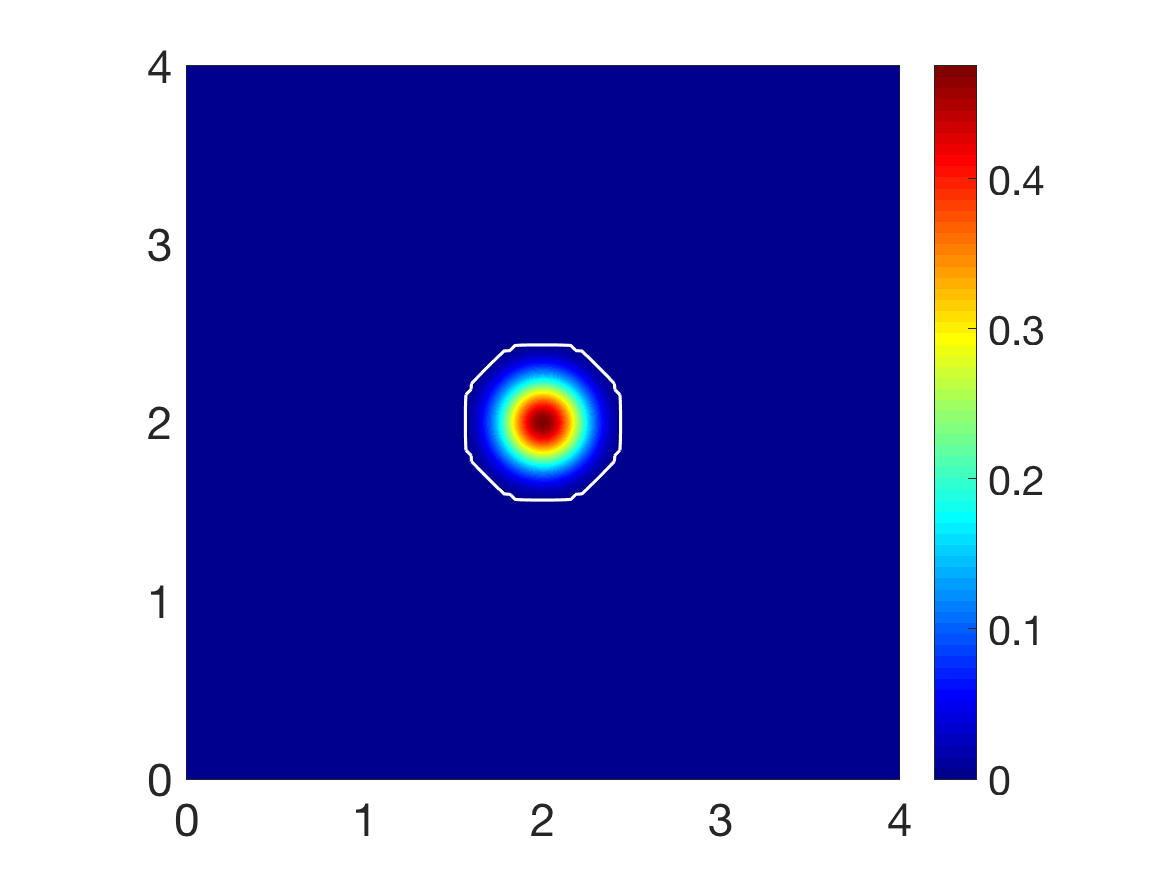}
\caption[Initial condition for the cancer cell population.]{\emph{Initial condition for the cancer cell population $c(x,0)$, illustrating the tumour boundary by the white contour.}}
\label{fig:iccancer}
\end{figure}

We consider the same initial conditions from \cite{Shutt_2018}, such that we consider the initial cancer cell population $c(x,0)$ to occupy the region $\Omega(0) = \textbf{B}((2,2),0.5)$ positioned in the centre of the tissue cube $Y$, Figure \ref{fig:iccancer},
\bequ
c(x,0)=0.5\left(\text{exp}\left(-\frac{||x-(2,2)||^2_2}{0.03}\right)-\text{exp}(-28.125)\right)\left(\chi_{_{\Bila((2,2),0.5-\gamma)}} \ast \psi_{\gamma}\right),
\label{eq:canceric}
\eequ
where $\psi$ is the standard mollifier detailed in Appendix \ref{standardmollifier} that acts within a radius $\gamma<<\frac{\Delta x}{3}$ from $\partial \textbf{B}((2,2),0.5-\gamma)$ to smooth out the characteristic function $\chi_{_{\Bila((2,2),0.5-\gamma)}}$.

\begin{figure}[ht!]
    \centering
    \begin{subfigure}{0.4\linewidth}
        \includegraphics[width=\linewidth]{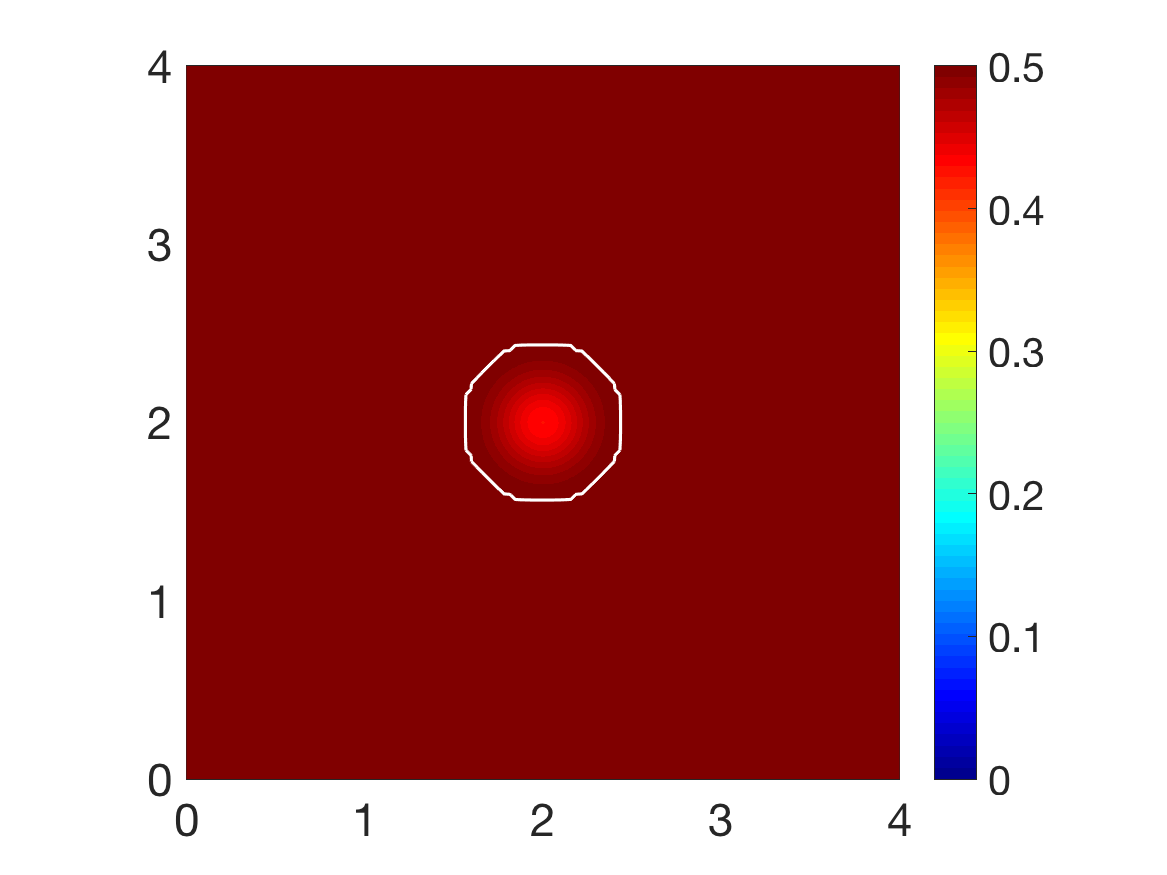}
        \caption{\emph{Initial ECM density - homogeneous}}
        \label{fig:ichomolf}
\end{subfigure}\hfil
    \begin{subfigure}{0.4\linewidth}
        \includegraphics[width=\linewidth]{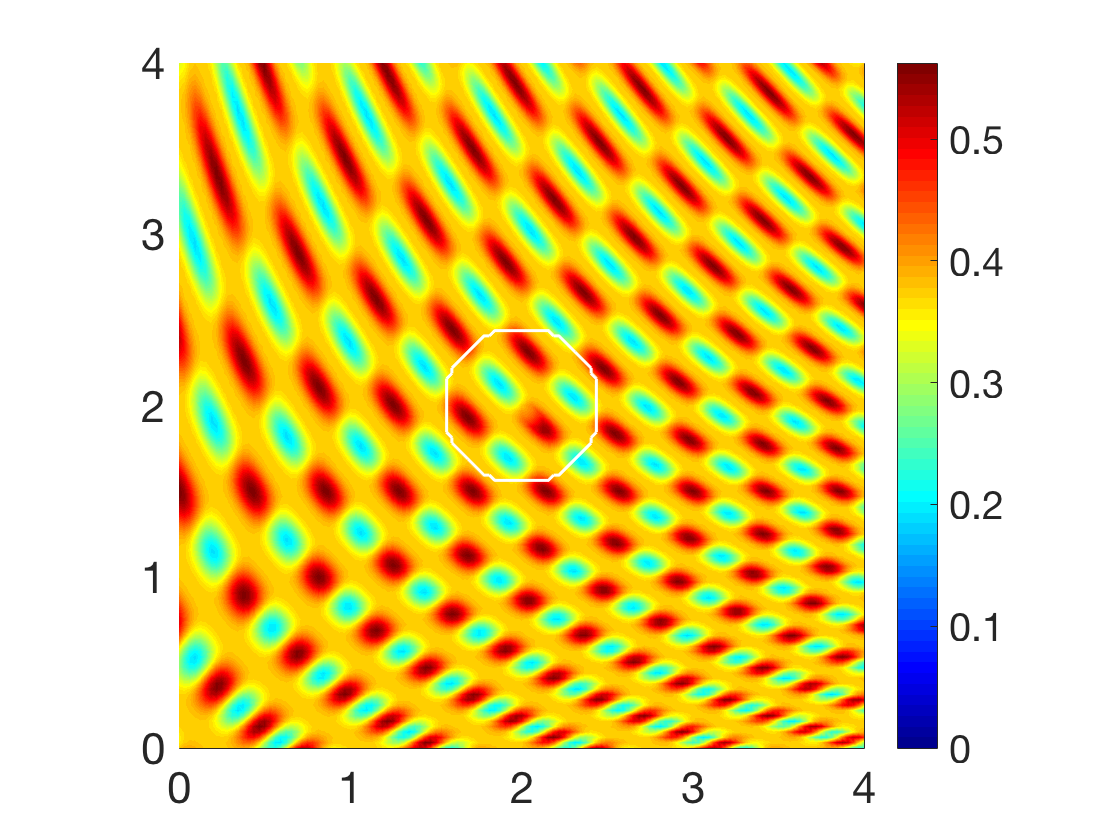}
        \caption{\emph{Initial ECM density - heterogeneous}}
        \label{fig:icheterolf}
    \end{subfigure}\hfil
    \caption[Initial conditions for the non-fibres ECM phase, illustrating both a homogeneous and heterogeneous example.]{\emph{Initial conditions for the non-fibres ECM phase $l(x,0)$, illustrating both a homogeneous (a) and heterogeneous (b) distribution.}}
    \label{fig:lphaseic}
\end{figure}

We explore the invasion of cancer within two different compositions of the non-fibres ECM phase, a homogeneous distribution given as 
\bequ
l(x,0)=\text{min}\{0.5,1-c(x,0)\}
\label{eq:lhomoic}
\eequ
and a heterogeneous distribution, previously explored in \cite{Shutt_2018} and defined below
\begin{equation}
l(x,0)=\text{min}\left\{ h(x_1,x_2), 1- c(x,0)\right\},
\label{eq:lheteroic}
\end{equation}
where 
\begin{align*}
h(x_1,x_2)&=\frac{1}{2}+\frac{1}{4}\text{sin}(\zeta x_1 x_2)^3 \cdot \text{sin}(\zeta \frac{x_2}{x_1}),  \\
(x_1,x_2)&= \frac{1}{3}(x+1.5) \ \in [0,1]^2 \ \text{for} \ x \in D, \quad \zeta = 7\pi.
\end{align*}
These initial conditions are shown in Figure \ref{fig:lphaseic}.

Finally, the fibres ECM phase will be initialised with both a homogeneous and heterogeneous macro-scale distribution. To this end, we first assume a random distribution of five pre-assigned micro-distributions containing patterns of five different micro-fibres (detailed in Appendix \ref{microfibres}) that are then randomly assigned onto $\delta Y(x) := x + \delta Y$. To calibrate the macro-fibre distributions, we use the initial conditions for the non-fibres ECM phase, for either a homogeneous or heterogeneous distribution, \eqref{eq:lhomoic} and \eqref{eq:lheteroic}, respectively, taking a percentage $p$ of the density of the non-fibres ECM at each spatio-temporal position $x$. This allows for the control of the maximal height of the micro-fibres in each $\delta Y(x)$, centred at macro-position $x$, so that the resulting macro-fibre distribution $F(x,0)$ represents the percentage $p$, which here will be $15\%, \ p=0.15$ or $20\%,\  p=0.2$, with the latter representing a denser collagen structure. All of the following simulations of the model have been performed on MATLAB.

\section{Results}
\begin{figure}[h!]
    \centering 
\begin{subfigure}{0.5\textwidth}
  \includegraphics[width=\linewidth]{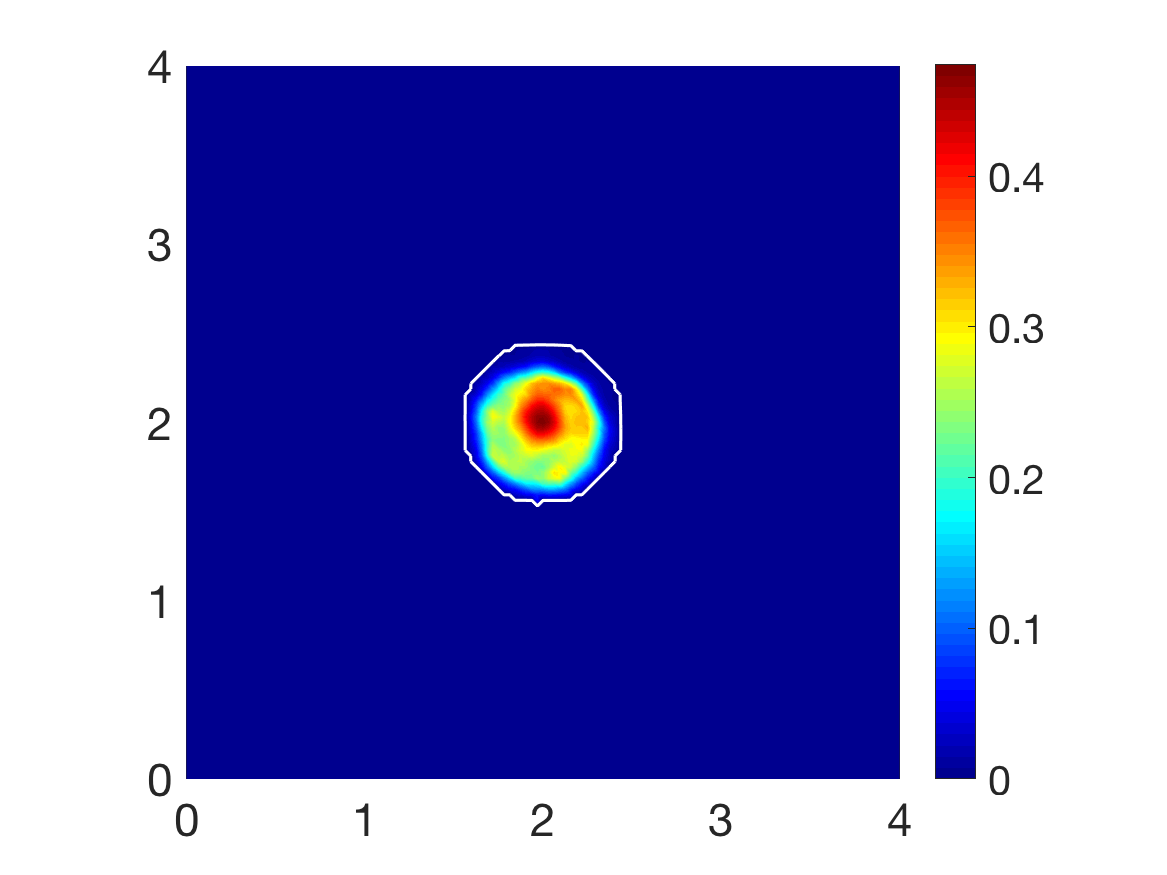}
  \caption{\emph{Cancer cell population}}
  \label{fig:fulldeghomo25a}
\end{subfigure}\hfil 
\begin{subfigure}{0.5\textwidth}
  \includegraphics[width=\linewidth]{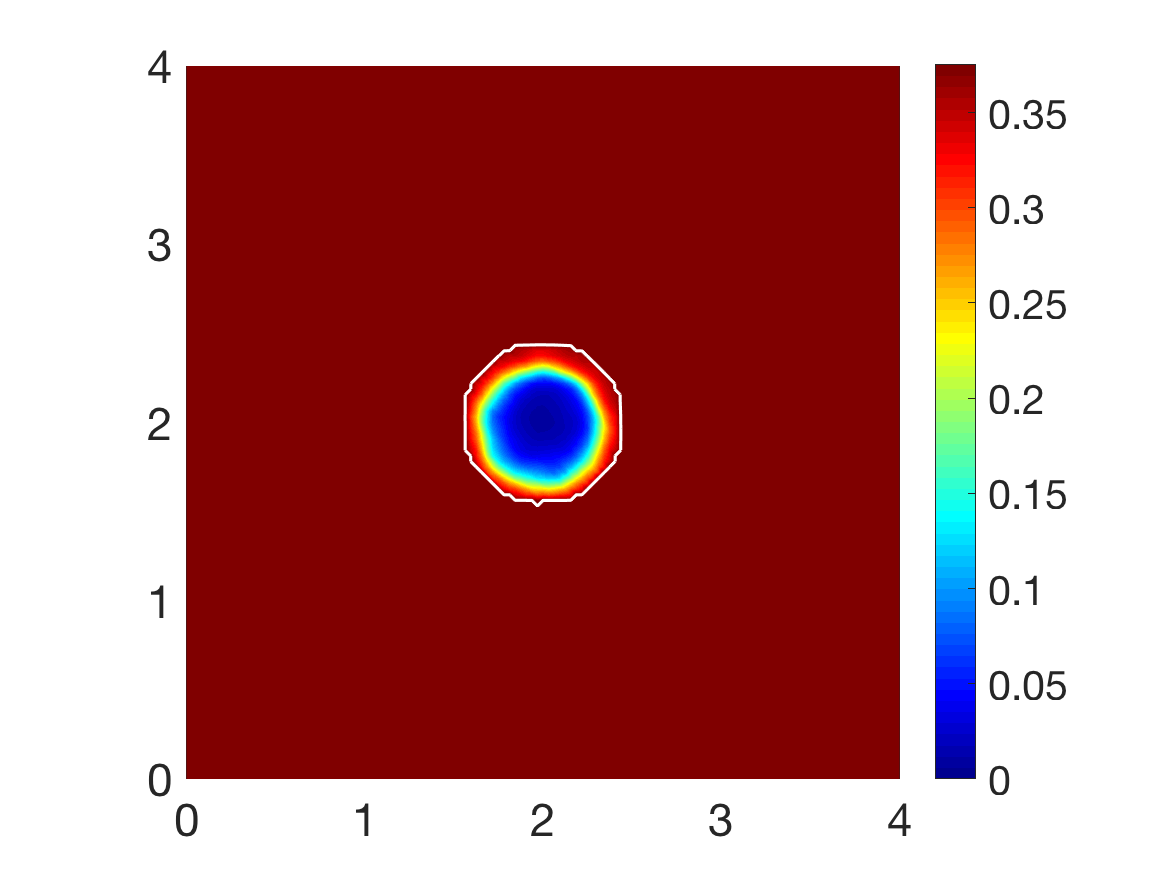}
  \caption{\emph{Non-fibres ECM distribution}}
  \label{fig:fulldeghomo25b}
\end{subfigure}\hfil 

\medskip
\begin{subfigure}{0.5\textwidth}
  \includegraphics[width=\linewidth]{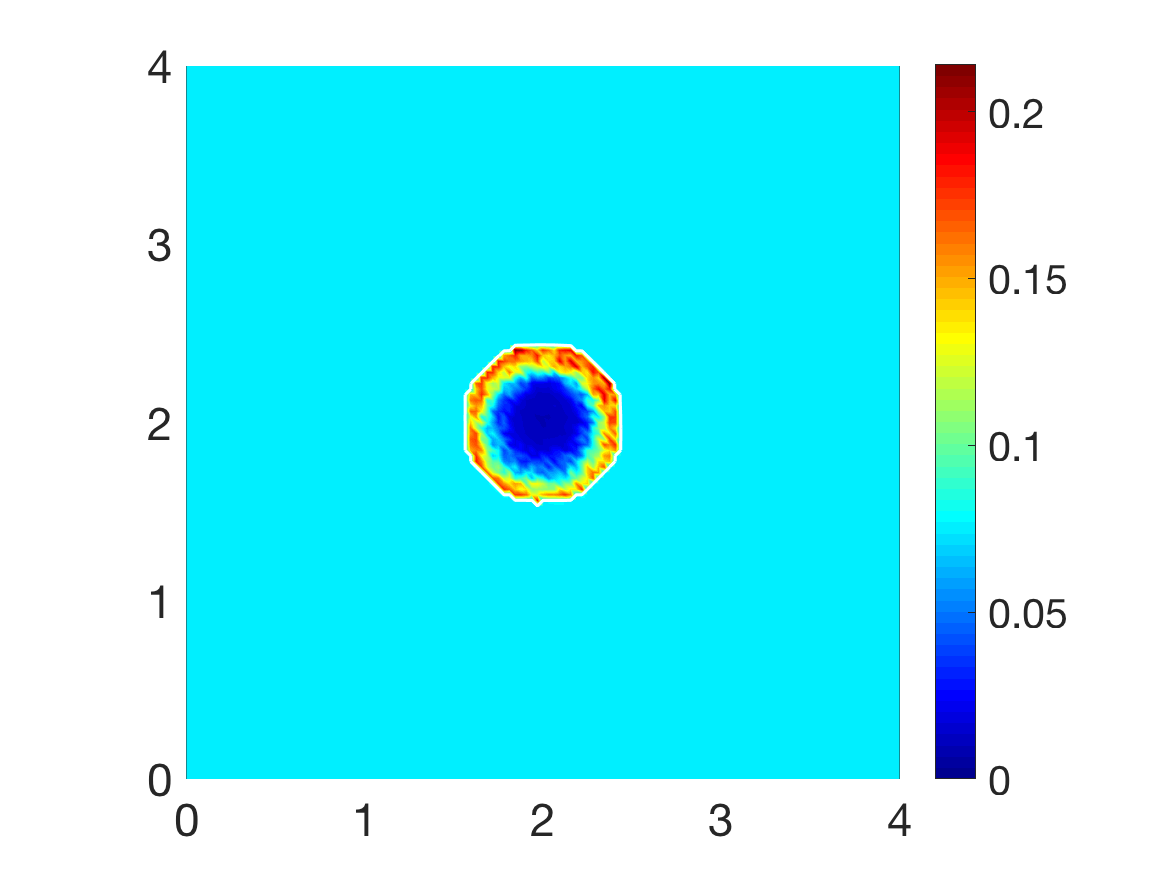}
  \caption{\emph{Fibre magnitude density}}
  \label{fig:fulldeghomo25c}
  \end{subfigure}\hfil 
\begin{subfigure}{0.5\textwidth}
  \includegraphics[width=\linewidth]{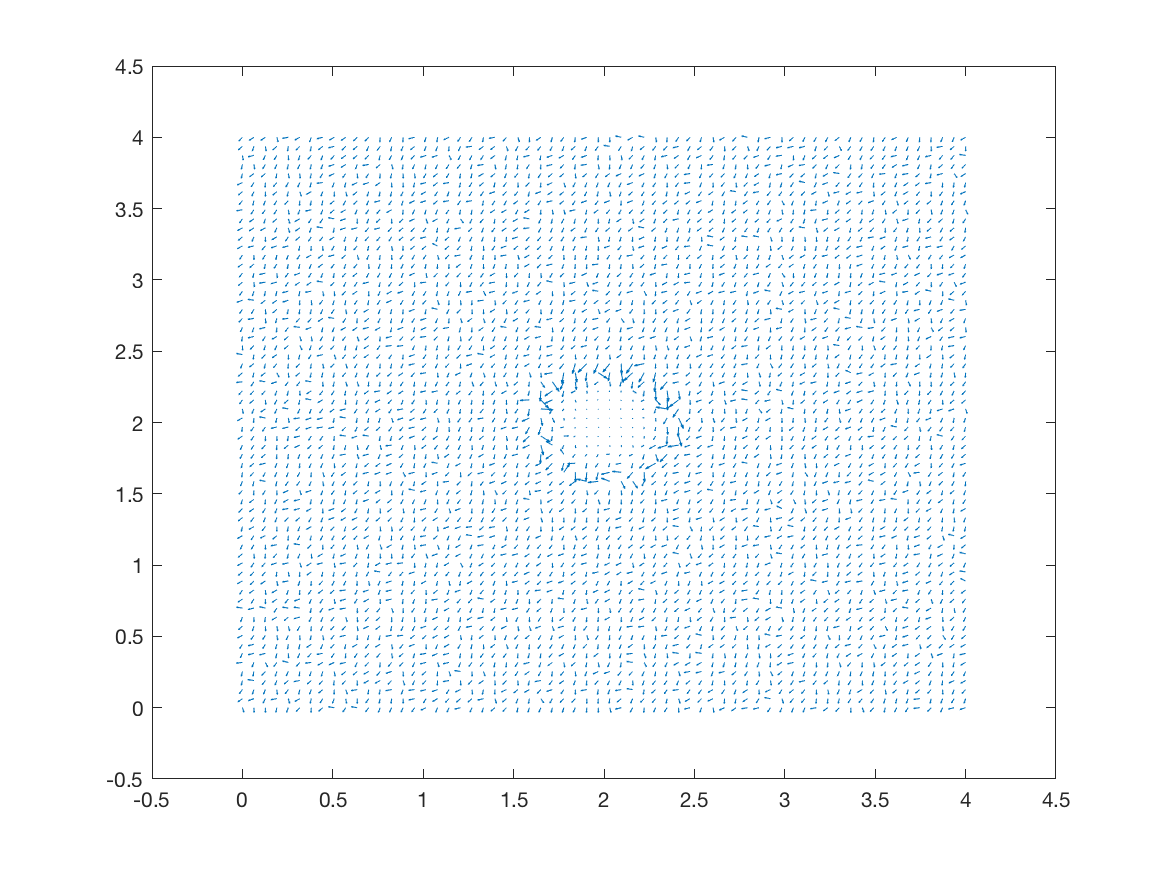}
  \caption{\emph{Fibre vector field - coarsened 2 fold}}
  \label{fig:fulldeghomo25d}
\end{subfigure}\hfil 

\medskip
\begin{subfigure}{0.5\textwidth}
  \includegraphics[width=\linewidth]{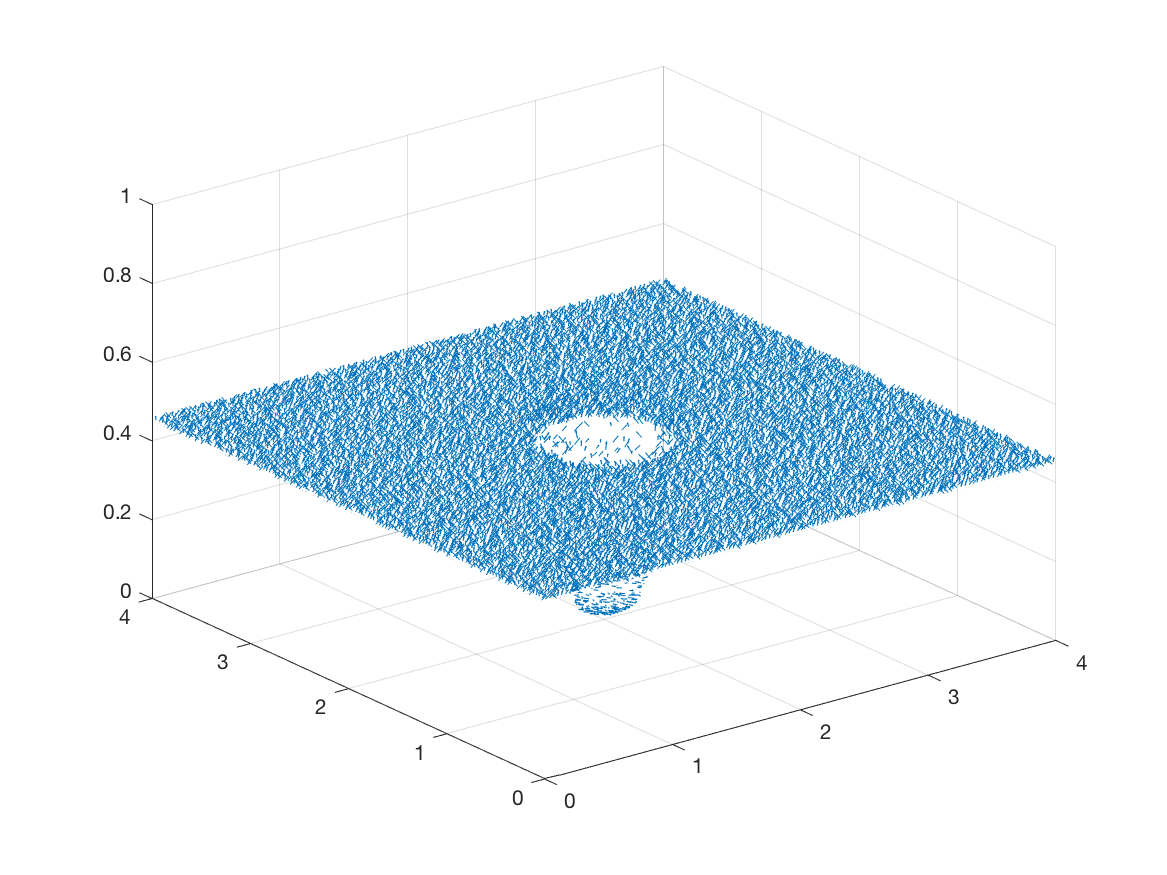}
  \caption{\emph{3D ECM vector field}}
  \label{fig:fulldeghomo25e}
  \end{subfigure}\hfil 
\begin{subfigure}{0.5\textwidth}
  \includegraphics[width=\linewidth]{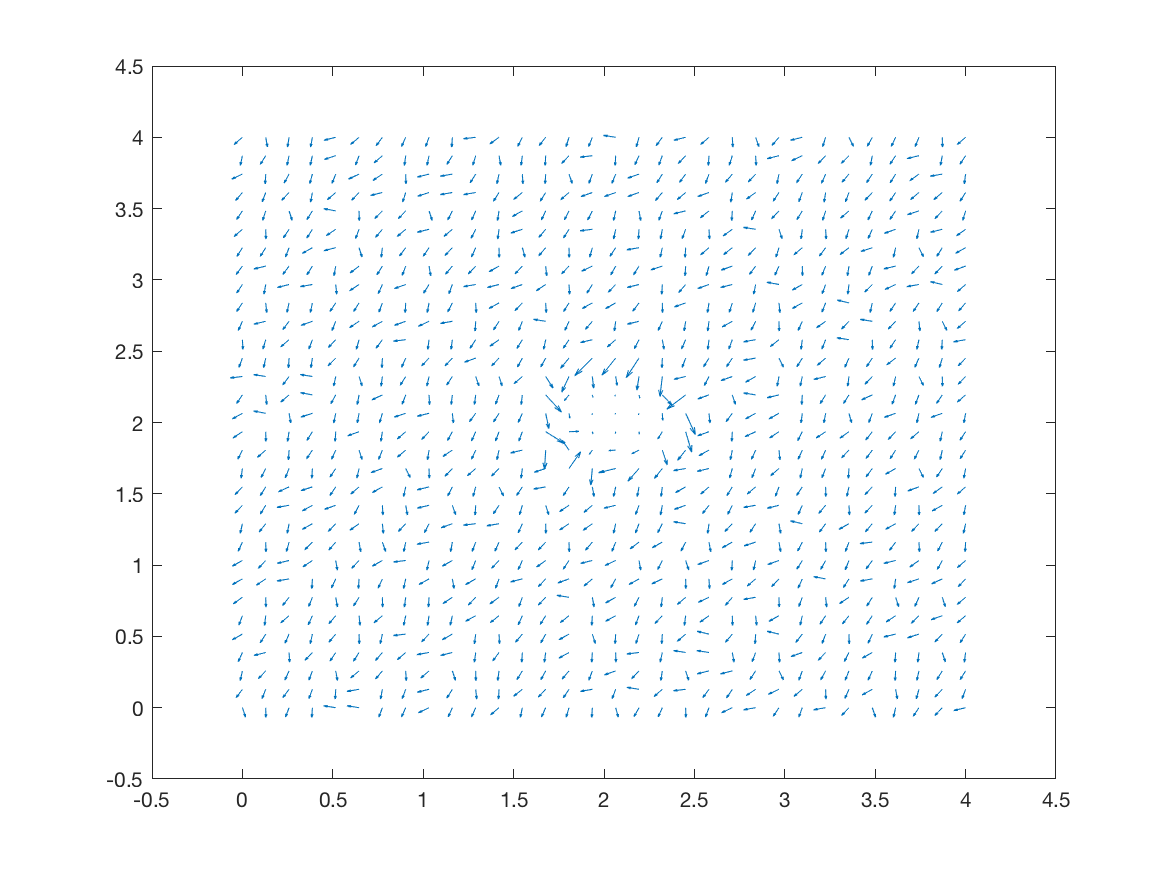}
  \caption{\emph{Fibre vector field - coarsened 4 fold}}
  \label{fig:fulldeghomo25f}
\end{subfigure}\hfil 

\caption[Simulations at stage $25\Delta t$ with a homogeneous distribution of the non-fibrous and fibres phase of the ECM and a micro-fibres degradation rate of $d_f = 1$.]{\emph{Simulations at stage $25\Delta t$ with a homogeneous distribution of the non-fibrous and fibres phase of the ECM and a micro-fibres degradation rate of $d_f = 1$.}}
\label{fig:fulldeghomo25}
\end{figure}

\begin{figure}[h!]
    \centering 
\begin{subfigure}{0.5\textwidth}
  \includegraphics[width=\linewidth]{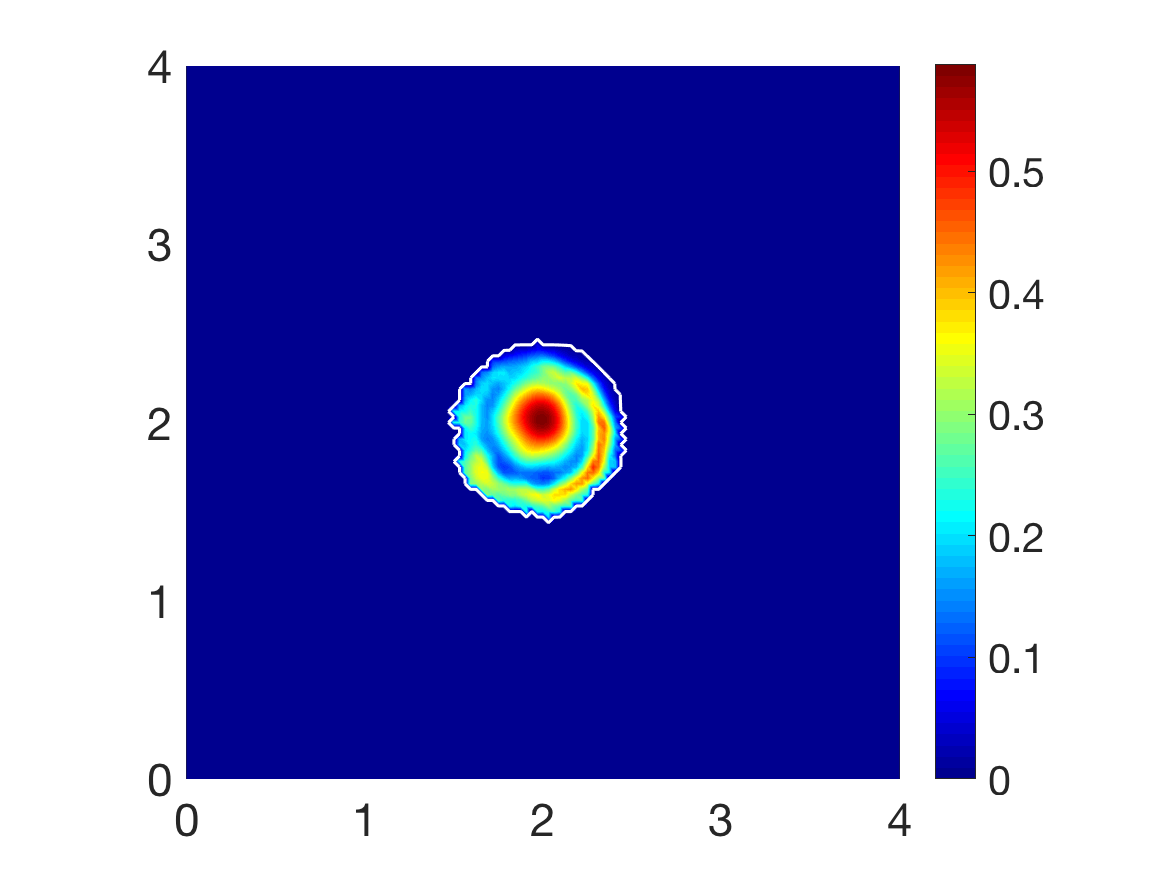}
  \caption{\emph{Cancer cell population}}
  \label{fig:fulldeghomo50a}
\end{subfigure}\hfil 
\begin{subfigure}{0.5\textwidth}
  \includegraphics[width=\linewidth]{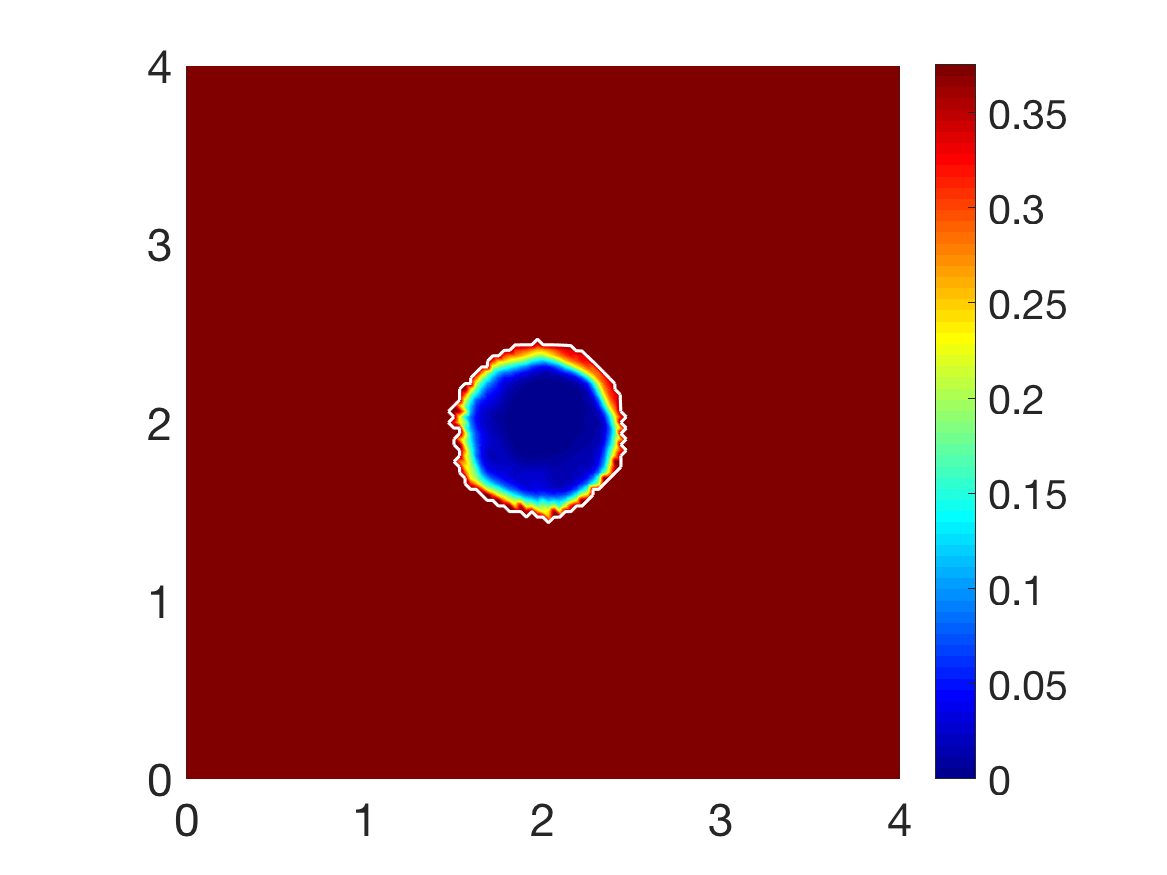}
  \caption{\emph{Non-fibres ECM distribution}}
  \label{fig:fulldeghomo50b}
\end{subfigure}\hfil 

\medskip
\begin{subfigure}{0.5\textwidth}
  \includegraphics[width=\linewidth]{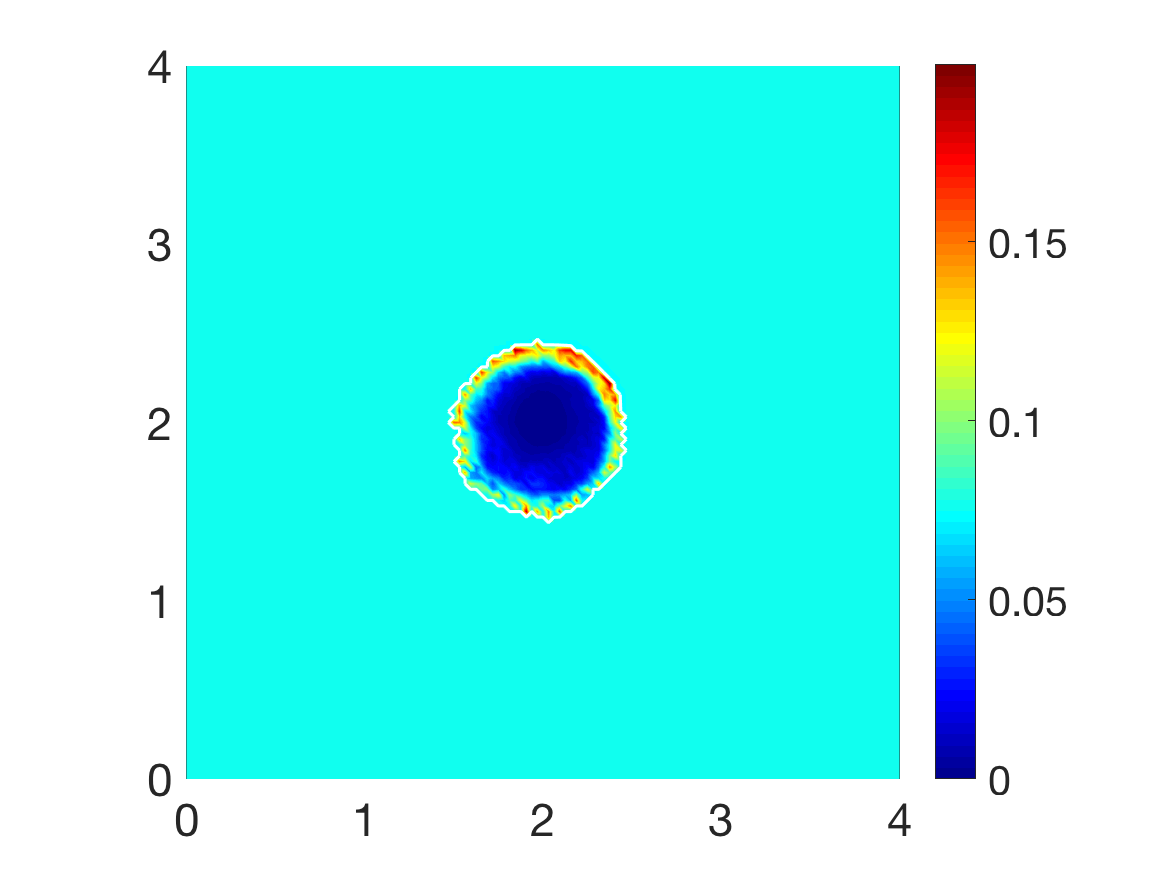}
  \caption{\emph{Fibre magnitude density}}
  \label{fig:fulldeghomo50c}
  \end{subfigure}\hfil 
\begin{subfigure}{0.5\textwidth}
  \includegraphics[width=\linewidth]{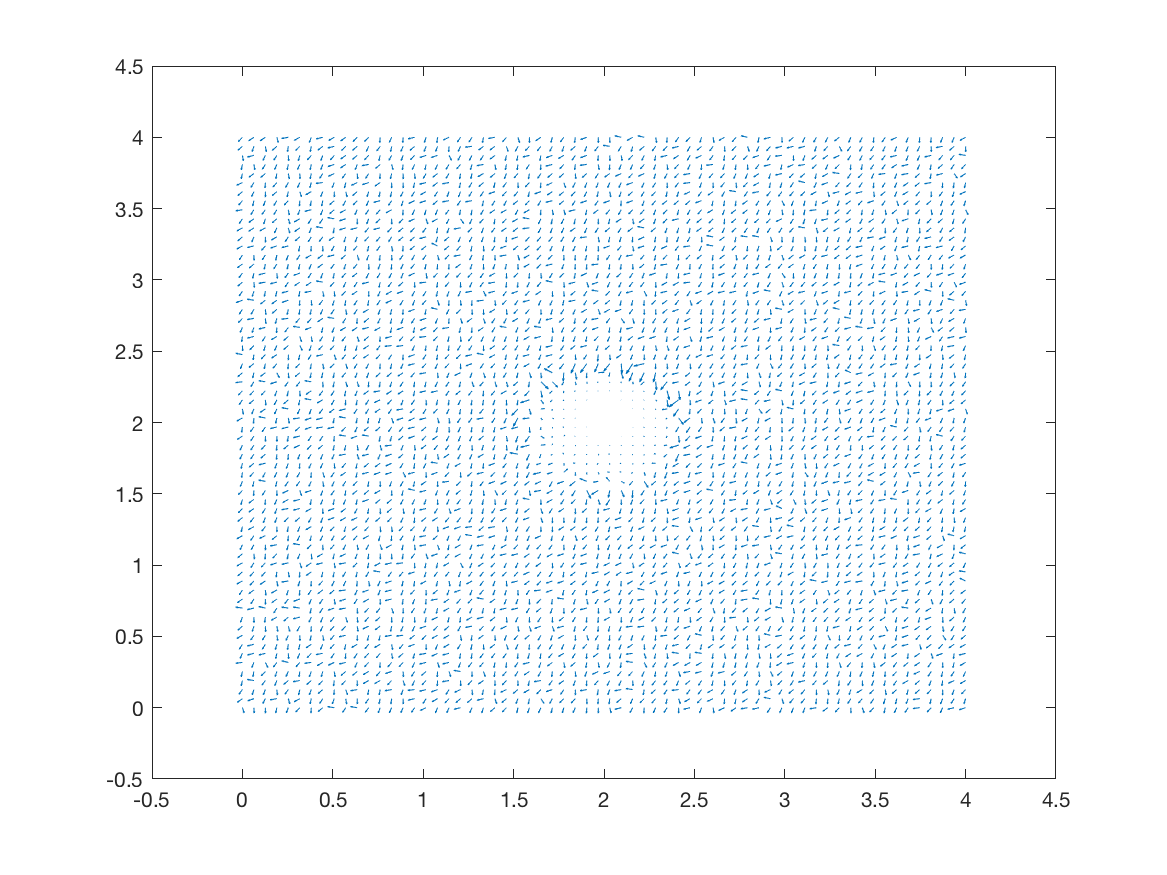}
  \caption{\emph{Fibre vector field - coarsened 2 fold}}
  \label{fig:fulldeghomo50d}
\end{subfigure}\hfil 

\medskip
\begin{subfigure}{0.5\textwidth}
  \includegraphics[width=\linewidth]{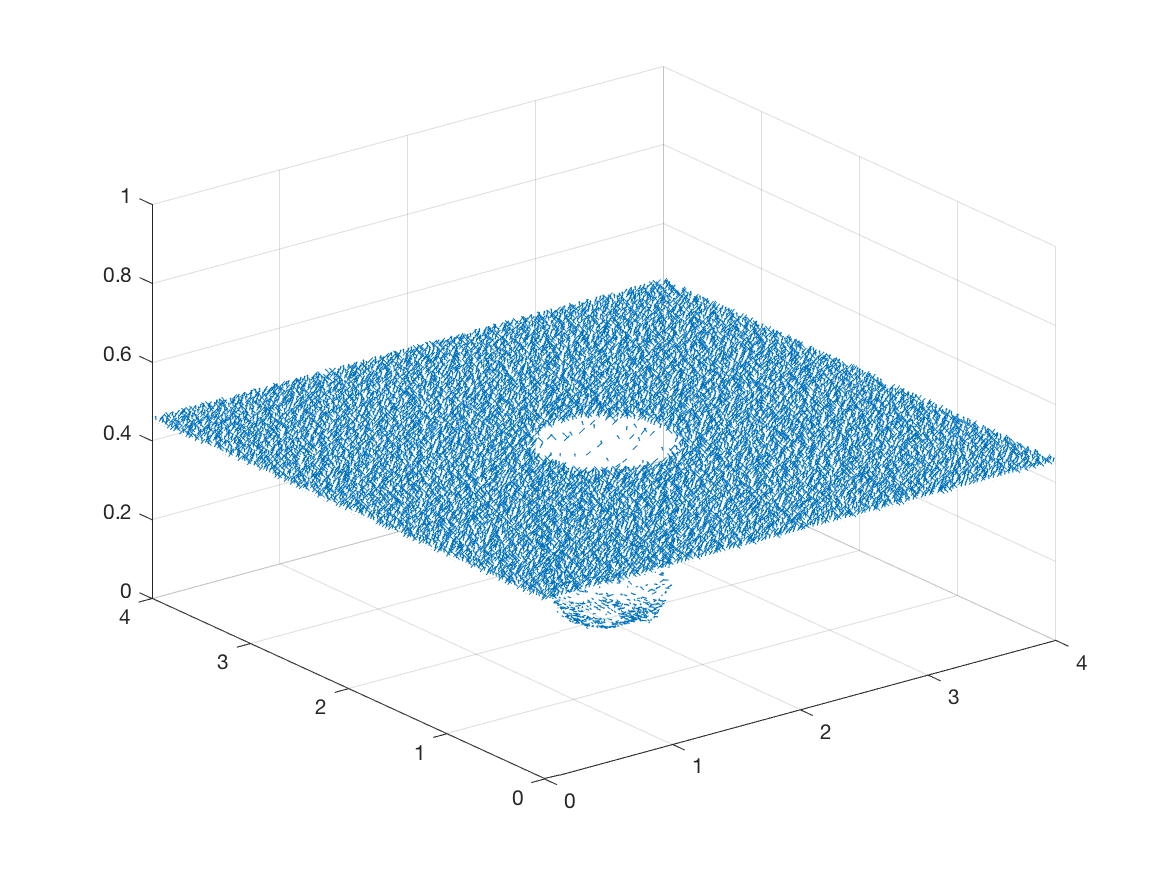}
  \caption{\emph{3D ECM vector field}}
  \label{fig:fulldeghomo50e}
  \end{subfigure}\hfil 
\begin{subfigure}{0.5\textwidth}
  \includegraphics[width=\linewidth]{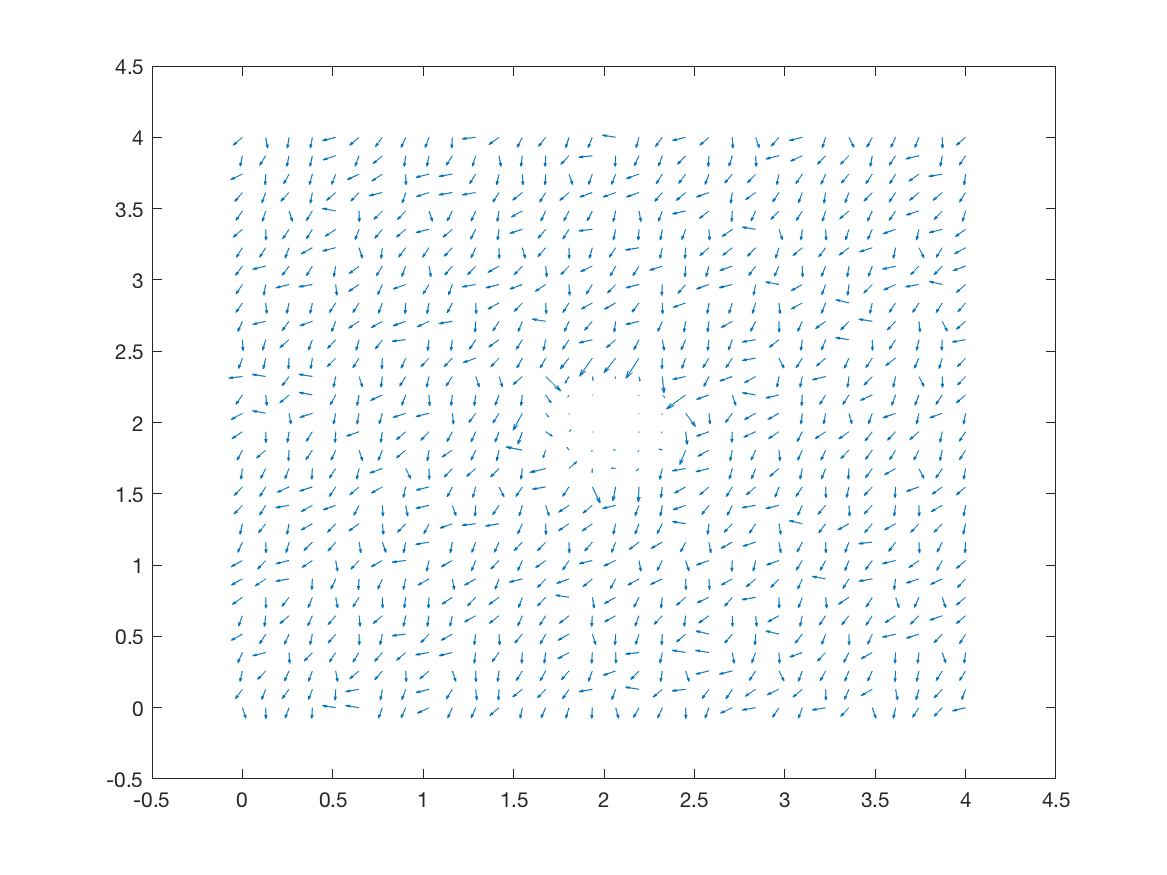}
  \caption{\emph{Fibre vector field - coarsened 4 fold}}
  \label{fig:fulldeghomo50f}
\end{subfigure}\hfil 

\caption[Simulations at stage $50\Delta t$ with a homogeneous distribution of the non-fibrous and fibres phase of the ECM and a micro-fibres degradation rate of $d_f = 1$.]{\emph{Simulations at stage $50\Delta t$ with a homogeneous distribution of the non-fibrous and fibres phase of the ECM and a micro-fibres degradation rate of $d_f = 1$.}}
\label{fig:fulldeghomo50}
\end{figure}

\begin{figure}[h!]
    \centering 
\begin{subfigure}{0.5\textwidth}
  \includegraphics[width=\linewidth]{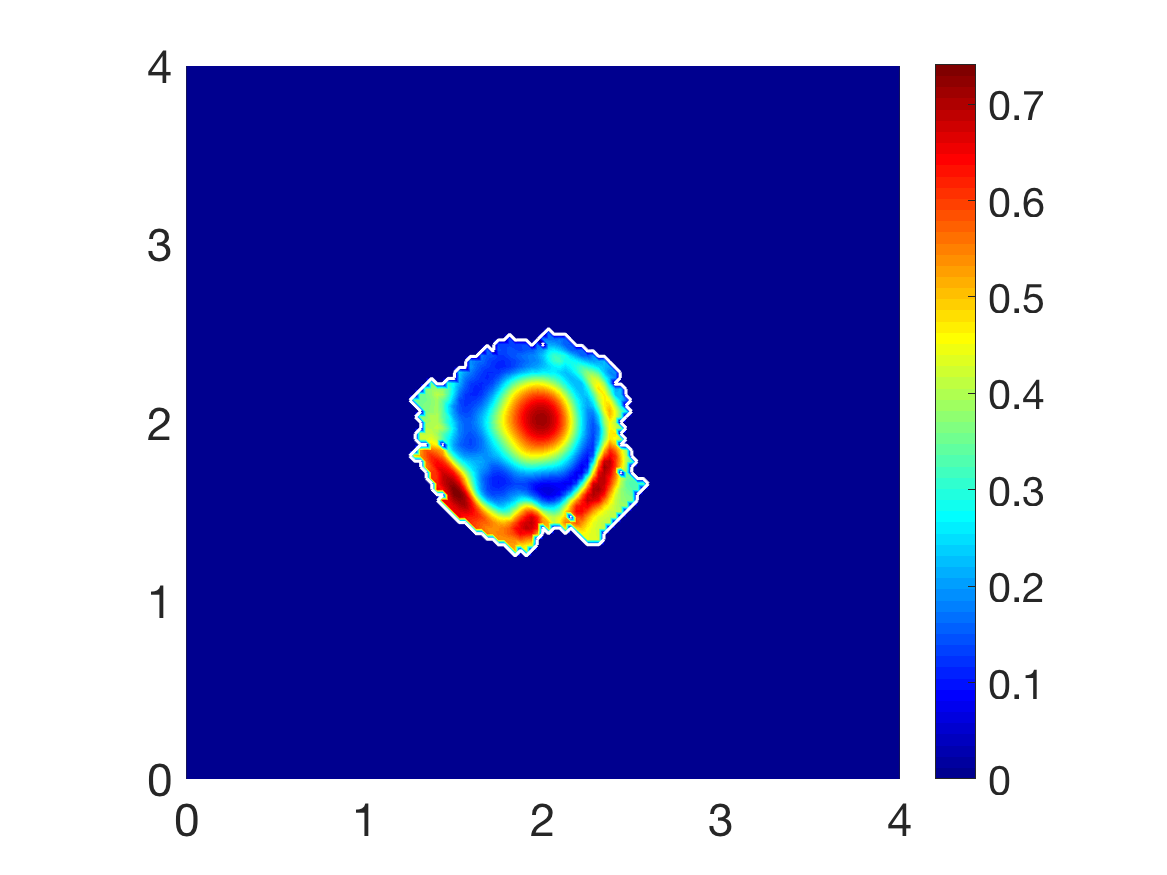}
  \caption{\emph{Cancer cell population}}
  \label{fig:fulldeghomo75a}
\end{subfigure}\hfil 
\begin{subfigure}{0.5\textwidth}
  \includegraphics[width=\linewidth]{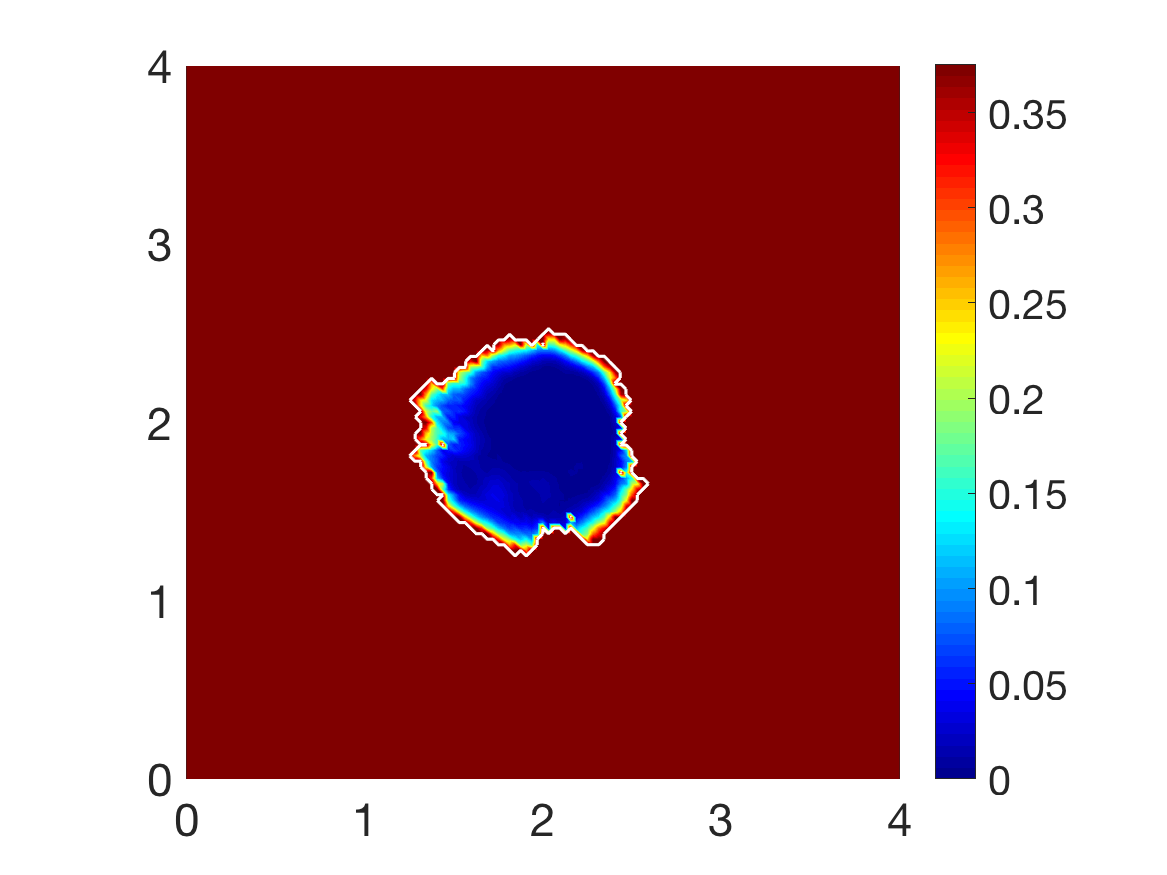}
  \caption{\emph{Non-fibres ECM distribution}}
  \label{fig:fulldeghomo75b}
\end{subfigure}\hfil 

\medskip
\begin{subfigure}{0.5\textwidth}
  \includegraphics[width=\linewidth]{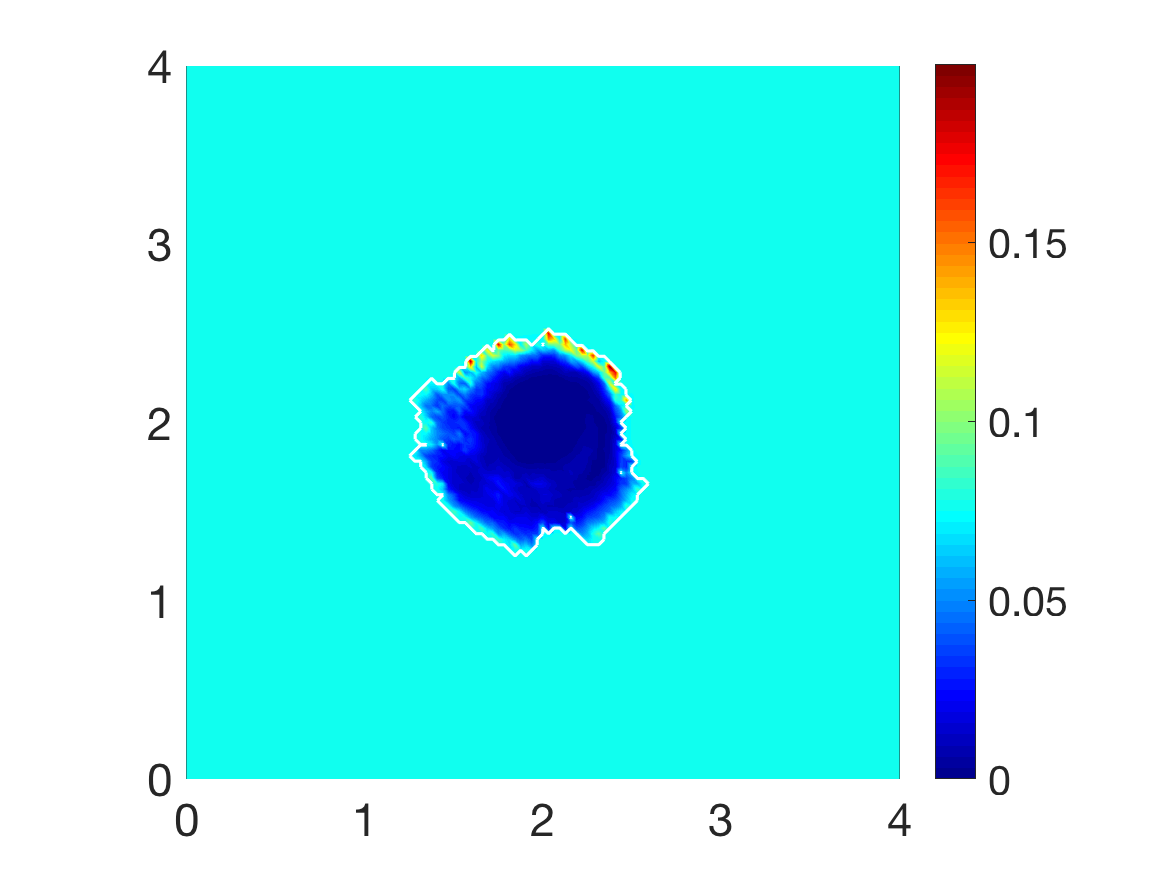}
  \caption{\emph{Fibre magnitude density}}
  \label{fig:fulldeghomo75c}
  \end{subfigure}\hfil 
\begin{subfigure}{0.5\textwidth}
  \includegraphics[width=\linewidth]{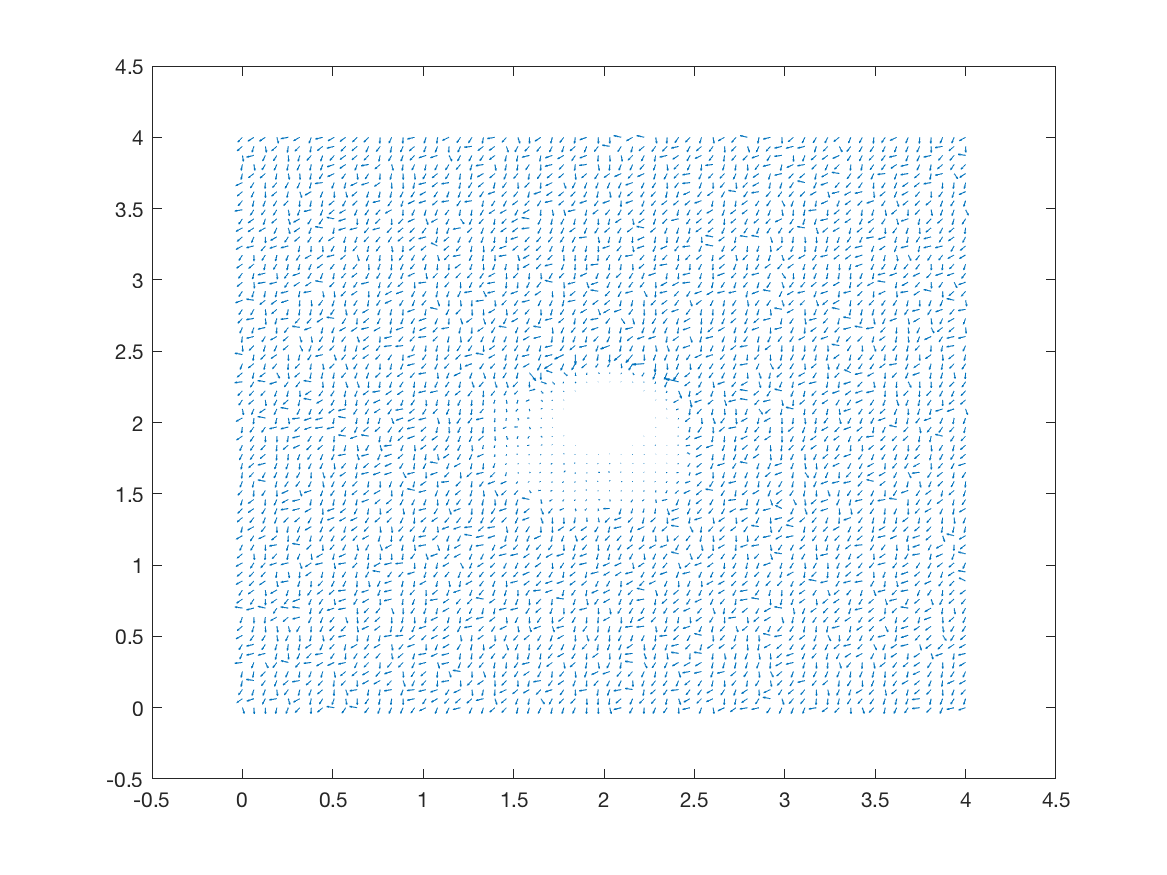}
  \caption{\emph{Fibre vector field - coarsened 2 fold}}
  \label{fig:fulldeghomo75d}
\end{subfigure}\hfil 

\medskip
\begin{subfigure}{0.5\textwidth}
  \includegraphics[width=\linewidth]{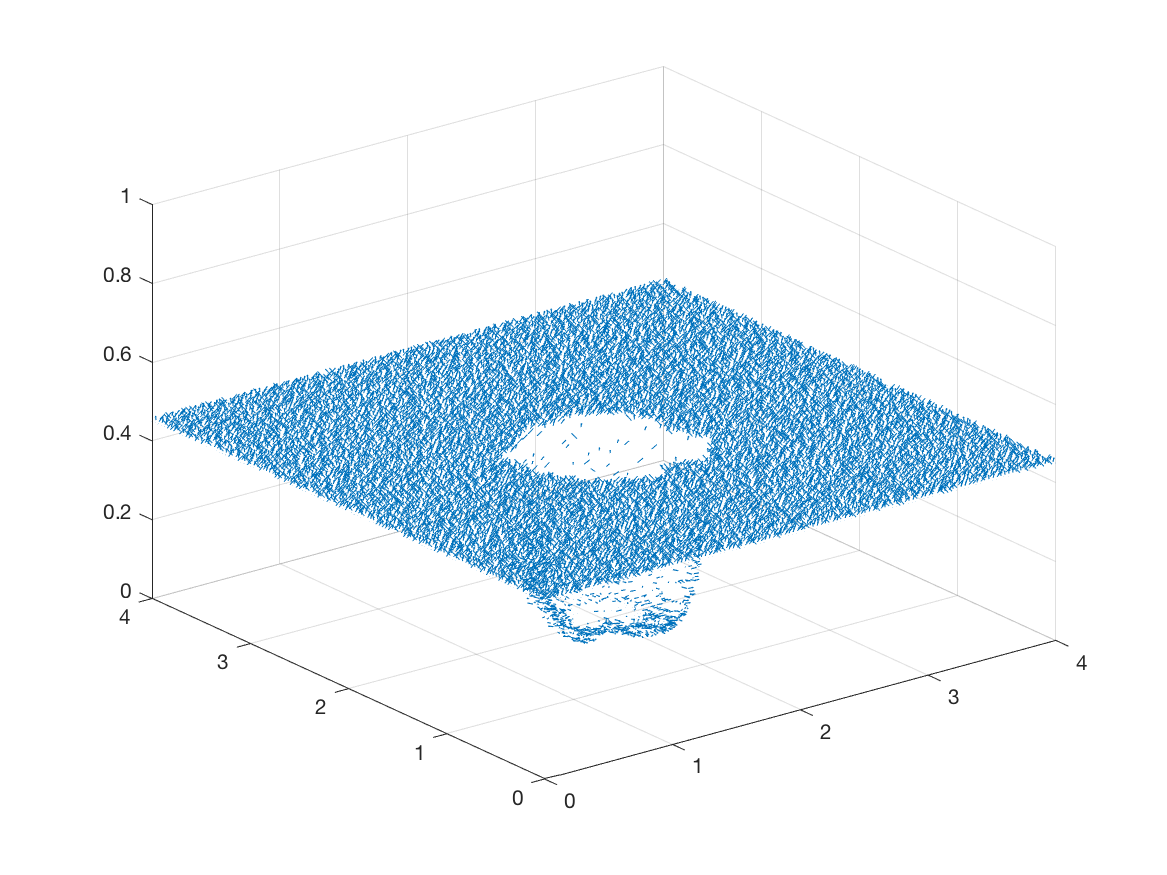}
  \caption{\emph{3D ECM vector field}}
  \label{fig:fulldeghomo75e}
  \end{subfigure}\hfil 
\begin{subfigure}{0.5\textwidth}
  \includegraphics[width=\linewidth]{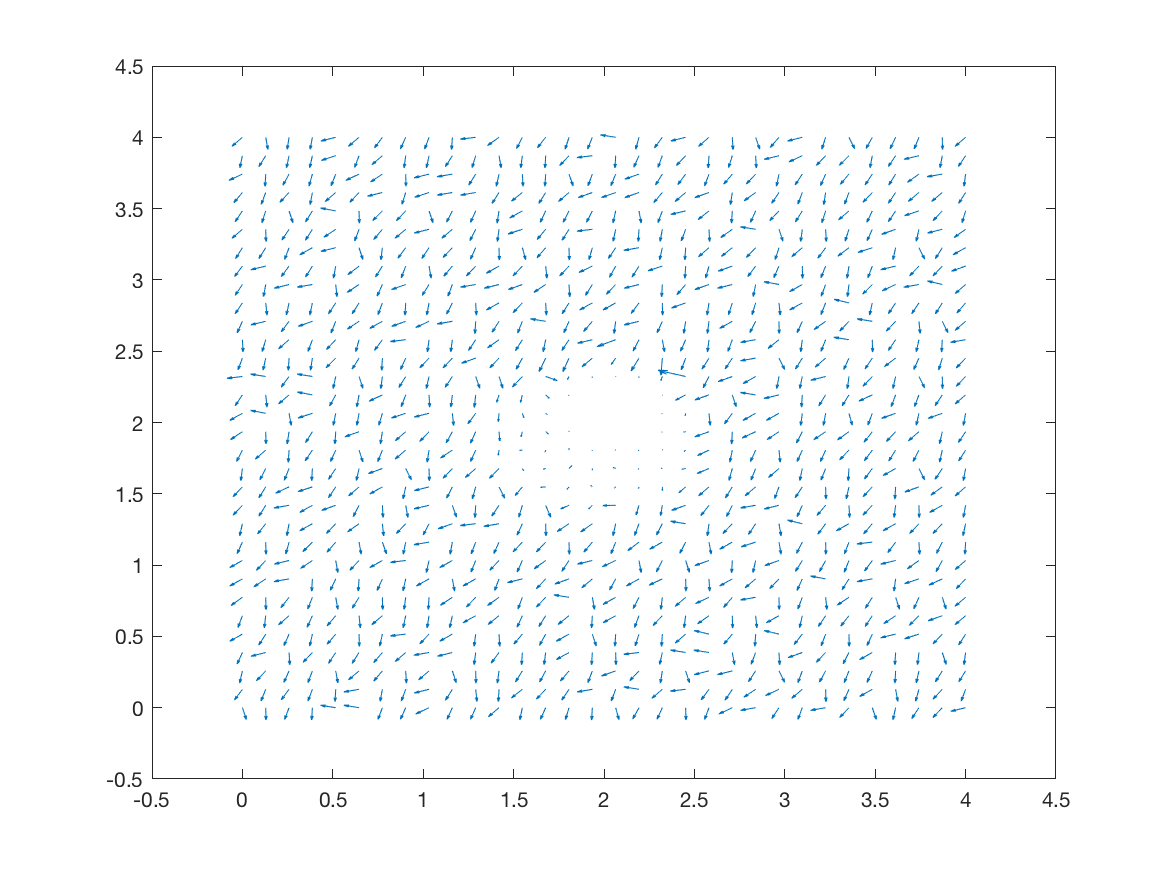}
  \caption{\emph{Fibre vector field - coarsened 4 fold}}
  \label{fig:fulldeghomo75f}
\end{subfigure}\hfil 

\caption[Simulations at stage $75\Delta t$ with a homogeneous distribution of the non-fibrous and fibres phase of the ECM and a micro-fibres degradation rate of $d_f = 1$.]{\emph{Simulations at stage $75\Delta t$ with a homogeneous distribution of the non-fibrous and fibres phase of the ECM and a micro-fibres degradation rate of $d_f = 1$.}}
\label{fig:fulldeghomo75}
\end{figure}

We first explore tumour invasion in the presence of a homogeneous ECM, where the non-fibres phase is the homogeneous distribution \eqref{eq:lhomoic} with the fibres phase taking the percentage $p=0.15$ of $l(x,0)$. Using the parameter set $\Sigma_{1}$ from Appendix \ref{paramSection}, the microscopic fibre degradation rate $d_f=1$, and the cell adhesion coefficients
\[
\textbf{S}_{max}=0.5, \quad \textbf{S}_{cF}=0.1 \quad \text{and} \quad \textbf{S}_{cl}=0.05,
\]
we investigate the effects micro-fibres degradation on tumour progression. Figures \ref{fig:fulldeghomo25}, \ref{fig:fulldeghomo50} and \ref{fig:fulldeghomo75} display the evolution of the tumour at stages $\Delta 25 t$, $\Delta 50 t$ and $\Delta 75 t$, respectively. Each figure contains the subfigures: \dt{(a)} cancer cell population\dt{; (b)} non-fibres ECM density\dt{; (c)} macroscopic fibre distribution\dt{;} vector fields of fibre orientations at two different resolutions, namely\dt{: (d)} coarsened two-fold\dt{; and (f)} coarsened four fold\dt{; and finally in (e)} a $3$D plot of the orientation of the complete ECM distribution $v(x,t)=F(x,t)+l(x,t)$.

After $25 \Delta t$ macro-stages, Figure \ref{fig:fulldeghomo25}, the boundary of the tumour has remained largely unchanged, subfigure \ref{fig:fulldeghomo25a}. Both the non-fibres and fibres ECM phase undergo degradation where the cancer cell distribution is highest, subfigures \ref{fig:fulldeghomo25b}, \ref{fig:fulldeghomo25c}, whilst minor degradation of the fibres has also occurred at the tumour boundary. The masses of micro-fibres have been rearranged such that the macroscopic orientation of fibres remains in line with the general fibre direction, subfigures \ref{fig:fulldeghomo25d}, \ref{fig:fulldeghomo25f} which points towards the origin of the space. However, there are some irregularities, particularly visible in subfigure \ref{fig:fulldeghomo25f} which has been magnified 4-fold, where the orientation of the fibres is mixed with some fibres realigned to near perpendicular of their original orientation. The cancer cells are pushing and rearranging the fibres in a direction opposite that of the initial fibre orientation, subfigure \ref{fig:fulldeghomo25c}, where a build up of fibre distributions occurs on the top right of the tumour region, situated close to the bulk of the tumour mass. As the tumour expands in size, the fibres are subsequently degraded during each stage of evolution, Figures \ref{fig:fulldeghomo50} and \ref{fig:fulldeghomo75}, illustrating the simulations after $50 \Delta t$ and $75 \Delta t$, respectively. At stage $50\Delta t$, some cells have detached and formed a region encircling the initial bulk of tumour cells, subfigure \ref{fig:fulldeghomo50a}. The boundary of the tumour is expanding into the surrounding tissue, subfigure \ref{fig:fulldeghomo50b}, exhibiting a ``rippling'' effect along the proliferating edge caused by the fibre mediated movement of the boundary. The macroscopic fibre distribution becomes depressed as the cancer cells increase in distribution and thus increase in their degradative behaviour, subfigure \ref{fig:fulldeghomo50c}. Moving on to final stage $75 \Delta t$, the tumour has largely increased in size and there are dense regions of cell distribution, subfigure \ref{fig:fulldeghomo75a}. The cells are gathering at the tumour interface, pulled in this direction by cell-fibre adhesion, following the direction of the fibres observed at the previous interval in subfigures \ref{fig:fulldeghomo50d}, \ref{fig:fulldeghomo50f}. The source of MMPs induced by the cancer cells are very high in these dense areas, therefore the degradation of fibres is higher, subfigure \ref{fig:fulldeghomo75c}, witnessed here by the absence of macroscopic fibre distribution.

\begin{figure}[h!]
    \centering 
\begin{subfigure}{0.5\textwidth}
  \includegraphics[width=\linewidth]{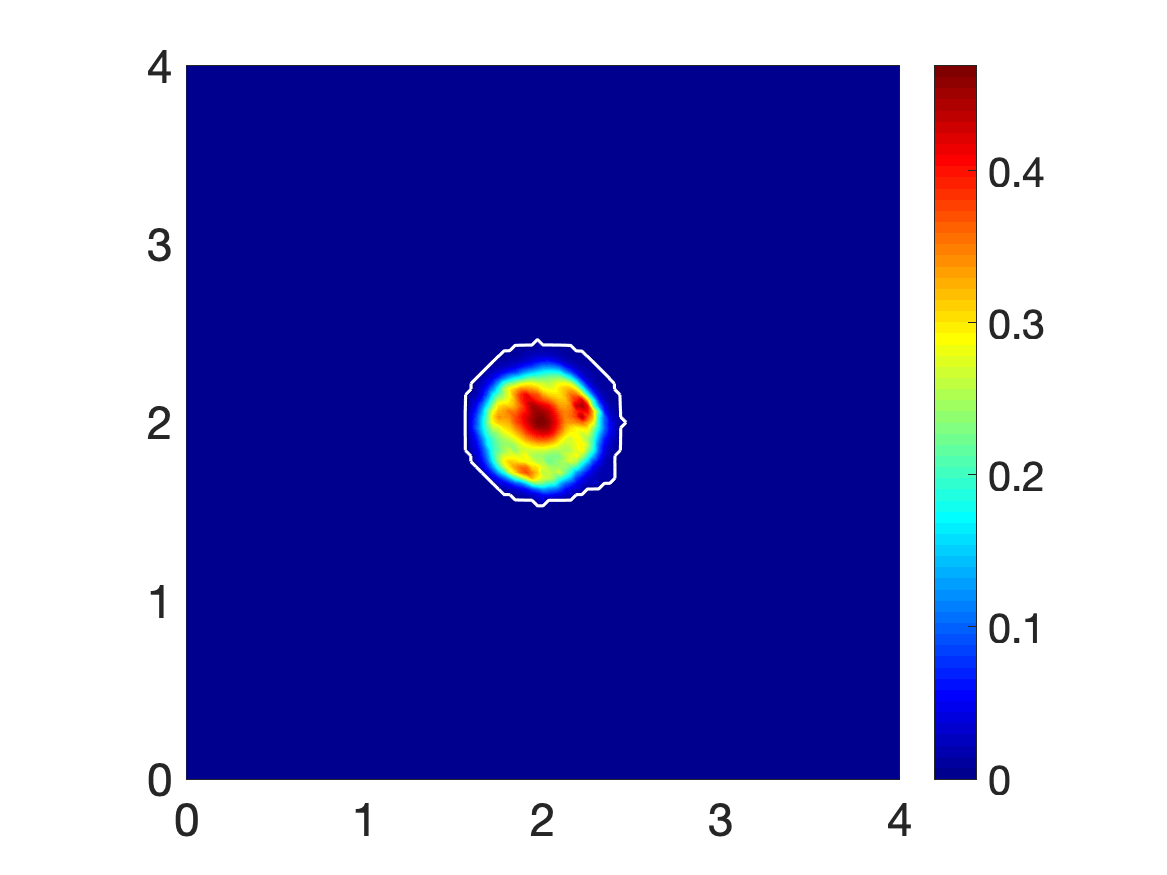}
  \caption{\emph{Cancer cell population}}
  \label{fig:fulldeghetero25a}
\end{subfigure}\hfil 
\begin{subfigure}{0.5\textwidth}
  \includegraphics[width=\linewidth]{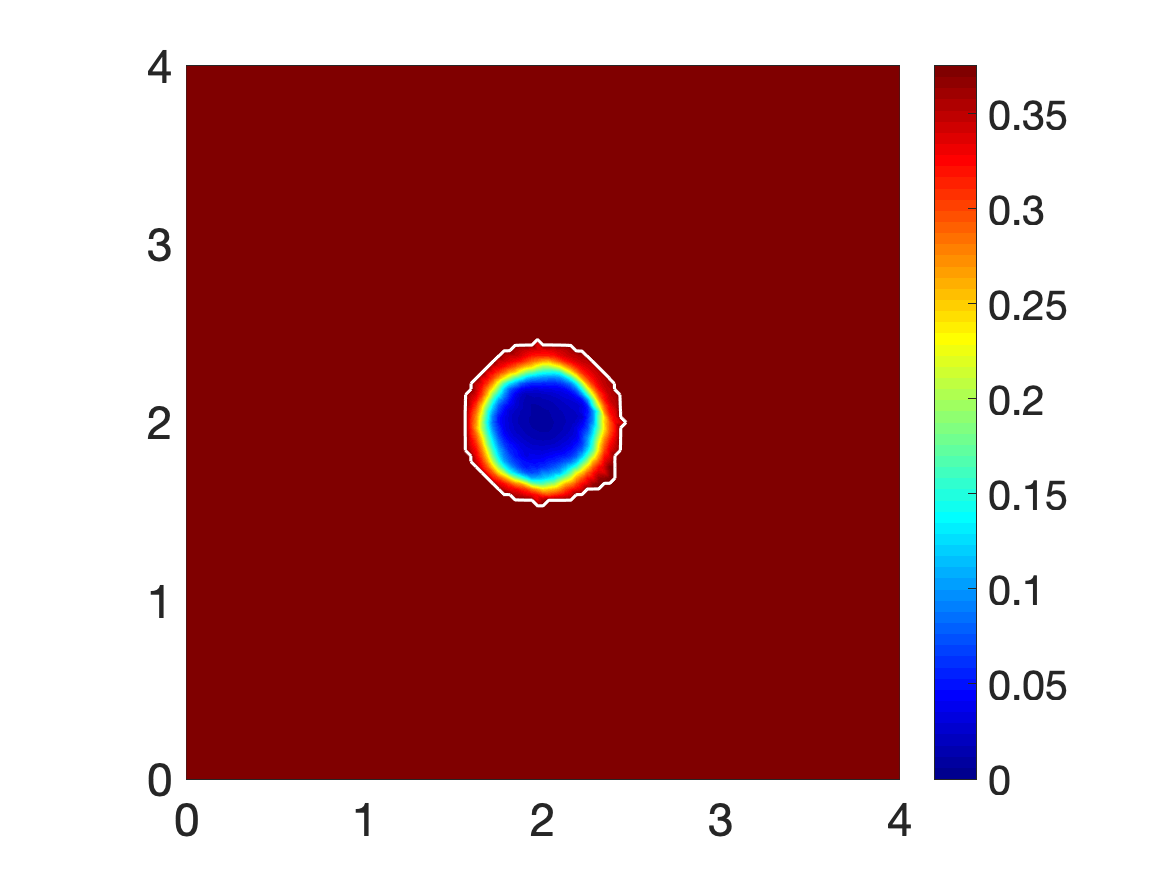}
  \caption{\emph{Non-fibres ECM distribution}}
  \label{fig:fulldeghetero25b}
\end{subfigure}\hfil 

\medskip
\begin{subfigure}{0.5\textwidth}
  \includegraphics[width=\linewidth]{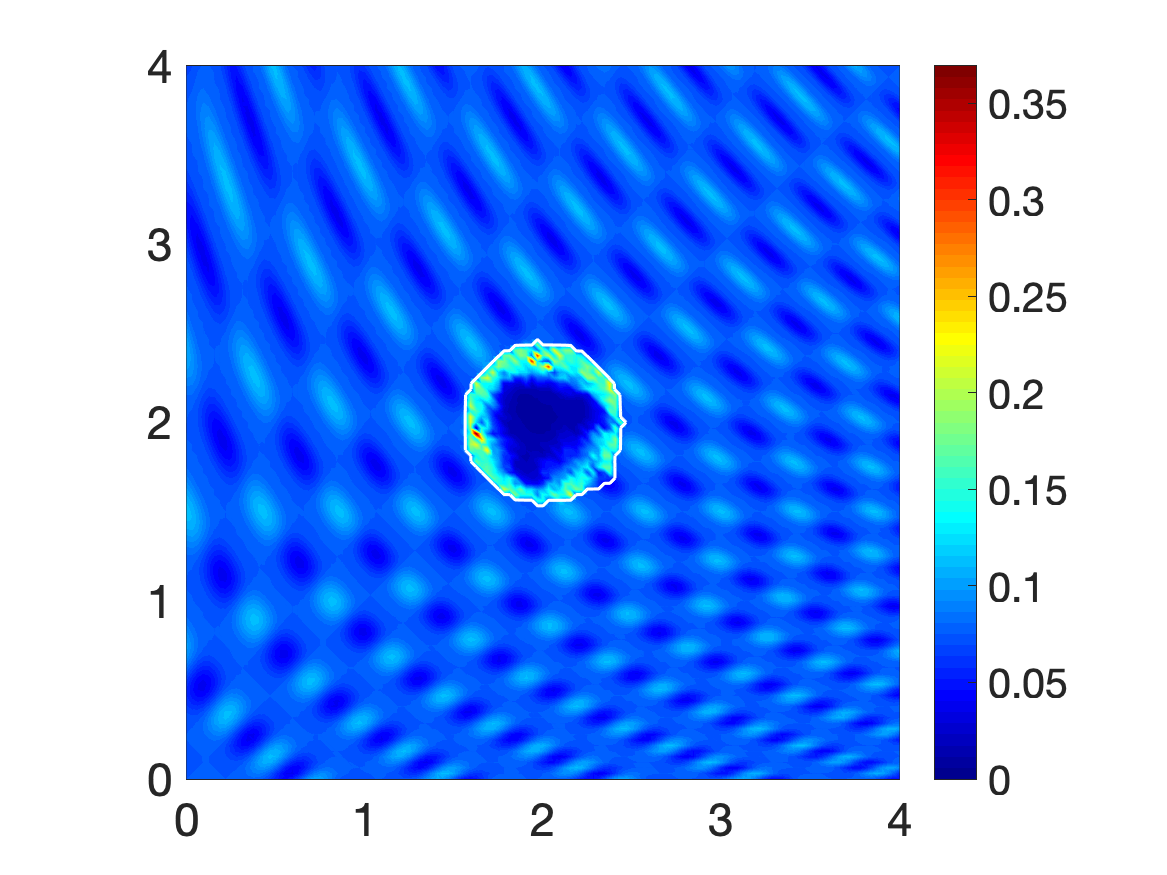}
  \caption{\emph{Fibre magnitude density}}
  \label{fig:fulldeghetero25c}
  \end{subfigure}\hfil 
\begin{subfigure}{0.5\textwidth}
  \includegraphics[width=\linewidth]{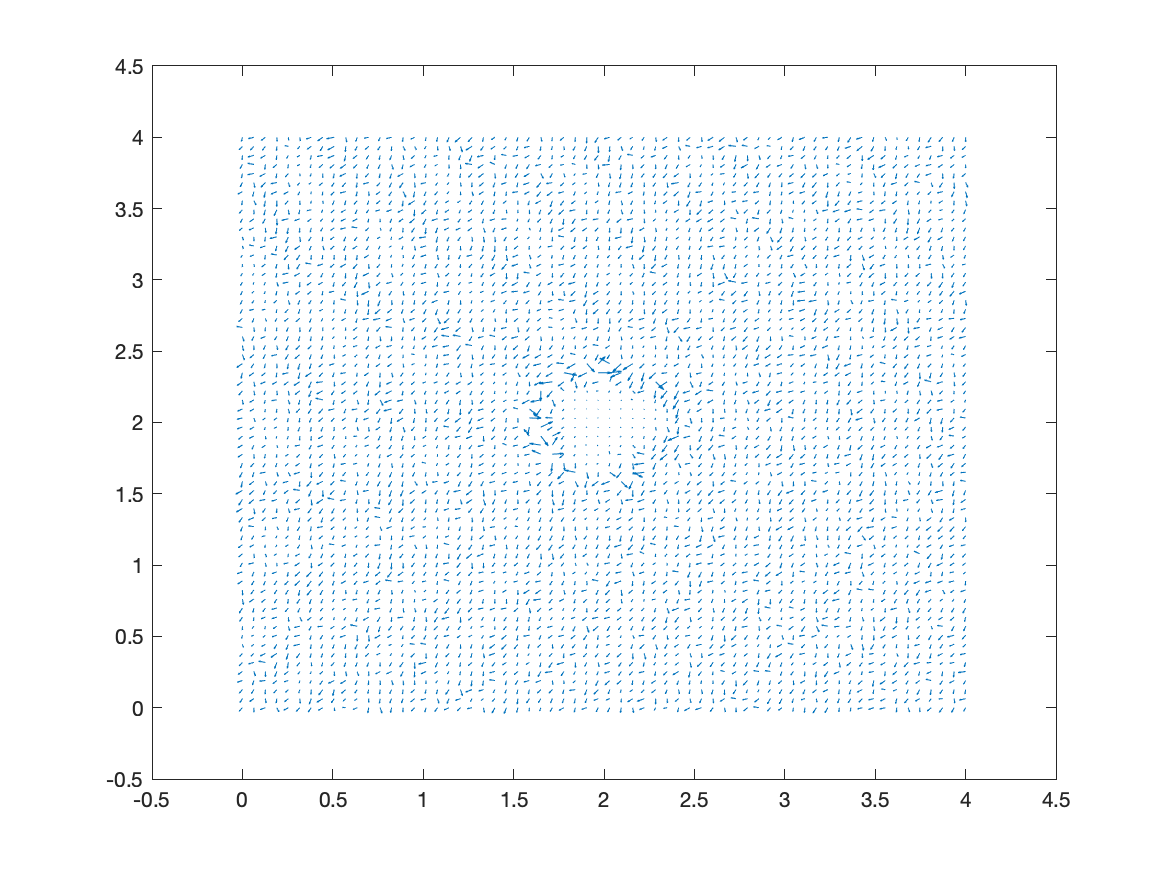}
  \caption{\emph{Fibre vector field - coarsened 2 fold}}
  \label{fig:fulldeghetero25d}
\end{subfigure}\hfil 

\medskip
\begin{subfigure}{0.5\textwidth}
  \includegraphics[width=\linewidth]{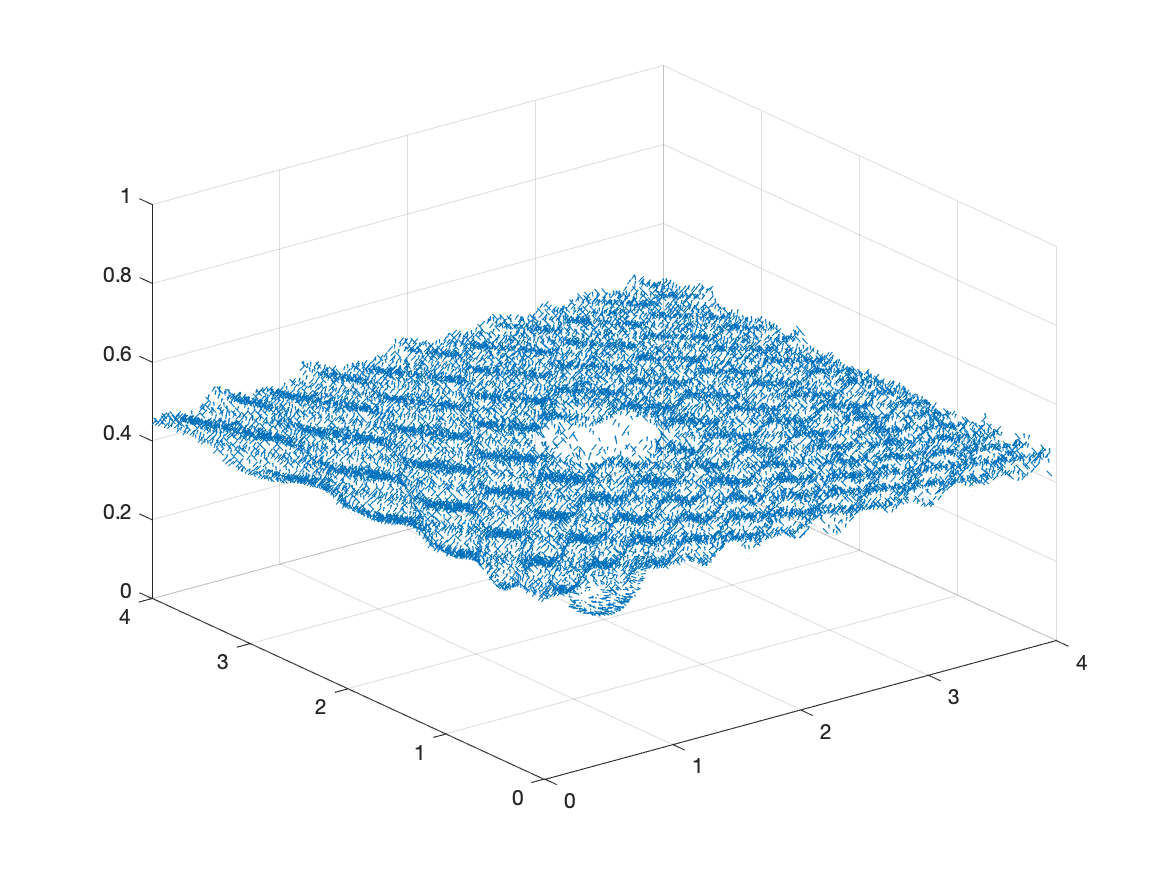}
  \caption{\emph{3D ECM vector field}}
  \label{fig:fulldeghetero25e}
  \end{subfigure}\hfil 
\begin{subfigure}{0.5\textwidth}
  \includegraphics[width=\linewidth]{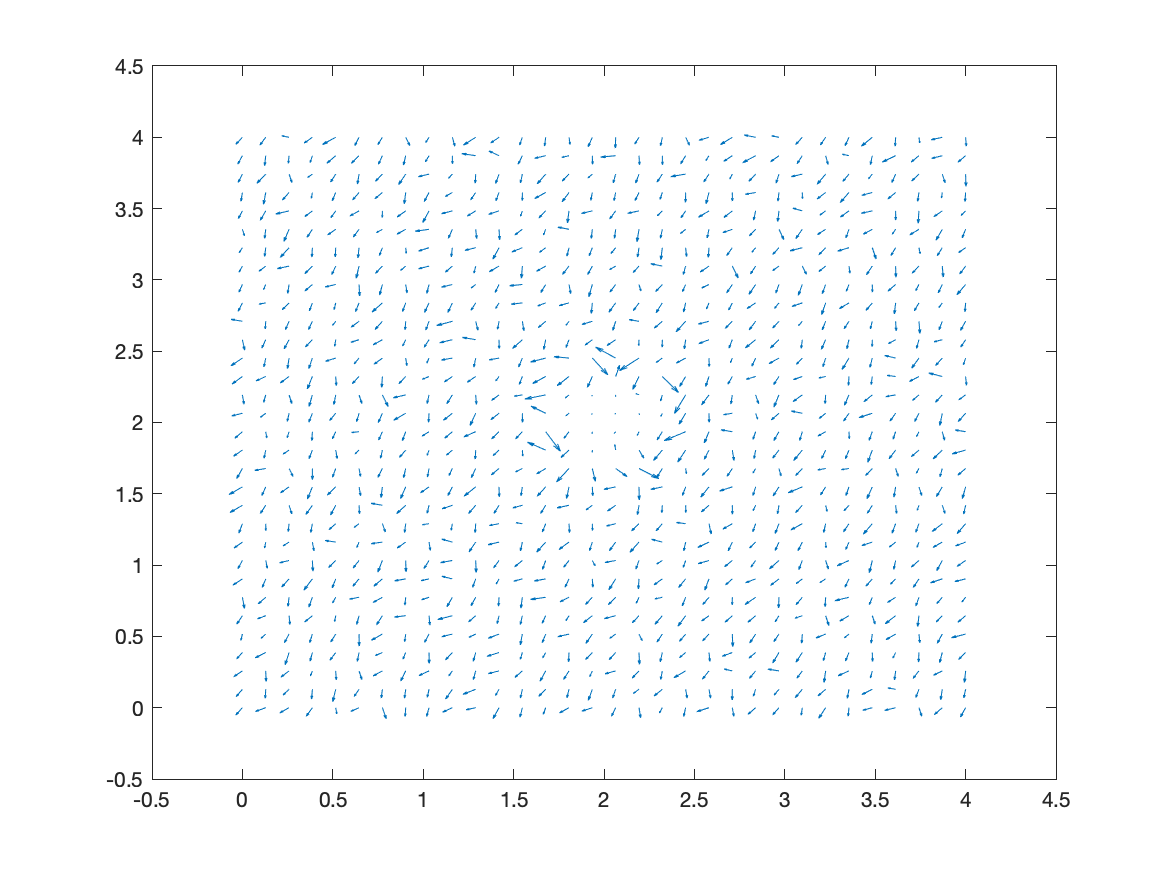}
  \caption{\emph{Fibre vector field - coarsened 4 fold}}
  \label{fig:fulldeghetero25f}
\end{subfigure}\hfil 

\caption[Simulations at stage $25\Delta t$ with a homogeneous distribution of the non-fibrous phase and $15\%$ heterogeneous fibres phase of the ECM with a micro-fibres degradation rate of $d_f = 1$.]{\emph{Simulations at stage $25\Delta t$ with a homogeneous distribution of the non-fibrous phase and $15\%$ heterogeneous fibres phase of the ECM with a micro-fibres degradation rate of $d_f = 1$.}}
\label{fig:fulldeghetero25}
\end{figure}

\begin{figure}[h!]
    \centering 
\begin{subfigure}{0.5\textwidth}
  \includegraphics[width=\linewidth]{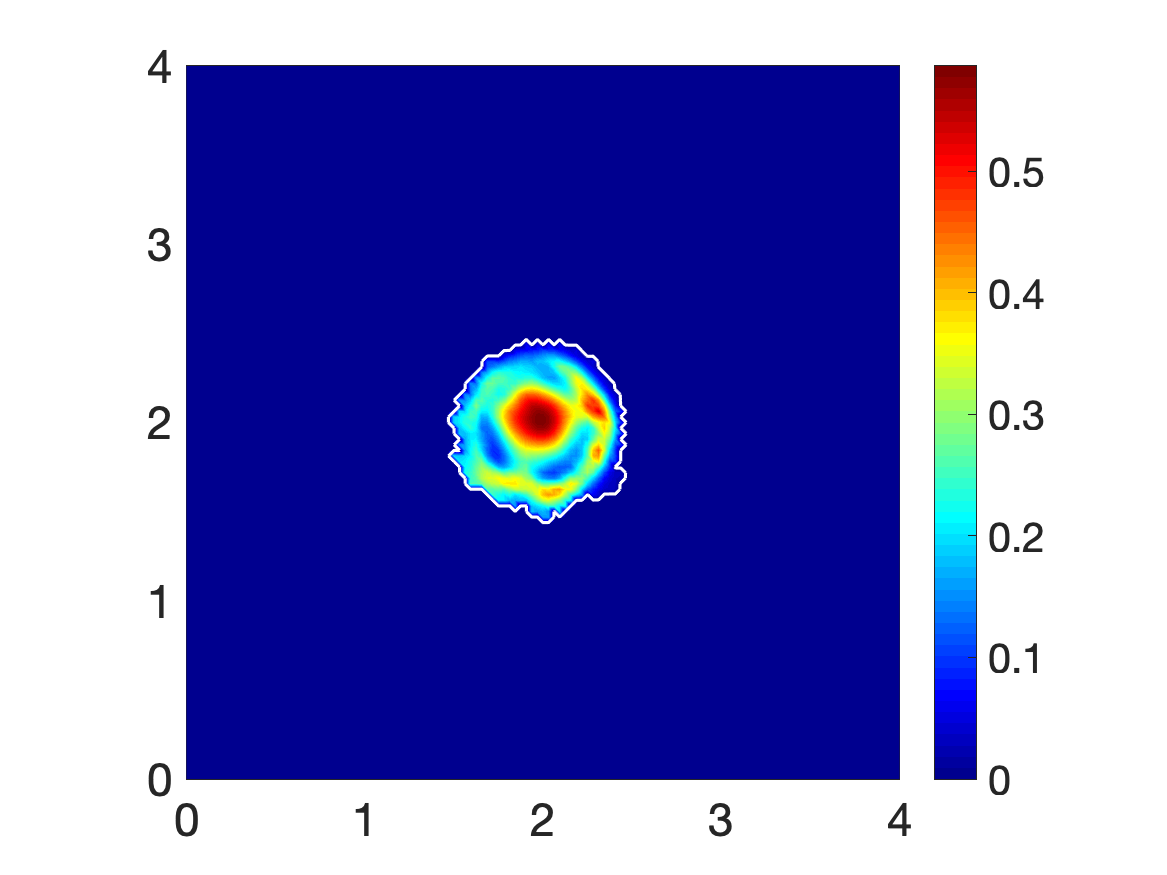}
  \caption{\emph{Cancer cell population}}
  \label{fig:fulldeghetero50a}
\end{subfigure}\hfil 
\begin{subfigure}{0.5\textwidth}
  \includegraphics[width=\linewidth]{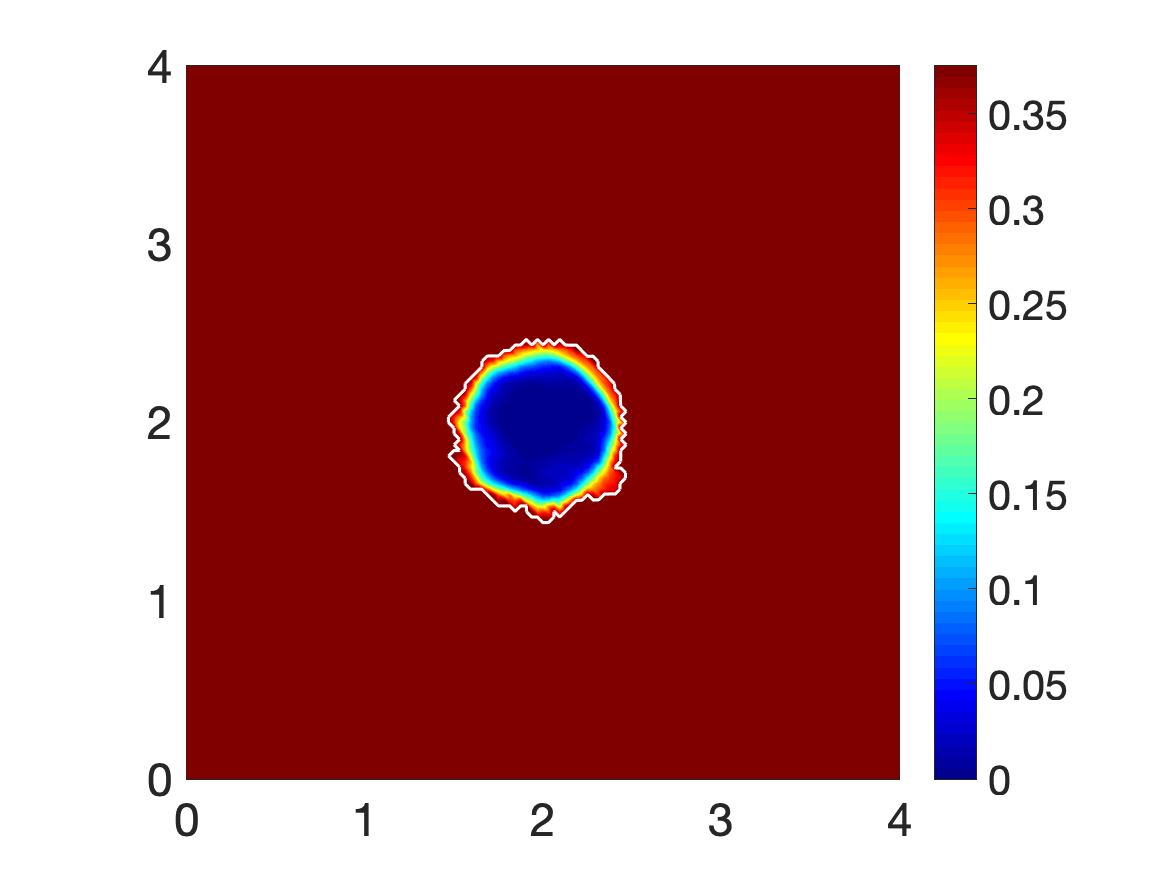}
  \caption{\emph{Non-fibres ECM distribution}}
  \label{fig:fulldeghetero50b}
\end{subfigure}\hfil 

\medskip
\begin{subfigure}{0.5\textwidth}
  \includegraphics[width=\linewidth]{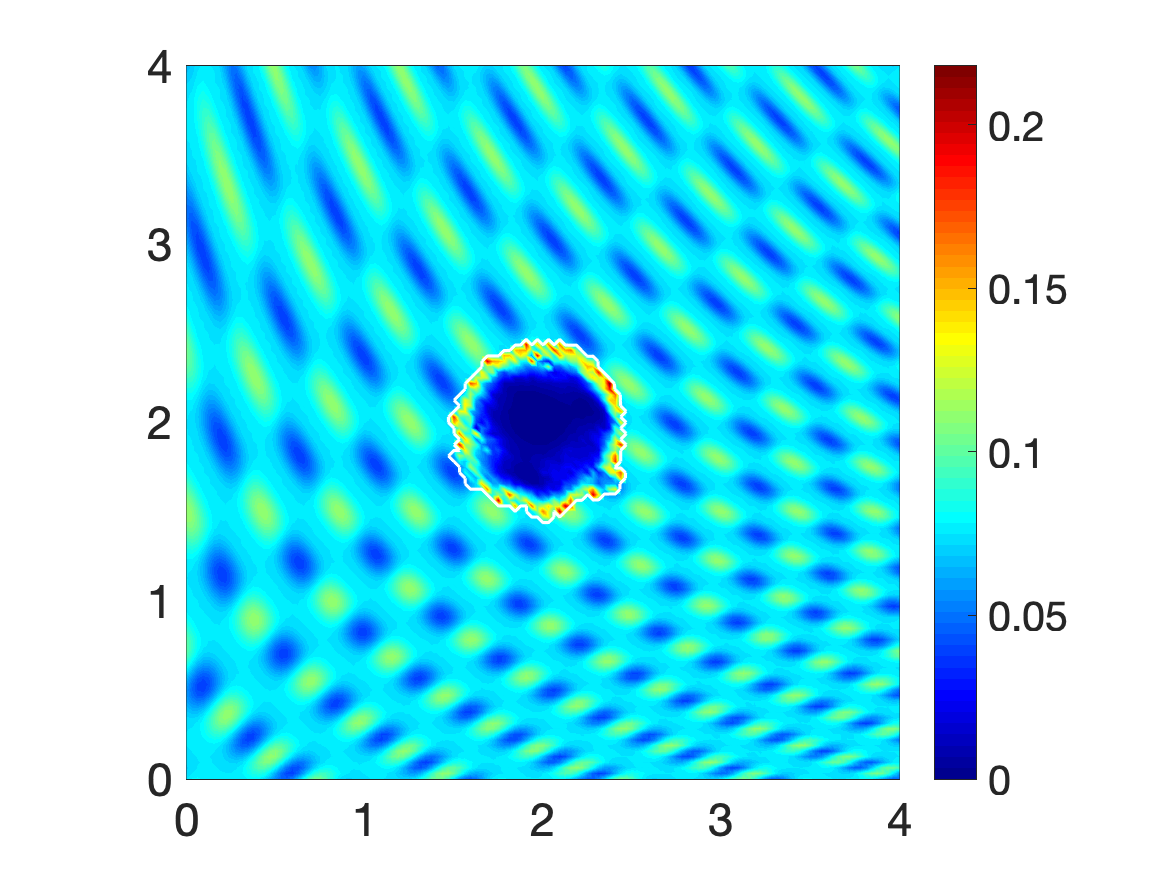}
  \caption{\emph{Fibre magnitude density}}
  \label{fig:fulldeghetero50c}
  \end{subfigure}\hfil 
\begin{subfigure}{0.5\textwidth}
  \includegraphics[width=\linewidth]{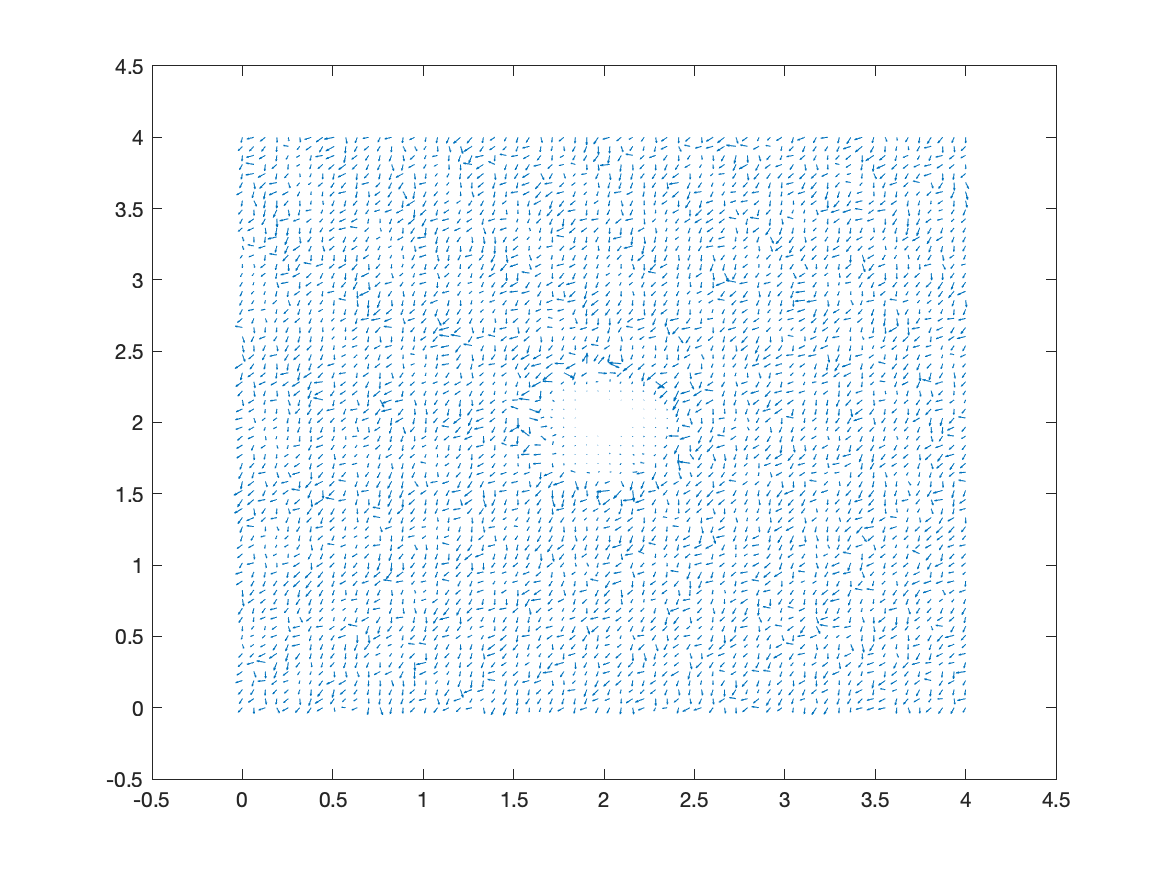}
  \caption{\emph{Fibre vector field - coarsened 2 fold}}
  \label{fig:fulldeghetero50d}
\end{subfigure}\hfil 

\medskip
\begin{subfigure}{0.5\textwidth}
  \includegraphics[width=\linewidth]{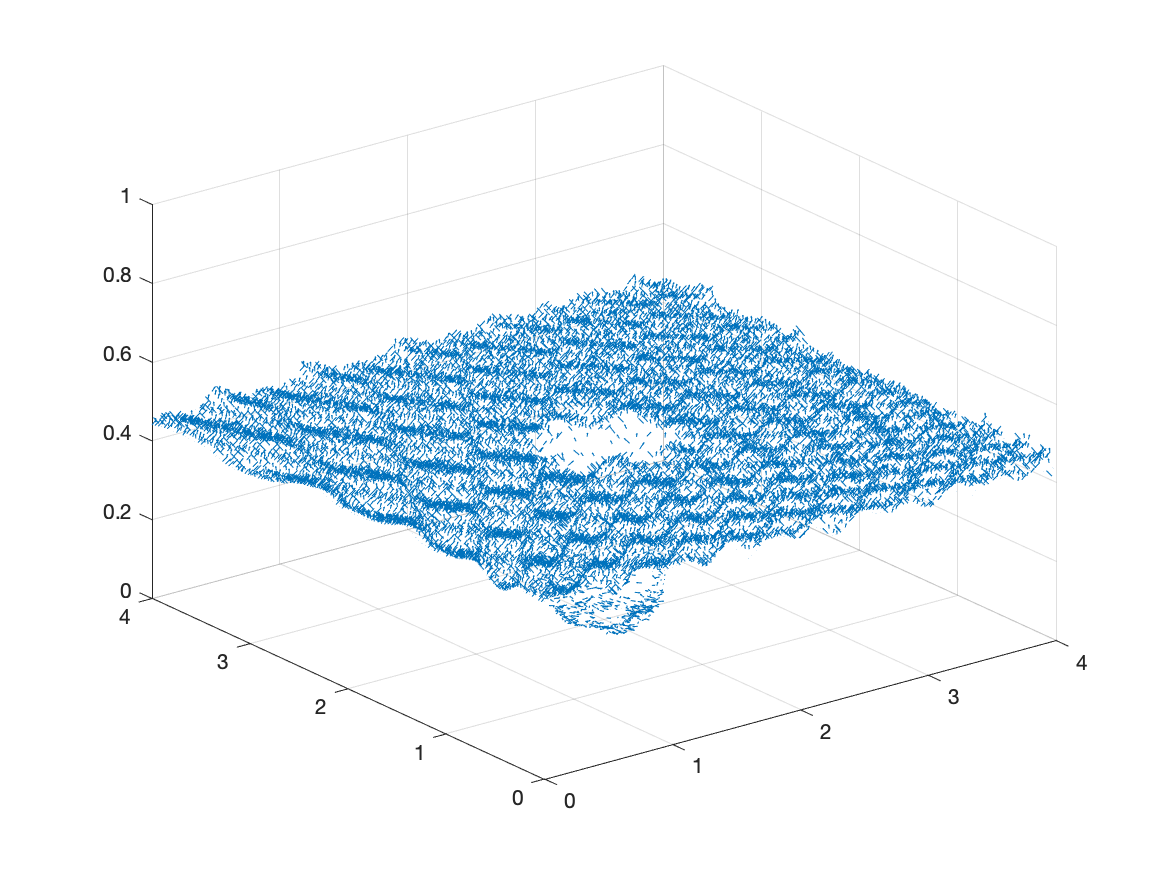}
  \caption{\emph{3D ECM vector field}}
  \label{fig:fulldeghetero50e}
  \end{subfigure}\hfil 
\begin{subfigure}{0.5\textwidth}
  \includegraphics[width=\linewidth]{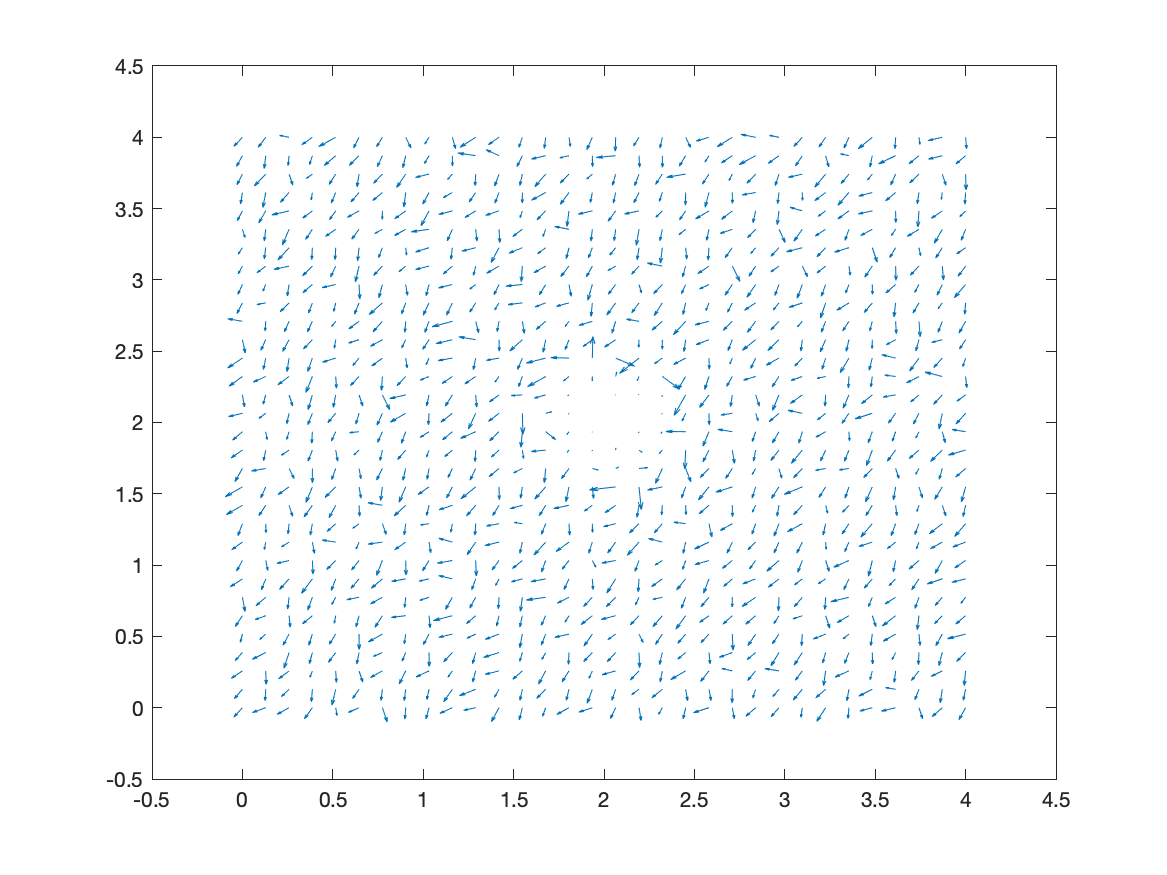}
  \caption{\emph{Fibre vector field - coarsened 4 fold}}
  \label{fig:fulldeghetero50f}
\end{subfigure}\hfil 

\caption[Simulations at stage $50\Delta t$ with a homogeneous distribution of the non-fibrous phase and $15\%$ heterogeneous fibres phase of the ECM with a micro-fibres degradation rate of $d_f = 1$.]{\emph{Simulations at stage $50\Delta t$ with a homogeneous distribution of the non-fibrous phase and $15\%$ heterogeneous fibres phase of the ECM with a micro-fibres degradation rate of $d_f = 1$.}}
\label{fig:fulldeghetero50}
\end{figure}

\begin{figure}[h!]
    \centering 
\begin{subfigure}{0.5\textwidth}
  \includegraphics[width=\linewidth]{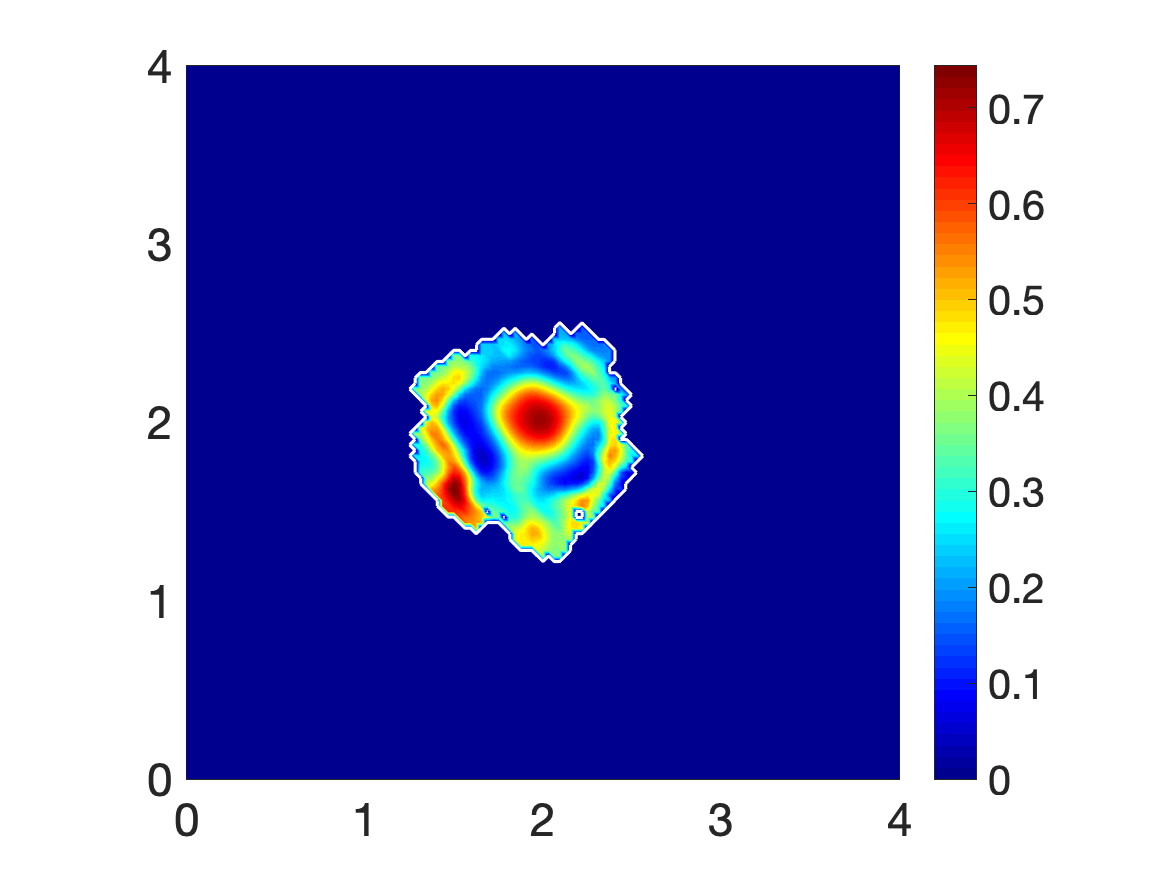}
  \caption{\emph{Cancer cell population}}
  \label{fig:fulldeghetero75a}
\end{subfigure}\hfil 
\begin{subfigure}{0.5\textwidth}
  \includegraphics[width=\linewidth]{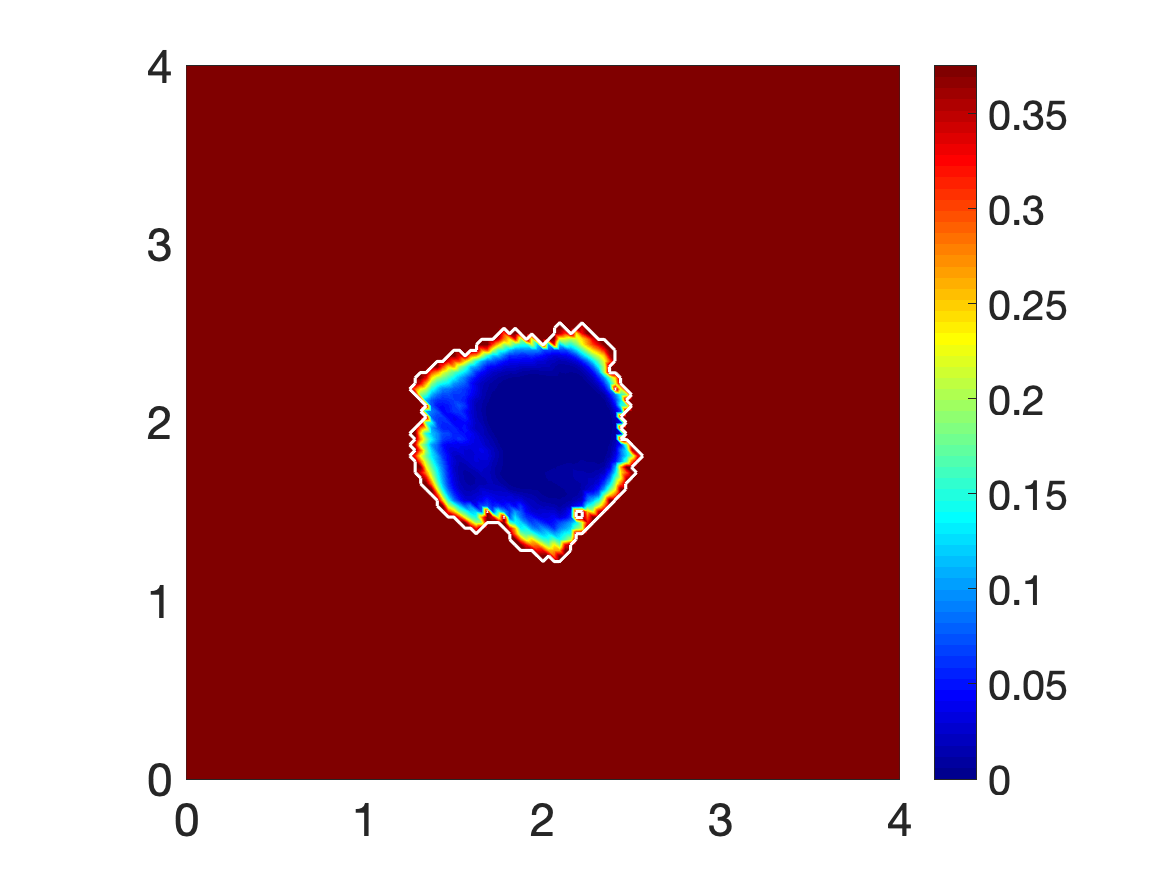}
  \caption{\emph{Non-fibres ECM distribution}}
  \label{fig:fulldeghetero75b}
\end{subfigure}\hfil 

\medskip
\begin{subfigure}{0.5\textwidth}
  \includegraphics[width=\linewidth]{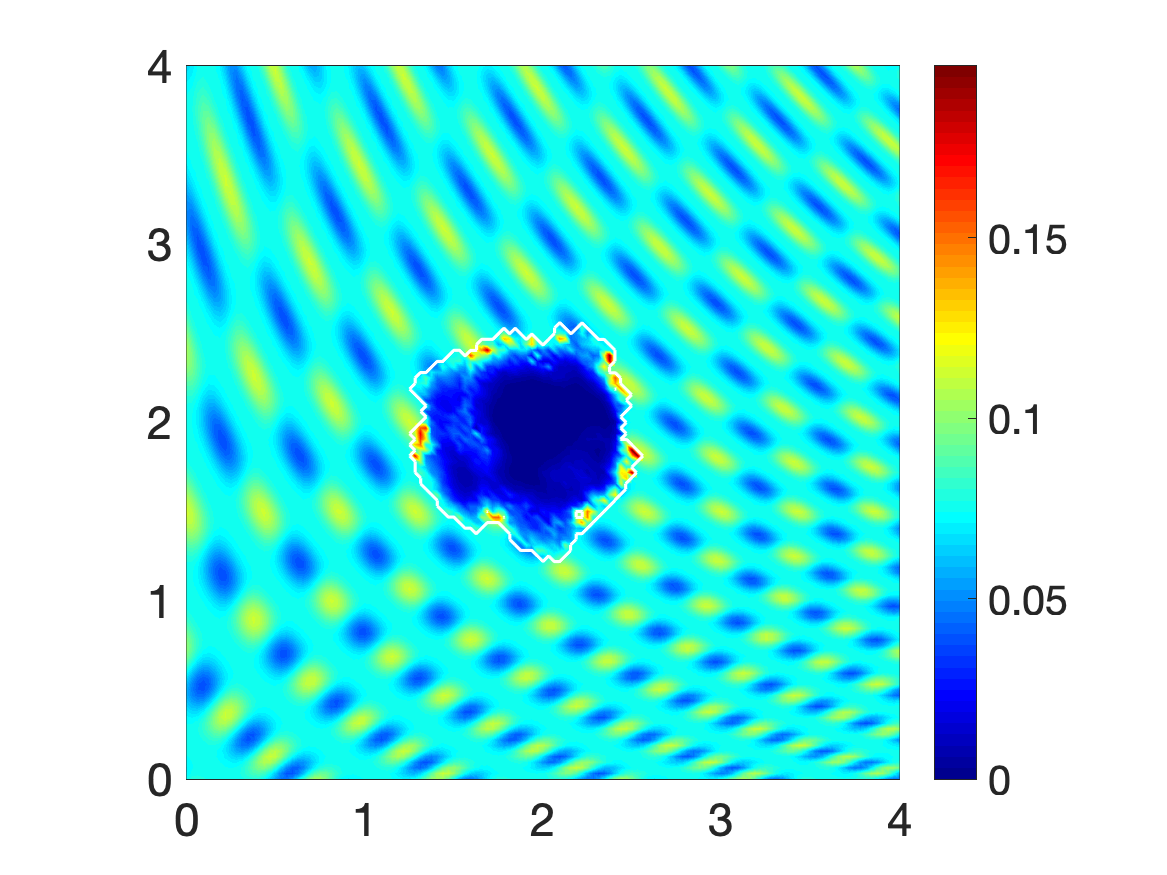}
  \caption{\emph{Fibre magnitude density}}
  \label{fig:fulldeghetero75c}
  \end{subfigure}\hfil 
\begin{subfigure}{0.5\textwidth}
  \includegraphics[width=\linewidth]{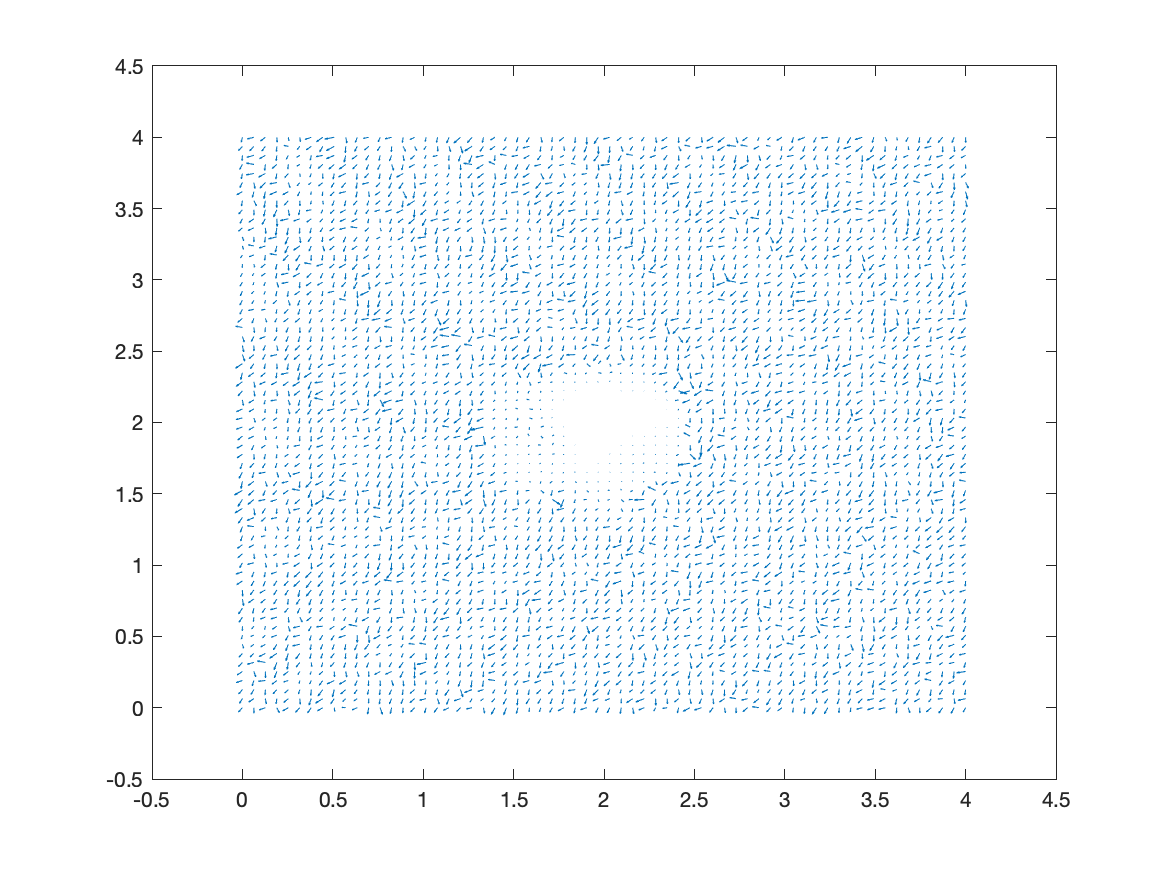}
  \caption{\emph{Fibre vector field - coarsened 2 fold}}
  \label{fig:fulldeghetero75d}
\end{subfigure}\hfil 

\medskip
\begin{subfigure}{0.5\textwidth}
  \includegraphics[width=\linewidth]{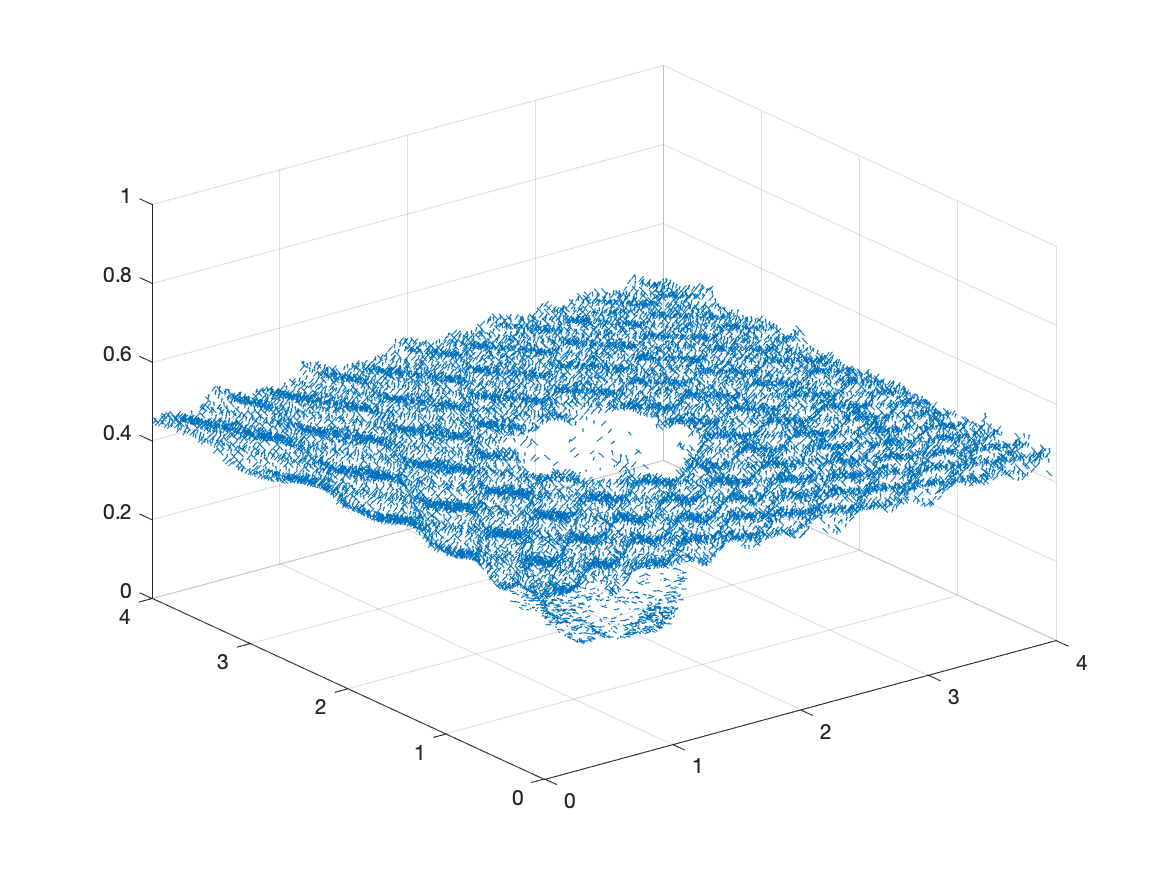}
  \caption{\emph{3D ECM vector field}}
  \label{fig:fulldeghetero75e}
  \end{subfigure}\hfil 
\begin{subfigure}{0.5\textwidth}
  \includegraphics[width=\linewidth]{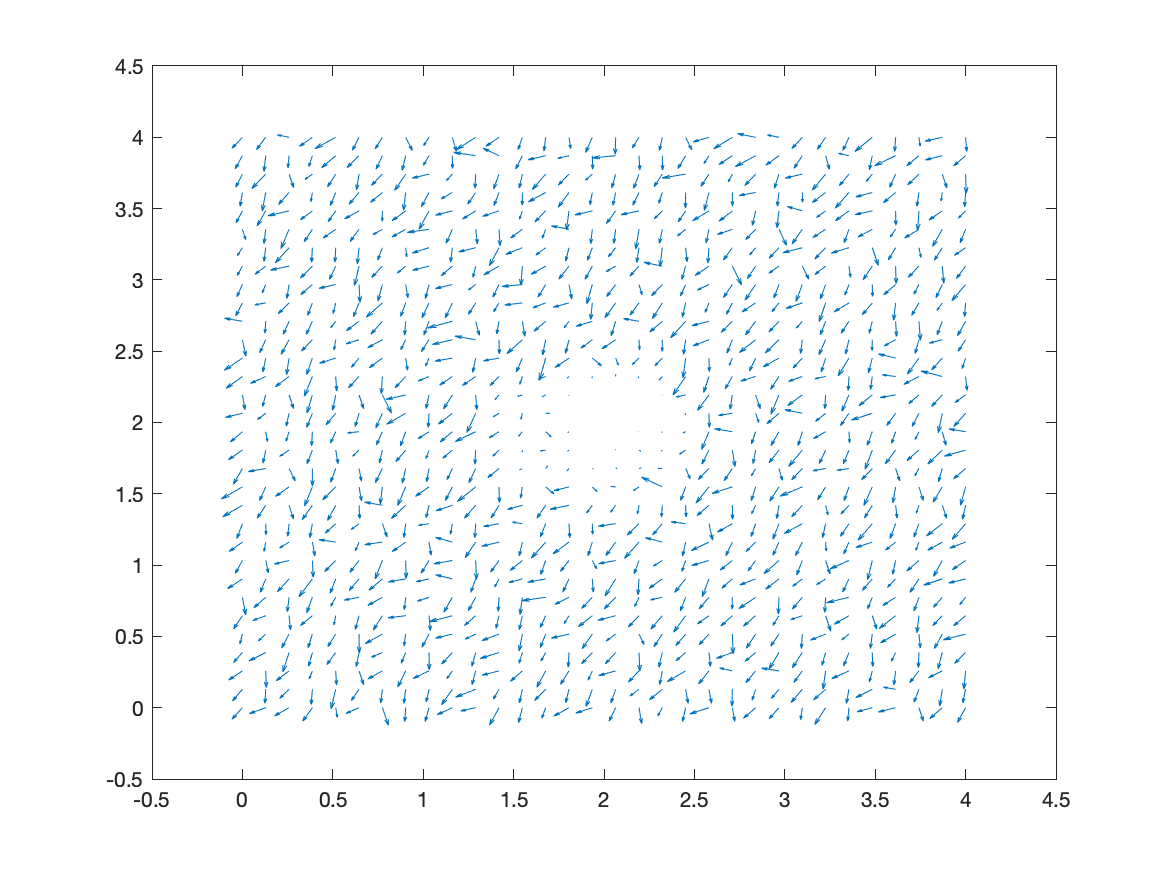}
  \caption{\emph{Fibre vector field - coarsened 4 fold}}
  \label{fig:fulldeghetero75f}
\end{subfigure}\hfil 

\caption[Simulations at stage $75\Delta t$ with a homogeneous distribution of the non-fibrous phase and $15\%$ heterogeneous fibres phase of the ECM with a micro-fibres degradation rate of $d_f = 1$.]{\emph{Simulations at stage $75\Delta t$ with a homogeneous distribution of the non-fibrous phase and $15\%$ heterogeneous fibres phase of the ECM with a micro-fibres degradation rate of $d_f = 1$.}}
\label{fig:fulldeghetero75}
\end{figure}

To investigate the effects of fibre distribution, whilst keeping the non-fibre ECM phase initially homogeneous \eqref{eq:lhomoic}, we initialise the fibre distribution $F(x,0)$ with $15\%$, or $p=0.15$, of the heterogeneous distribution \eqref{eq:lheteroic}. The simulations at stage $25 \Delta t$, Figure \ref{fig:fulldeghetero25}, indicate the initial distribution of fibres is significant during the evolution of a tumour. The primary bulk of cancer cells have dispersed within the boundary into regions of high cell distribution, subfigure \ref{fig:fulldeghetero25a}, these coinciding with low density regions of the fibres ECM phase. The boundary of the tumour is circular with small defects in the direction of the fibre orientation, subfigure \ref{fig:fulldeghomo25f}. The fibre magnitude density in subfigure \ref{fig:fulldeghetero25c} has small regions of high fibre distributions, again pushed outwards towards the proliferating edge. Due to the heterogeneity of the fibre distributions, the fibre orientations are subject to an increased degree of realignment, subfigures \ref{fig:fulldeghetero25d}, \ref{fig:fulldeghetero25f}, attributed to the initially different levels of density at each macro-spatial position. The fibres are realigned towards the higher density regions, where ``frenzied'' groups of fibre orientations can be observed. Moving on to simulations at stage $50 \Delta t$, Figure \ref{fig:fulldeghetero50}, many of the behaviours previously observed at stage $25 \Delta t$ are magnified. The cancer bundle has increased in size and notably spread further into the areas of initially low fibre density, subfigure \ref{fig:fulldeghetero50a} with the tumour boundary also increasing in size and irregularity. The fibres are being rearranged and pushed further outwards towards the tumour boundary, subfigure \ref{fig:fulldeghetero50c}, whilst simultaneously undergoing macroscopic degradation at the tissue scale \eqref{eq:fib}, thus resulting is areas of very low to no fibre density. Finally, the simulations in Figure \ref{fig:fulldeghetero75} show the evolution of the tumour at stage $75 \Delta t$. The cancer cells are forming patterns within the tumour boundary, subfigure \ref{fig:fulldeghetero75a}, in the areas of low ECM density, and cells are migrating towards the tumour boundary collecting in high distribution bundles that more rapidly degrade the surrounding ECM. Furthermore, the cells are flooding areas where there is very low to no fibre density, subfigure \ref{fig:fulldeghetero75c}, and forming dense bundles of cells. This behaviour is in accordance with the conclusions presented in \cite{Shutt_2018} that cancer cells can more freely invade areas of no ECM density and will progress upon these areas first before engulfing the higher density regions.

\subsection{Increased collagen density}
\begin{figure}[h!]
    \centering 
\begin{subfigure}{0.5\textwidth}
  \includegraphics[width=\linewidth]{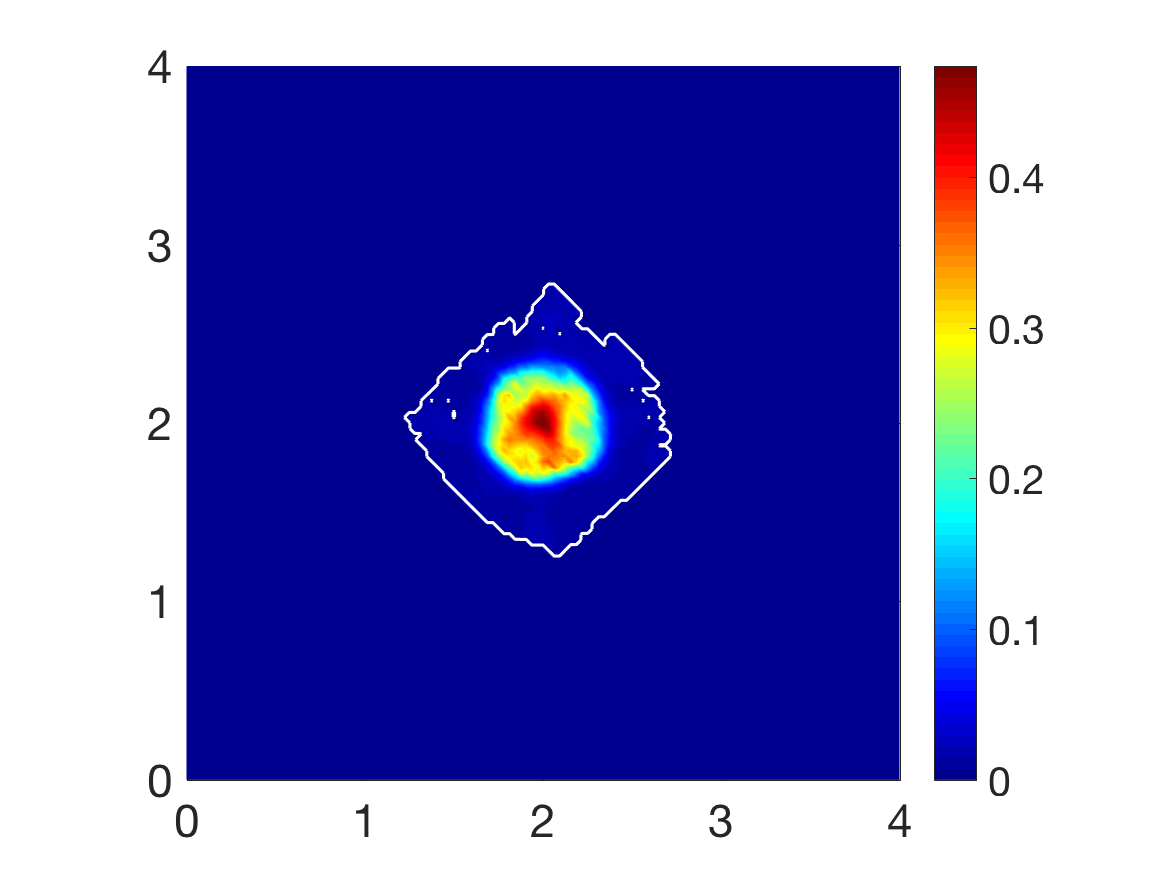}
  \caption{\emph{Cancer cell population}}
  \label{fig:fulldeghighhomo25a}
\end{subfigure}\hfil 
\begin{subfigure}{0.5\textwidth}
  \includegraphics[width=\linewidth]{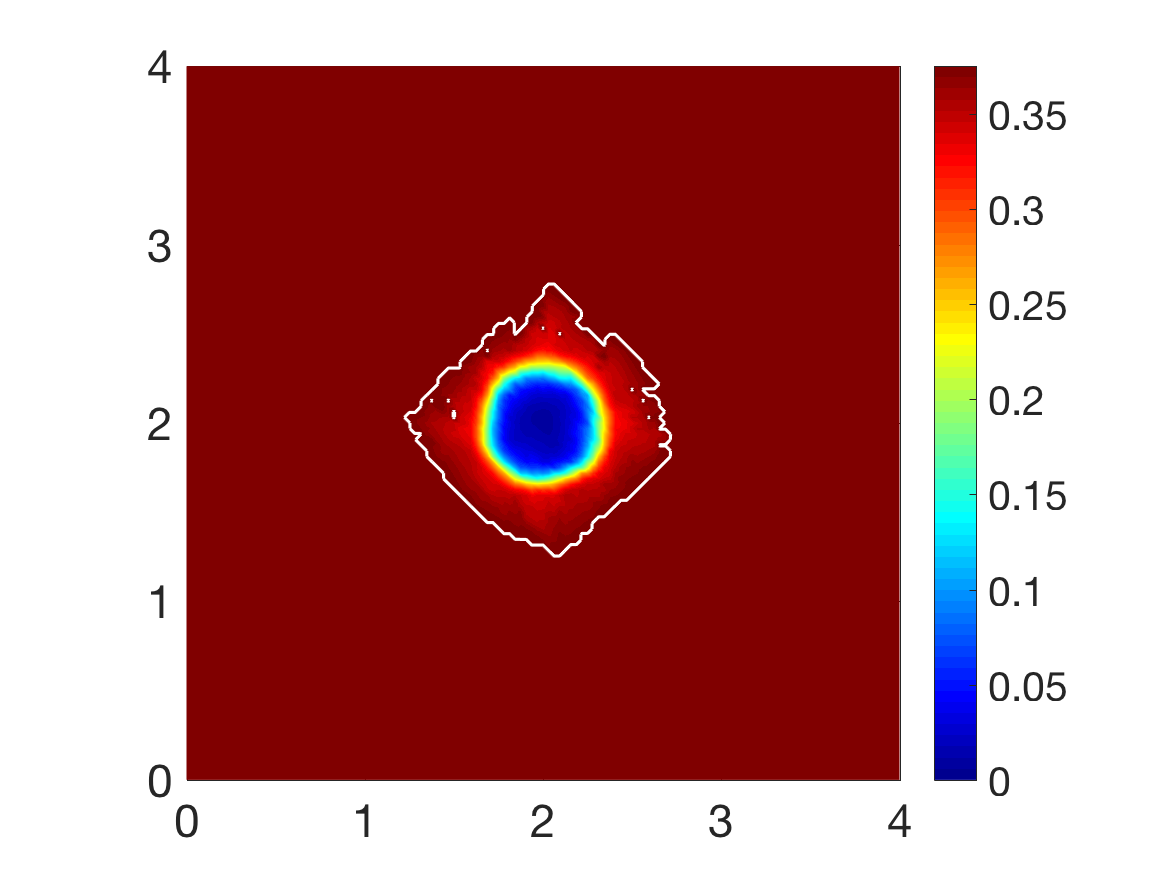}
  \caption{\emph{Non-fibres ECM distribution}}
  \label{fig:fulldeghighhomo25b}
\end{subfigure}\hfil 

\medskip
\begin{subfigure}{0.5\textwidth}
  \includegraphics[width=\linewidth]{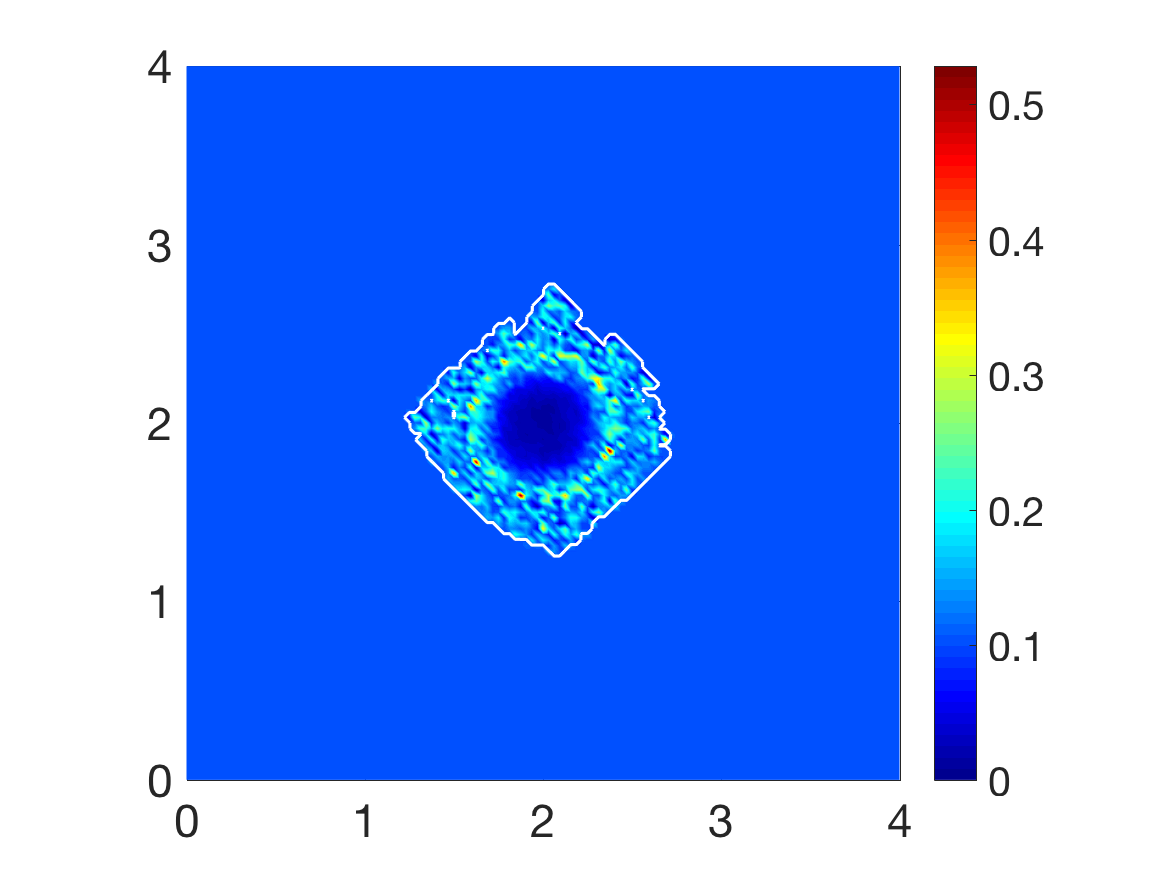}
  \caption{\emph{Fibre magnitude density}}
  \label{fig:fulldeghighhomo25c}
  \end{subfigure}\hfil 
\begin{subfigure}{0.5\textwidth}
  \includegraphics[width=\linewidth]{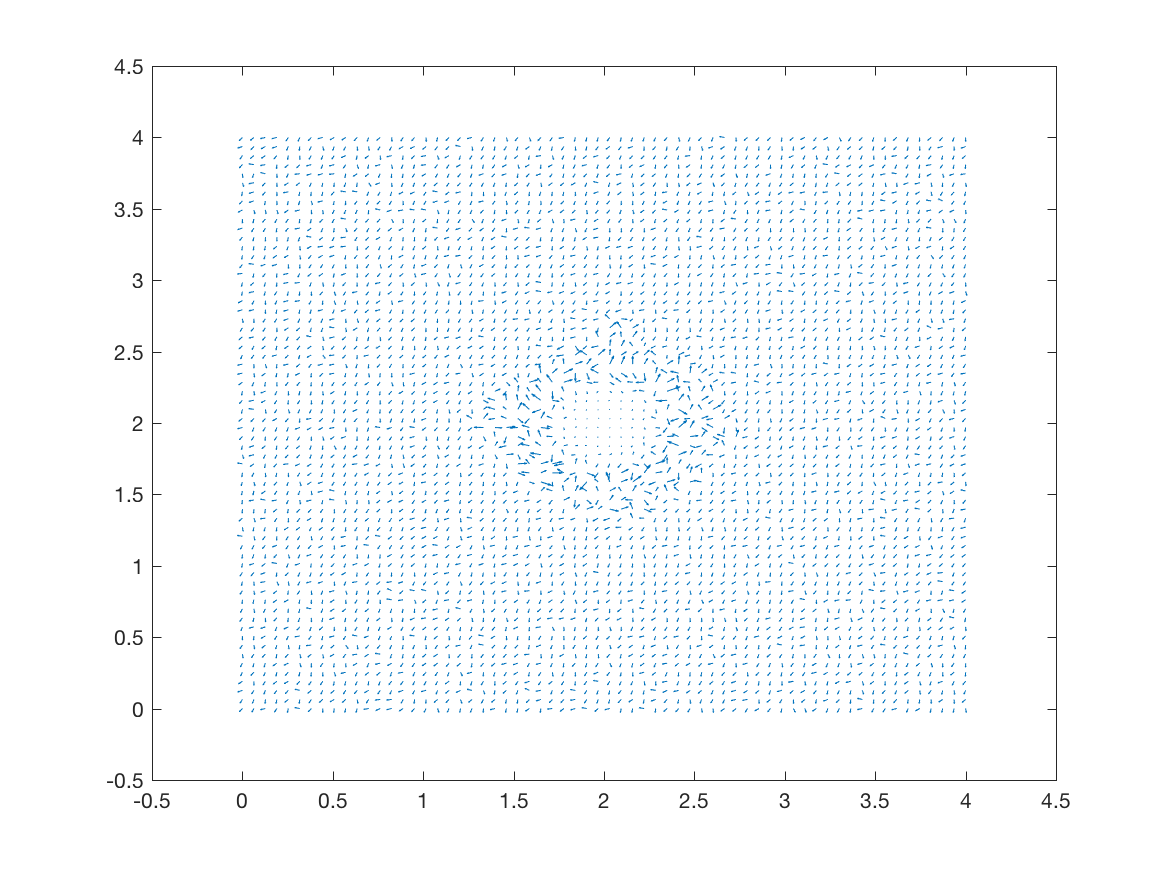}
  \caption{\emph{Fibre vector field - coarsened 2 fold}}
  \label{fig:fulldeghighhomo25d}
\end{subfigure}\hfil 

\medskip
\begin{subfigure}{0.5\textwidth}
  \includegraphics[width=\linewidth]{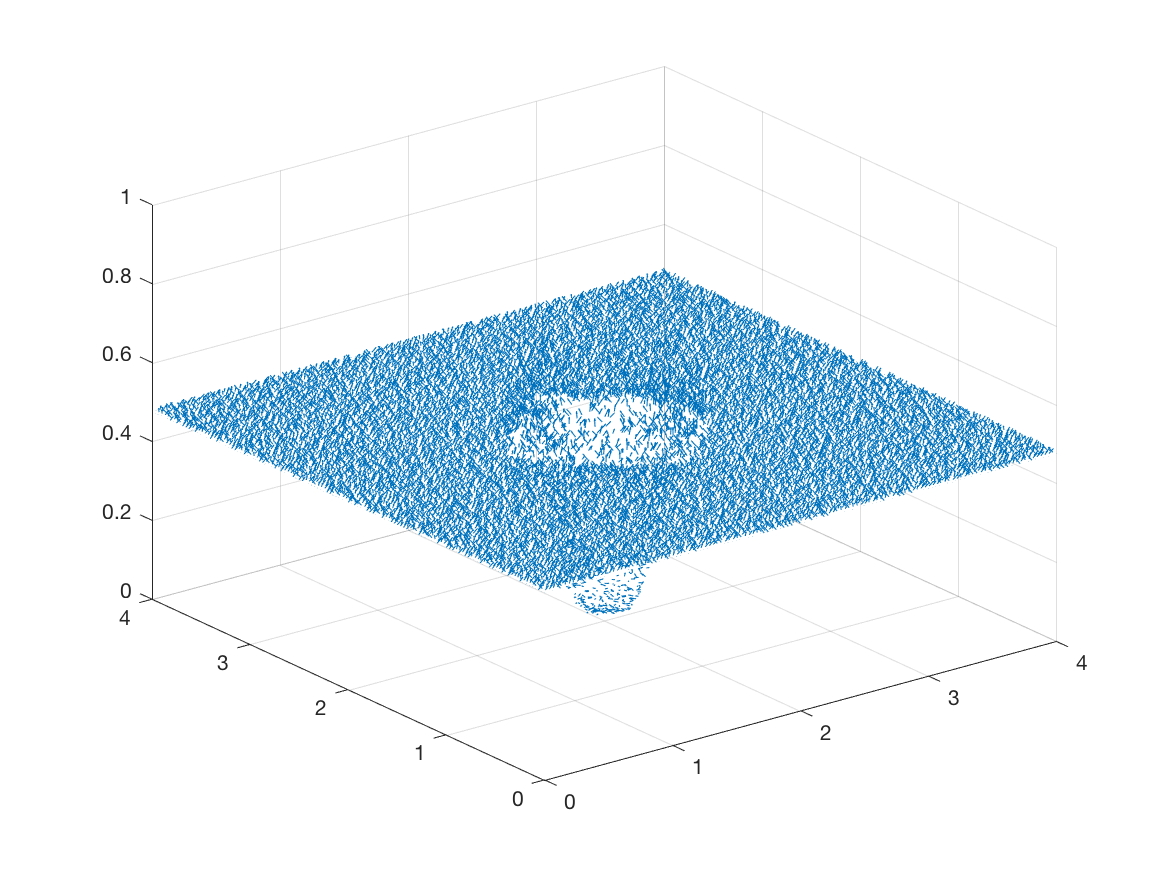}
  \caption{\emph{3D ECM vector field}}
  \label{fig:fulldeghighhomo25e}
  \end{subfigure}\hfil 
\begin{subfigure}{0.5\textwidth}
  \includegraphics[width=\linewidth]{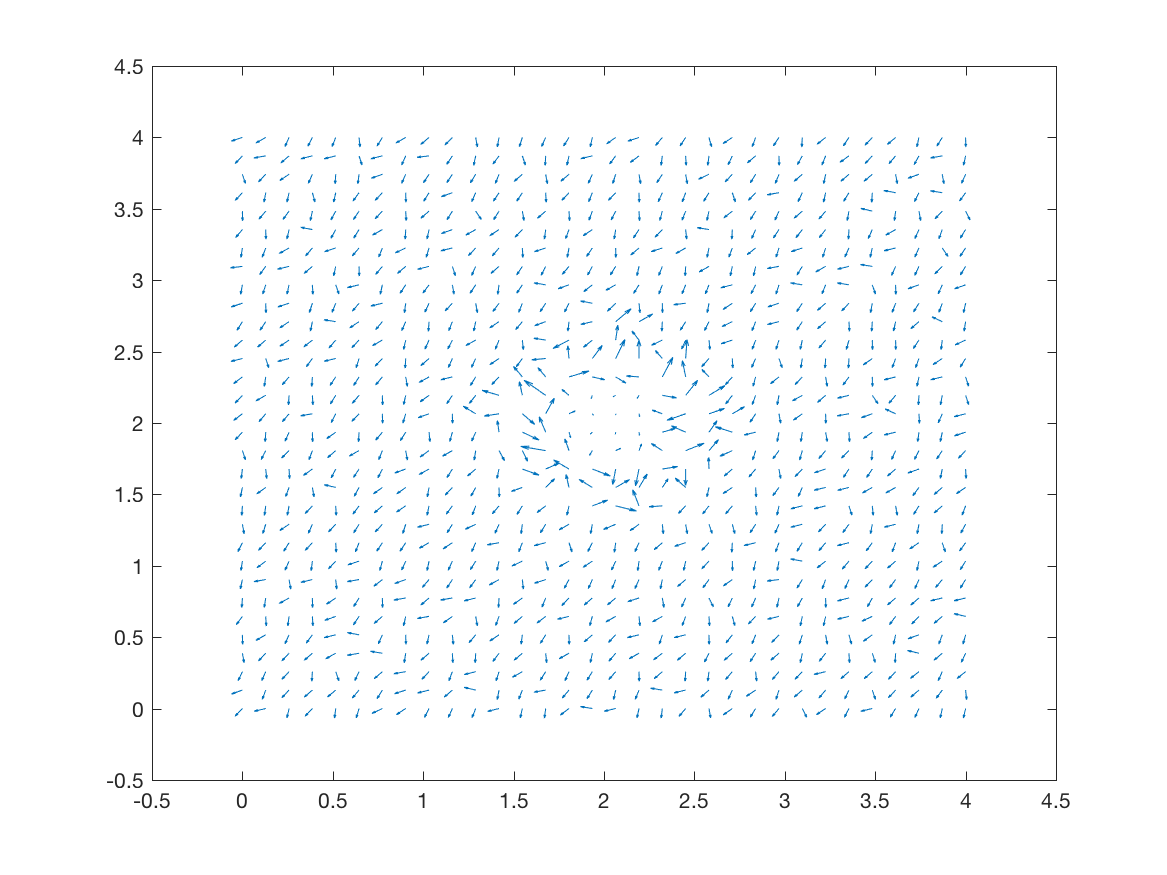}
  \caption{\emph{Fibre vector field - coarsened 4 fold}}
  \label{fig:fulldeghighhomo25f}
\end{subfigure}\hfil 

\caption[Simulations at stage $25\Delta t$ with a homogeneous distribution of the non-fibrous phase and $20\%$ homogeneous fibres phase of the ECM with a micro-fibres degradation rate of $d_f = 0.5$.]{\emph{Simulations at stage $25\Delta t$ with a homogeneous distribution of the non-fibrous phase and $20\%$ homogeneous fibres phase of the ECM with a micro-fibres degradation rate of $d_f = 0.5$.}}
\label{fig:fulldeghighhomo25}
\end{figure}

\begin{figure}[h!]
    \centering 
\begin{subfigure}{0.5\textwidth}
  \includegraphics[width=\linewidth]{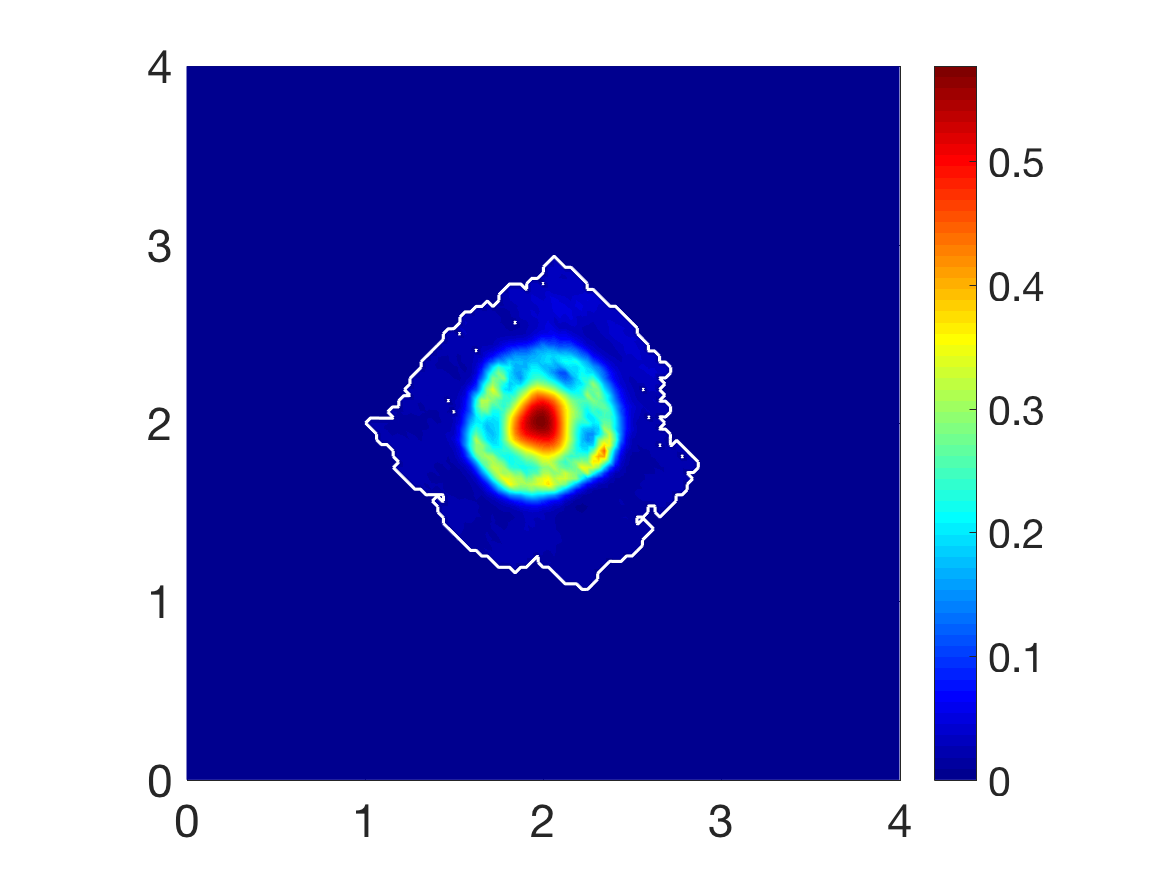}
  \caption{\emph{Cancer cell population}}
  \label{fig:fulldeghighhomo50a}
\end{subfigure}\hfil 
\begin{subfigure}{0.5\textwidth}
  \includegraphics[width=\linewidth]{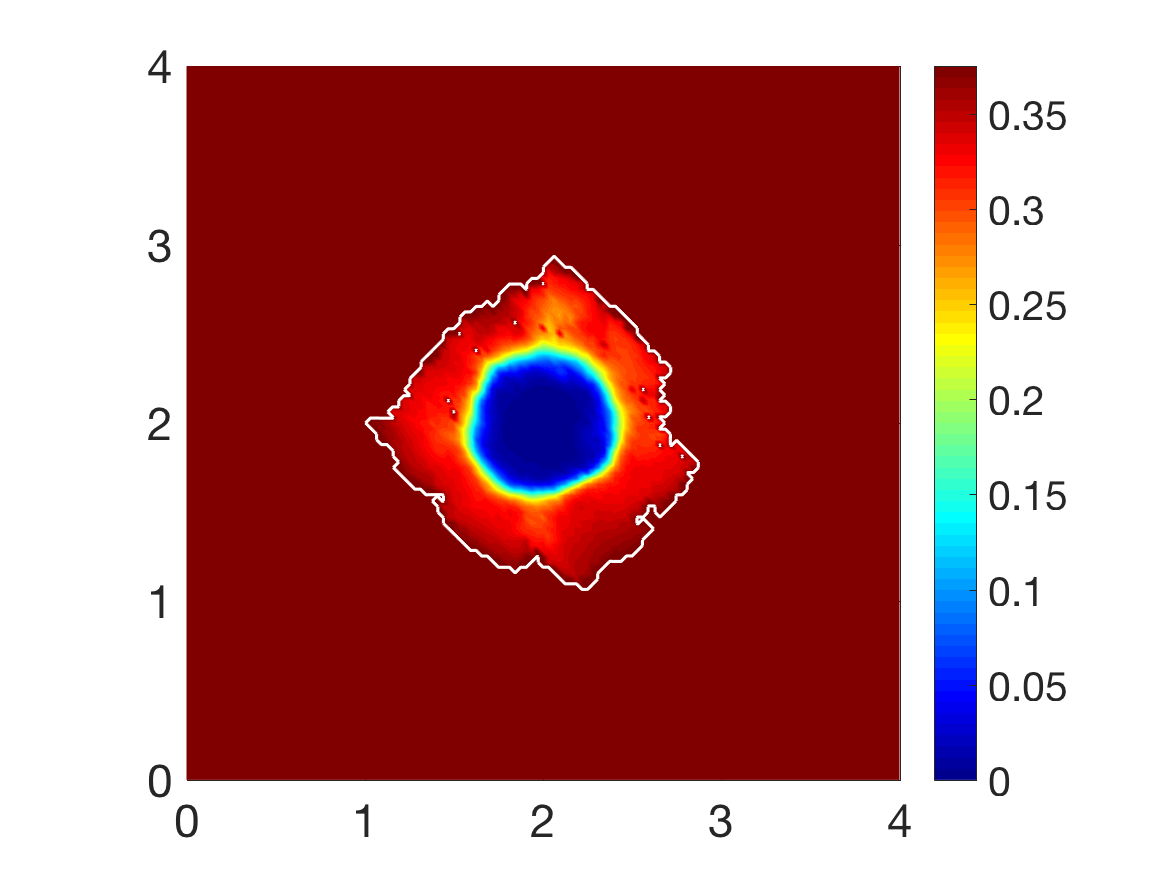}
  \caption{\emph{Non-fibres ECM distribution}}
  \label{fig:fulldeghighhomo50b}
\end{subfigure}\hfil 

\medskip
\begin{subfigure}{0.5\textwidth}
  \includegraphics[width=\linewidth]{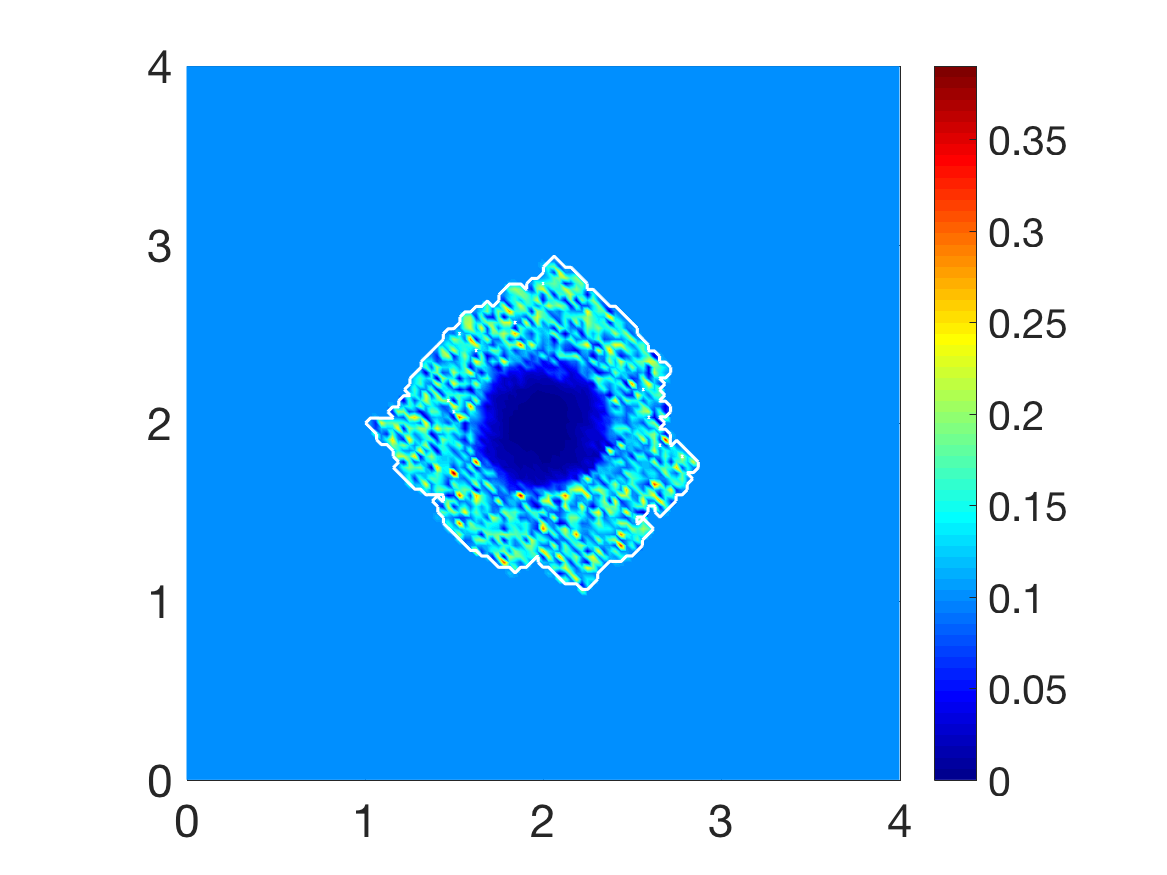}
  \caption{\emph{Fibre magnitude density}}
  \label{fig:fulldeghighhomo50c}
  \end{subfigure}\hfil 
\begin{subfigure}{0.5\textwidth}
  \includegraphics[width=\linewidth]{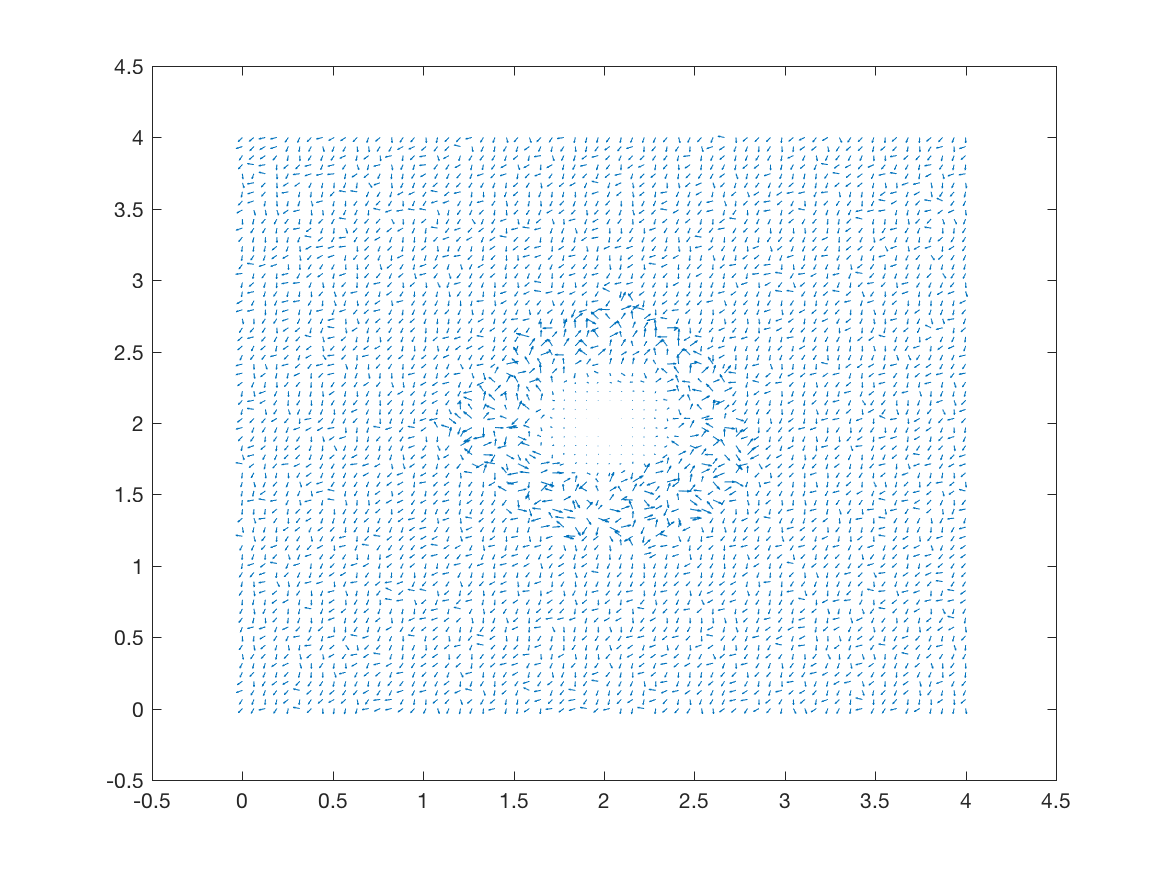}
  \caption{\emph{Fibre vector field - coarsened 2 fold}}
  \label{fig:fulldeghighhomo50d}
\end{subfigure}\hfil 

\medskip
\begin{subfigure}{0.5\textwidth}
  \includegraphics[width=\linewidth]{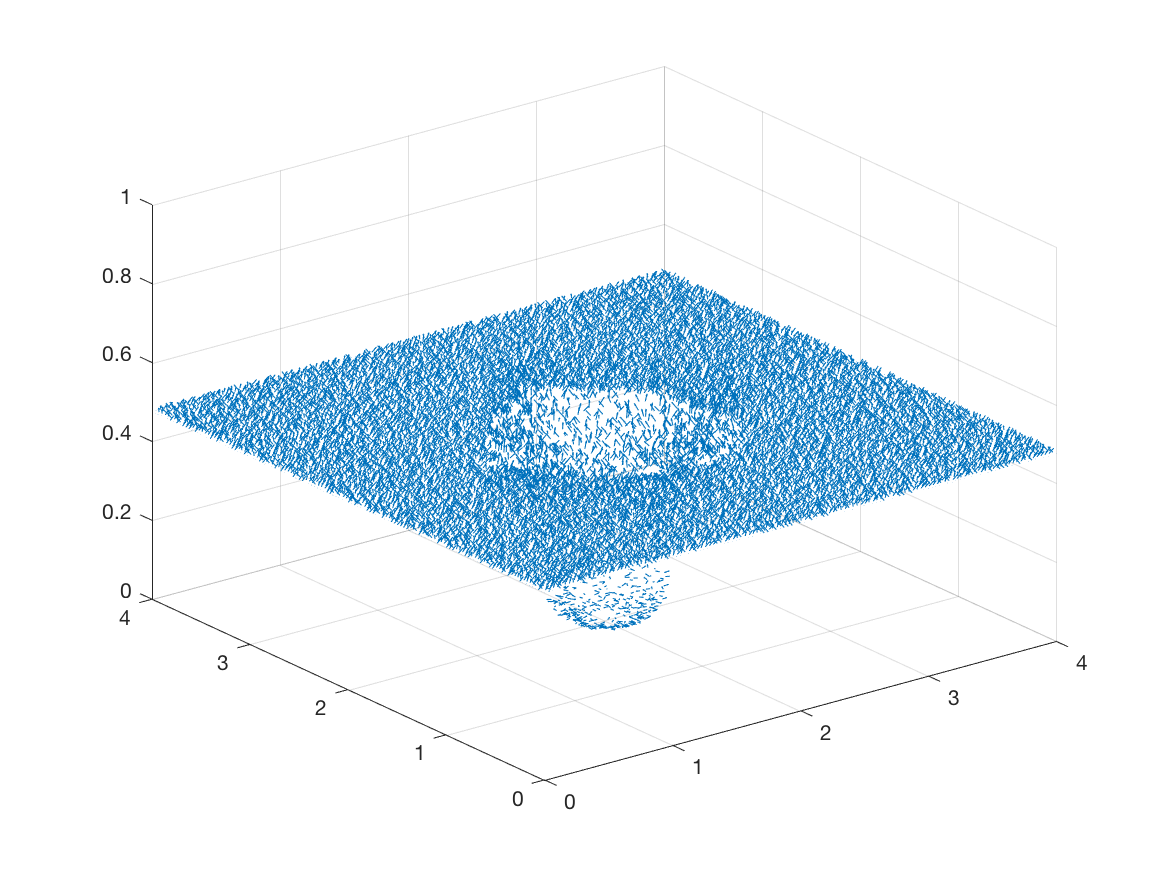}
  \caption{\emph{3D ECM vector field}}
  \label{fig:fulldeghighhomo50e}
  \end{subfigure}\hfil 
\begin{subfigure}{0.5\textwidth}
  \includegraphics[width=\linewidth]{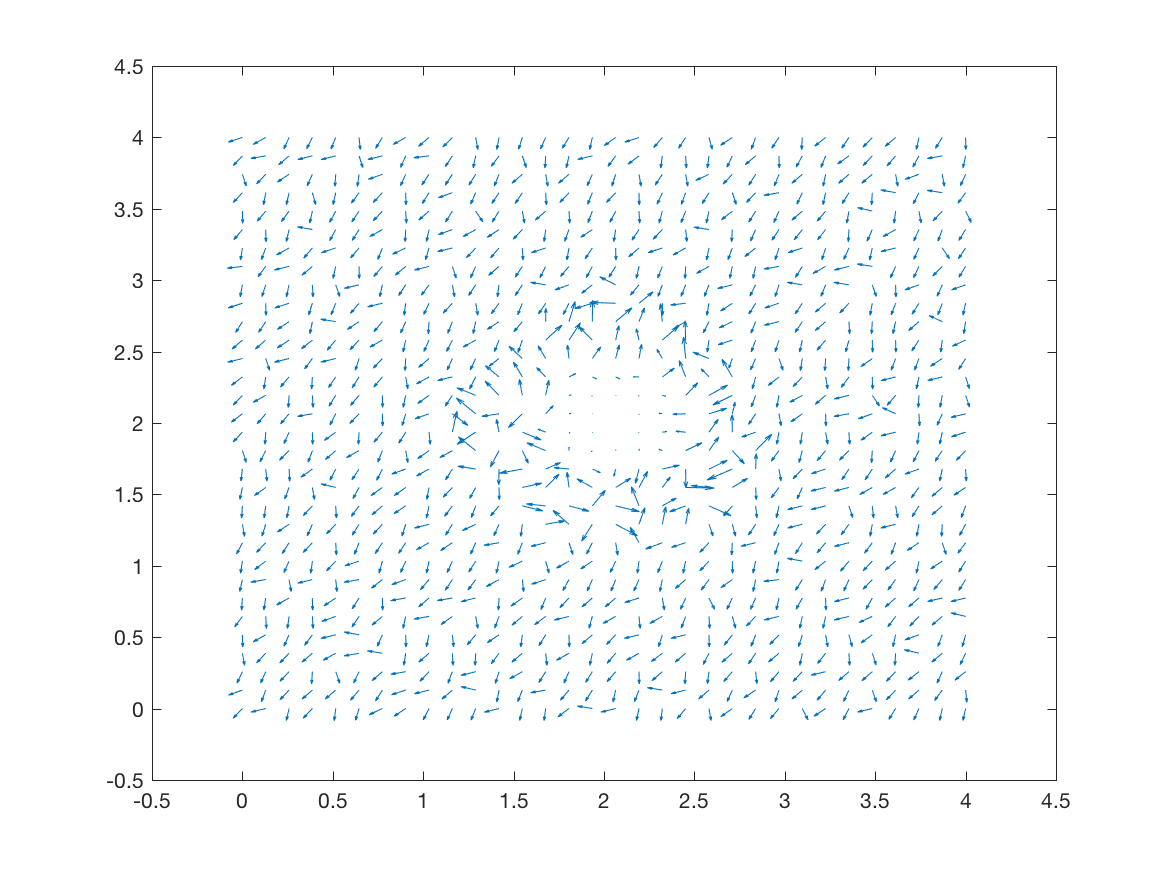}
  \caption{\emph{Fibre vector field - coarsened 4 fold}}
  \label{fig:fulldeghighhomo50f}
\end{subfigure}\hfil 

\caption[Simulations at stage $50\Delta t$ with a homogeneous distribution of the non-fibrous phase and $20\%$ homogeneous fibres phase of the ECM with a micro-fibres degradation rate of $d_f = 0.5$.]{\emph{Simulations at stage $50\Delta t$ with a homogeneous distribution of the non-fibrous phase and $20\%$ homogeneous fibres phase of the ECM with a micro-fibres degradation rate of $d_f = 0.5$.}}
\label{fig:fulldeghighhomo50}
\end{figure}

\begin{figure}[h!]
    \centering 
\begin{subfigure}{0.5\textwidth}
  \includegraphics[width=\linewidth]{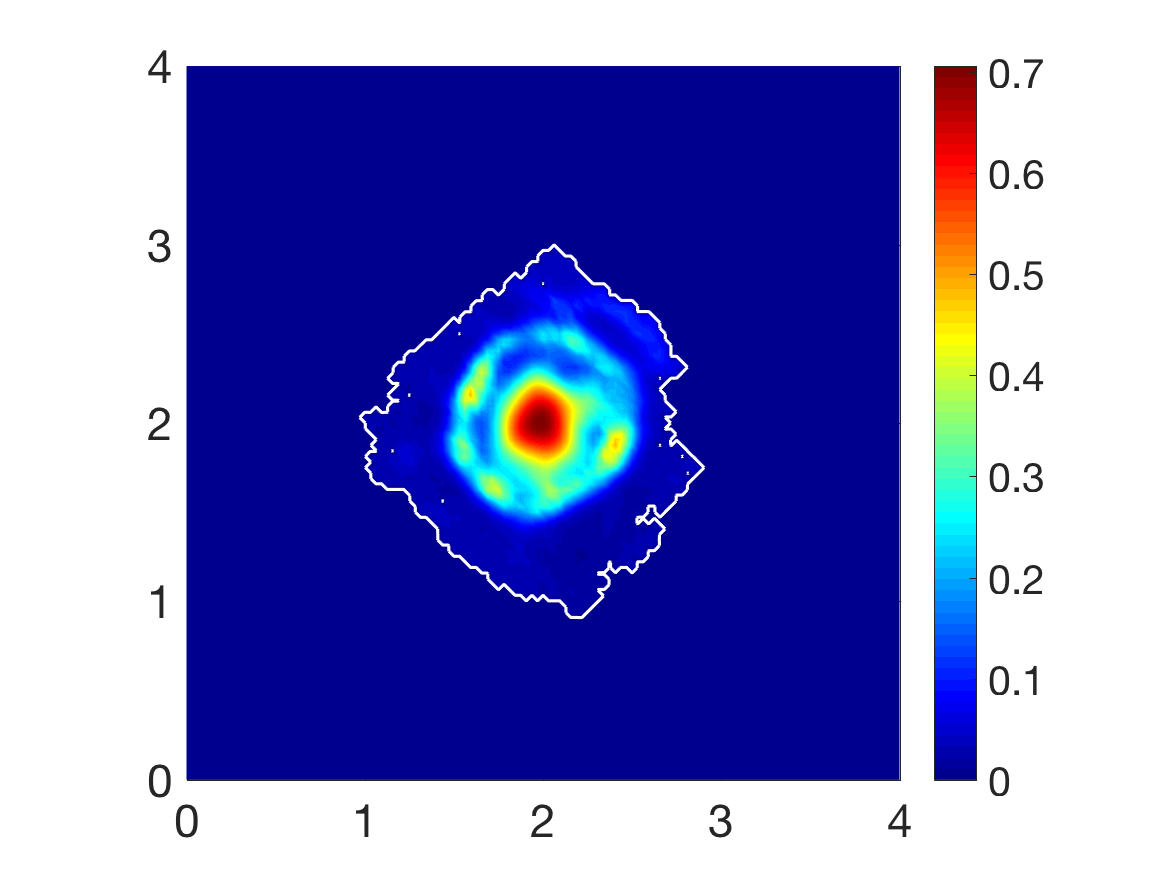}
  \caption{\emph{Cancer cell population}}
  \label{fig:fulldeghighhomo75a}
\end{subfigure}\hfil 
\begin{subfigure}{0.5\textwidth}
  \includegraphics[width=\linewidth]{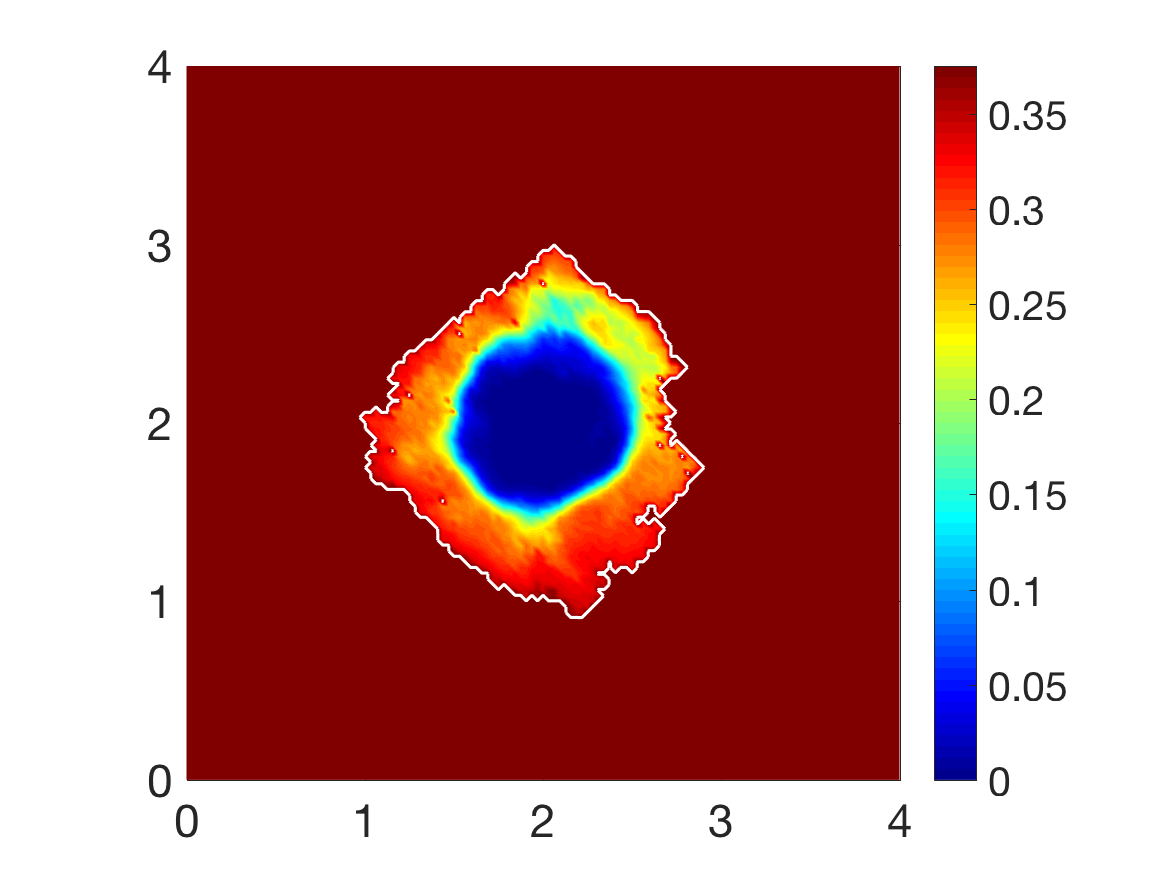}
  \caption{\emph{Non-fibres ECM distribution}}
  \label{fig:fulldeghighhomo75b}
\end{subfigure}\hfil 

\medskip
\begin{subfigure}{0.5\textwidth}
  \includegraphics[width=\linewidth]{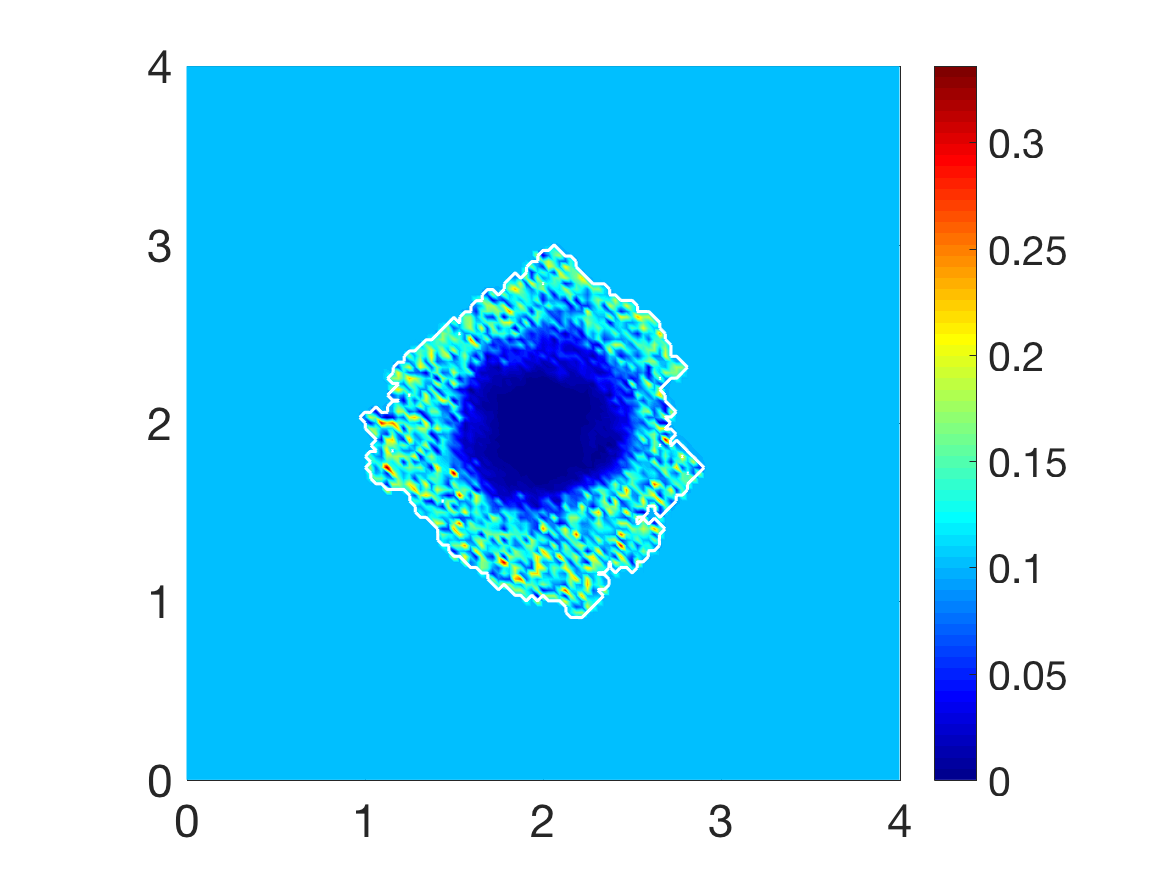}
  \caption{\emph{Fibre magnitude density}}
  \label{fig:fulldeghighhomo75c}
  \end{subfigure}\hfil 
\begin{subfigure}{0.5\textwidth}
  \includegraphics[width=\linewidth]{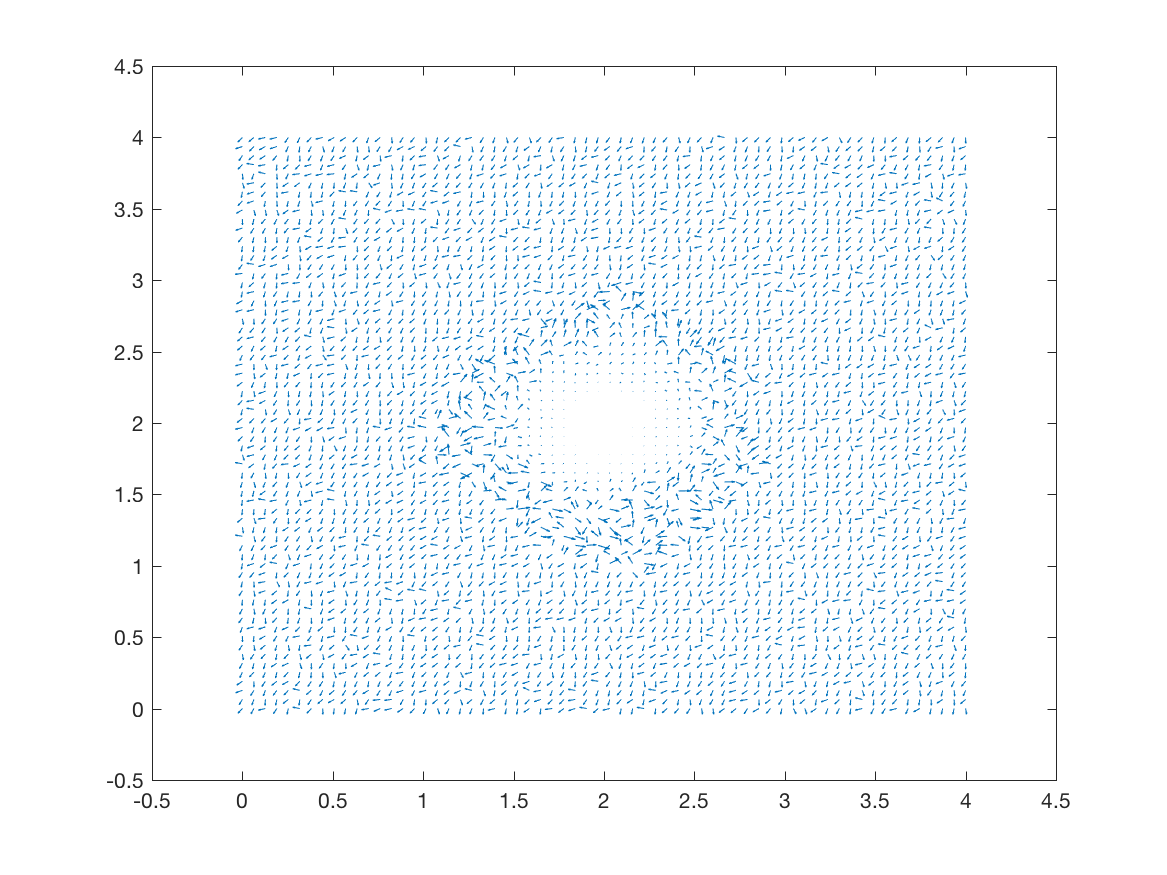}
  \caption{\emph{Fibre vector field - coarsened 2 fold}}
  \label{fig:fulldeghighhomo75d}
\end{subfigure}\hfil 

\medskip
\begin{subfigure}{0.5\textwidth}
  \includegraphics[width=\linewidth]{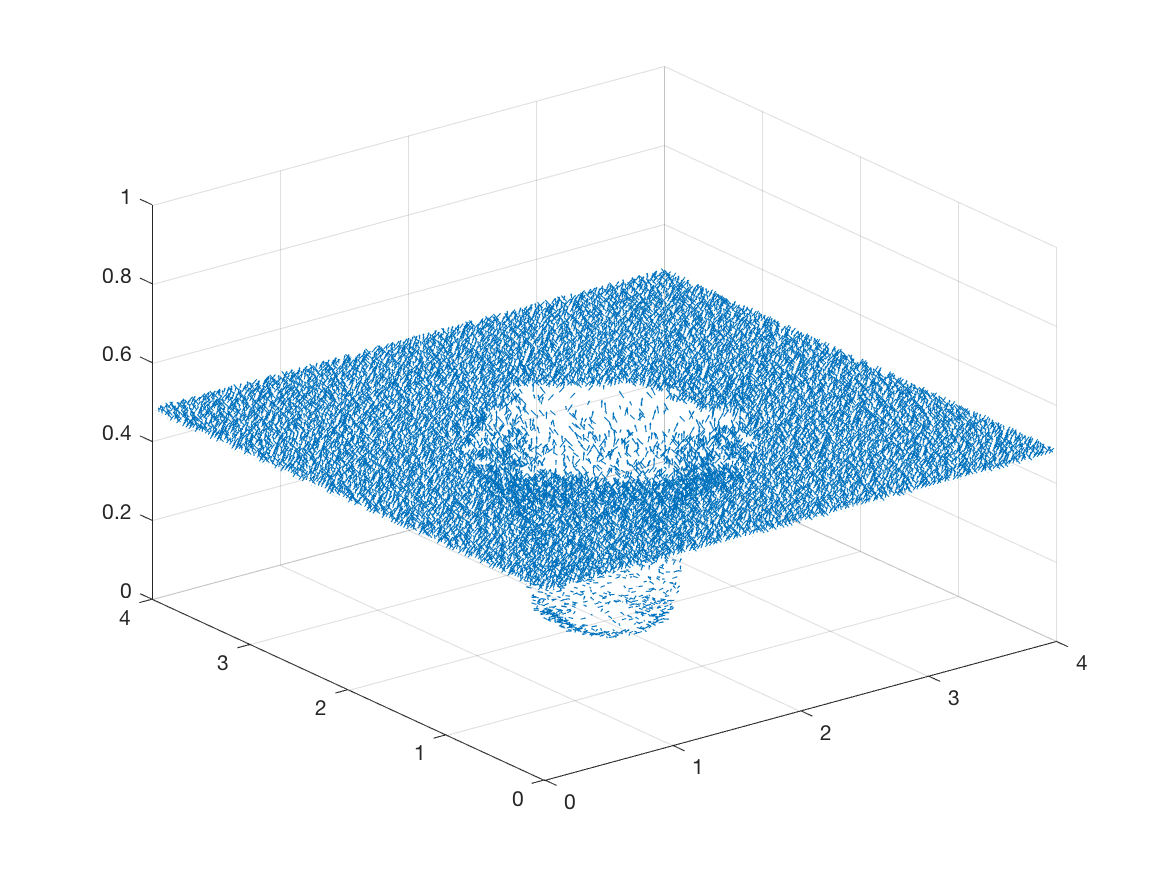}
  \caption{\emph{3D ECM vector field}}
  \label{fig:fulldeghighhomo75e}
  \end{subfigure}\hfil 
\begin{subfigure}{0.5\textwidth}
  \includegraphics[width=\linewidth]{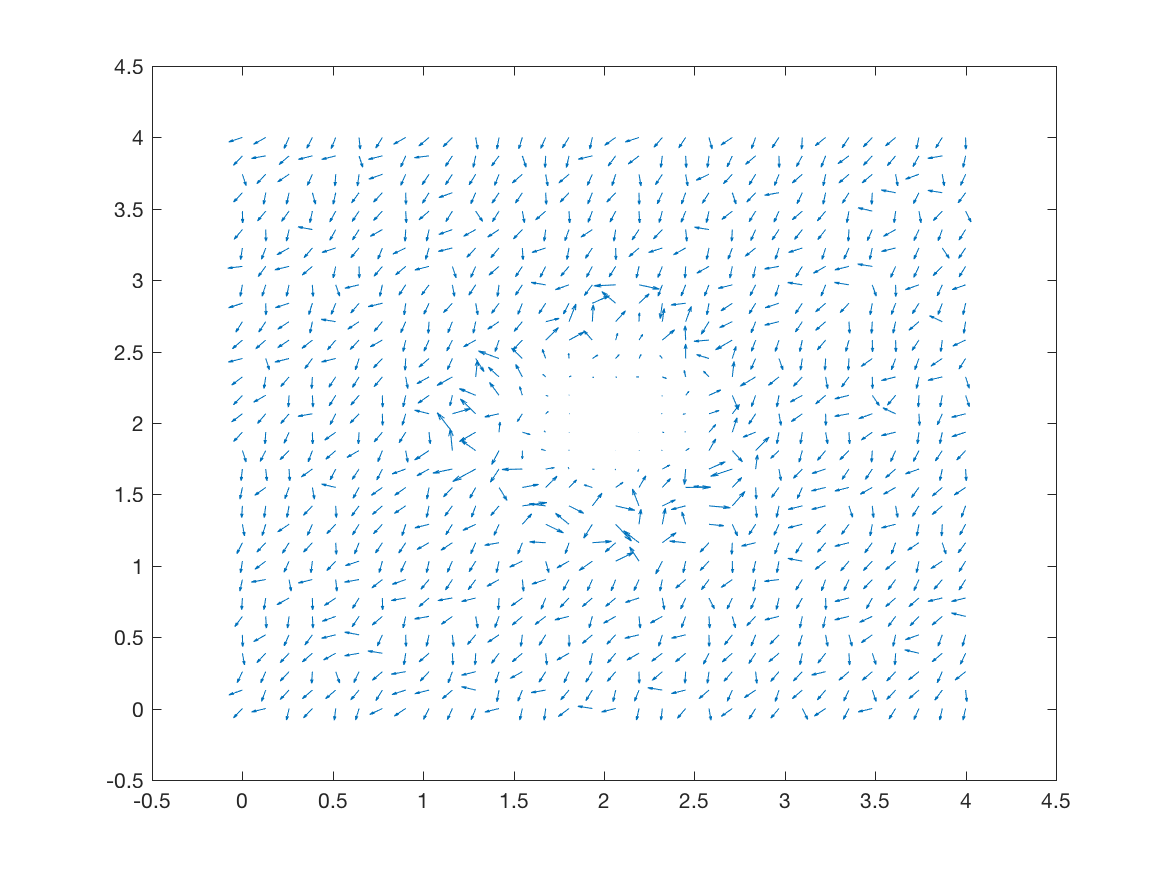}
  \caption{\emph{Fibre vector field - coarsened 4 fold}}
  \label{fig:fulldeghighhomof75f}
\end{subfigure}\hfil 

\caption[Simulations at stage $75\Delta t$ with a homogeneous distribution of the non-fibrous phase and $20\%$ homogeneous fibres phase of the ECM with a micro-fibres degradation rate of $d_f = 0.5$.]{\emph{Simulations at stage $75\Delta t$ with a homogeneous distribution of the non-fibrous phase and $20\%$ homogeneous fibres phase of the ECM with a micro-fibres degradation rate of $d_f = 0.5$.}}
\label{fig:fulldeghighhomo75}
\end{figure}

We proceed by exploring the cancer dynamics within an initial $20\%$ homogeneous fibre distribution, \rs{taken as $p=0.2$ of the non-fibres ECM phase $l(x,0)$, and in the presence of the micro-fibre degradation rate $d_{f}=0.5$}. At stage $25 \Delta t$, Figure \ref{fig:fulldeghighhomo25}, the non-fibres ECM phase has been degraded by the cancer cells, subfigure \ref{fig:fulldeghighhomo25b}, with the highest degradation occurring in the regions of highest cell distribution.  \rs{The tumour boundary at this stage is larger than in previous simulations, this correlating with results in \cite{Shutt_twopop,Shutt_2018} where an initially higher density fibre ECM phase resulted in an accelerated spread of the tumour. Additionally, the macroscopic fibre density, subfigure \ref{fig:fulldeghighhomo25c}, also exhibits different behaviour than in previous simulations, where a similar pattern of fibres was noted in \cite{Shutt_twopop,Shutt_2018}. This pattern of fibres is witnessed because the tumour boundary is expanding faster than the micro-fibres are being rearranged, thus the distributions are not found to build on the proliferating edge.} Moving on to stage $50 \Delta t$, the cancer cell distribution is increasing and spreading within the tumour region, subfigure \ref{fig:fulldeghighhomo50a}. There is significant non-fibres \rs{ECM} degradation stretching the entire area of the tumour, subfigure \ref{fig:fulldeghighhomo50b}, with this only becoming more pronounced at later stages, subfigure \ref{fig:fulldeghighhomo75b}.\rs{The macroscopic fibre orientations, subfigures \ref{fig:fulldeghighhomo50d}, \ref{fig:fulldeghighhomo50f}, display high levels of reorientation with a similar ``frenzied'' appearance to figures in \cite{Shutt_twopop}.} At the final stage $75 \Delta t$, Figure \ref{fig:fulldeghighhomo75}, the cancer cells are dispersed further into the matrix, subfigure \ref{fig:fulldeghighhomo75a} and noticeably in the pattern of degradation of the non-fibres ECM phase, subfigure \ref{fig:fulldeghighhomo75b}. Overall, we conclude that in the presence of a homogeneous ECM, with a high initial homogeneous fibre distribution, the cancer cells have increased opportunity for adhesion and thus they are able to easier invade the surrounding matrix, resulting in a larger tumour region and a higher level of ECM degradation.

\begin{figure}[h!]
    \centering 
\begin{subfigure}{0.5\textwidth}
  \includegraphics[width=\linewidth]{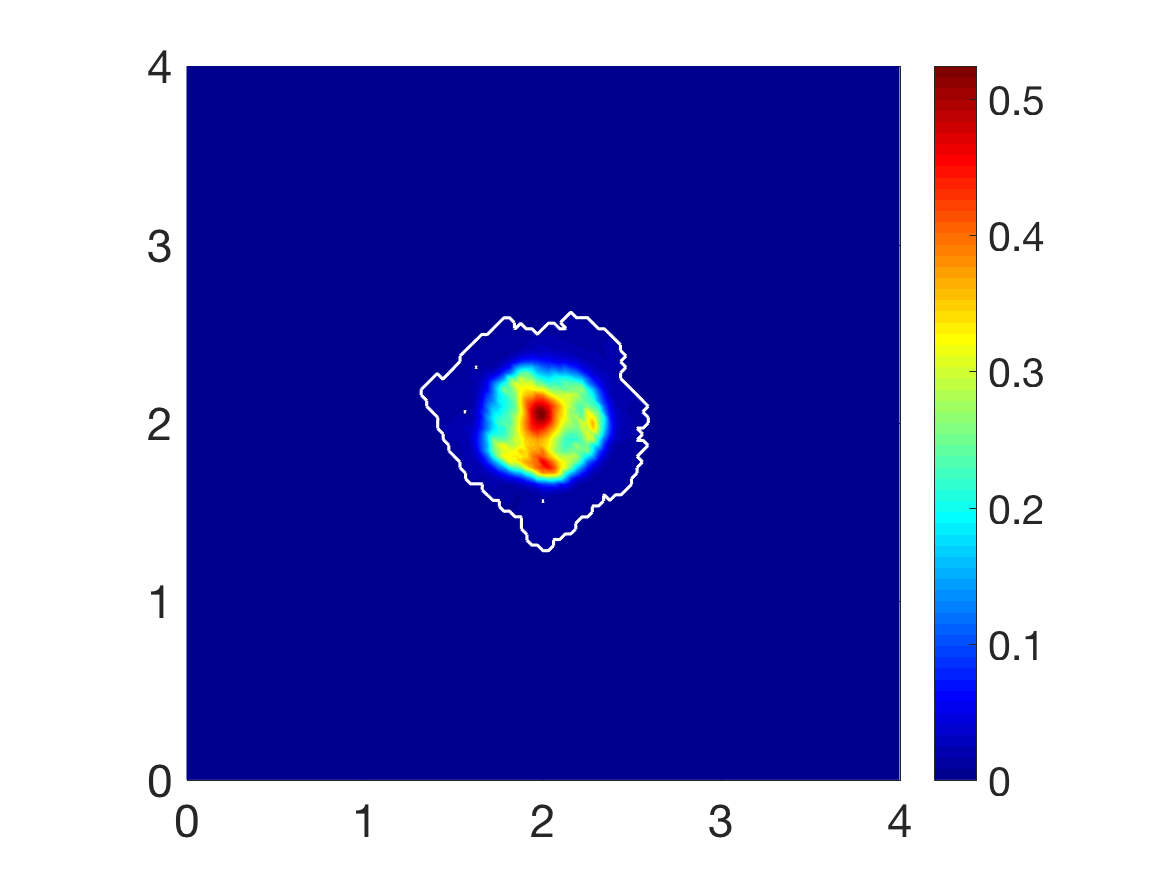}
  \caption{\emph{Cancer cell population}}
  \label{fig:fulldeghighhetero25a}
\end{subfigure}\hfil 
\begin{subfigure}{0.5\textwidth}
  \includegraphics[width=\linewidth]{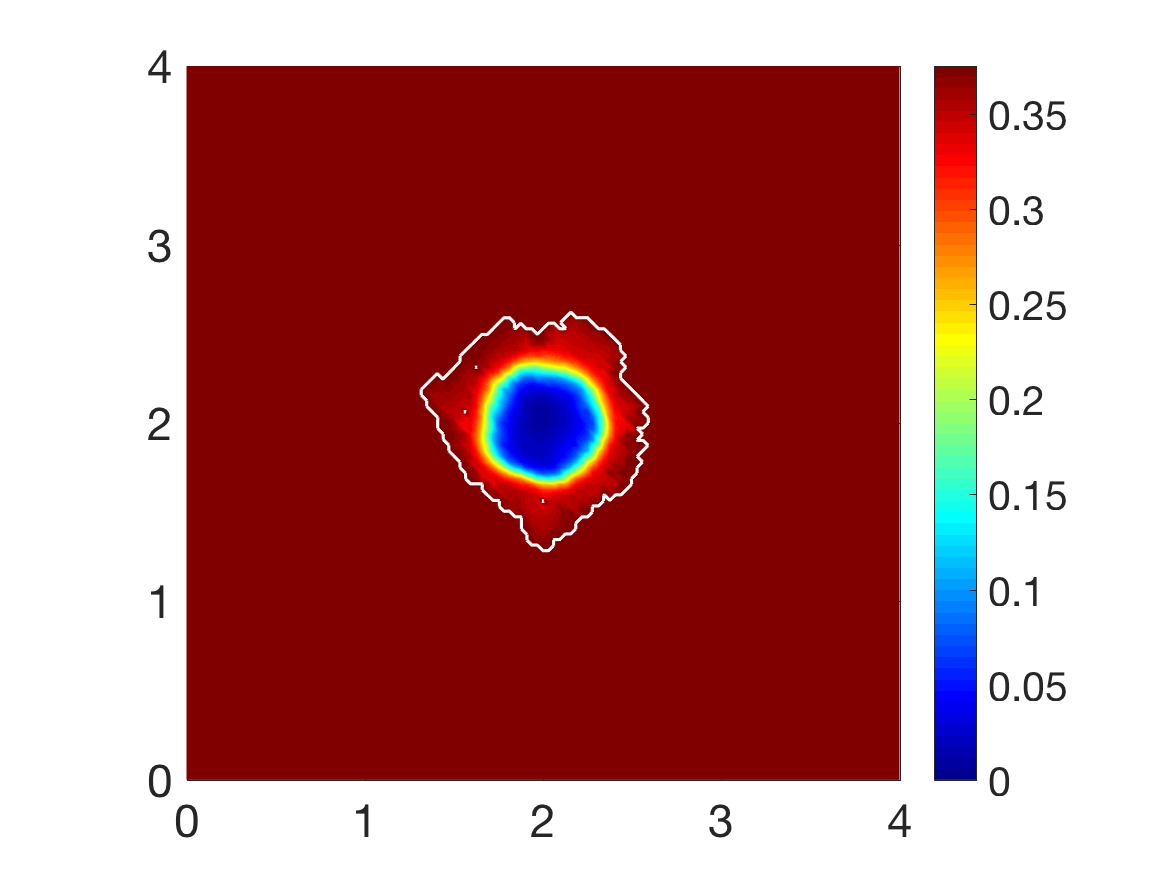}
  \caption{\emph{Non-fibres ECM distribution}}
  \label{fig:fulldeghighhetero25b}
\end{subfigure}\hfil 

\medskip
\begin{subfigure}{0.5\textwidth}
  \includegraphics[width=\linewidth]{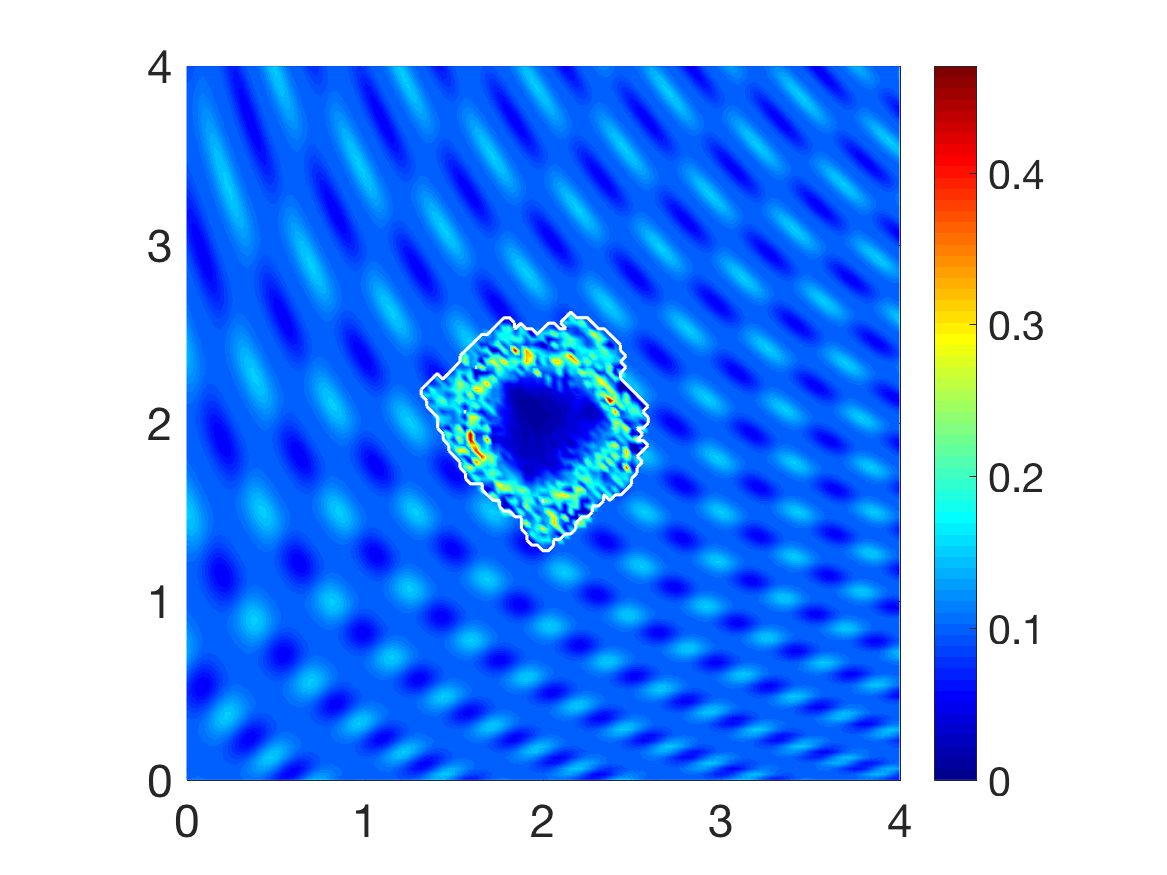}
  \caption{\emph{Fibre magnitude density}}
  \label{fig:fulldeghighhetero25c}
  \end{subfigure}\hfil 
\begin{subfigure}{0.5\textwidth}
  \includegraphics[width=\linewidth]{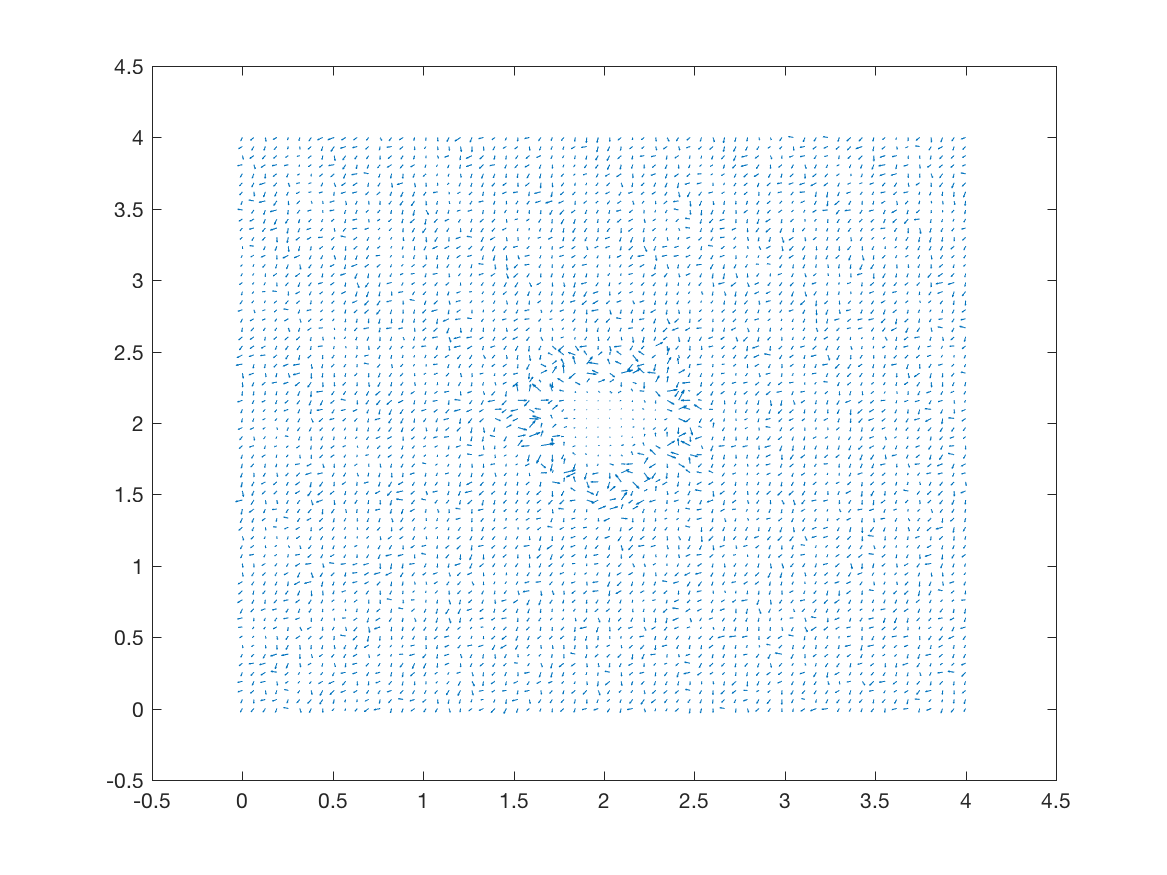}
  \caption{\emph{Fibre vector field - coarsened 2 fold}}
  \label{fig:fulldeghighhetero25d}
\end{subfigure}\hfil 

\medskip
\begin{subfigure}{0.5\textwidth}
  \includegraphics[width=\linewidth]{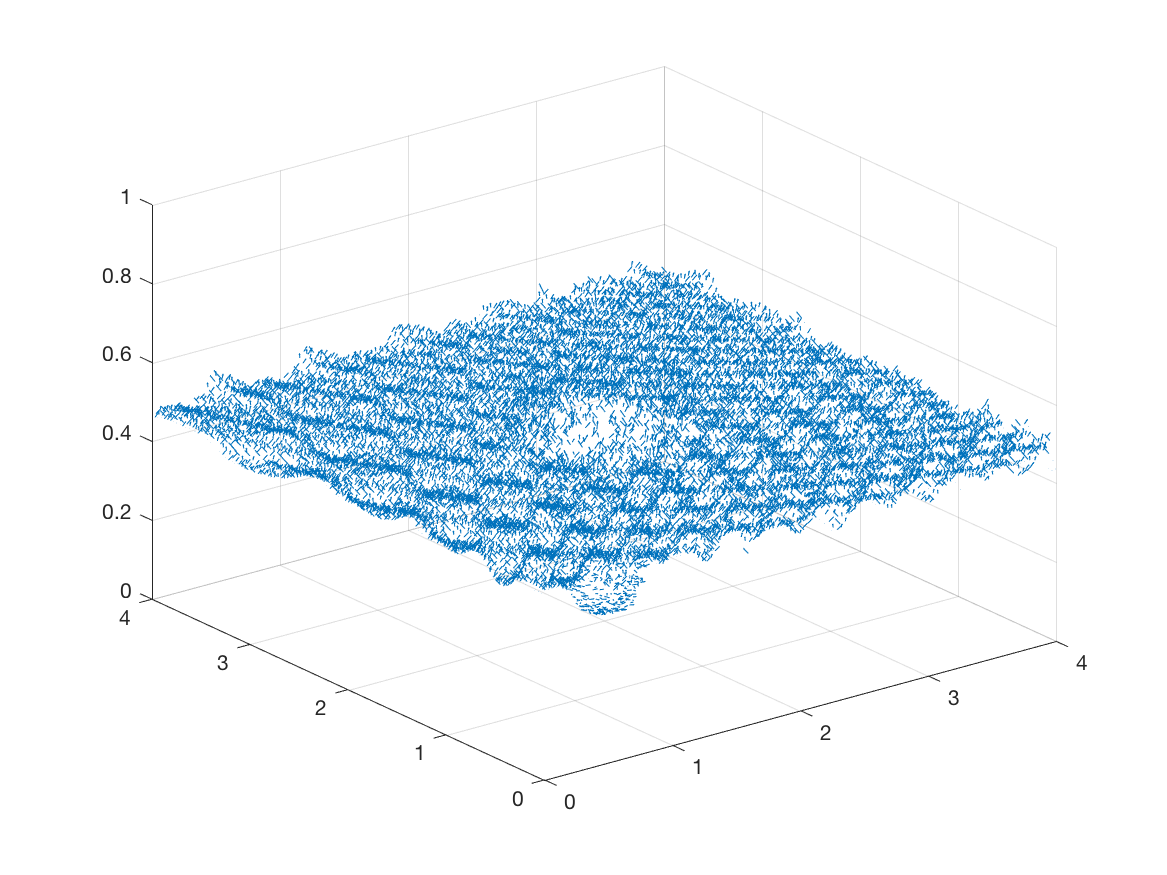}
  \caption{\emph{3D ECM vector field}}
  \label{fig:fulldeghighhetero25e}
  \end{subfigure}\hfil 
\begin{subfigure}{0.5\textwidth}
  \includegraphics[width=\linewidth]{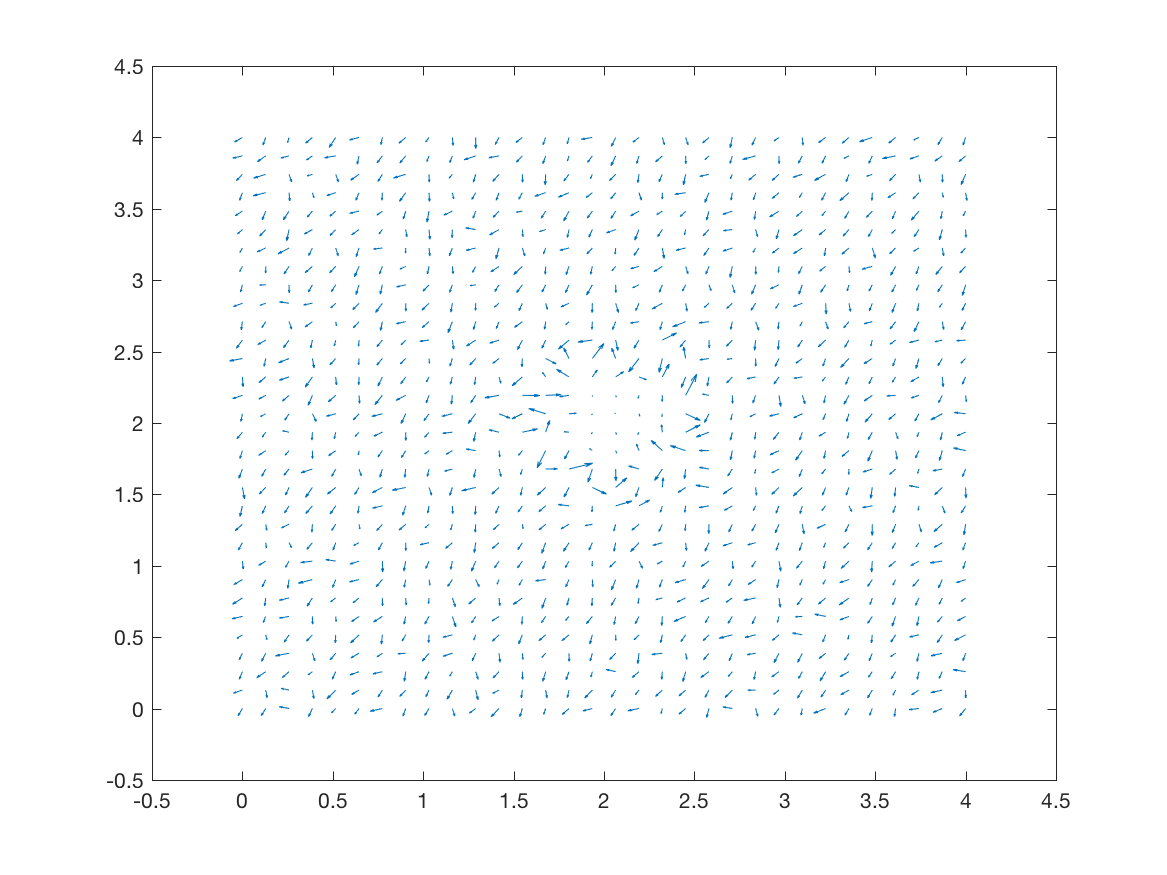}
  \caption{\emph{Fibre vector field - coarsened 4 fold}}
  \label{fig:fulldeghighhetero25f}
\end{subfigure}\hfil 

\caption[Simulations at stage $25\Delta t$ with a homogeneous distribution of the non-fibrous phase and $20\%$ homogeneous fibres phase of the ECM with a micro-fibres degradation rate of $d_f = 0.5$.]{\emph{Simulations at stage $25\Delta t$ with a homogeneous distribution of the non-fibrous phase and $20\%$ homogeneous fibres phase of the ECM with a micro-fibres degradation rate of $d_f = 0.5$.}}
\label{fig:fulldeghighhetero25}
\end{figure}

\begin{figure}[h!]
    \centering 
\begin{subfigure}{0.5\textwidth}
  \includegraphics[width=\linewidth]{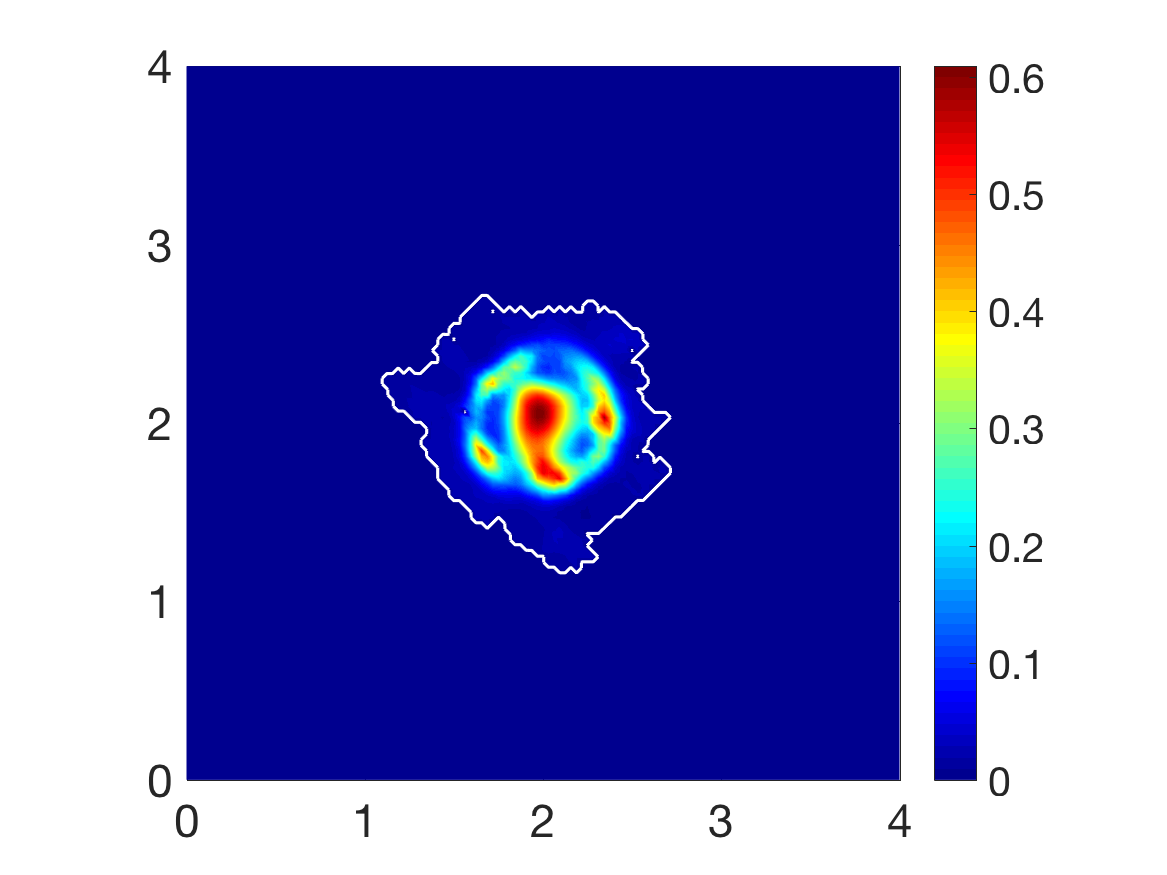}
  \caption{\emph{Cancer cell population}}
  \label{fig:fulldeghighhetero50a}
\end{subfigure}\hfil 
\begin{subfigure}{0.5\textwidth}
  \includegraphics[width=\linewidth]{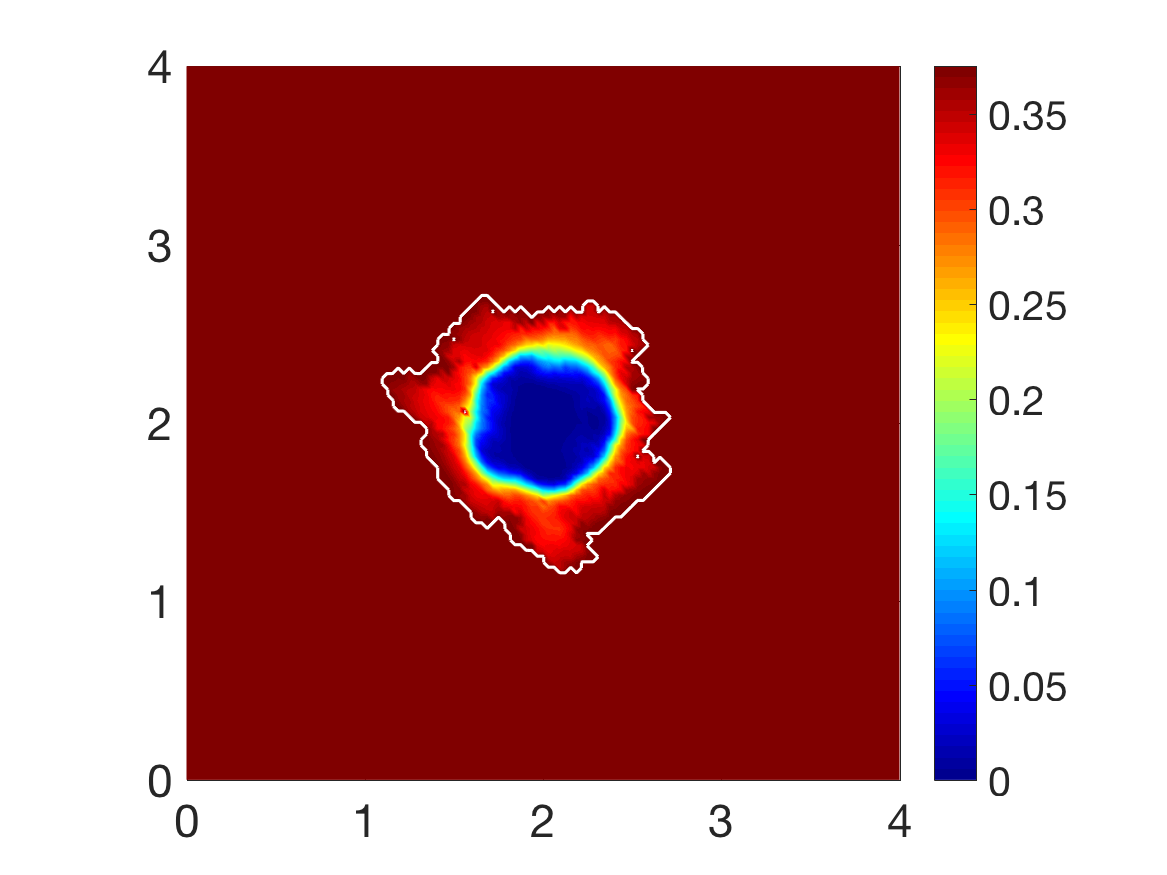}
  \caption{\emph{Non-fibres ECM distribution}}
  \label{fig:fulldeghighhetero50b}
\end{subfigure}\hfil 

\medskip
\begin{subfigure}{0.5\textwidth}
  \includegraphics[width=\linewidth]{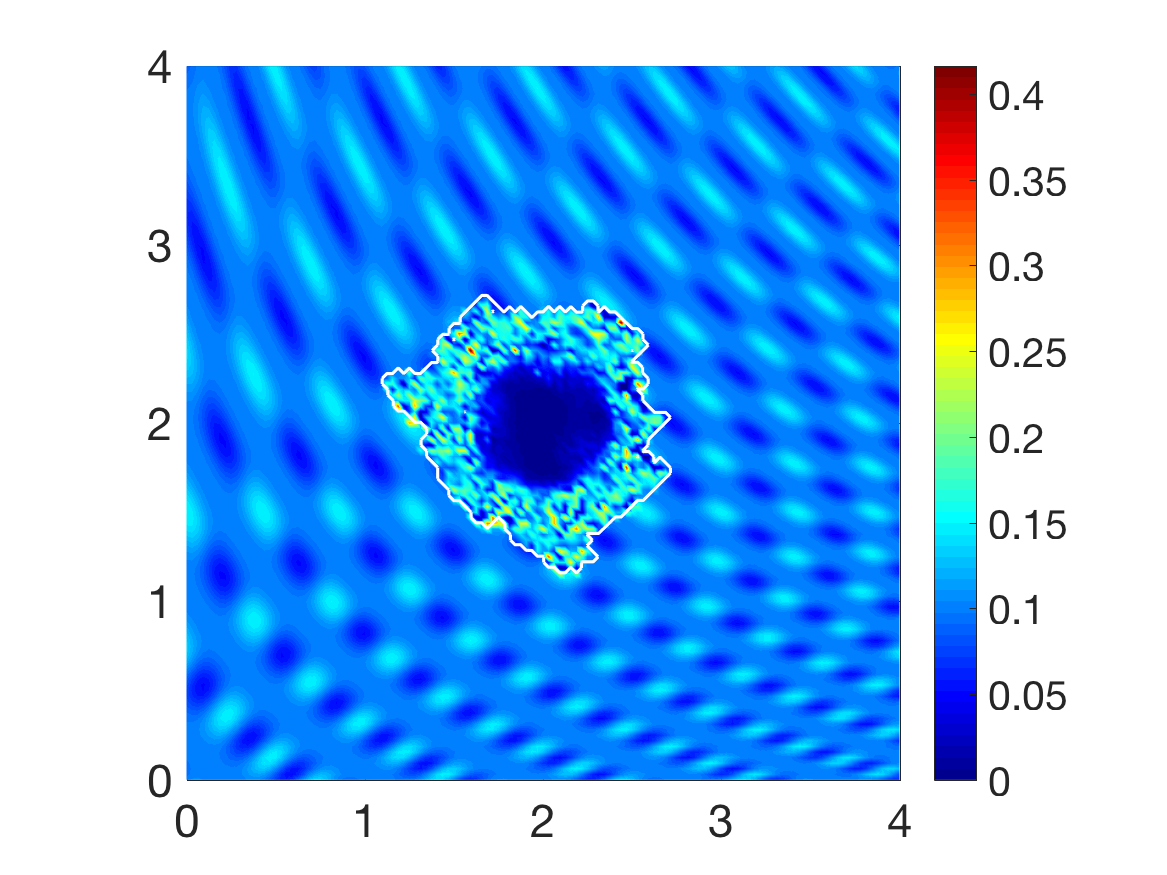}
  \caption{\emph{Fibre magnitude density}}
  \label{fig:fulldeghighhetero50c}
  \end{subfigure}\hfil 
\begin{subfigure}{0.5\textwidth}
  \includegraphics[width=\linewidth]{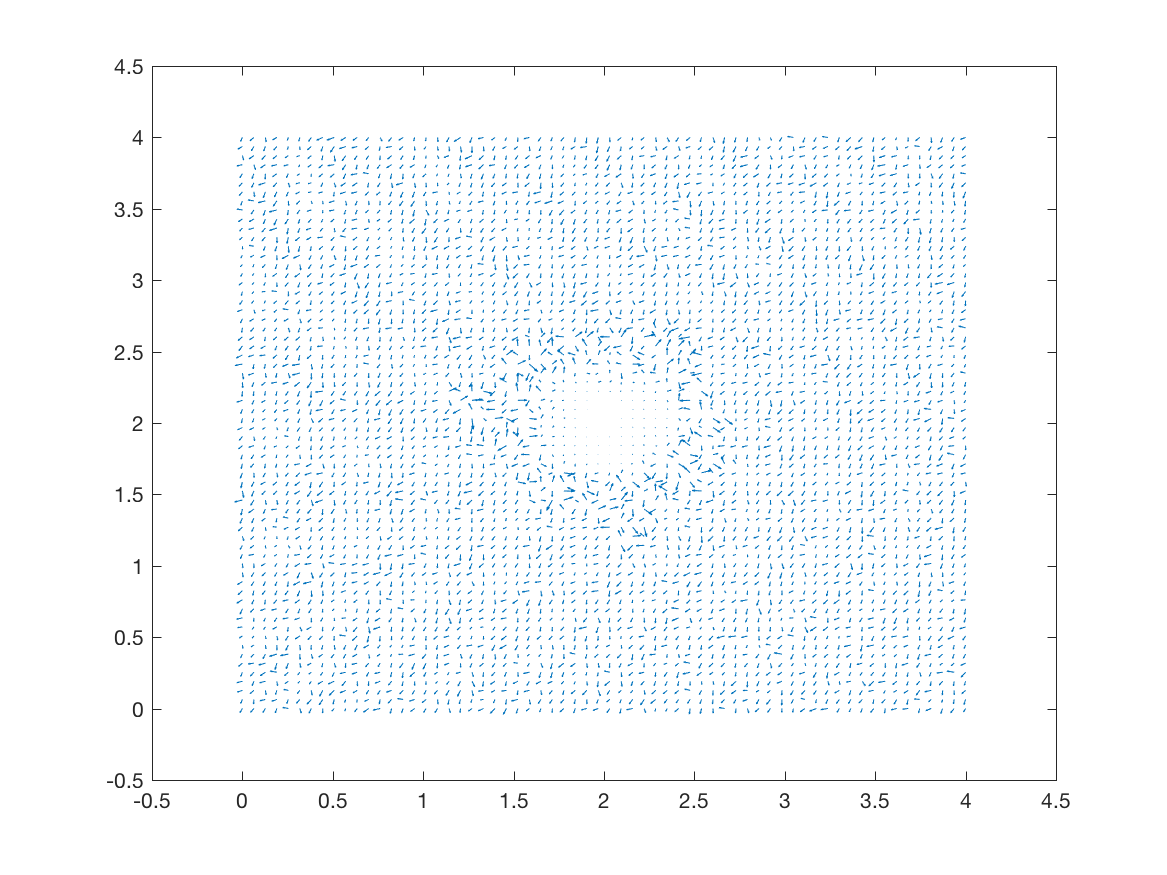}
  \caption{\emph{Fibre vector field - coarsened 2 fold}}
  \label{fig:fulldeghighhetero50d}
\end{subfigure}\hfil 

\medskip
\begin{subfigure}{0.5\textwidth}
  \includegraphics[width=\linewidth]{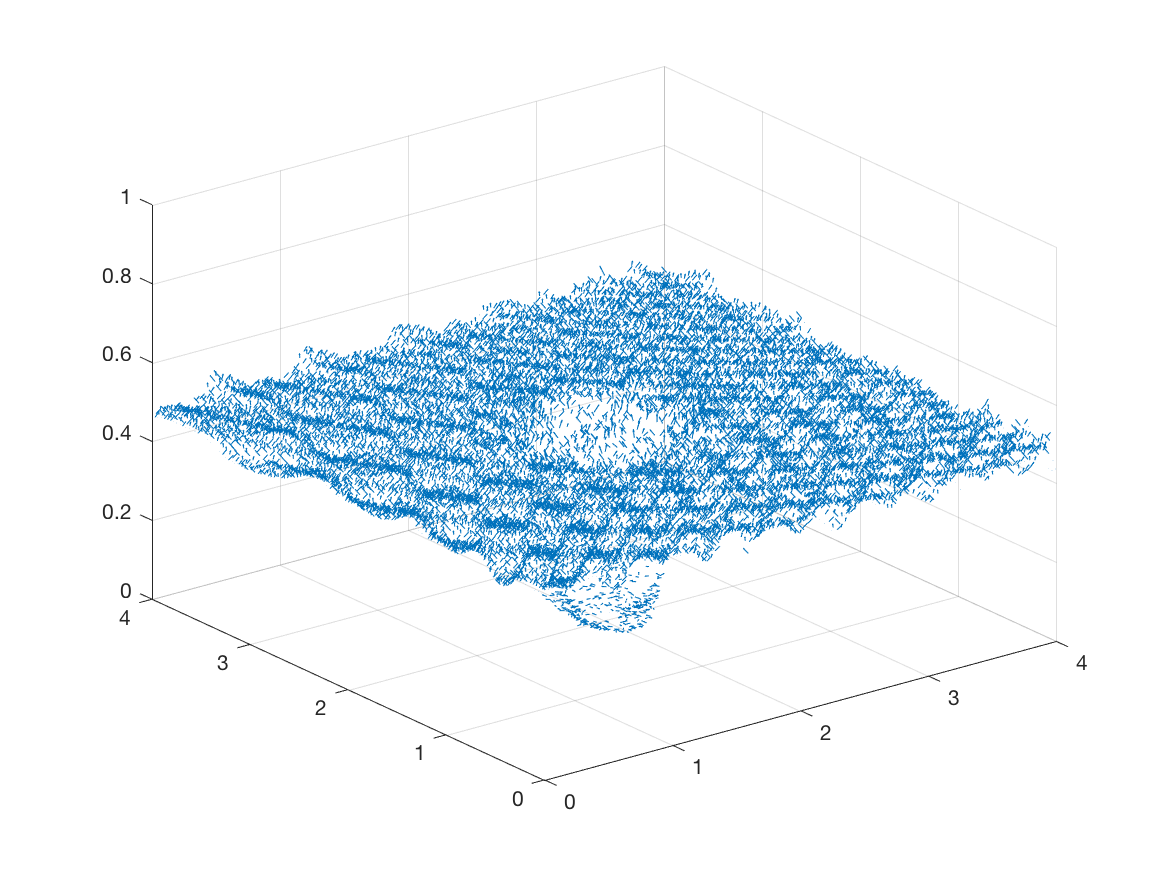}
  \caption{\emph{3D ECM vector field}}
  \label{fig:fulldeghighhetero50e}
  \end{subfigure}\hfil 
\begin{subfigure}{0.5\textwidth}
  \includegraphics[width=\linewidth]{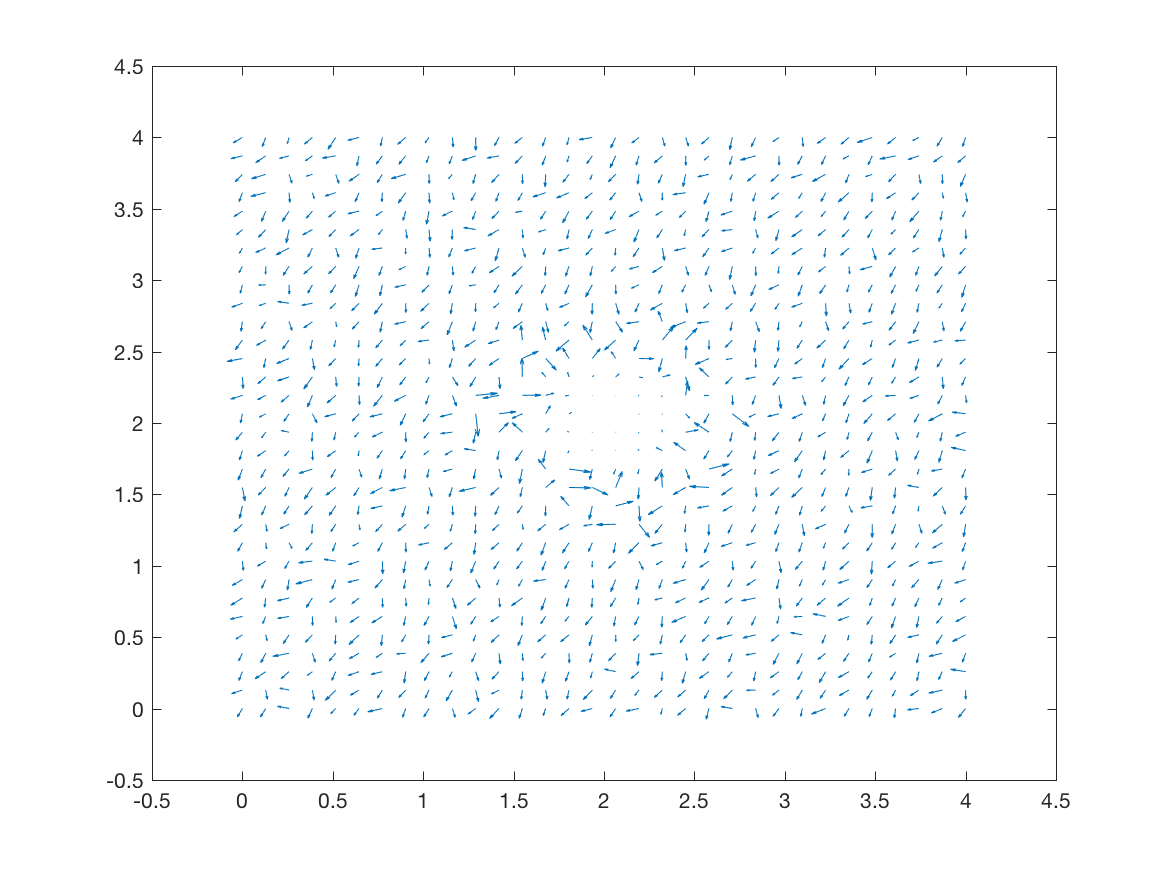}
  \caption{\emph{Fibre vector field - coarsened 4 fold}}
  \label{fig:fulldeghighhetero50f}
\end{subfigure}\hfil 

\caption[Simulations at stage $50\Delta t$ with a homogeneous distribution of the non-fibrous phase and $20\%$ homogeneous fibres phase of the ECM with a micro-fibres degradation rate of $d_f = 0.5$.]{\emph{Simulations at stage $50\Delta t$ with a homogeneous distribution of the non-fibrous phase and $20\%$ homogeneous fibres phase of the ECM with a micro-fibres degradation rate of $d_f = 0.5$.}}
\label{fig:fulldeghighhetero50}
\end{figure}

\begin{figure}[h!]
    \centering 
\begin{subfigure}{0.5\textwidth}
  \includegraphics[width=\linewidth]{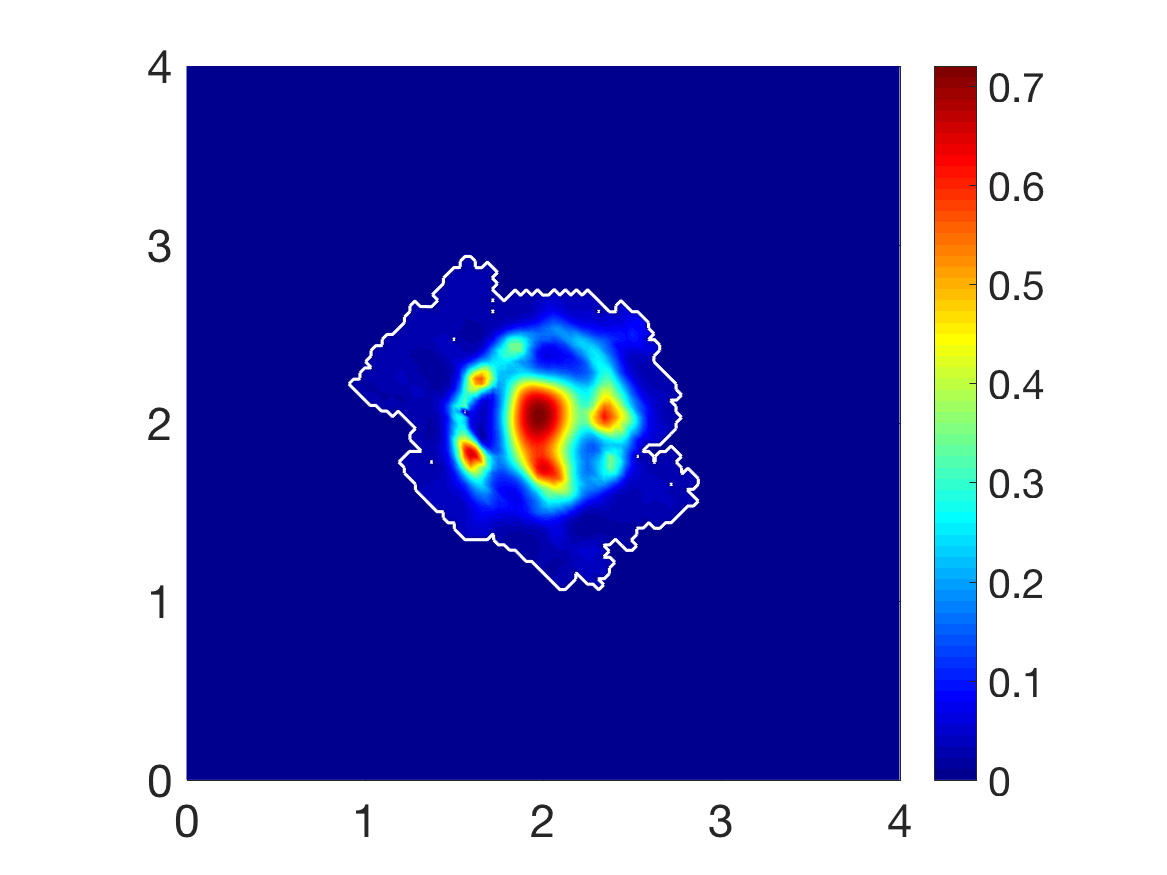}
  \caption{\emph{Cancer cell population}}
  \label{fig:fulldeghighhetero75a}
\end{subfigure}\hfil 
\begin{subfigure}{0.5\textwidth}
  \includegraphics[width=\linewidth]{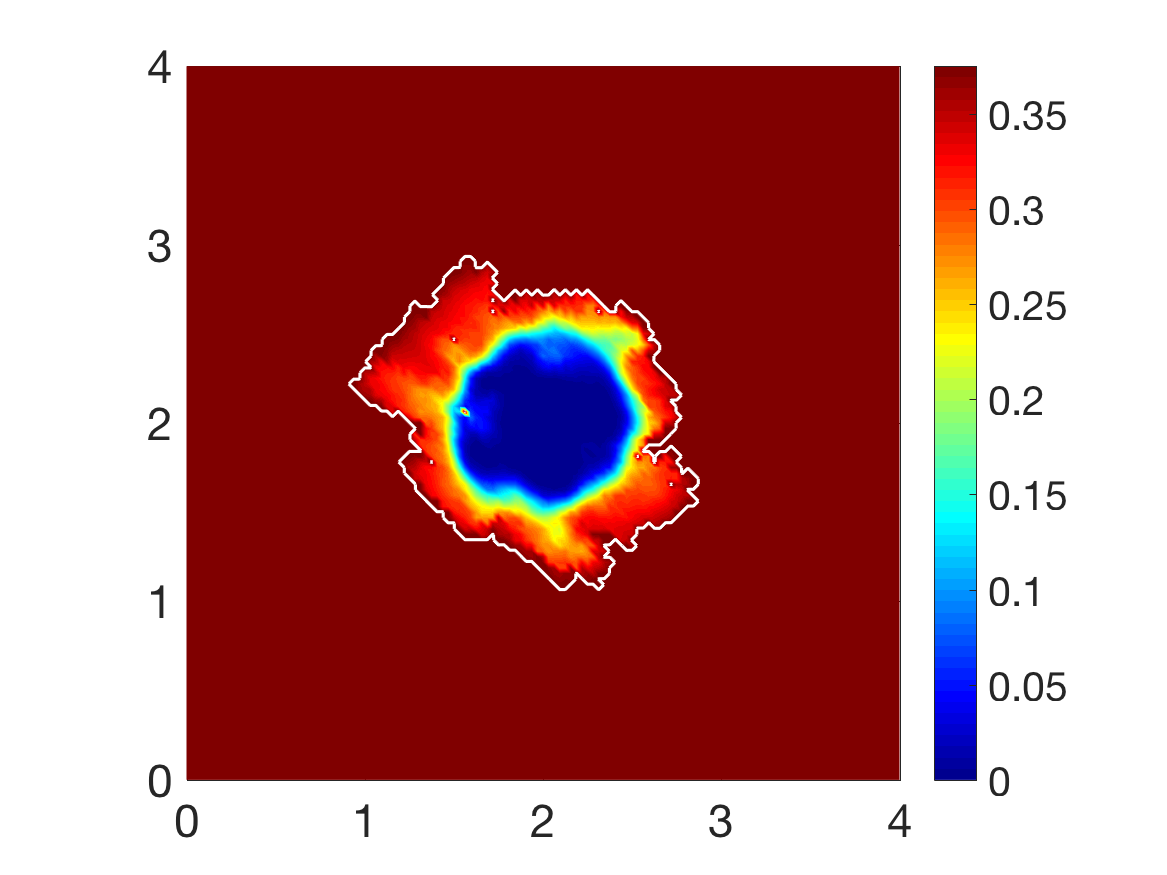}
  \caption{\emph{Non-fibres ECM distribution}}
  \label{fig:fulldeghighhetero75b}
\end{subfigure}\hfil 

\medskip
\begin{subfigure}{0.5\textwidth}
  \includegraphics[width=\linewidth]{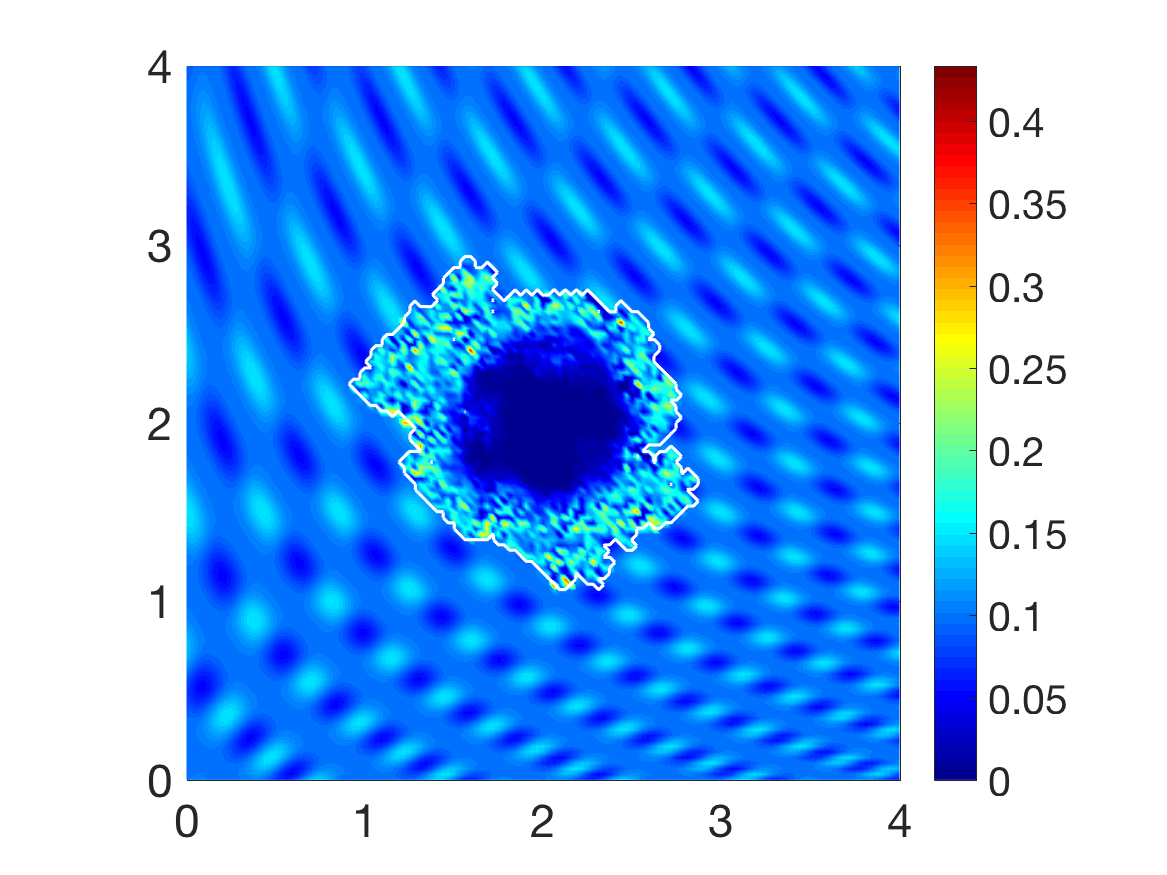}
  \caption{\emph{Fibre magnitude density}}
  \label{fig:fulldeghighhetero75c}
  \end{subfigure}\hfil 
\begin{subfigure}{0.5\textwidth}
  \includegraphics[width=\linewidth]{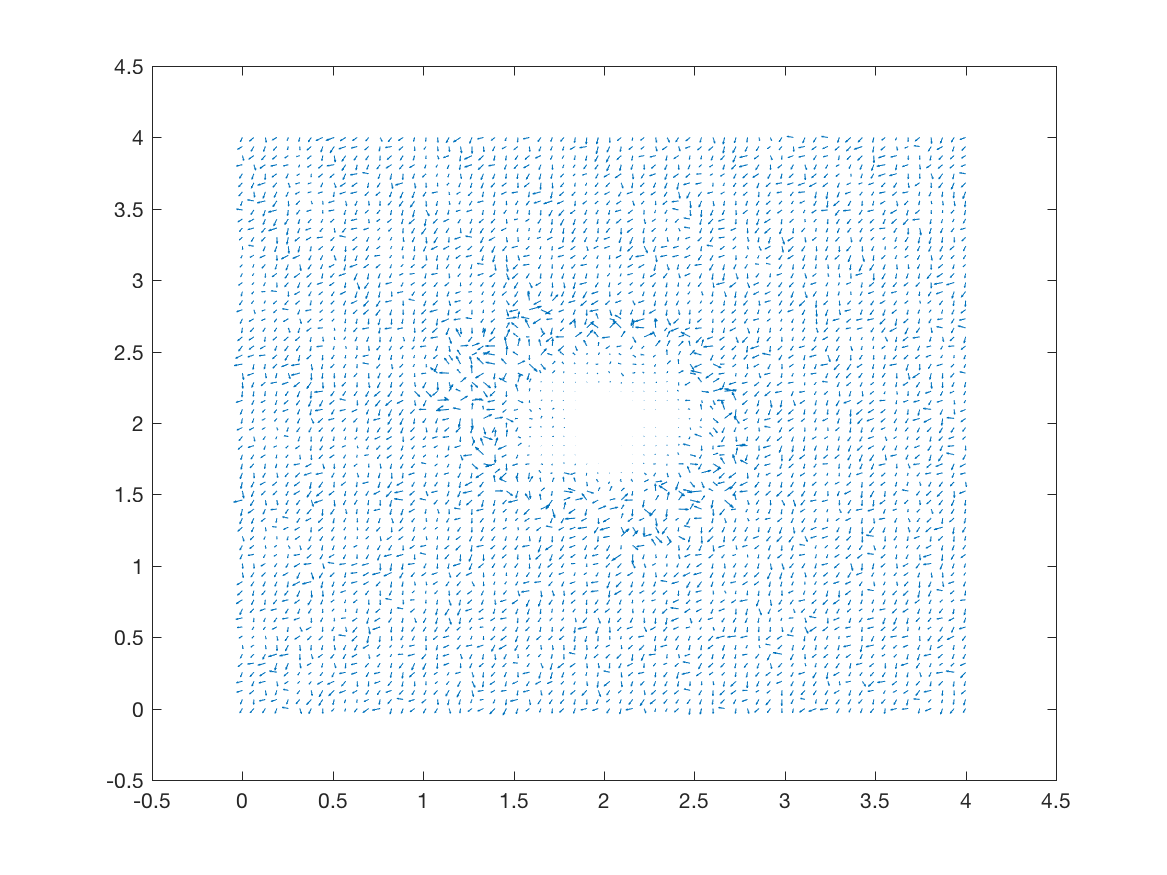}
  \caption{\emph{Fibre vector field - coarsened 2 fold}}
  \label{fig:fulldeghighhetero75d}
\end{subfigure}\hfil 

\medskip
\begin{subfigure}{0.5\textwidth}
  \includegraphics[width=\linewidth]{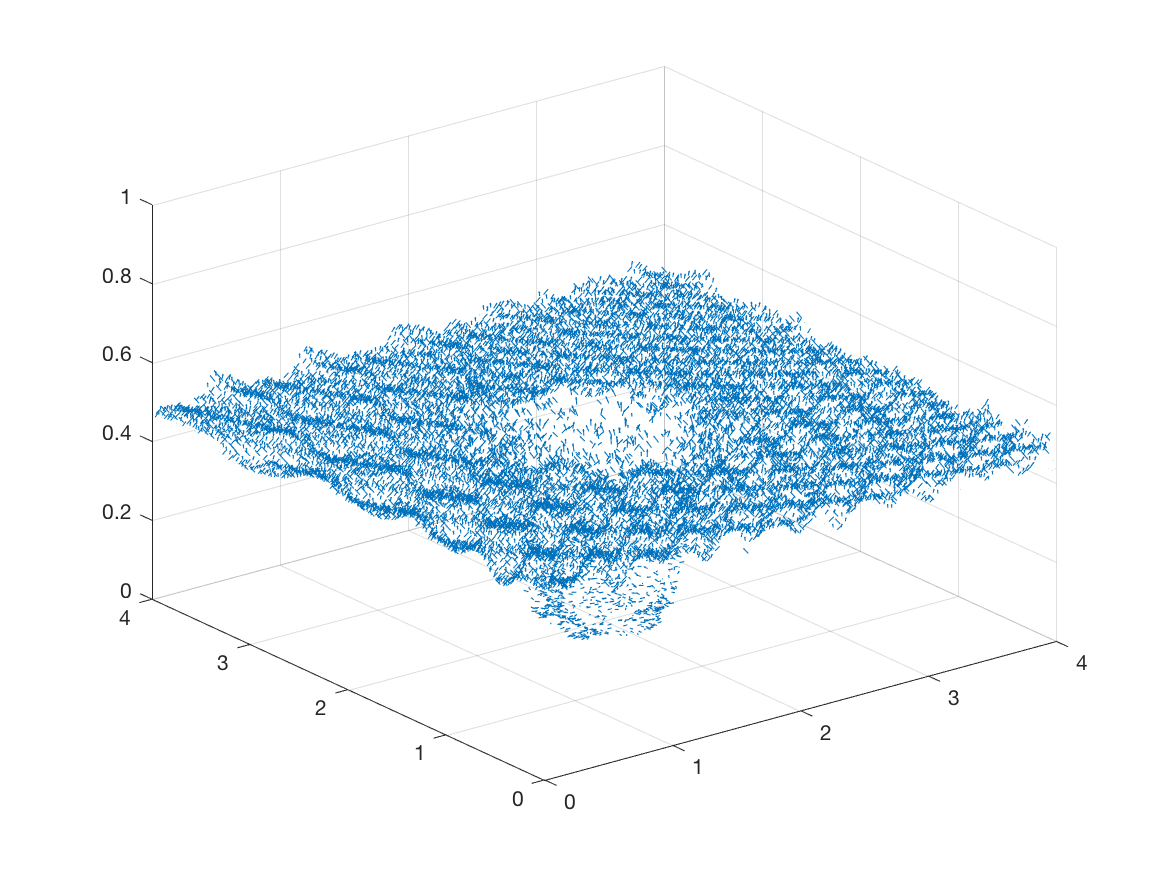}
  \caption{\emph{3D ECM vector field}}
  \label{fig:fulldeghighhetero75e}
  \end{subfigure}\hfil 
\begin{subfigure}{0.5\textwidth}
  \includegraphics[width=\linewidth]{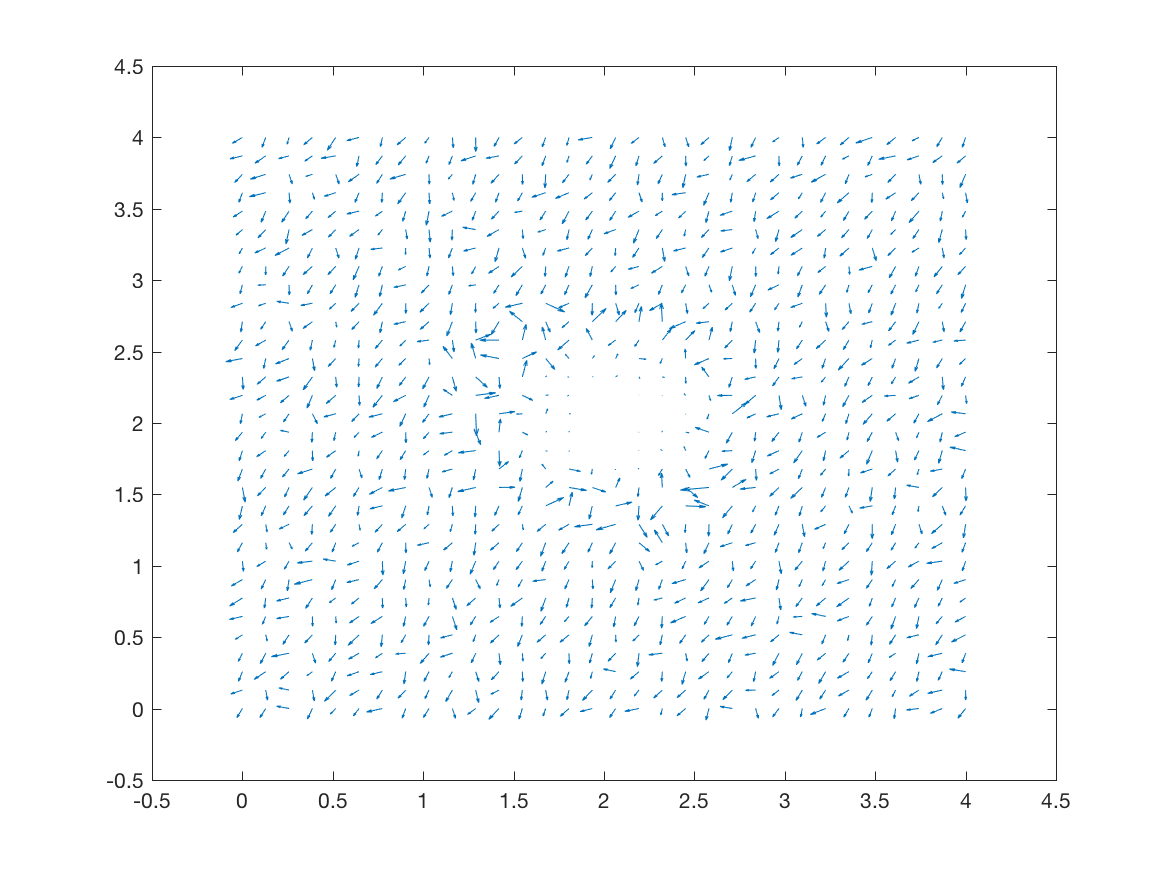}
  \caption{\emph{Fibre vector field - coarsened 4 fold}}
  \label{fig:fulldeghighheterof75f}
\end{subfigure}\hfil 

\caption[Simulations at stage $75\Delta t$ with a homogeneous distribution of the non-fibrous phase and $20\%$ homogeneous fibres phase of the ECM with a micro-fibres degradation rate of $d_f = 0.5$.]{\emph{Simulations at stage $75\Delta t$ with a homogeneous distribution of the non-fibrous phase and $20\%$ homogeneous fibres phase of the ECM with a micro-fibres degradation rate of $d_f = 0.5$.}}
\label{fig:fulldeghighhetero75}
\end{figure}

\rs{We continue our investigation of cancer invasion within a collagen dense environment by exploring the multiscale model within an initially $20\%$ heterogeneous fibre ECM phase embedded in a homogeneous non-fibres ECM phase. Figures \ref{fig:fulldeghighhetero25}-\ref{fig:fulldeghighhetero75}} display simulation results at stages $25 \Delta t$, $50 \Delta t$ and final stage $75 \Delta t$. When comparing with the results in Figures \ref{fig:fulldeghighhomo25}-\ref{fig:fulldeghighhomo75}, the significant differences when in a homogeneous or heterogeneous fibre environment is the shape and density of the main tumour bulk and the pattern of the proliferating boundary. The tumour boundary in Figure \ref{fig:fulldeghighhetero25} is spreading first to the low density regions of fibres, this behaviour consistent with previous results in \cite{Shutt_2018}, \rs{where the} cancer cells can more easily migrate to close-by areas of low fibre density due to the physical space available, thus the boundary of the tumour becomes lobular as the cells migrate outwards, subfigure \ref{fig:fulldeghighhetero25c}. The main bulk of tumour cells are also exhibiting this behaviour as they have formed a high distribution region of cells in the area of lowest fibre density, subfigure \ref{fig:fulldeghighhetero25a}. The macroscopic fibre orientations have been rearranged to \rs{direct} the movement of the tumour boundary, witnessed in Figure \ref{fig:fulldeghighhetero50} at the next time stage interval $50 \Delta t$. The main body of the tumour has distributed the cells into several small high distribution bundles, subfigure \ref{fig:fulldeghighhetero50a} particularly in the low density regions of fibres, subfigure \ref{fig:fulldeghighhetero50c}. The microscopic fibres are continuously rearranged and \rs{due to} a higher rate of fibres degradation where the cancer cell distribution is highest, we observe a very low central region of fibre density with them being both degraded and pushed outwards from the central region of the tumour, subfigures \ref{fig:fulldeghighhetero50d}, \ref{fig:fulldeghighhetero50f}. The orientation of the fibres become increasingly \rs{erratic} in areas of higher fibre density, witnessed in the \rs{protrusions of the tumour boundary, whereby they are} orientated in opposing directions. This trait is exaggerated at final stage $75 \Delta t$, Figure \ref{fig:fulldeghighhetero75}, where \rs{the} tumour has spread further into the ECM and the \rs{protrusions have} increased in size. \rs{Furthermore,} the bulk of cancer cells have become separated and \rs{formed distinct regions within the tumour boundary.} When comparing the cancer cell distribution and the macro-fibre\rs{s} density, subfigures \ref{fig:fulldeghighhetero75a} and \ref{fig:fulldeghighhetero75c} respectively, it can be noted that the cancer cells have bypassed the higher regions of fibres and formed bundles of \rs{cells} around these areas. This behaviour can also be seen in the homogeneous case, however it is a more prominent feature when in the presence of a heterogeneous fibre ECM phase. \rs{In conclusion, in the presence of} an initially high fibre ECM density, tumour progression is \rs{accelerated} and encourages a more aggressively spreading tumour, in both the case of either a homogeneous or heterogeneous \rs{fibre} distribution. Additionally, when the initial fibre density is high, the boundary of the tumour spreads further away from the main body of the tumour \rs{at a rate which the cells cannot maintain}. Hence, the main body of the tumour stays centralised, compared to a lower initial fibre density, whereby the cancer cells spread at a \rs{consistent} rate within the tumour region and stay close to the boundary.

\subsection{Invasion patterns of a heterotypic cancer cell population}
To conclude our exploration of cancer invasion within a heterogeneous microenvironment, we \dt{focus now on} the invasion patterns of a heterotypic cell population using the two cell sub-populations macro-dynamics introduced in \cite{shutt_chapter} and expanded upon in \cite{Shutt_twopop}. The macro-dynamics of the two cell sub-populations are similar in flavour and can be mathematically expressed as
\begin{align} 
\frac{\partial c_1}{\partial t} &= \nabla \cdot [D_{1} \nabla c_1 - c_1 \mathcal{A}_{1}(x,t,\textbf{u}(t,\cdot),\theta_{f}(\cdot,t))] +\mu_{1}c_1(1-\rho(\textbf{u})) - M_{c}(\textbf{u},t)c_1, \label{eq:c1} \\ 
\frac{\partial c_2}{\partial t} &= \nabla \cdot [D_{2} \nabla c_2 - c_2 \mathcal{A}_{2}(x,t,\textbf{u}(t,\cdot),\theta_{f}(\cdot,t))] +\mu_{2}c_2(1-\rho(\textbf{u})) + M_{c}(\textbf{u},t)c_1,
\end{align}
where the individual terms retain their exact meaning from Section \ref{macrodynamics}, \dt{while} $M_{c}(\textbf{u},t)$ represents the mutation of cells from cell population $c_{1}$ to cell sub-population $c_{2}$, defined in \cite{Shutt_twopop},
\[
M_{c}(x,t):=
\left\{
\begin{array}{l}
\frac{\text{exp}\left(\frac{-1}{\kappa^{2}-(1-v(x,t))^{2}}\right)}{\text{exp}\left(\frac{1}{\kappa^{2}}\right)}  \cdot H(t-t_m) \quad \text{if} \ 1-\kappa<v(x,t)<1, \\
0, \quad  \quad \quad \textrm{otherwise}.
\end{array}
\right.
\label{mutation}
\]
\dt{Here} the mutations are considered to be dependent on the underlying fibre densities, \dt{with} $\kappa$ \dt{representing} a certain level of ECM beyond which mutations can occur, $H(\cdot)$ \dt{being} the usual Heaviside function, with $t_m$ being the time at which mutations begin. Furthermore, we consider both cancer cells sub-populations to contribute to the source of MMPs on the tumour boundary, hence the microscopic source term \eqref{eq:sourceMDEs} will be readdressed as
\begin{align}
\begin{split}
1. \quad &g_{\epsilon Y}(y,\tau) = \frac{\int\limits_{\textbf{B}(y,\gamma)\cap\Omega(t_0)} (\alpha_{1} c_{1}(x,t_0 + \tau) + \alpha_{2} c_{2}(x,t_0 + \tau)) \ \widetilde{F}(x,t_0 + \tau) \ dx}{\widetilde{F} \cdot \lambda (\textbf{B}(y,\gamma)\cap\Omega(t_0))}, \\
&\qquad \qquad \qquad \qquad \qquad \qquad \qquad \qquad \qquad \qquad \qquad \qquad \qquad \qquad  y \in \epsilon Y \cap \Omega(t_0), \\[0.7cm]
2. \quad &g_{\epsilon Y}(y,\tau) = 0,\quad  y \in \epsilon Y \setminus \big( \Omega(t_0)+\{ y \in Y|  \ ||y||_2 < \gamma\}), 
\end{split}
\label{eq:sourceMDEs_twopop}
\end{align}
where $\alpha_{1}, \alpha_{2}$ are the MMP secretion rates for cell populations $c_{1}$ and $c_{2}$, respectively, $\widetilde{F}(x,t_0 + \tau):=1+F(x,t)$ and $\widetilde{F} \cdot \lambda$ are defined as in Section \ref{microdynamics}. Using parameter set $\Sigma_{1}$ from Appendix \ref{paramSection}, the micro-fibre degradation rate $d_{f}=1$ and the cell adhesion matrices 
\begin{equation*}
\textbf{S}_{max}=
\left( \begin{array}{cc}
0.5 & 0  \\
0 & 0.3  \end{array} \right)
\quad
\text{and} \quad \textbf{S}_{cF}=
\left( \begin{array}{cc}
0.1 & 0  \\
0 & 0.2  \end{array} \right),
\quad
\textbf{S}_{cl}=
\left( \begin{array}{cc}
0.05 & 0  \\
0 & 0.05  \end{array} \right),
\label{norm_matrices}
\end{equation*}
\dt{in following Figures \ref{fig:homotwopop25} - \ref{fig:heterotwopop75}} we present the computational results for the evolution of\dt{: (a)} the primary cancer cell sub-population\dt{; (b)} the secondary cell sub-population\dt{; (c)} the fibre magnitude density\dt{; (d)} the non-fibres ECM distribution\dt{; and of} the vector field of orientated fibres at two different resolutions, namely\dt{: (e)} coarsened twice\dt{; and (f)} coarsened fourfold. Considering an initially homogeneous fibre ECM phase, when comparing these results directly with the \rs{results} in \cite{Shutt_twopop}, we observe no differences between simulations performed in the absence of micro-fibre degradation at the tumour interface at the first interval $25 \Delta t$, Figure \ref{fig:homotwopop25}, with the fibre orientations consistent between the results, subfigures \ref{fig:homotwopop25e}, \ref{fig:homotwopop25f}. Proceeding to later stages, $50 \Delta t$ and $75 \Delta t$, Figures \ref{fig:homotwopop50} and \ref{fig:homotwopop75}, respectively, we begin to witness changes, \rs{whereby} the boundary of the tumour is visibly smaller than in previous results, \rs{thus} implying a slower progression of the tumour. The macroscopic fibre orientations are aligned differently in subfigures \ref{fig:homotwopop50e}, \ref{fig:homotwopop50f}, \rs{where} the fibres are directed inwards to the centre of the tumour, confining the tumour to the centre of the domain, subfigure \ref{fig:homotwopop25a}. Moving on to the final stage, Figure \ref{fig:homotwopop75}, the tumour region is smaller than in previous results in \cite{Shutt_twopop}. The bulk of the cancer cells stick closely to the tumour boundary and both the fibre and non-fibre ECM phase have undergone a higher level of degradation within the tumour region.

\rs{Finally, to further our understanding} of the effects of micro-fibre degradation at the tumour interface, we explore the evolution of a heterotypic cell population within a heterogeneous fibre ECM phase, whilst the non-fibre ECM phase remains homogeneous, Figures \ref{fig:heterotwopop25}-\ref{fig:homotwopop75}. As with an initially homogeneous fibre density, at stage $25 \Delta t$, Figure \ref{fig:heterotwopop25}, there is very little difference when compared to simulations presented in \cite{Shutt_twopop}. Moving on to later stages it is obvious that the process of boundary micro-fibre degradation causes a slower rate of tumour progression. The tumour region is considerably smaller in Figures \ref{fig:heterotwopop50} and \ref{fig:heterotwopop75} when compared with previous results \cite{Shutt_twopop}. The bulk of tumour cells display a similar pattern, however much closer to the tumour boundary, particularly cancer cell population $c_{2}$ where the cells have formed high distribution bundles \rs{at the leading edge.} Much like in the presence of a homogeneous fibre density, both the fibre and non-fibre ECM phase \rs{has been subject} to a higher level of degradation in the tumour region. This is attributed to the micro-fibre degradation at the tumour boundary, \rs{as degradation continuously occurs as the boundary expands we witness a lower fibre density as the tumour evolves.} It can be concluded from these simulations and comparisons with previous results that the process of micro-fibre degradation at the tumour interface is disadvantageous to tumour progression, inhibiting the full invasive capabilities of the tumour. This is \rs{due to} the lower levels of fibre density at the tumour interface, \rs{as degradation of the micro-fibres occurs both inside and within the peritumoural region of the tumour,} the cell-fibre adhesion rate is reduced in line with low fibre levels, therefore the cancer cells do not have the same opportunities for adhesion and thus their migration is greatly reduced.

\begin{figure}[h!]
    \centering 
\begin{subfigure}{0.5\textwidth}
  \includegraphics[width=\linewidth]{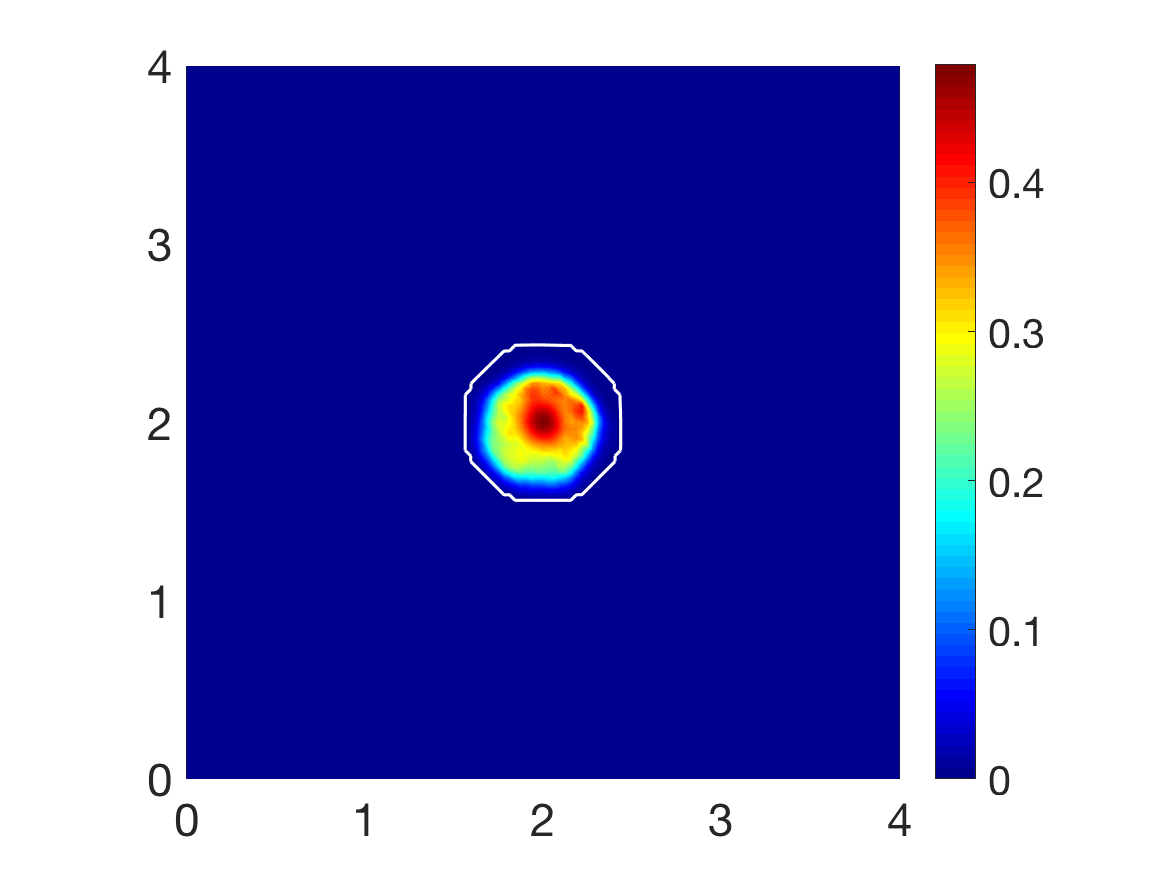}
  \caption{\emph{Cancer cell population 1}}
  \label{fig:homotwopop25a}
\end{subfigure}\hfil 
\begin{subfigure}{0.5\textwidth}
  \includegraphics[width=\linewidth]{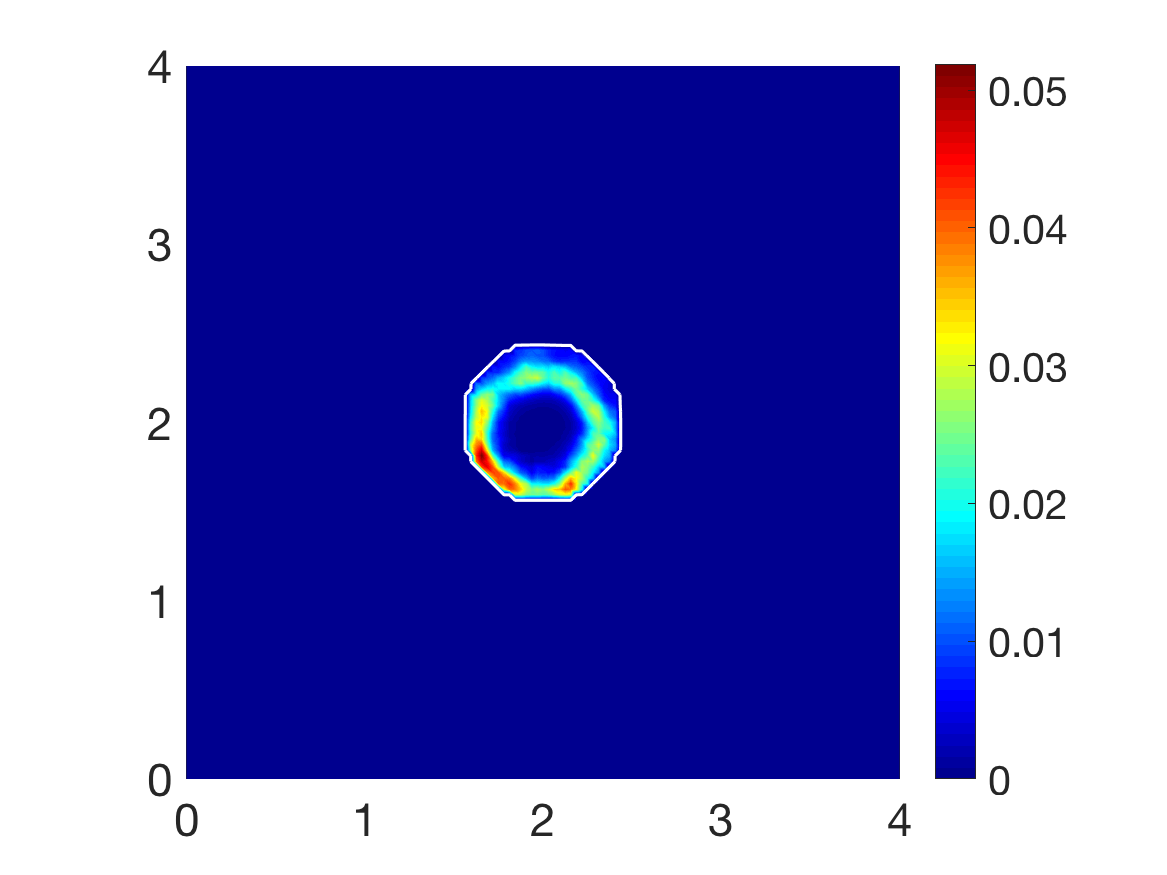}
  \caption{\emph{Cancer cell population 2}}
  \label{fig:homotwopop25b}
\end{subfigure}\hfil 

\medskip
\begin{subfigure}{0.5\textwidth}
  \includegraphics[width=\linewidth]{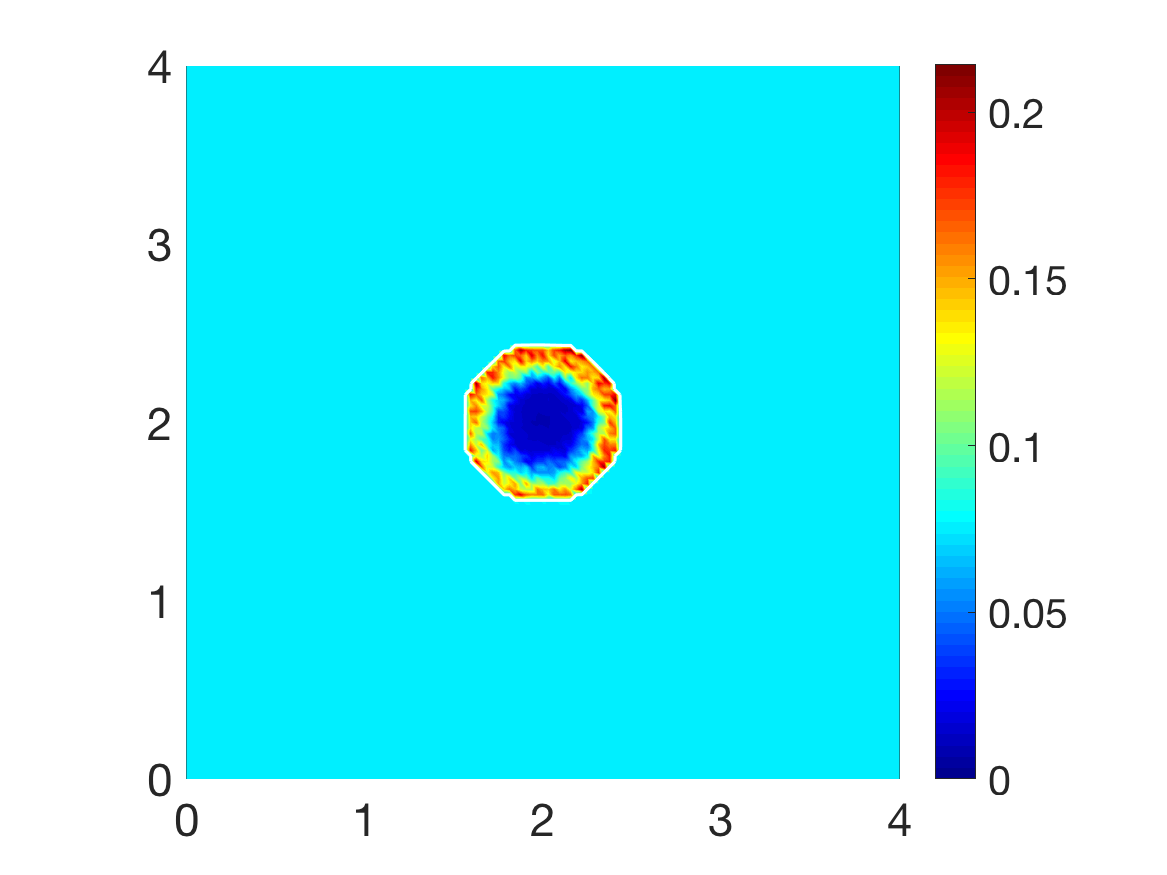}
  \caption{\emph{Fibre magnitude density}}
  \label{fig:homotwopop25c}
  \end{subfigure}\hfil 
\begin{subfigure}{0.5\textwidth}
  \includegraphics[width=\linewidth]{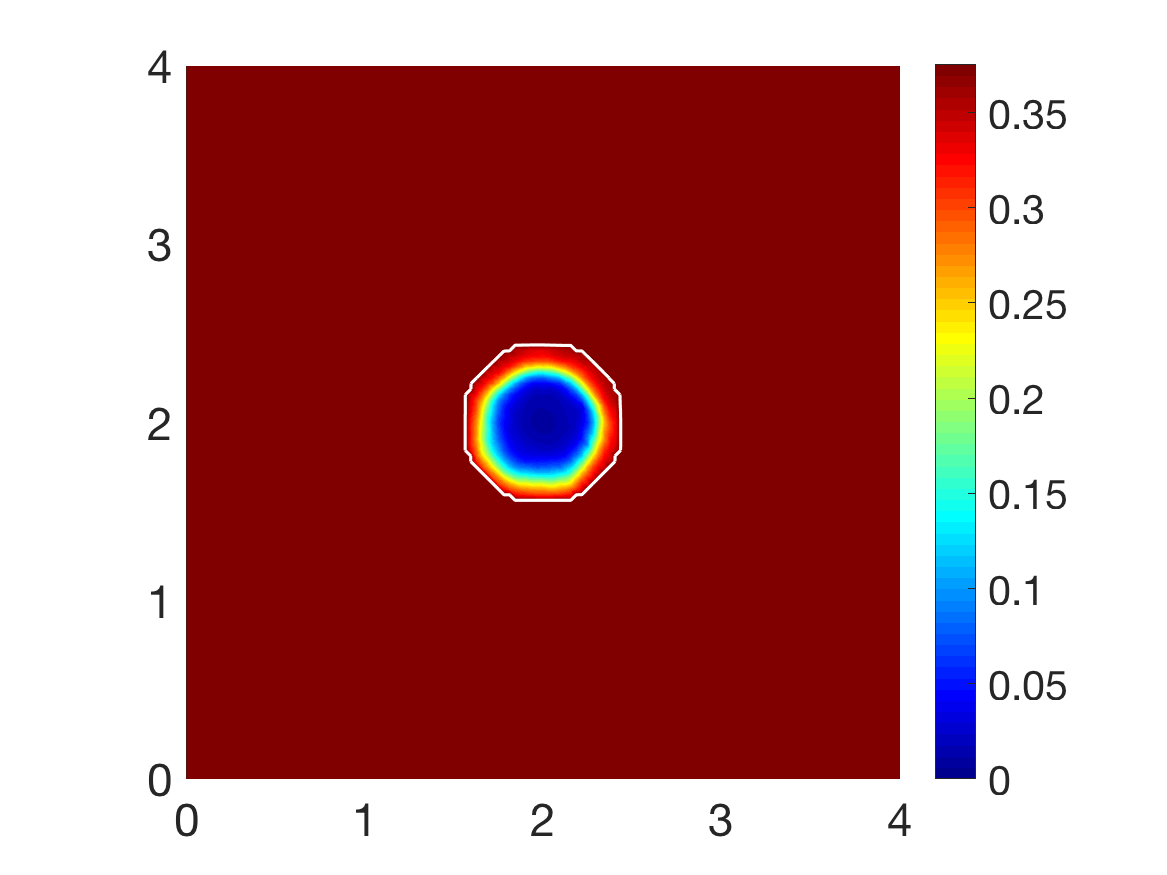}
  \caption{\emph{Non-fibres ECM distribution}}
  \label{fig:homotwopop25d}
\end{subfigure}\hfil 

\medskip
\begin{subfigure}{0.5\textwidth}
  \includegraphics[width=\linewidth]{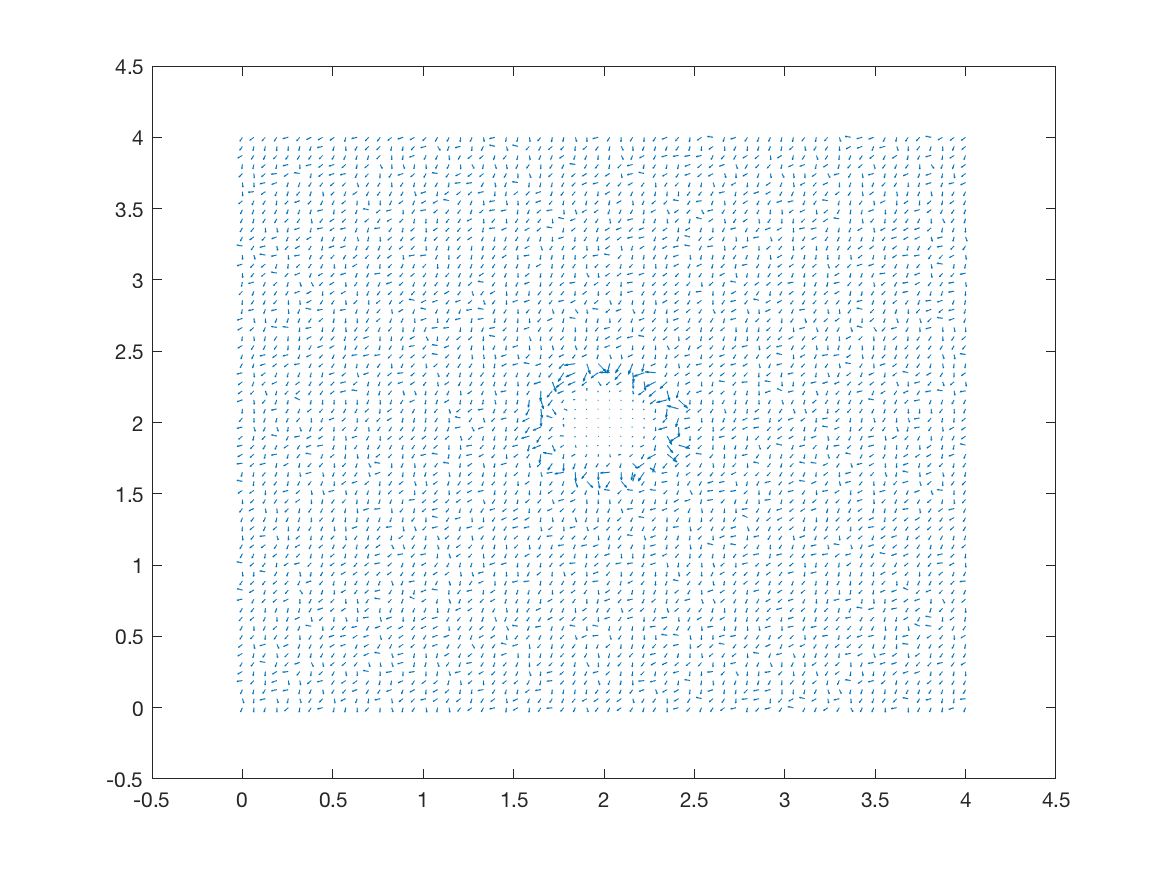}
  \caption{\emph{Fibre vector field - coarsened 2 fold}}
  \label{fig:homotwopop25e}
  \end{subfigure}\hfil 
\begin{subfigure}{0.5\textwidth}
  \includegraphics[width=\linewidth]{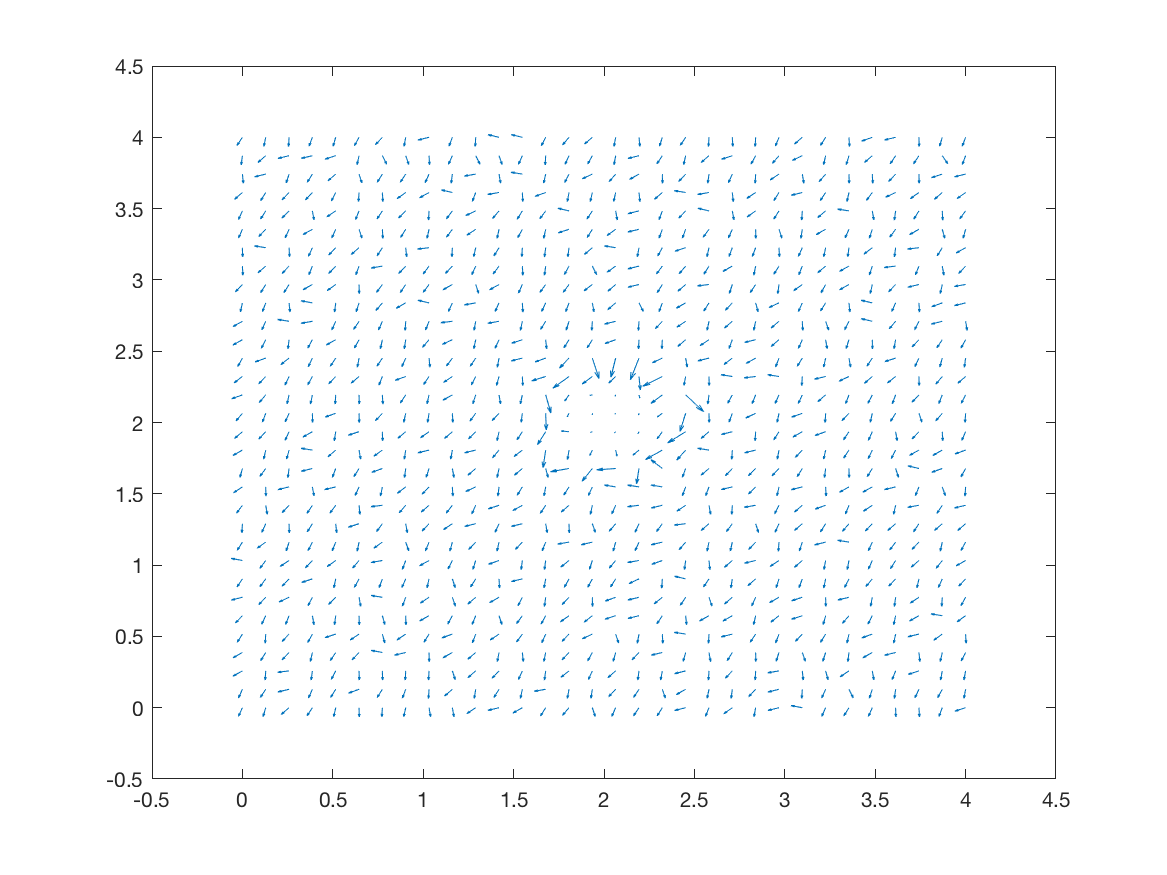}
  \caption{\emph{Fibre vector field - coarsened 4 fold}}
  \label{fig:homotwopop25f}
\end{subfigure}\hfil 

\caption[Simulations at stage $25\Delta t$ with a homogeneous distribution of the non-fibrous phase and $15\%$ homogeneous fibres phase of the ECM with a micro-fibres degradation rate of $d_f = 1$.]{\emph{Simulations at stage $25\Delta t$ with a homogeneous distribution of the non-fibrous phase and $15\%$ homogeneous fibres phase of the ECM with a micro-fibres degradation rate of $d_f = 1$.}}
\label{fig:homotwopop25}
\end{figure}

 \begin{figure}[h!]
    \centering 
\begin{subfigure}{0.5\textwidth}
  \includegraphics[width=\linewidth]{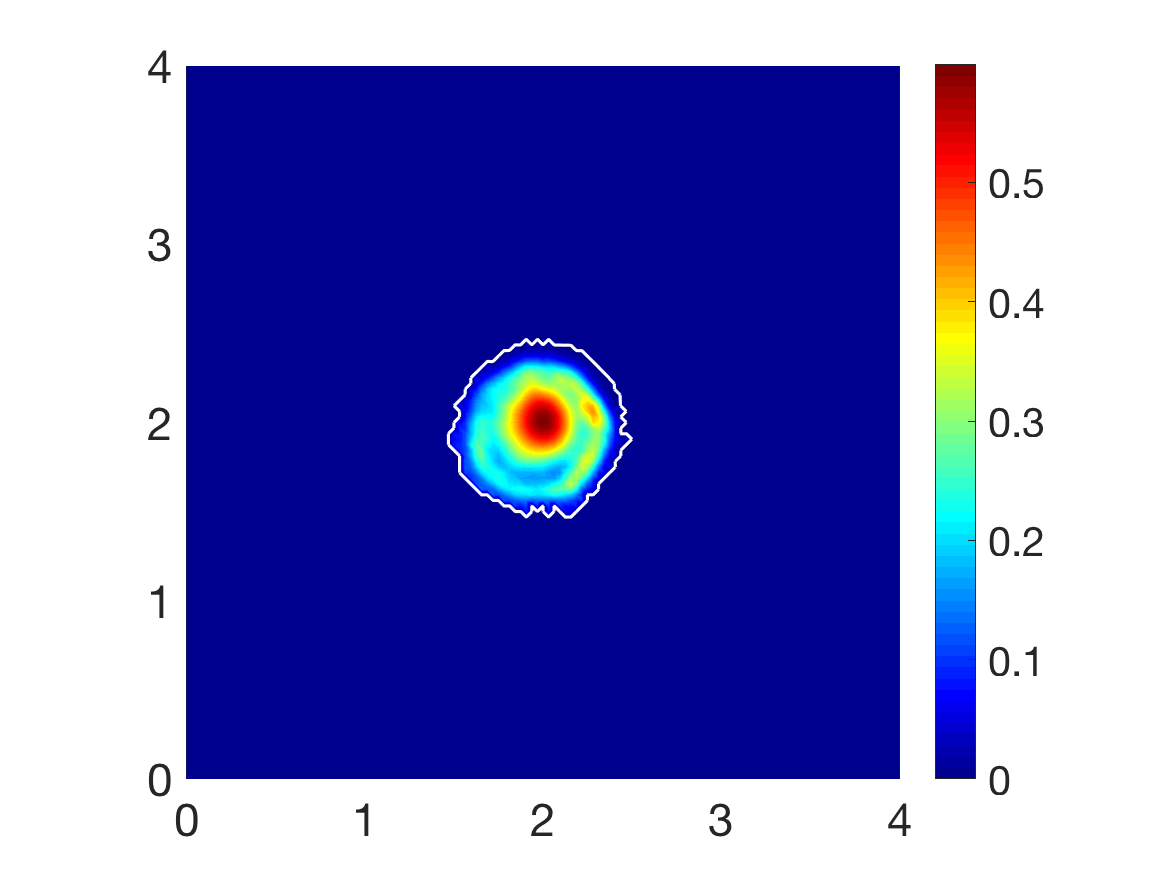}
  \caption{\emph{Cancer cell population}}
  \label{fig:homotwopop50a}
\end{subfigure}\hfil 
\begin{subfigure}{0.5\textwidth}
  \includegraphics[width=\linewidth]{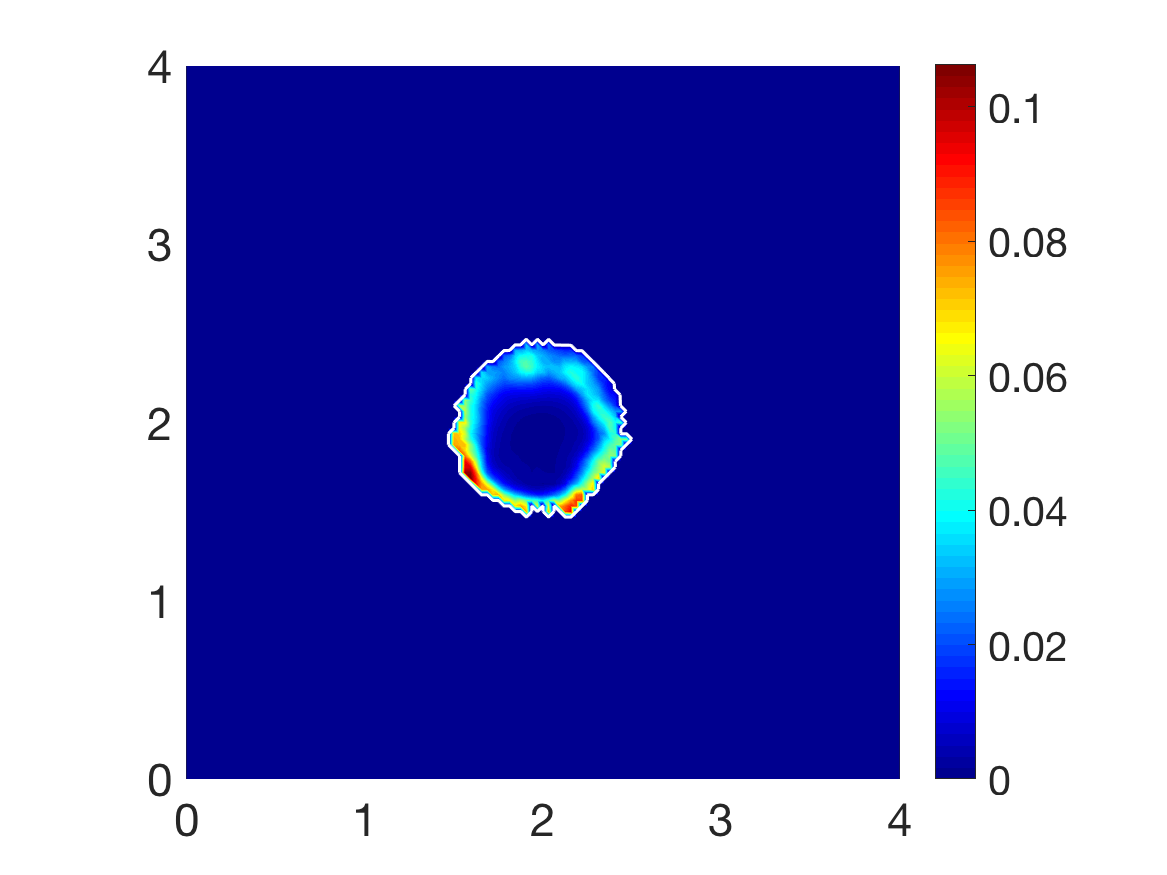}
  \caption{\emph{Cancer cell population 2}}
  \label{fig:homotwopop50b}
\end{subfigure}\hfil 

\medskip
\begin{subfigure}{0.5\textwidth}
  \includegraphics[width=\linewidth]{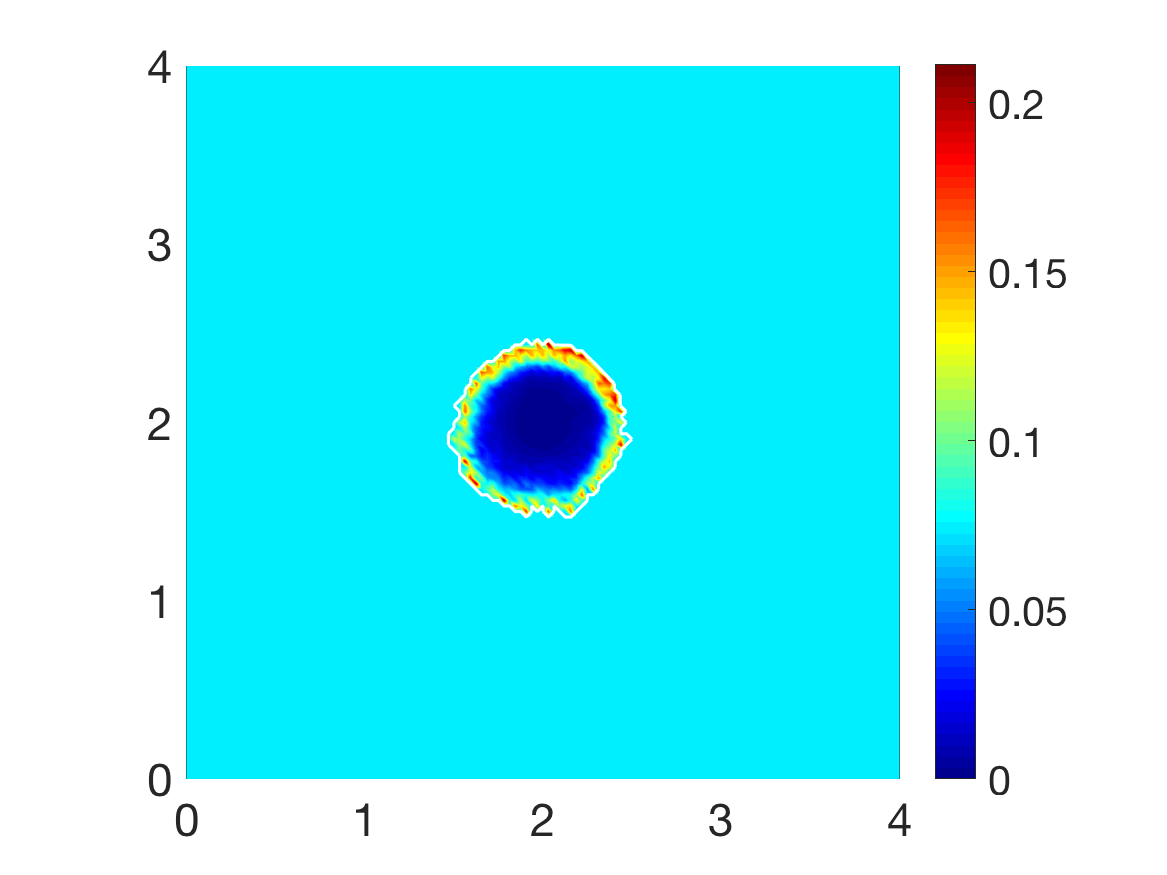}
  \caption{\emph{Fibre magnitude density}}
  \label{fig:homotwopop50c}
  \end{subfigure}\hfil 
\begin{subfigure}{0.5\textwidth}
  \includegraphics[width=\linewidth]{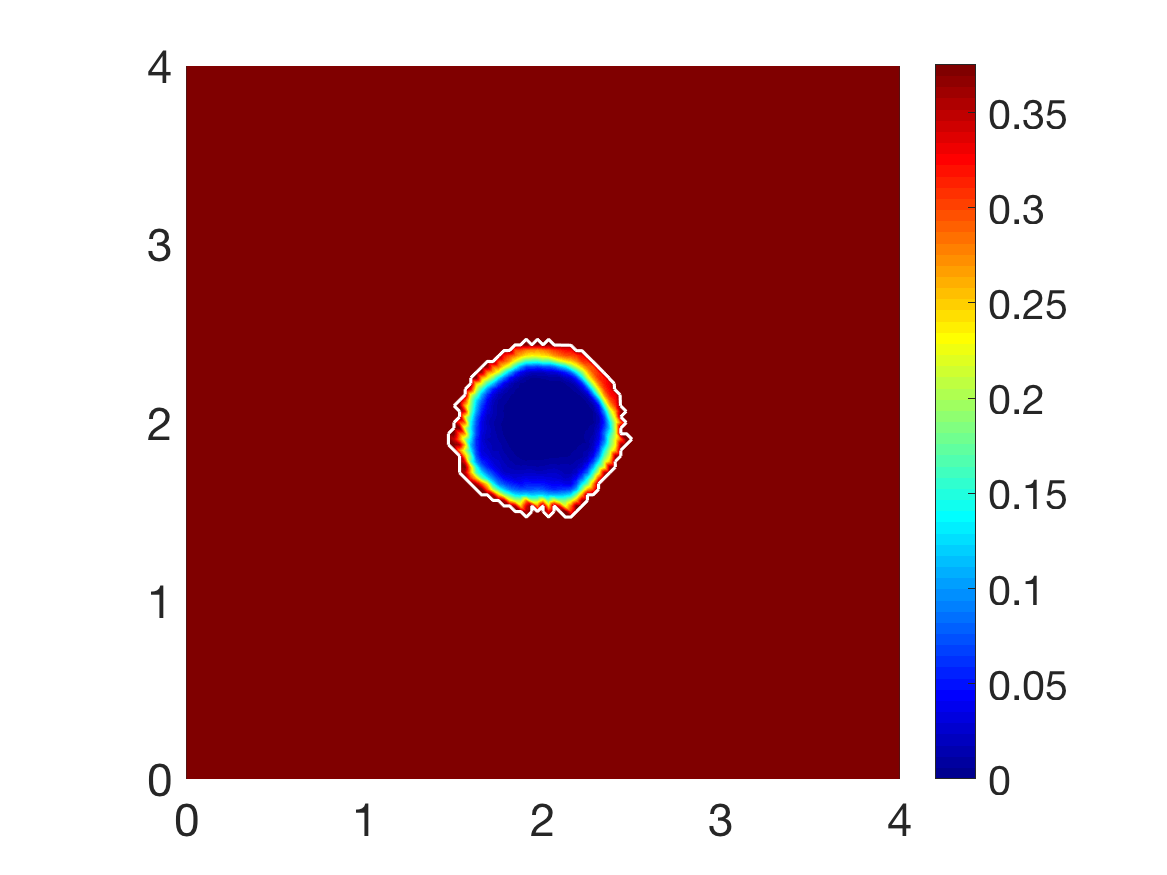}
  \caption{\emph{Non-fibres ECM distribution}}
  \label{fig:homotwopop50d}
\end{subfigure}\hfil 

\medskip
\begin{subfigure}{0.5\textwidth}
  \includegraphics[width=\linewidth]{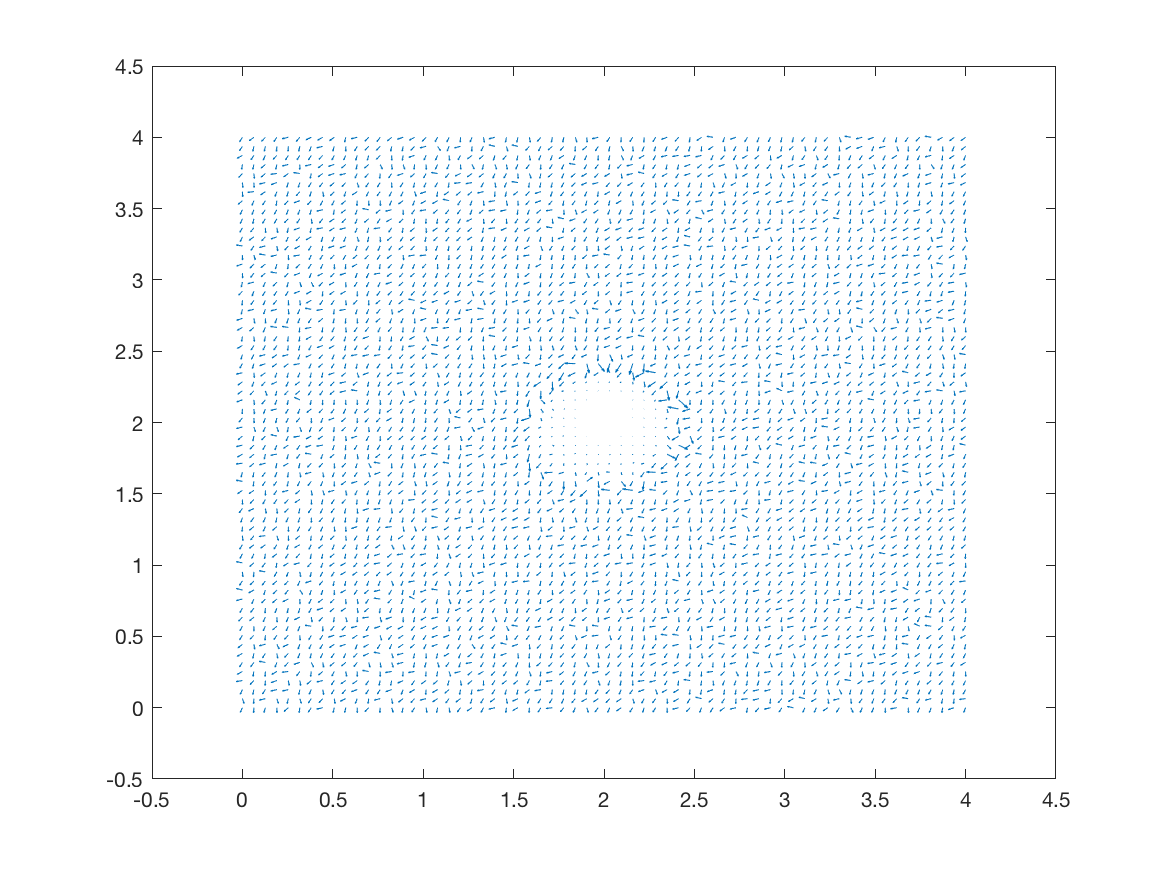}
  \caption{\emph{Fibre vector field - coarsened 2 fold}}
  \label{fig:homotwopop50e}
  \end{subfigure}\hfil 
\begin{subfigure}{0.5\textwidth}
  \includegraphics[width=\linewidth]{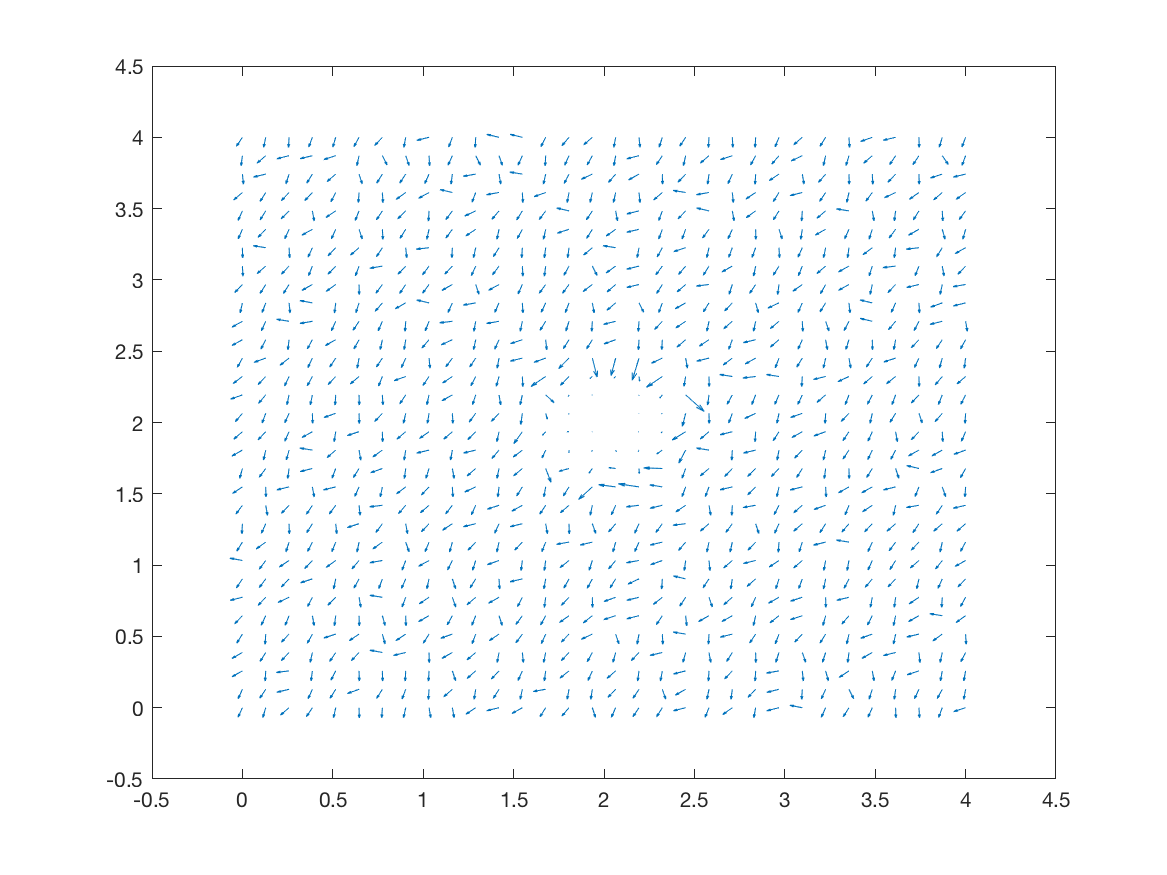}
  \caption{\emph{Fibre vector field - coarsened 4 fold}}
  \label{fig:homotwopop50f}
\end{subfigure}\hfil 

\caption[Simulations at stage $50\Delta t$ with a homogeneous distribution of the non-fibrous phase and $15\%$ homogeneous fibres phase of the ECM with a micro-fibres degradation rate of $d_f = 1$.]{\emph{Simulations at stage $50\Delta t$ with a homogeneous distribution of the non-fibrous phase and $15\%$ homogeneous fibres phase of the ECM with a micro-fibres degradation rate of $d_f = 1$.}}
\label{fig:homotwopop50}
\end{figure}

\begin{figure}[h!]
    \centering 
\begin{subfigure}{0.5\textwidth}
  \includegraphics[width=\linewidth]{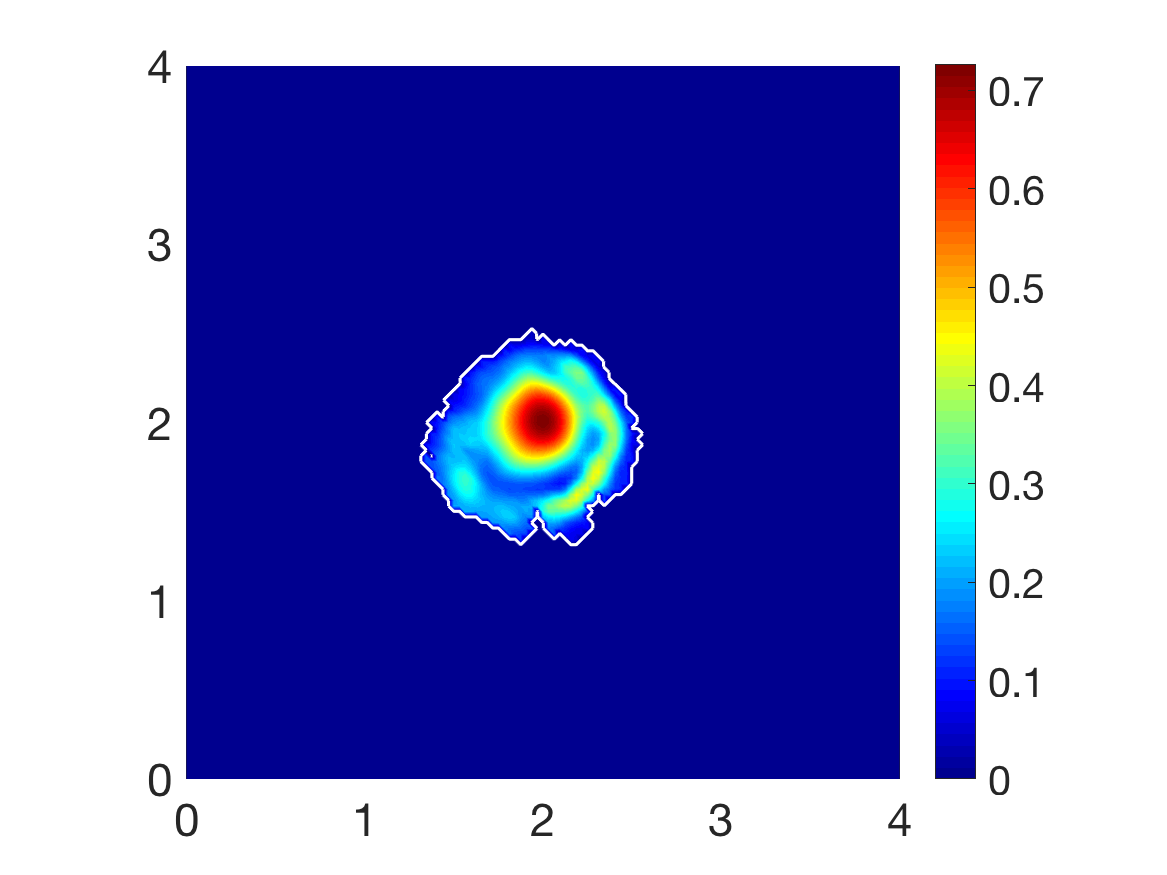}
  \caption{\emph{Cancer cell population}}
  \label{fig:homotwopop75a}
\end{subfigure}\hfil 
\begin{subfigure}{0.5\textwidth}
  \includegraphics[width=\linewidth]{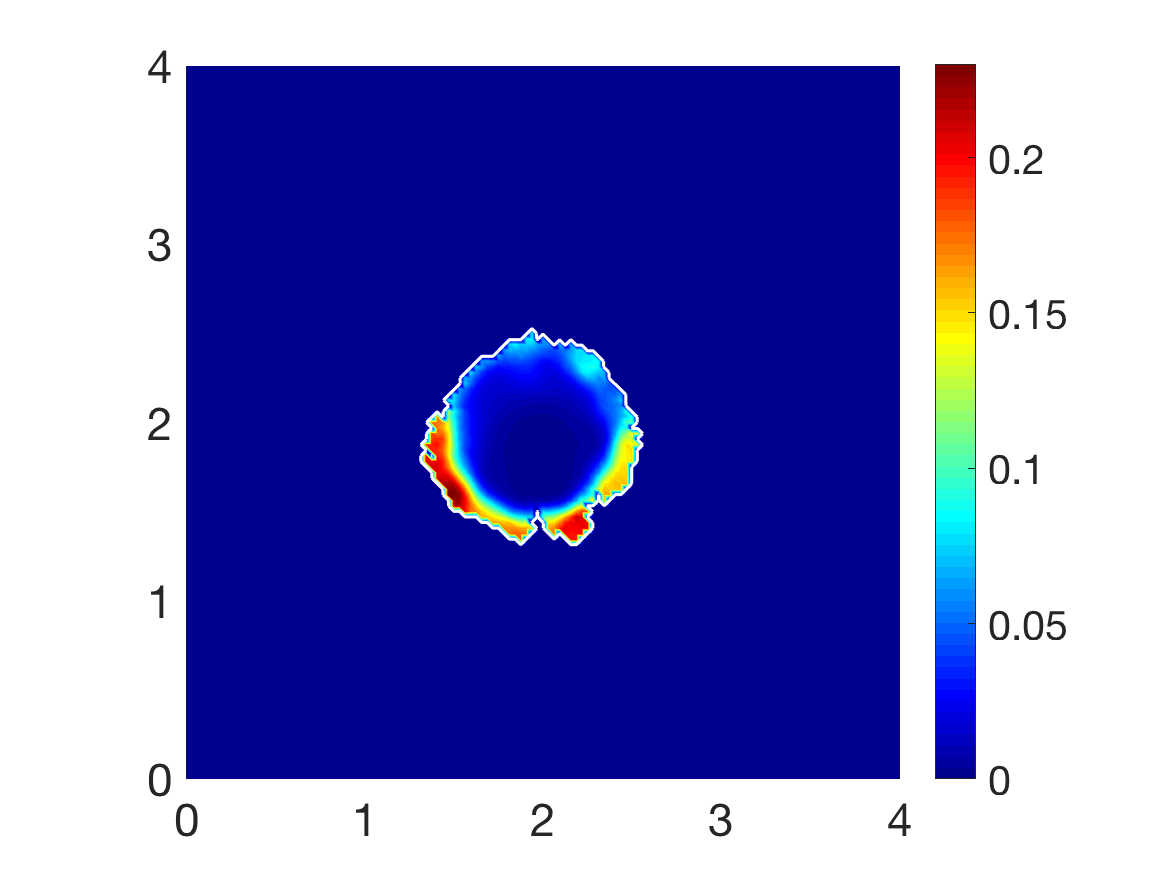}
  \caption{\emph{cancer cell population 2}}
  \label{fig:homotwopop75b}
\end{subfigure}\hfil 

\medskip
\begin{subfigure}{0.5\textwidth}
  \includegraphics[width=\linewidth]{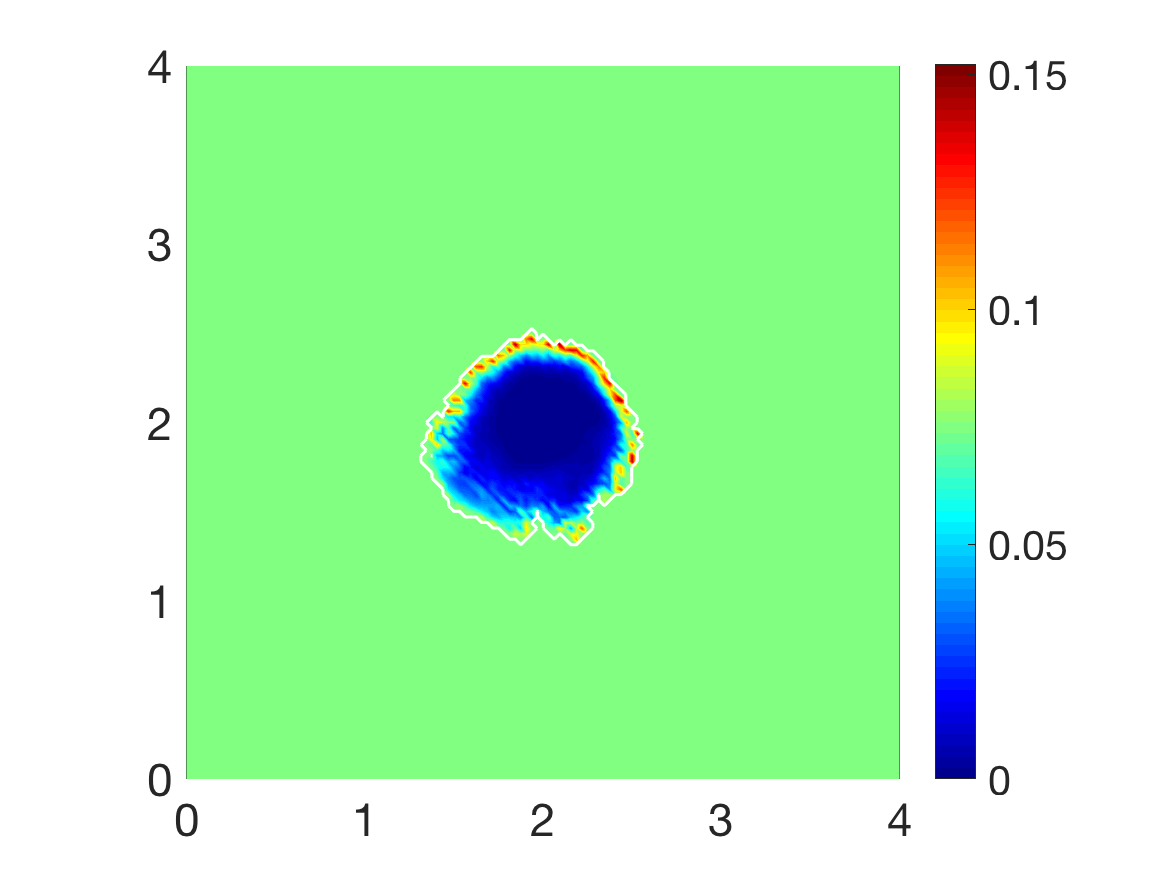}
  \caption{\emph{Fibre magnitude density}}
  \label{fig:homotwopop75c}
  \end{subfigure}\hfil 
\begin{subfigure}{0.5\textwidth}
  \includegraphics[width=\linewidth]{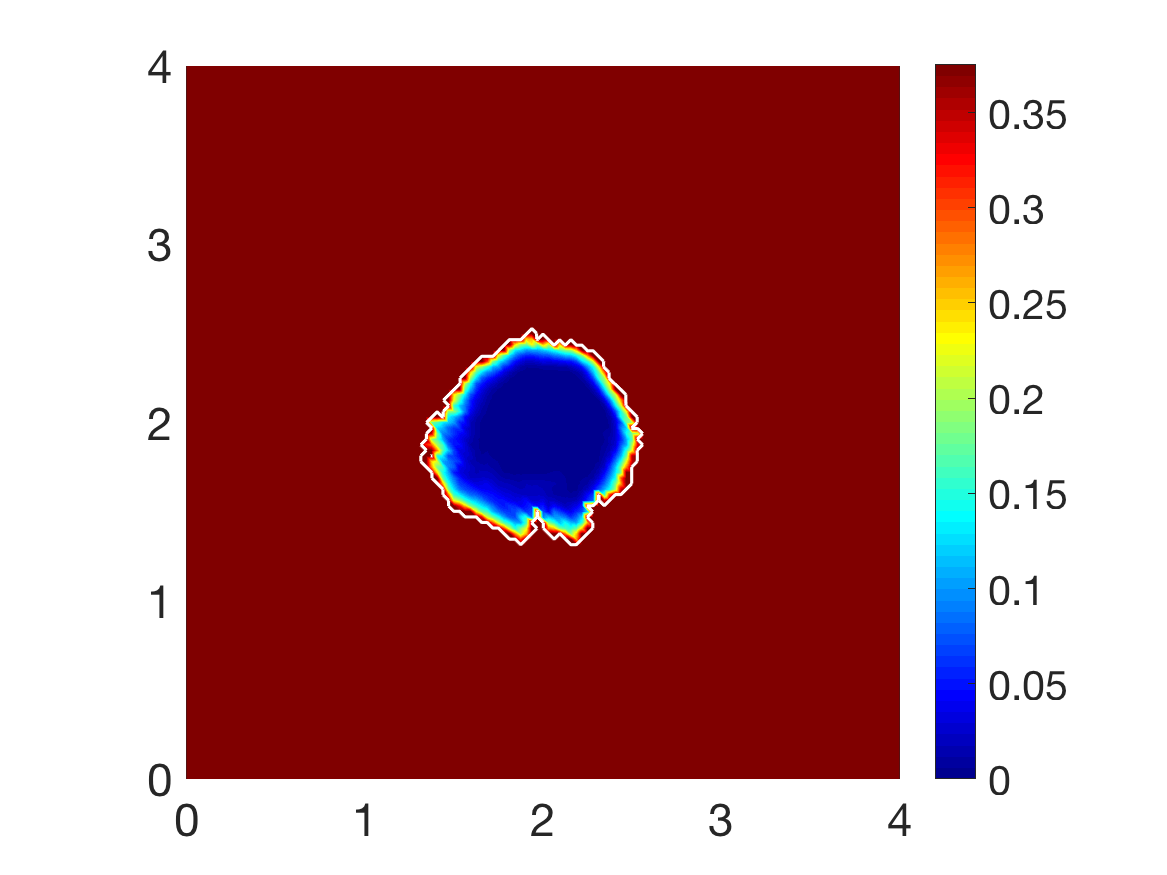}
  \caption{\emph{Non-fibres ECM distribution}}
  \label{fig:homotwopop75d}
\end{subfigure}\hfil 

\medskip
\begin{subfigure}{0.5\textwidth}
  \includegraphics[width=\linewidth]{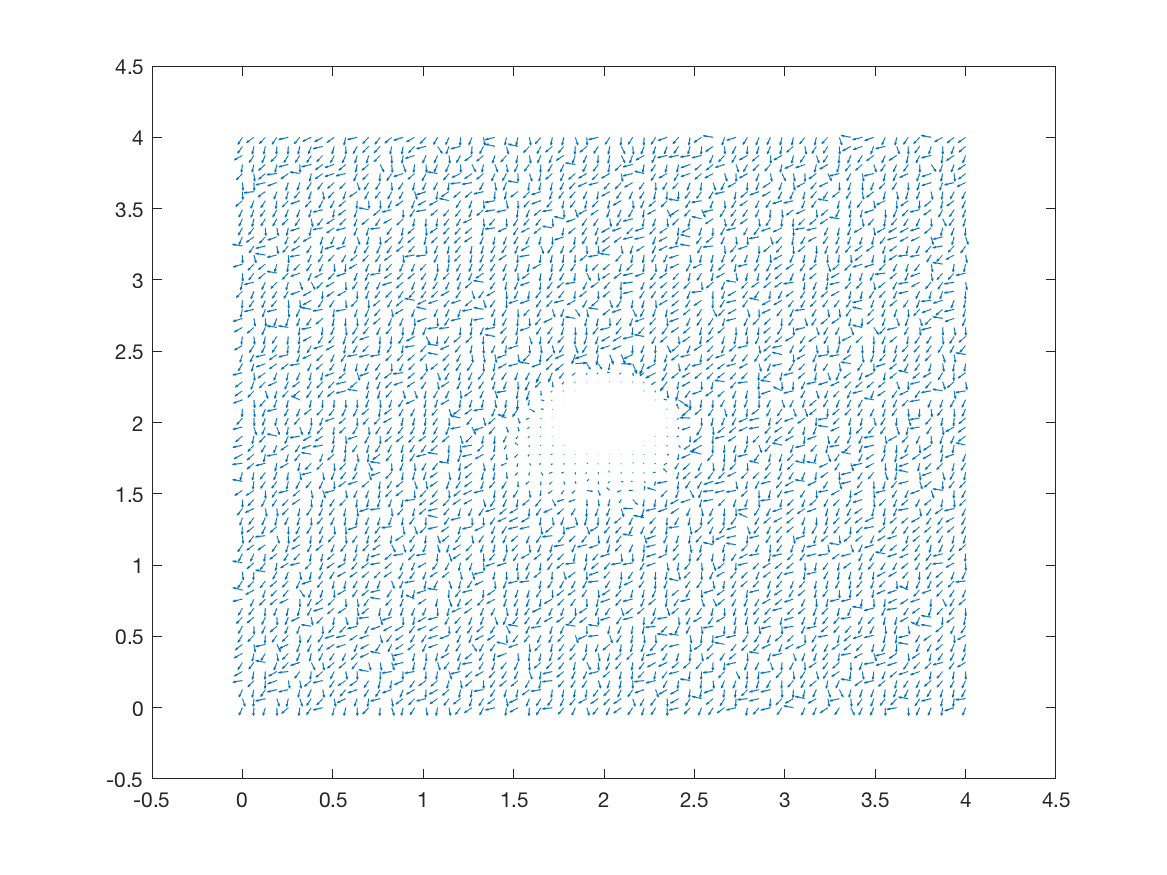}
  \caption{\emph{Fibre vector field - coarsened 2 fold}}
  \label{fig:homotwopop75e}
  \end{subfigure}\hfil 
\begin{subfigure}{0.5\textwidth}
  \includegraphics[width=\linewidth]{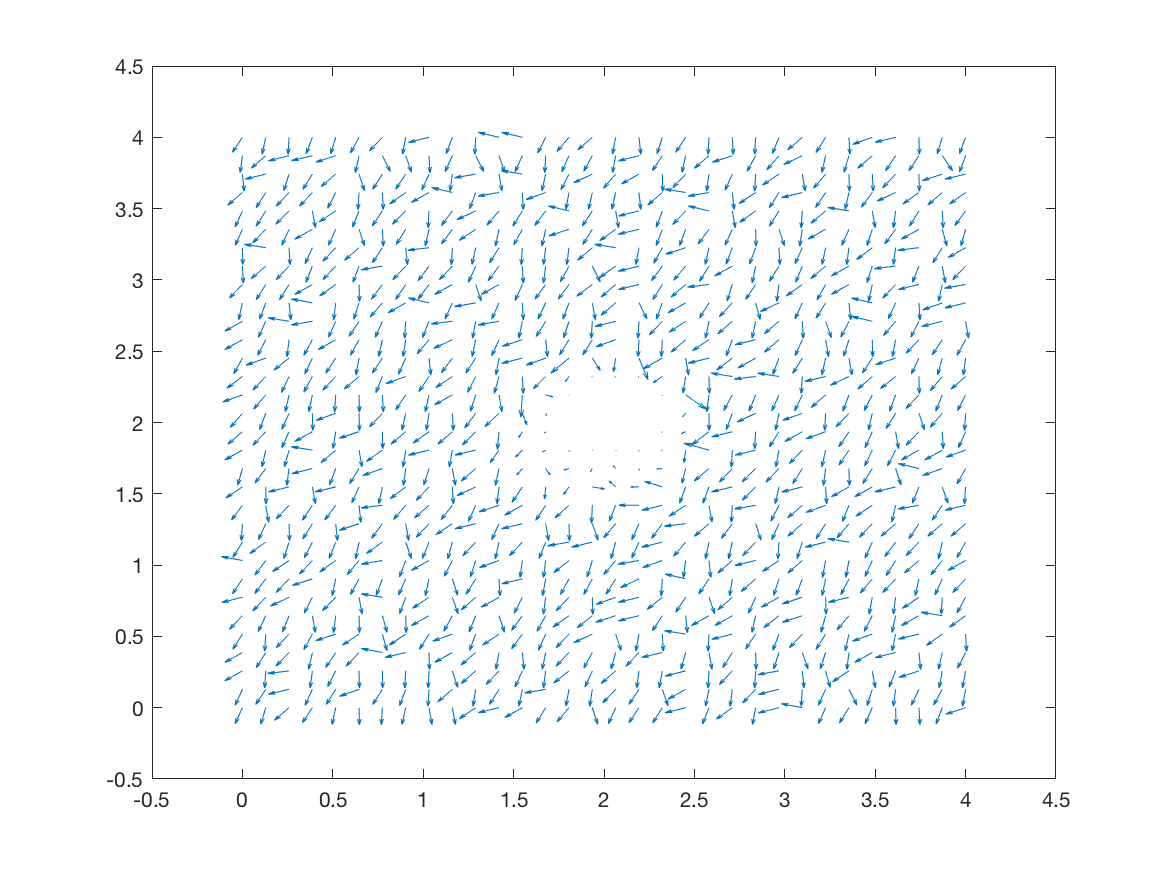}
  \caption{\emph{Fibre vector field - coarsened 4 fold}}
  \label{fig:homotwopop75f}
\end{subfigure}\hfil 

\caption[Simulations at stage $75\Delta t$ with a homogeneous distribution of the non-fibrous phase and $15\%$ homogeneous fibres phase of the ECM with a micro-fibres degradation rate of $d_f = 1$.]{\emph{Simulations at stage $75\Delta t$ with a homogeneous distribution of the non-fibrous phase and $15\%$ homogeneous fibres phase of the ECM with a micro-fibres degradation rate of $d_f = 1$.}}
\label{fig:homotwopop75}
\end{figure}

\begin{figure}[h!]
    \centering 
\begin{subfigure}{0.5\textwidth}
  \includegraphics[width=\linewidth]{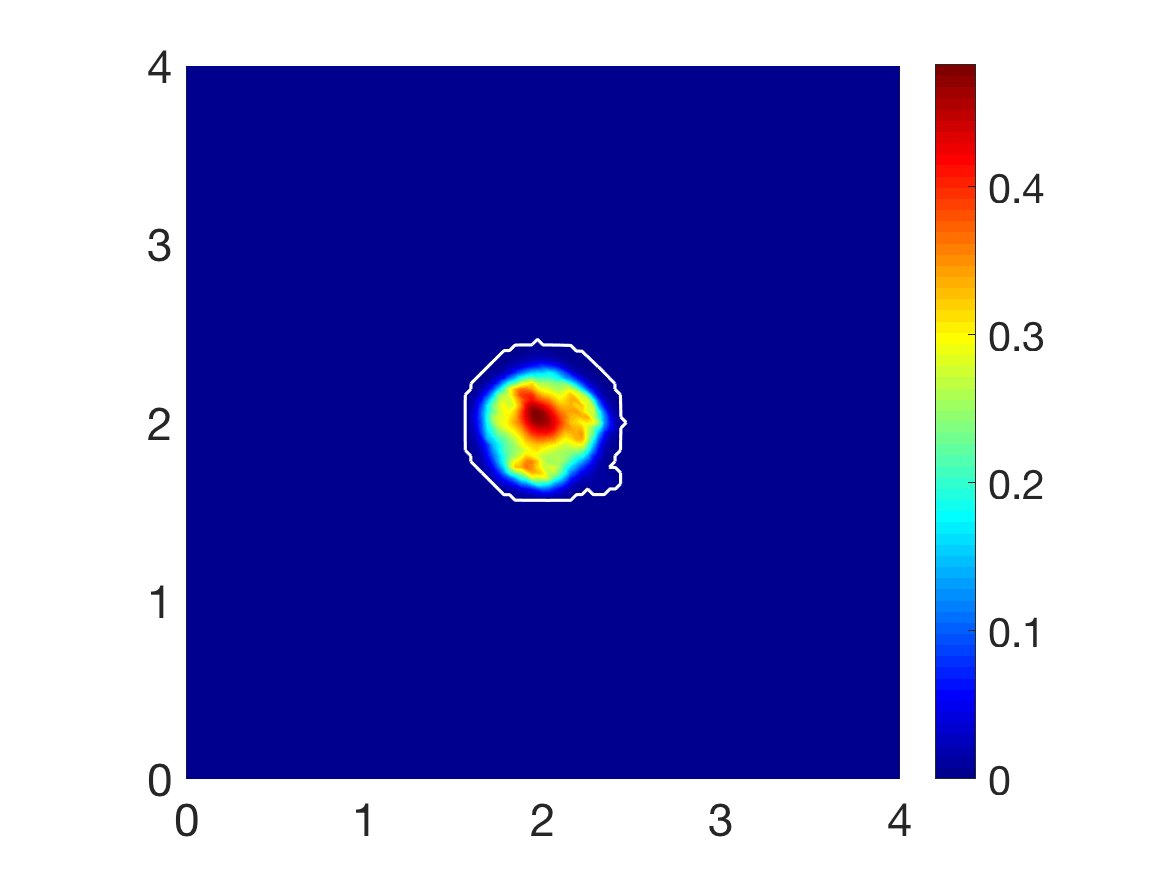}
  \caption{\emph{Cancer cell population}}
  \label{fig:heterotwopop25a}
\end{subfigure}\hfil 
\begin{subfigure}{0.5\textwidth}
  \includegraphics[width=\linewidth]{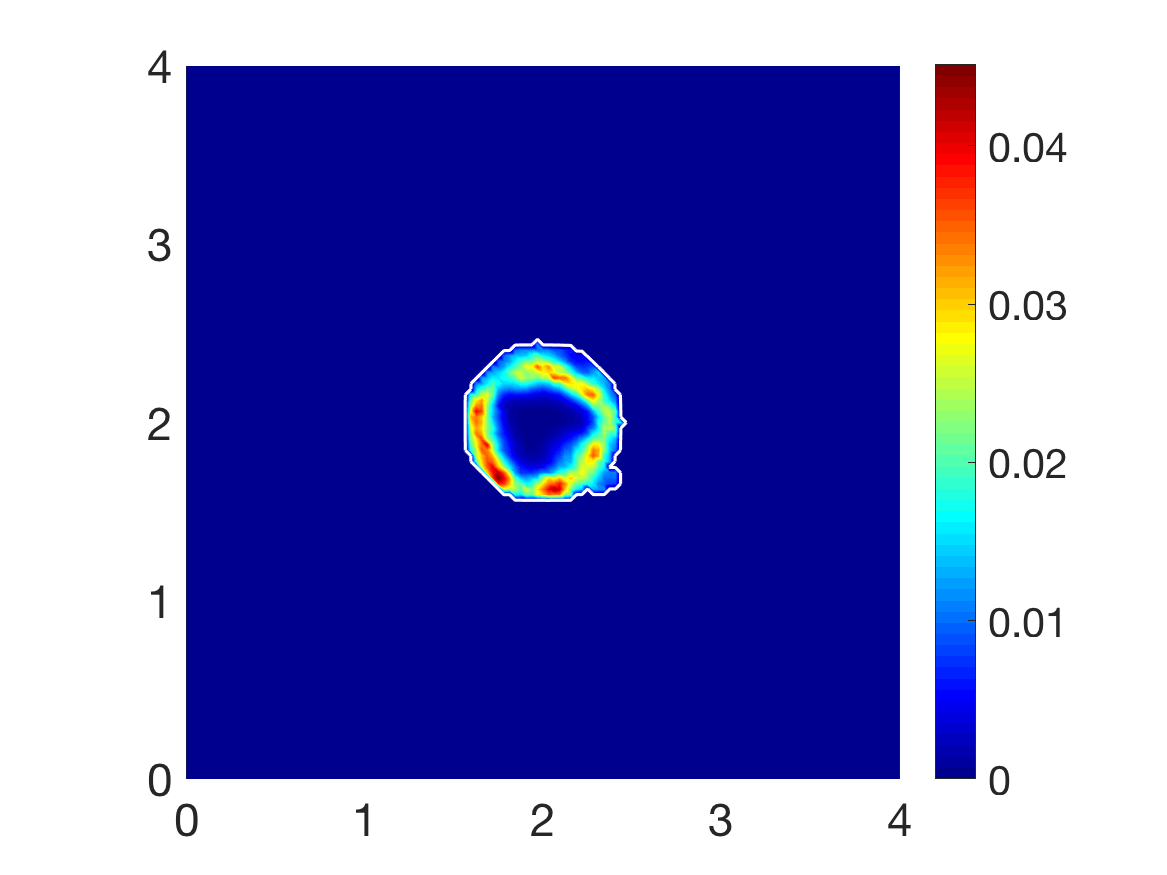}
  \caption{\emph{Cancer cell population 2}}
  \label{fig:heterotwopop25b}
\end{subfigure}\hfil 

\medskip
\begin{subfigure}{0.5\textwidth}
  \includegraphics[width=\linewidth]{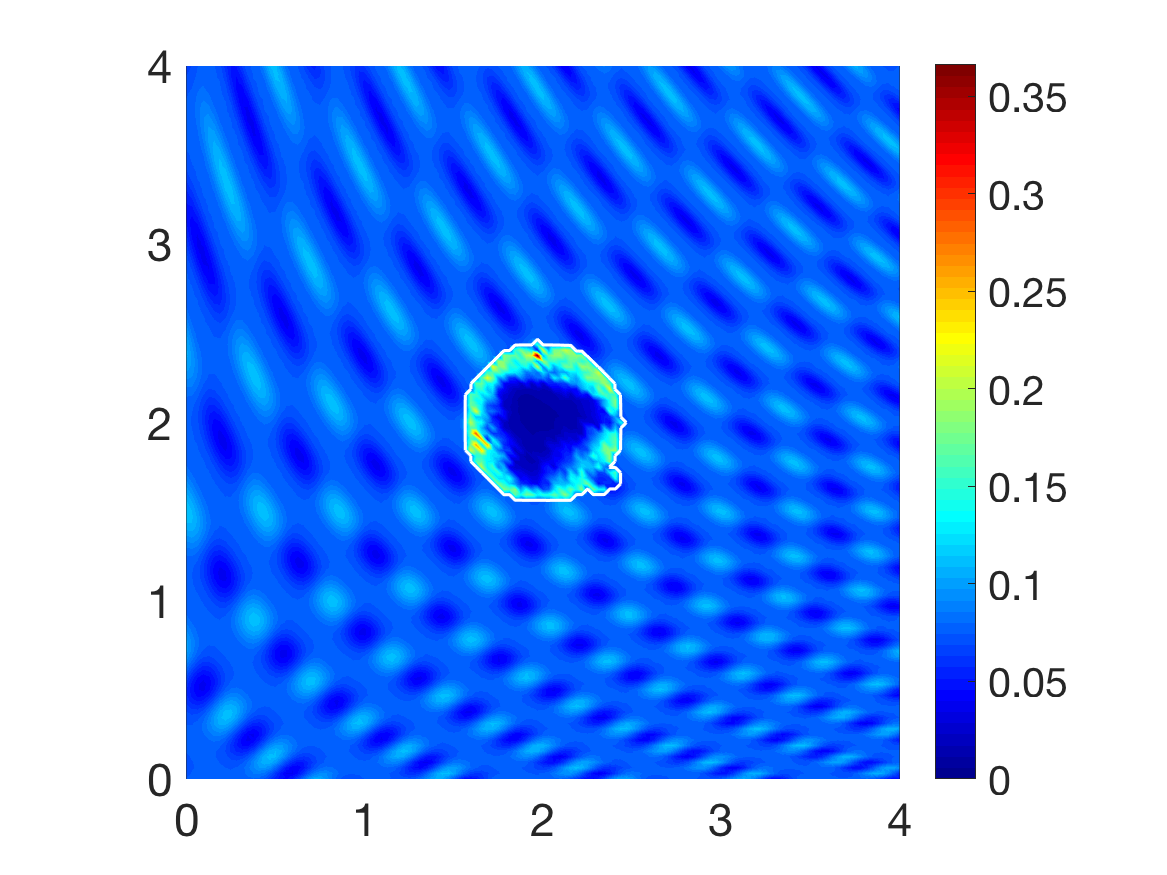}
  \caption{\emph{Fibre magnitude density}}
  \label{fig:heterotwopop25c}
  \end{subfigure}\hfil 
\begin{subfigure}{0.5\textwidth}
  \includegraphics[width=\linewidth]{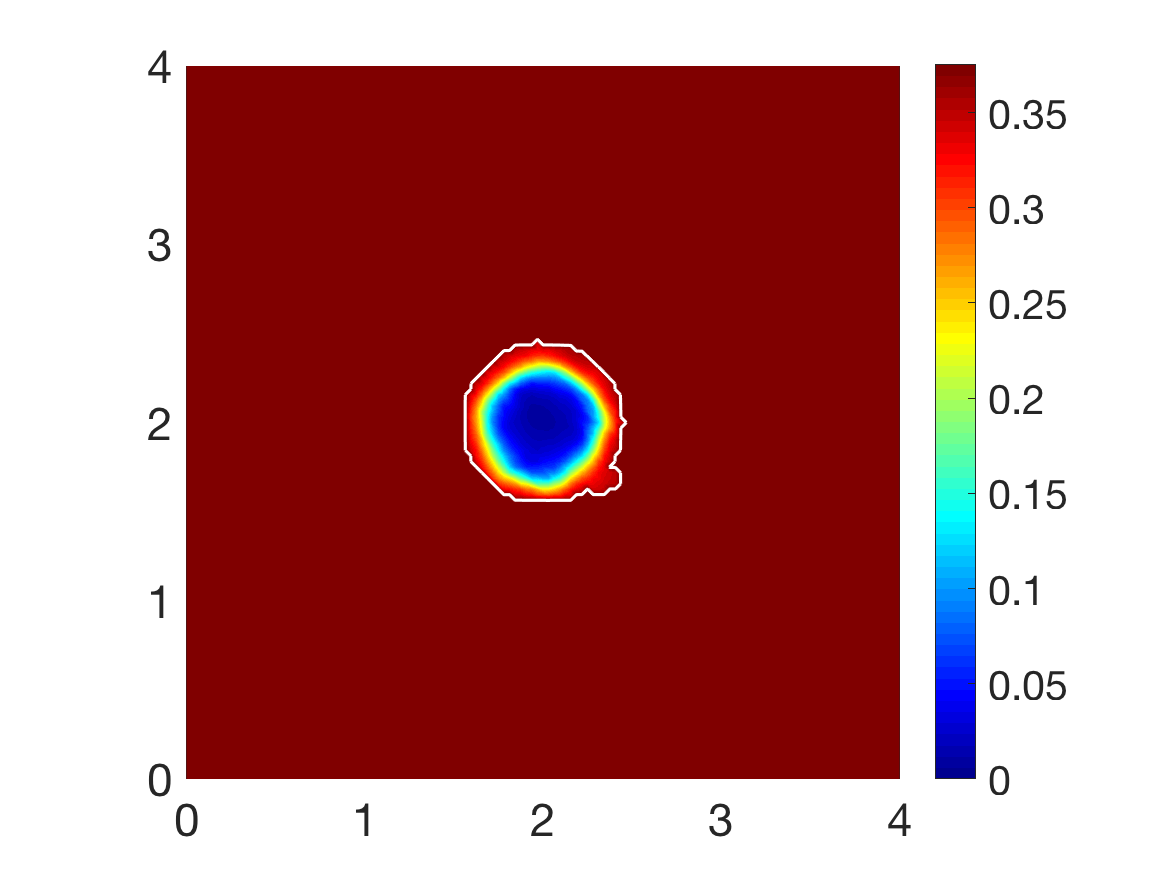}
  \caption{\emph{Non-fibres ECM distribution}}
  \label{fig:heterotwopop25d}
\end{subfigure}\hfil 

\medskip
\begin{subfigure}{0.5\textwidth}
  \includegraphics[width=\linewidth]{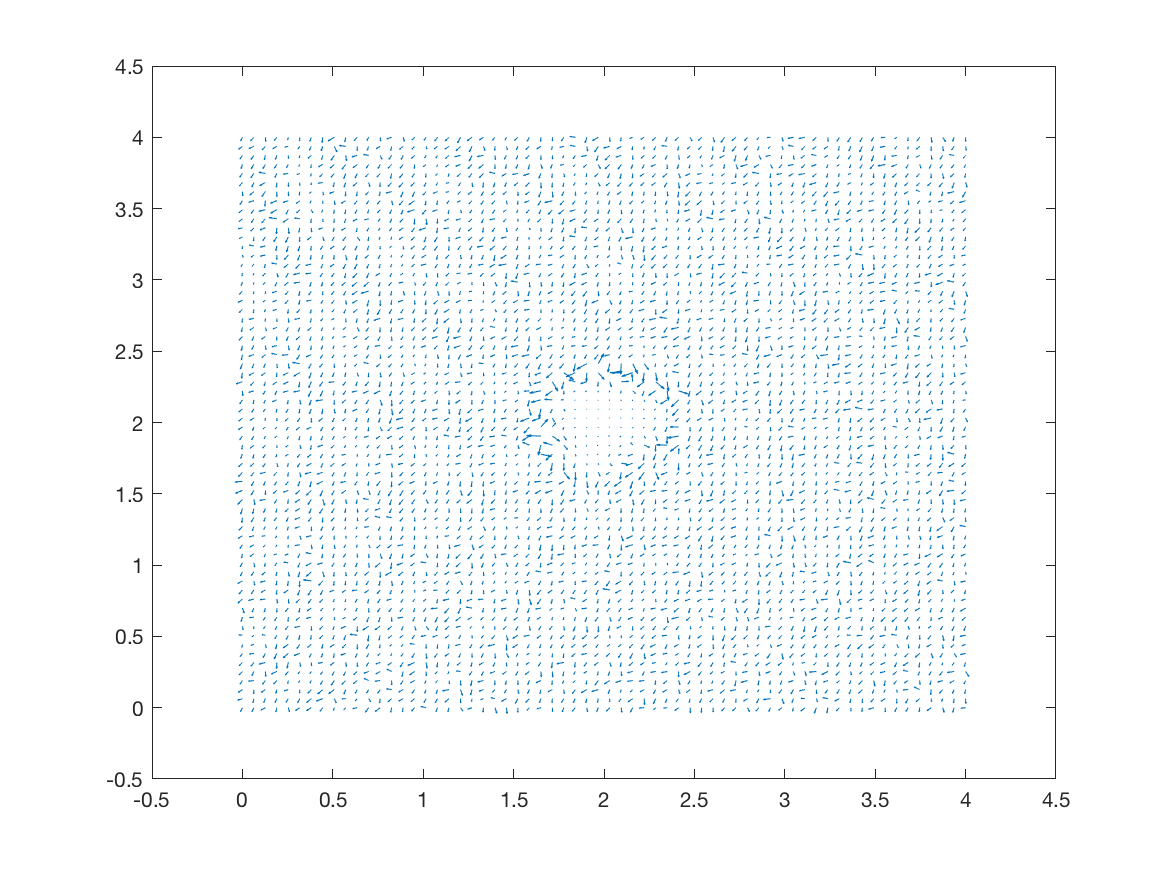}
  \caption{\emph{Fibre vector field - coarsened 2 fold}}
  \label{fig:heterotwopop25e}
  \end{subfigure}\hfil 
\begin{subfigure}{0.5\textwidth}
  \includegraphics[width=\linewidth]{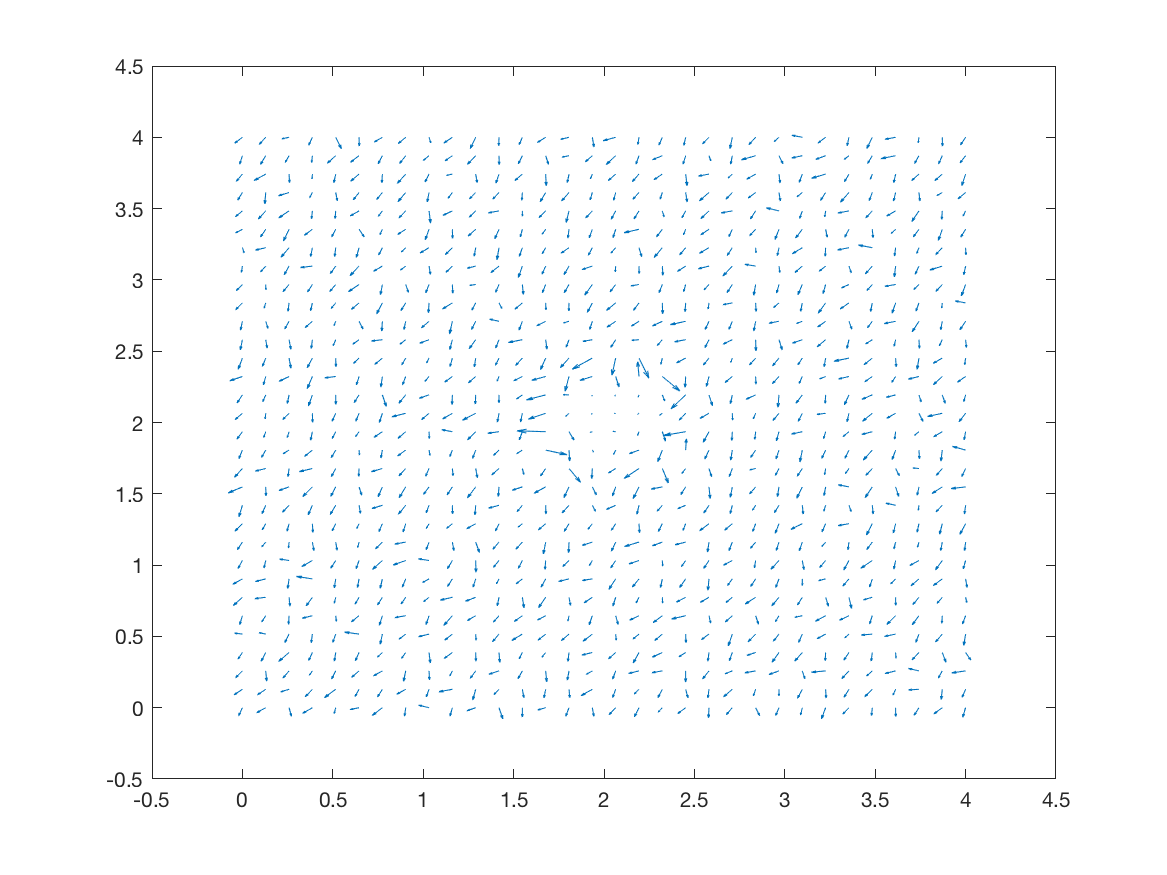}
  \caption{\emph{Fibre vector field - coarsened 4 fold}}
  \label{fig:heterotwopop25f}
\end{subfigure}\hfil 

\caption[Simulations at stage $25\Delta t$ with a homogeneous distribution of the non-fibrous phase and $15\%$ heterogeneous fibres phase of the ECM with a micro-fibres degradation rate of $d_f = 1$.]{\emph{Simulations at stage $25\Delta t$ with a homogeneous distribution of the non-fibrous phase and $15\%$ heterogeneous fibres phase of the ECM with a micro-fibres degradation rate of $d_f = 1$.}}
\label{fig:heterotwopop25}
\end{figure}

 \begin{figure}[h!]
    \centering 
\begin{subfigure}{0.5\textwidth}
  \includegraphics[width=\linewidth]{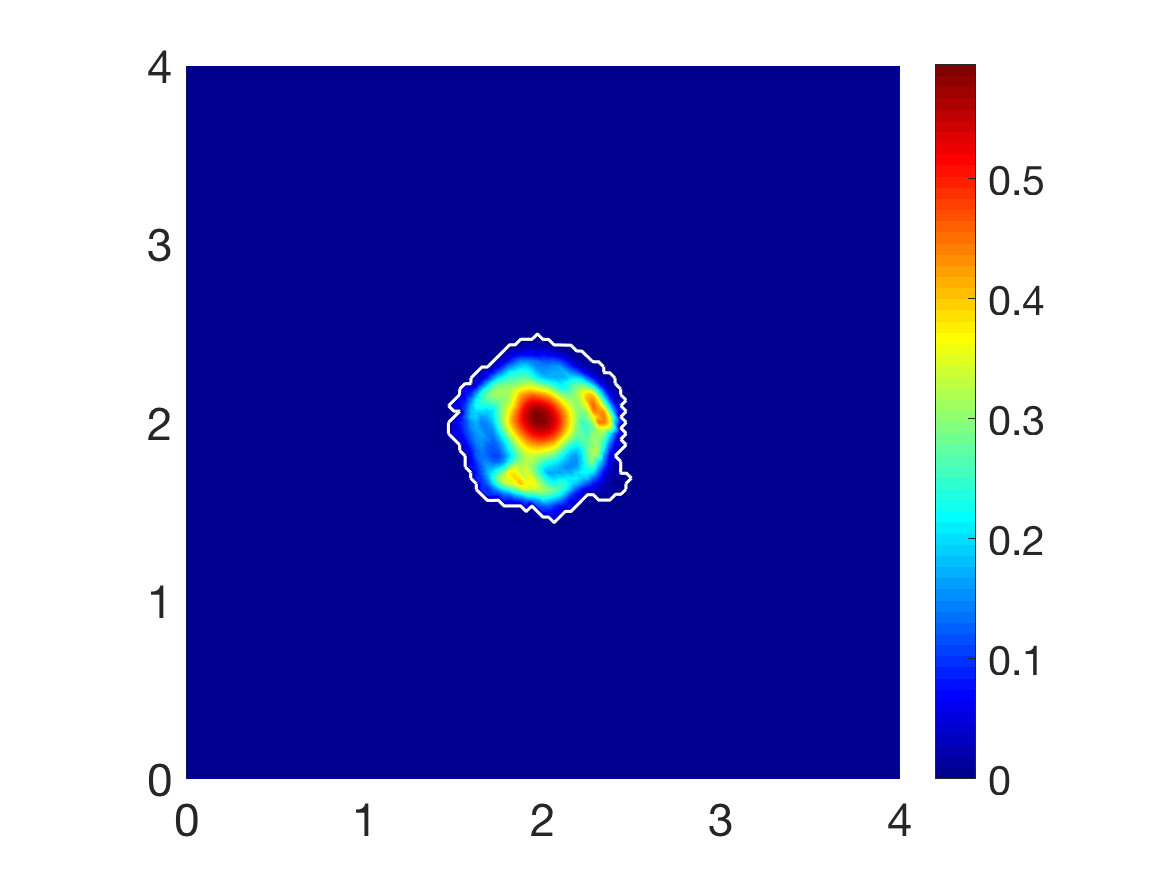}
  \caption{\emph{Cancer cell population}}
  \label{fig:heterotwopop50a}
\end{subfigure}\hfil 
\begin{subfigure}{0.5\textwidth}
  \includegraphics[width=\linewidth]{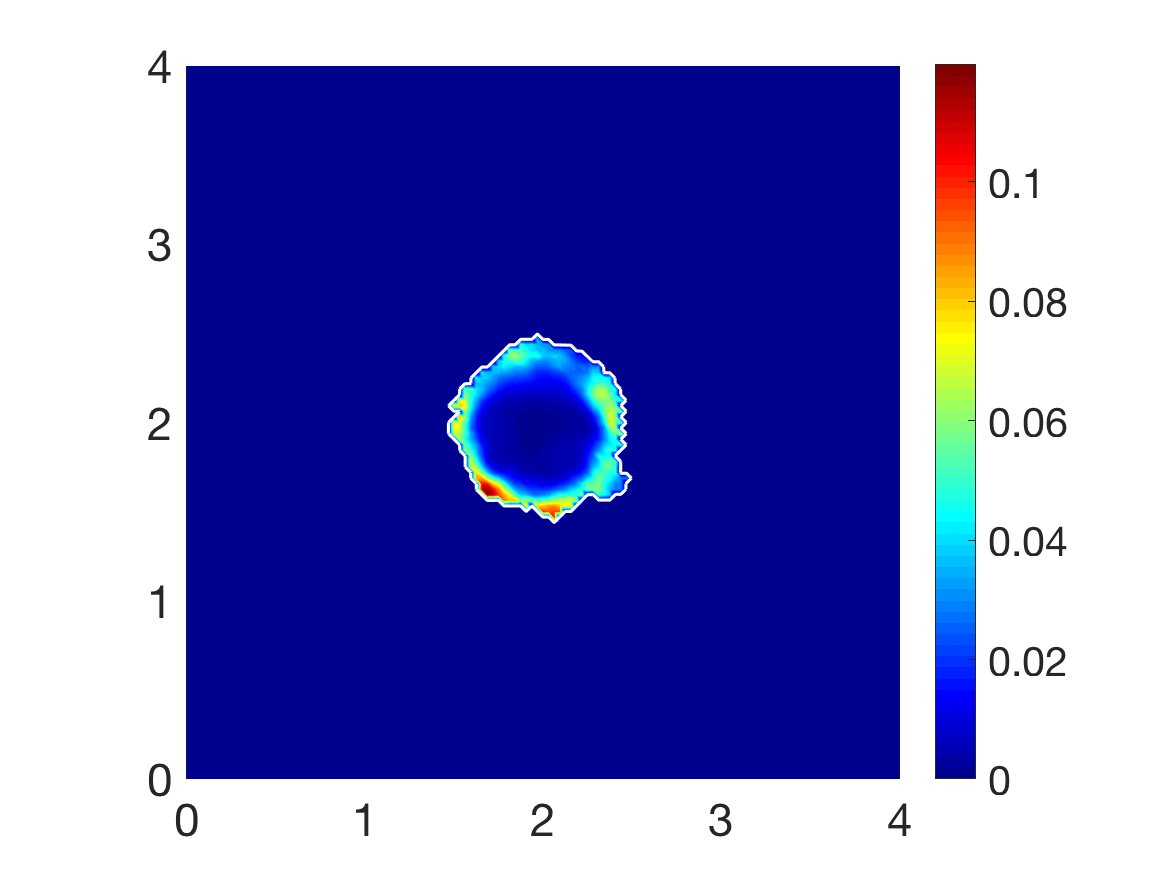}
  \caption{\emph{Cancer cell population 2}}
  \label{fig:heterotwopop50b}
\end{subfigure}\hfil 

\medskip
\begin{subfigure}{0.5\textwidth}
  \includegraphics[width=\linewidth]{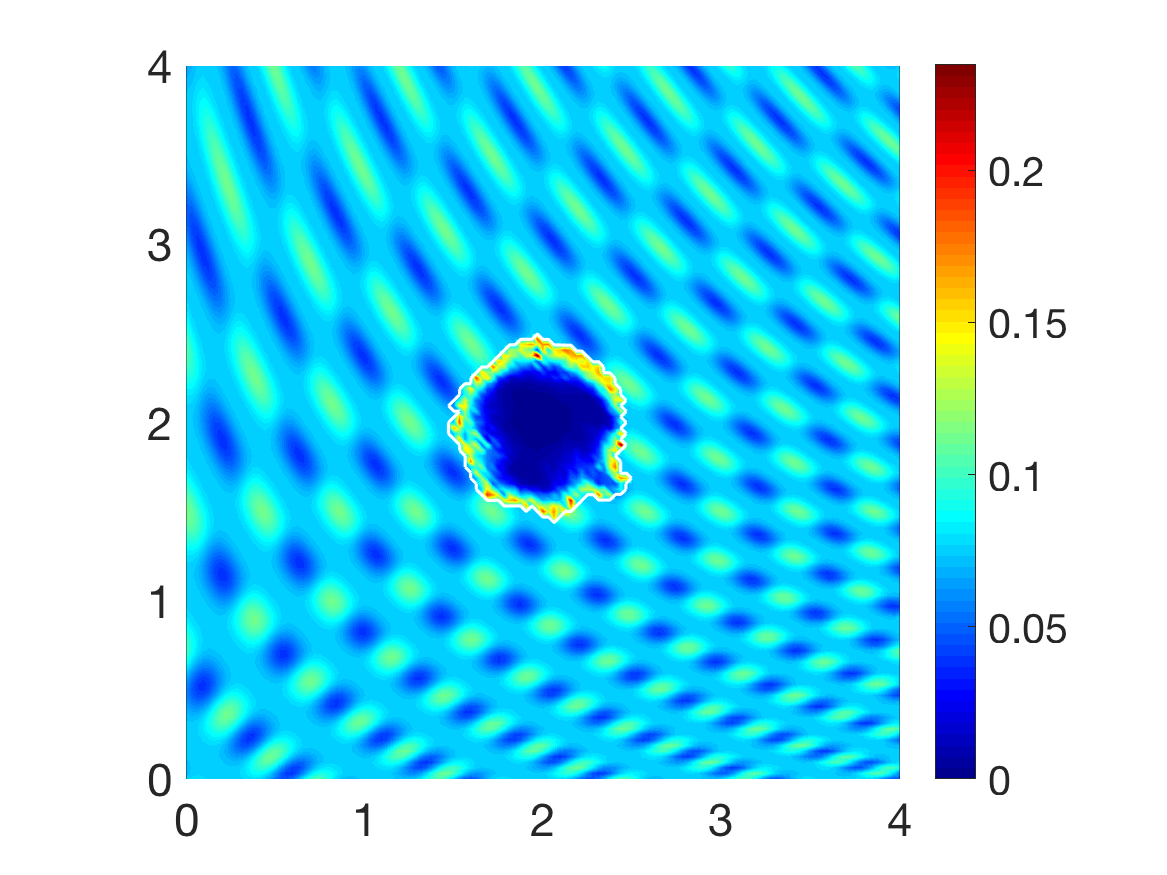}
  \caption{\emph{Fibre magnitude density}}
  \label{fig:heterotwopop50c}
  \end{subfigure}\hfil 
\begin{subfigure}{0.5\textwidth}
  \includegraphics[width=\linewidth]{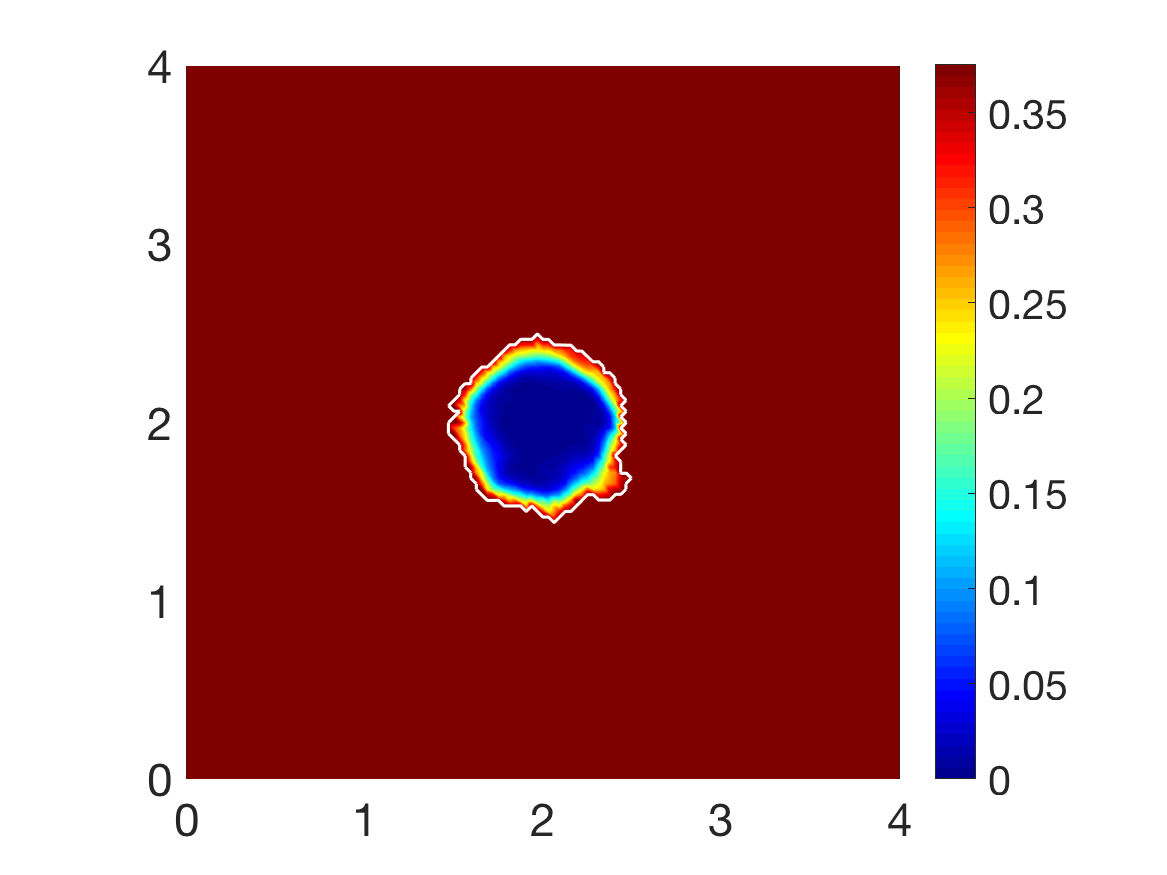}
  \caption{\emph{Non-fibres ECM distribution}}
  \label{fig:heterotwopop50d}
\end{subfigure}\hfil 

\medskip
\begin{subfigure}{0.5\textwidth}
  \includegraphics[width=\linewidth]{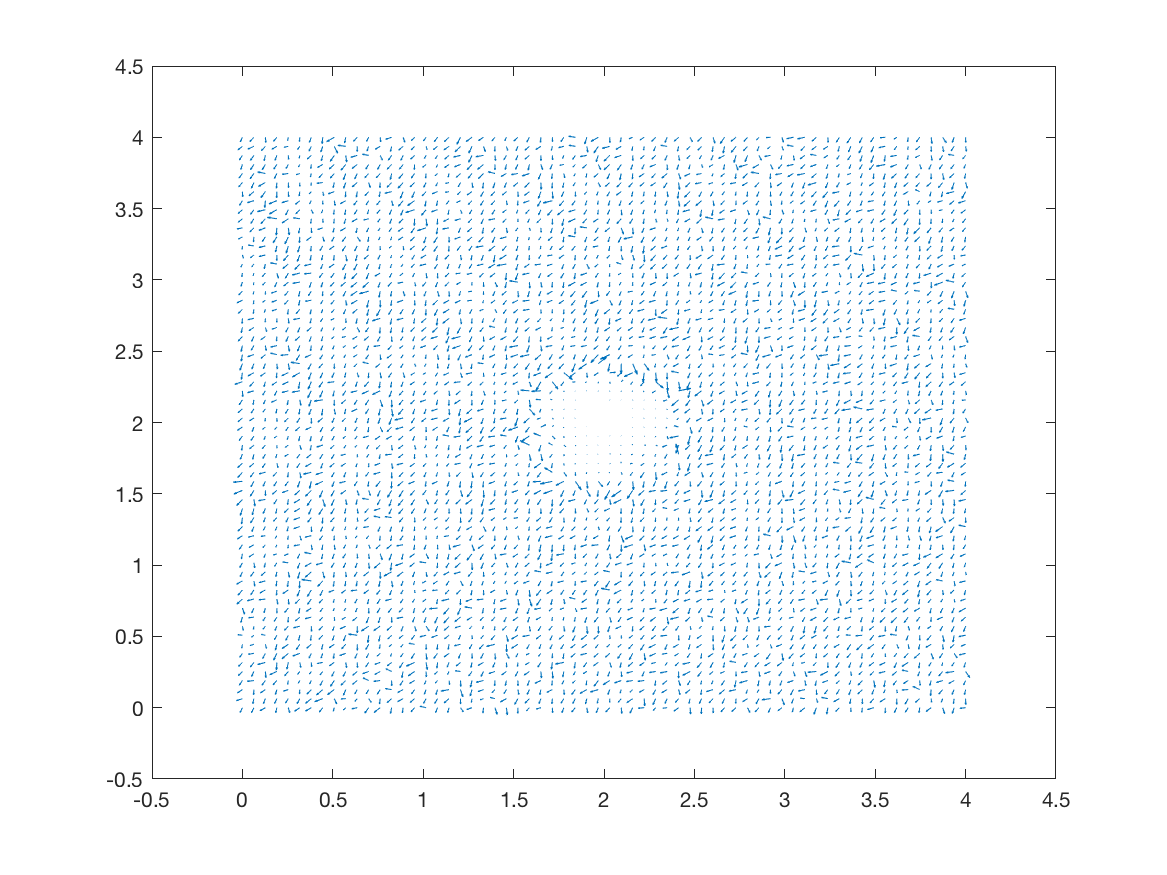}
  \caption{\emph{Fibre vector field - coarsened 2 fold}}
  \label{fig:heterotwopop50e}
  \end{subfigure}\hfil 
\begin{subfigure}{0.5\textwidth}
  \includegraphics[width=\linewidth]{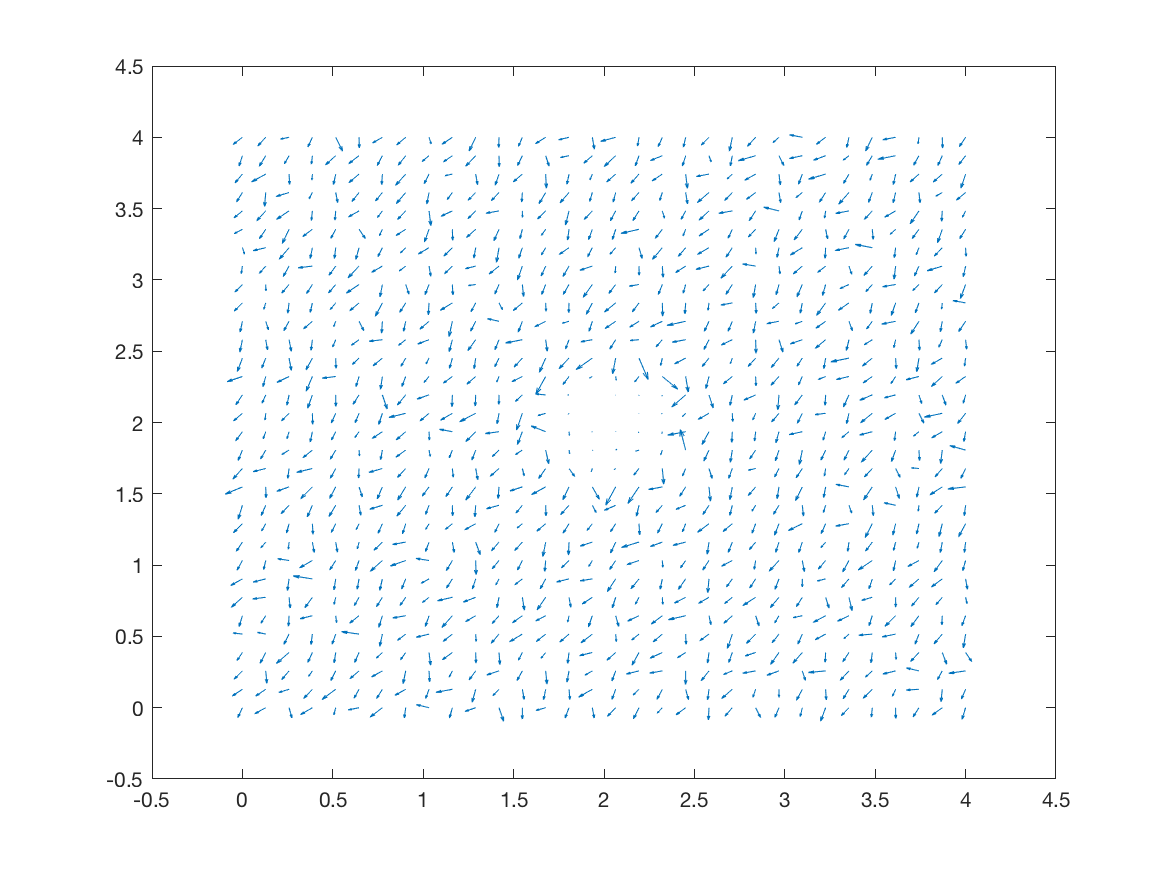}
  \caption{\emph{Fibre vector field - coarsened 4 fold}}
  \label{fig:heterotwopop50f}
\end{subfigure}\hfil 

\caption[Simulations at stage $50\Delta t$ with a homogeneous distribution of the non-fibrous phase and $15\%$ heterogeneous fibres phase of the ECM with a micro-fibres degradation rate of $d_f = 1$.]{\emph{Simulations at stage $50\Delta t$ with a homogeneous distribution of the non-fibrous phase and $15\%$ heterogeneous fibres phase of the ECM with a micro-fibres degradation rate of $d_f = 1$.}}
\label{fig:heterotwopop50}
\end{figure}

\begin{figure}[h!]
    \centering 
\begin{subfigure}{0.5\textwidth}
  \includegraphics[width=\linewidth]{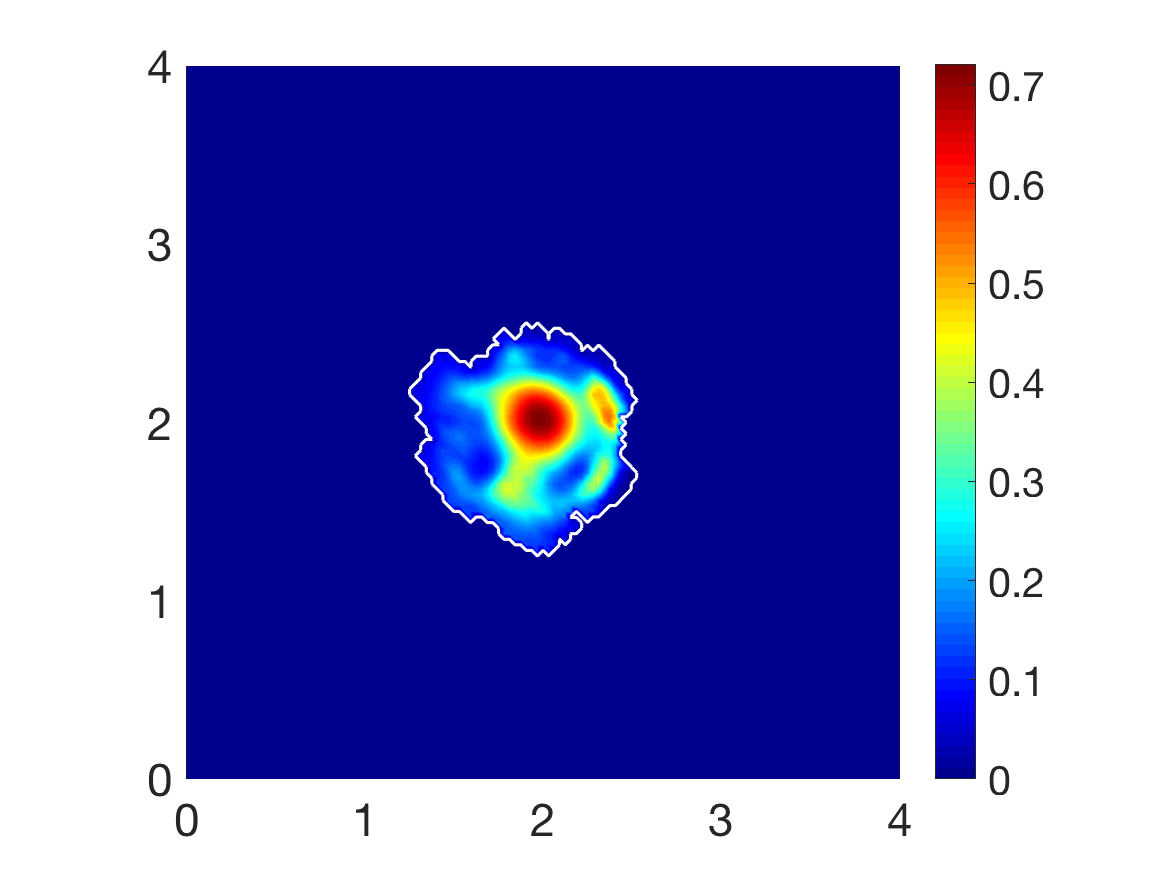}
  \caption{\emph{Cancer cell population}}
  \label{fig:heterotwopop75a}
\end{subfigure}\hfil 
\begin{subfigure}{0.5\textwidth}
  \includegraphics[width=\linewidth]{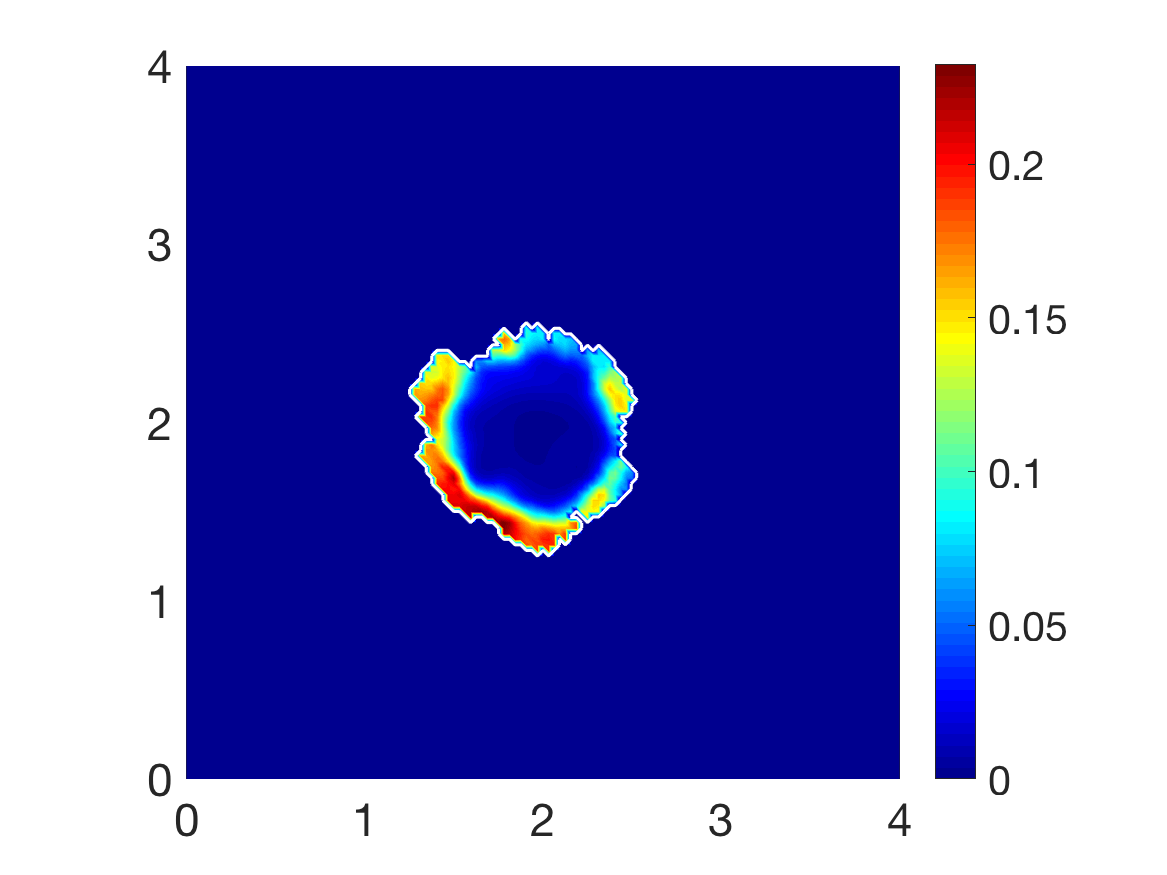}
  \caption{\emph{Cancer cell population 2}}
  \label{fig:heterotwopop75b}
\end{subfigure}\hfil 

\medskip
\begin{subfigure}{0.5\textwidth}
  \includegraphics[width=\linewidth]{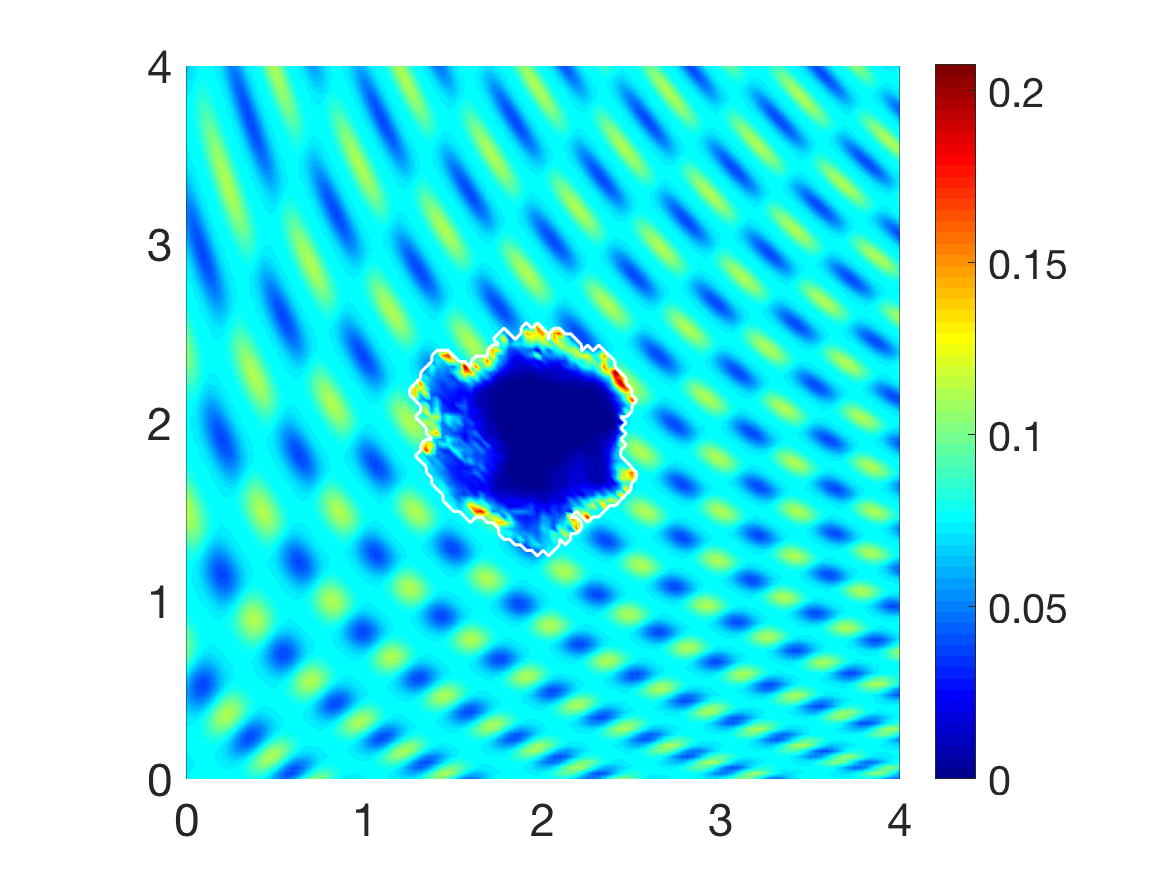}
  \caption{\emph{Fibre magnitude density}}
  \label{fig:heterotwopop75c}
  \end{subfigure}\hfil 
\begin{subfigure}{0.5\textwidth}
  \includegraphics[width=\linewidth]{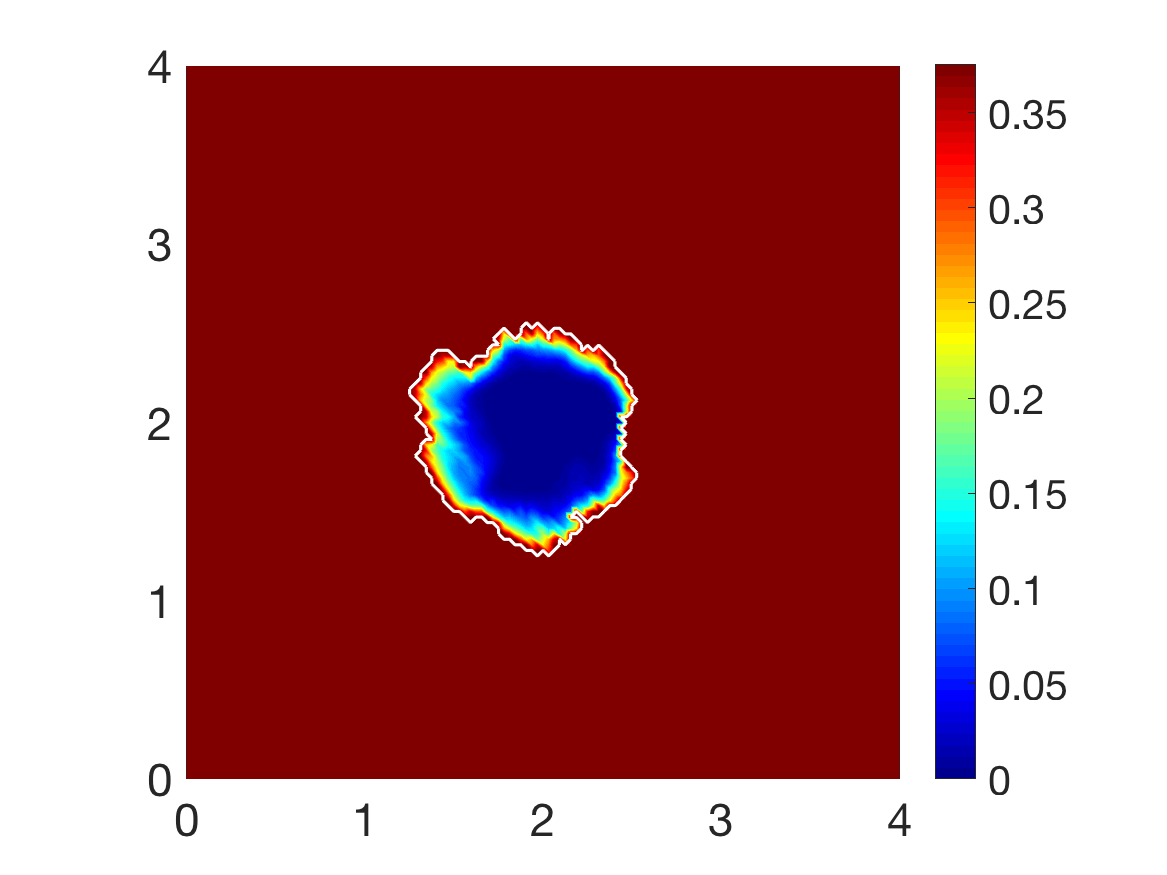}
  \caption{\emph{Non-fibres ECM distribution}}
  \label{fig:heterotwopop75d}
\end{subfigure}\hfil 

\medskip
\begin{subfigure}{0.5\textwidth}
  \includegraphics[width=\linewidth]{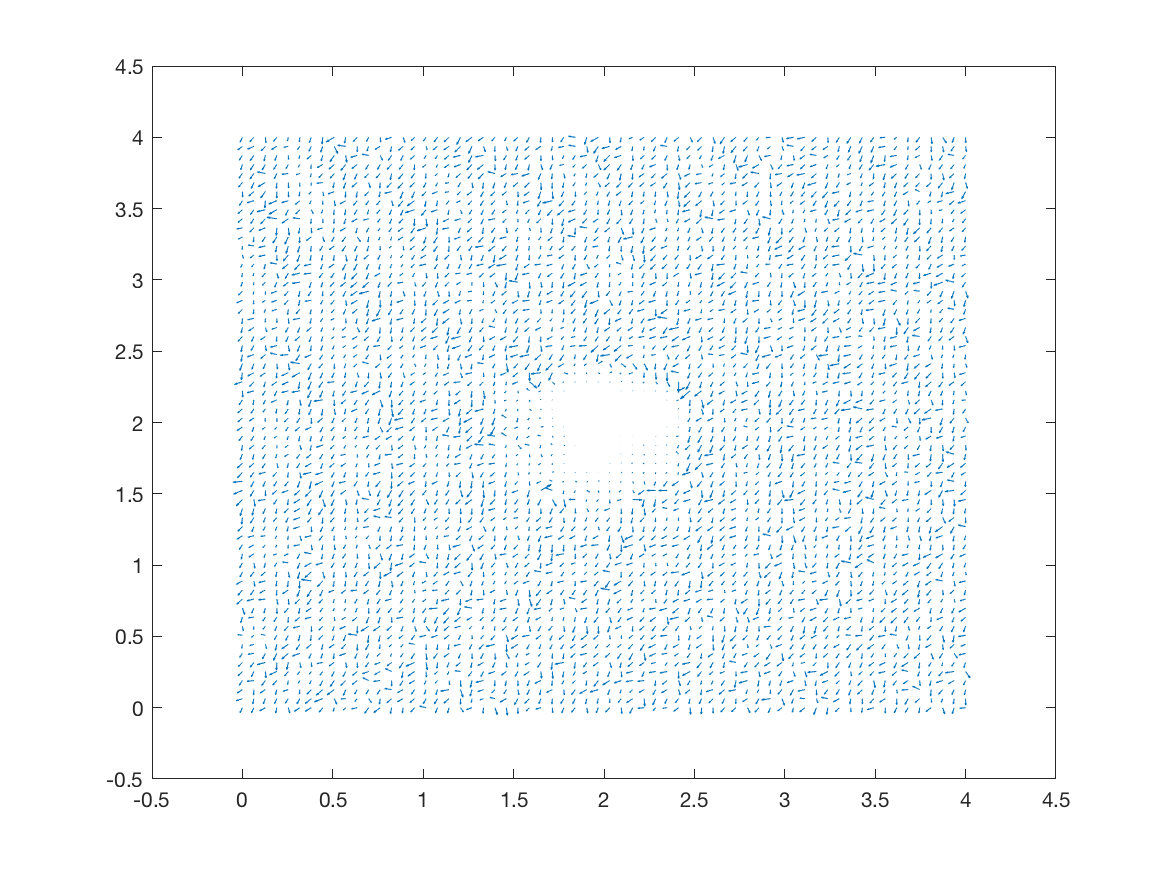}
  \caption{\emph{Fibre vector field - coarsened 2 fold}}
  \label{fig:heterotwopop75e}
  \end{subfigure}\hfil 
\begin{subfigure}{0.5\textwidth}
  \includegraphics[width=\linewidth]{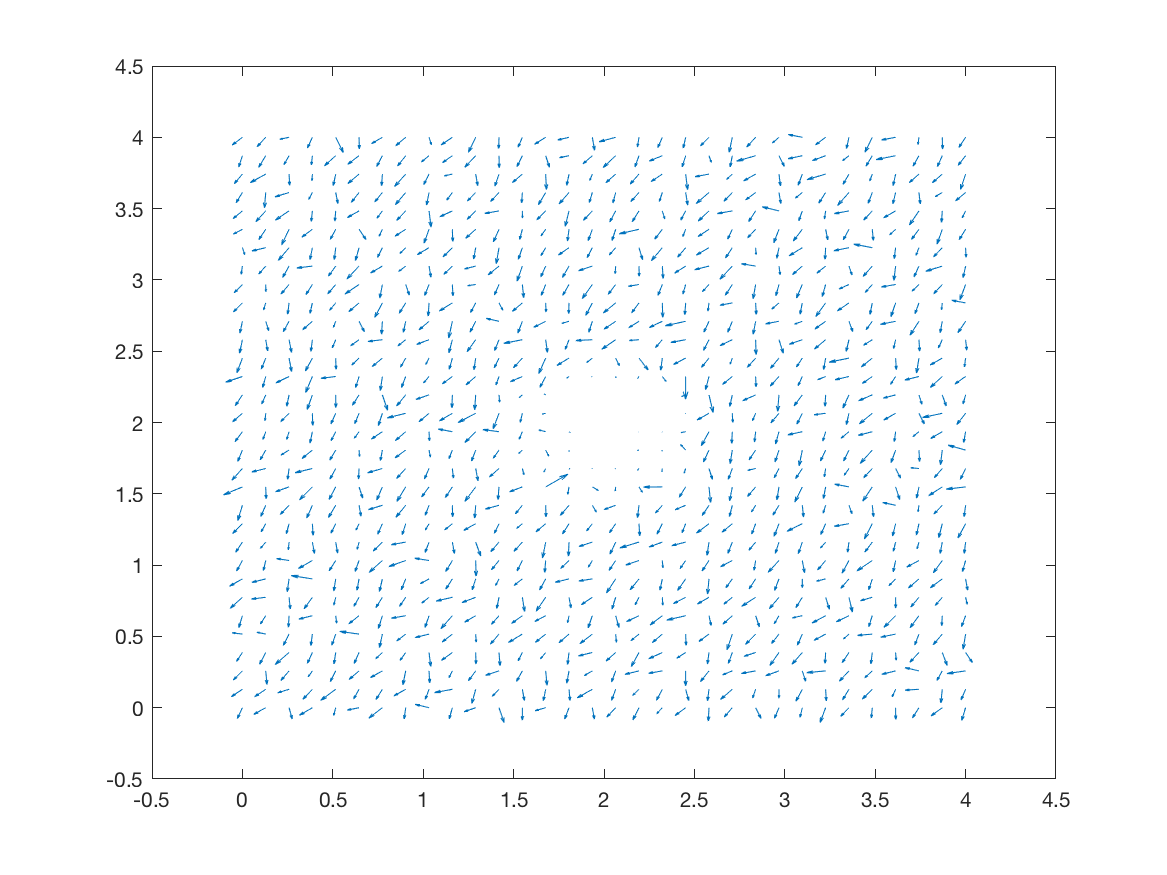}
  \caption{\emph{Fibre vector field - coarsened 4 fold}}
  \label{fig:heterotwopop75f}
\end{subfigure}\hfil 

\caption[Simulations at stage $75\Delta t$ with a homogeneous distribution of the non-fibrous phase and $15\%$ heterogeneous fibres phase of the ECM with a micro-fibres degradation rate of $d_f = 1$.]{\emph{Simulations at stage $75\Delta t$ with a homogeneous distribution of the non-fibrous phase and $15\%$ heterogeneous fibres phase of the ECM with a micro-fibres degradation rate of $d_f = 1$.}}
\label{fig:heterotwopop75}
\end{figure}

\section{Discussion}
In this paper we have presented \rs{an integrated} two-part multiscale model of cancer invasion, \dt{which builds on the approach introduced in} \cite{Shutt_2018} \dt{and extends that to capture explicitly the} \rs{dynamic} \dt{cell-scale interaction between the} MDE boundary \dt{micro-dynamics and the peritumoural mass distribution of micro-fibres}. 

\dt{Structured largely similar to the modelling framework introduced in \cite{Shutt_2018}}, the model proposed here combines two multiscale systems that \dt{share} the same tissue (macro-) scale dynamics while having separate cell- (micro-) scale processes that are simultaneously connected via two double feedback loops. \dt{Specifically, while this new modelling framework inherits completely the multiscale dynamics of naturally oriented ECM fibres (induced by the mass distribution of micro-fibres) occurring on the topological closure of the invading tumour (and including the dynamic rearrangement of fibres under the incidence of the macro-scale flux of cancer cells), this shares its macro-dynamics with a significantly extended multiscale moving-boundary modelling for the proteolytic dynamics at the tumour invasive edge that explicitly considers the interaction with the peritumoural fibres.} 

This model \dt{expands and takes forward both the initial multiscale moving-boundary framework introduced in \cite{Dumitru_et_al_2013} and its further development into the two-part multiscale modelling introduced in \cite{Shutt_2018}} by bringing in and exploring the \dt{cell-scale interactions} between \dt{the cross-interface} diffusion of MMPs \dt{and the micro-fibre distributions} in the peritumoural region, \dt{with direct impact upon} \rs{microscopic peritumoural degradation of micro-fibres} \dt{that results in a continuously altered macroscopic vector field of oriented ECM fibres at the tumour boundary. Moreover, these altered peritumoural ECM fibres have major relevance within the macroscopic dynamics of the cancer cells as this affects the cell-fibres adhesion properties at the leading edge of the tumour, impacting this way not only the tumour mechanics close to the tumour interface but the entire-tissue scale dynamics of the tumour}. 

\dt{To that end, we first explore mathematically the positive feedback that the macroscopic distribution of ECM fibres close to the tumour interface has upon the emerging cell-scale source of MMP-2 for the cross-interface micro-dynamics that MMP-2 exercise at the invading edge of the tumour. Specifically, in this new formulation we are able to capture the enhanced sources of MMP-2 in a relevant cell-scale neighbourhood which are enabled non-locally by the presence of elevated distributions of ECM fibres within neighbouring active regions from within the outer proliferating rim of the tumour where cancer cells arrive during their macro-dynamics and produce MMPs.}

\dt{Further, in the presence of the cell-scale MMP-2 source induced by the macro-dynamics, a cross-interface diffusion of MMP-2 occurs at the invasive edge of the tumour. However, as the MMP-2 find it easier to diverge along their gradient directions in regions with lower micro-fibres levels, by accounting for the presence micro-scale mass distribution, we finally obtain that the diffusion rate of this diffusive molecular transport process of MMP-2 naturally depend on the micro-fibres density.} \dt{Thus, this} \dt{cell-scale MMP-2 dynamics focuses the cross-interface molecular transport towards the regions of lower mass-distributions of micro-fibres, taking \emph{on-the-fly} advantage on the potential ``micro-fibres valleys" created by the multiscale dynamic rearrangement of fibres induced by the macro-scale flux of cancer cells, which was derived and explored with full details in \cite{Shutt_2018}}.  

\dt{Since the MMP-2 cross-interface molecular transport leads to peritumoural micro-fibres degradation at the micro-scale, we explored this degradation explicitly at the cell scale, remarking here at the same time that this complements the previous modelling framework introduced and discussed in \cite{Shutt_2018} where the fibres degradation was only considered on the bulk of the tumour at the macro-scale. To that end, we considered the correlation between the micro-fibres degradation and the incidence angle that the MMP-2 molecular flux makes with the regions of significant levels of micro-fibres distributed with any given $\epsilon Y$ from the covering bundle of $\{\epsilon Y\}_{\P}$ boundary micro-domains. This enabled us to derive mathematically a micro-fibres degradation law occurring on each fibre micro-domain $\delta Y(x)$ that has non-empty intersection with at least one of the boundary micro-domains $\epsilon Y$, in which maximum fibre degradation occurs when the angle between the fibres and MMPs flux is perpendicular, while the degradation decreases with increasing alignment of the fibres with the direction of the MMP-2 molecular flux.} This suggests that the highly aligned collagen fibrils will act as a pathway for invasion rather than a barrier against it, \dt{this being consistent with the biological evidence presented in \cite{Provenzano_06}}.\dt{Finally, this degradation of peritumoural mass distribution micro-fibres at the cell-scale is continuously in time translated back at macro-scale, having a natural and major impact upon \emph{on-the-fly} changes in the orientation and magnitudes of macro-scale ECM fibres from the peritumoural region.} 

\rs{This new modelling framework has been explored} in several scenarios, within both a homogeneous and heterogeneous initial distribution of fibres, and varying the initial ratio of macroscopic fibre distribution in relation to the non-fibres ECM phase. The non-fibres ECM phase was kept as a homogeneous density throughout the paper for the purpose of exploring only the influence of ECM fibres during cancer invasion. These scenarios were explored through randomly allocated \dt{distributions of micro-fibres over the fibres micro-domain, as considered} \dt{already} in \cite{Shutt_twopop} and defined in Appendix \ref{microfibres}.

We explored the differences between a homogeneous and heterogeneous initial distribution of fibres where we varied the initial percentage of fibres density from $15\%$ to $20\%$, as well as investigating the morphology of a heterotypic cancer cell population whose macroscopic dynamics were developed in \cite{Shutt_twopop}. We conclude from these simulations that a heterogeneous distribution of fibres induces a more lobular, fingering pattern of the tumour boundary, and an increase in initial fibre density promotes a more aggressive tumour spreading further into the surrounding tissue, a behaviour which is mirrored in the biological experiments performed in \cite{Provenzano_2008}. These remarks are in line with previous results in \cite{Shutt_2018,Shutt_twopop} where the same conclusions are drawn regarding the initial condition of the fibres ECM phase. The simulations performed in this paper exhibit an overall larger tumour spread than the simulations in \cite{Shutt_2018}, implying that the degradation of fibres at the tumour interface promotes tumour invasion. The final simulations, which consider the invasive behaviour of a heterotypic cell population, suggest that although the cancer cells migrate \rs{more easily} into low density regions of ECM, at the tumour interface a lack of fibre density is detrimental to the progression of the tumour. The low levels of fibre density inhibit the migration of cancer cells and thus the movement of the tumour boundary by reducing the opportunities for cell-fibre adhesion. In general, we conclude that the invasion of a heterotypic cancer cell population is accelerated in the presence of a high fibres ECM density, however within a low fibres ECM the tumour undergoes slower progression and the bulk of the cancer cells remain closer to the gradually expanding tumour boundary.

\rs{Looking forward, to advance with this model we would look to investigate the full MT1-MMP/MMP-2 cascade, namely, to include within the model the tissue-inhibitor of matrix-metalloproteinases-2, TIMP-2, a key molecule required for the activation of pro-MMP-2 \cite{Seiki_2003}, considering the role and regulation of these molecules and the resulting effects on peritumoural tissue degradation. Additionally, further exploration of the fibres network and its structure would permit better, more realistic modelling of human tissue, allowing the model to be compared with current biological experiments, for example, with the experiments performed in \cite{Provenzano_2008} that investigate the effects of increasing collagen density in the surrounding matrix.}

\section*{Appendix}
\appendix

\section{The radial kernel $\K(\cdot)$}\label{kernelAppendix}
To explore the influence on adhesion-driven migration decreases as the distance from $x+y$ to $x$ within the sensing region $B(x,r)$ increases, the expression of the radial dependent spatial kernel $\mathcal{K}(\cdot)$ appearing in \eqref{adhesionterm} is taken here to be: 
\vspace{-0.1cm}
\bequ
\mathcal{K}(r):=\frac{2\pi R^2}{3}\left(1-\frac{r}{2R}\right).
\eequ

\section{The mollifier $\psi_{\gamma}$} \label{standardmollifier}

The standard mollifier $\psi_{\gamma}:\R^{N}\to\R_{+}$ (which was used also in \cite{Shutt_2018,Dumitru_et_al_2013}) is defined as usual, namely
\[
\psi_{\gamma}(x):=\frac{1}{\gamma^{N}}\psi\big(\frac{x}{\gamma}\big),
\]
where $\psi$ is the smooth compact support function given by
\bequd
\psi(x):=
\left\{
\begin{array}{cll}
\frac{exp\frac{1}{\nor{x}^{2}_{_{2}}-1}}{\int\limits_{\Bila(0,1)}exp\frac{1}{\nor{z}^{2}_{_{2}}-1}dz},& \quad if & x\in \Bila(0,1),\\[0.3cm]
0, & \quad if & x\not\in \Bila(0,1).
\end{array}
\right.
\eequd

\section{Microscopic fibre domains}\label{microfibres}

For the fibres initial conditions,, on the micro-domains $\delta Y(x)$, we consider a family of five distinctive micro-fibres patterns, $\{P^{1}_{i}\}_{i\in J}$, which are defined by the union of paths $P^{1}_{i}=\bigcup\limits_{j=1..5} h^{1}_{i,j}$, which are given as follows.
\paragraph{For the family of fibre paths $P^{1}$, we have:}
\bequd
h^{1}_{1,1}: z_{1}=z_{2};  \quad
h^{1}_{1,2}: z_{1}=\frac{1}{2}; \quad
h^{1}_{1,3}: z_{1}=\frac{1}{5}; \quad
h^{1}_{1,4}: z_{2}=\frac{2}{5};\quad
h^{1}_{1,5}: z_{2}=\frac{4}{5}.
\eequd
\vspace{-0.6cm}
\bequd
h^{1}_{2,1}: z_{1}=z_{2};  \quad
h^{1}_{2,2}: z_{1}=\frac{1}{2}; \quad
h^{1}_{2,3}: z_{1}=\frac{1}{5}; \quad
h^{1}_{2,4}: z_{2}=\frac{2}{5};\quad
h^{1}_{2,5}: z_{2}=\frac{1}{10}.
\eequd
\vspace{-0.6cm}
\bequd
h^{1}_{3,1}: z_{1}=z_{2};  \quad
h^{1}_{3,2}: z_{1}=\frac{1}{10}; \quad
h^{1}_{3,3}: z_{1}=\frac{9}{10}; \quad
h^{1}_{3,4}: z_{2}=\frac{2}{5};\quad
h^{1}_{3,5}: z_{2}=\frac{1}{10}.
\eequd
\vspace{-0.6cm}
\bequd
h^{1}_{4,1}: z_{1}=z_{2};  \quad
h^{1}_{4,2}: z_{1}=\frac{1}{10}; \quad
h^{1}_{4,3}: z_{1}=\frac{3}{10}; \quad
h^{1}_{4,4}: z_{2}=\frac{2}{5};\quad
h^{1}_{4,5}: z_{2}=\frac{1}{10}.
\eequd
\vspace{-0.6cm}
\bequd
h^{1}_{5,1}: z_{1}=z_{2};  \quad
h^{1}_{5,2}: z_{1}=\frac{4}{5}; \quad
h^{1}_{5,3}: z_{1}=\frac{3}{10}; \quad
h^{1}_{5,4}: z_{2}=\frac{1}{10};\quad
h^{1}_{5,5}: z_{2}=\frac{7}{10}.
\eequd

For this family of fibre paths $P^{1}$, as described in \cite{Shutt_2018}, the micro-scale fibres pattern within each micro-domain $\delta Y(x)$ is given as 
\bequ
f(z,t):=\sum\limits_{j=1}^{5}\psi_{h^{l}_{i,j}}(z)(\chi_{_{(\delta-2\gamma_{_{0}})Y(x)}}\ast \psi_{\gamma_{_{0}}})(z)
\eequ
where $\{\psi_{h^{l}_{i,j}}\}_{i,j=1..5}$ are smooth compact support functions of the form
\bequ
\begin{array}{l}
\hspace{3.0cm}\psi_{h_{j}}: \delta Y(x) \rightarrow \mathbb{R}\\[0.3cm]
\textrm{defined as follows:}\\[0.6cm]
\textrm{$Case\,1:$ if $h^{l}_{i,j}$ is not parallel to $z_{1}-$axis} \\
\textrm{(i.e., $h^{l}_{i,j}$ is identified as the graph of a function of $z_{2}$)}\\[0.2cm]
\textrm{we have:}\\[0.2cm]
\psi_{h^{l}_{i,j}}(z_1,z_2):=
\left\{
\begin{array}{ll}
C_{h^{l}_{i,j}} e^{-{\frac{1}{r^2-(h_{j}(z_2) - z_1)^2}}}, &\quad \text{if} \ z_1 \in [h^{l}_{i,j}(z_2)-r, h^{l}_{i,j}(z_2)+r], \\[0.6cm]
0, &\quad \text{if} \ z_1 \not\in [h^{l}_{i,j}(z_2)-r, h^{l}_{i,j}(z_2)+r];
\end{array}
\right.\\[0.8cm]
\textrm{$Case\,2:$ if $h^{l}_{i,j}$ is parallel to $z_{1}-$axis}\\ 
\textrm{(i.e., $h^{l}_{i,j}$ is identified as the graph of a constant function of $z_{1}$)}\\[0.2cm]
\textrm{we have:}\\[0.2cm]
\psi_{h^{l}_{i,j}}(z_1,z_2):=
\left\{
\begin{array}{ll}
C_{h^{l}_{i,j}} e^{-{\frac{1}{r^2-(h^{l}_{i,j}(z_1) - z_2)^2}}}, &\quad \text{if} \ z_2 \in [h^{l}_{i,j}(z_1)-r, h^{l}_{i,j}(z_1)+r], \\[0.6cm]
0, &\quad \text{if} \ z_2 \not\in [h^{l}_{i,j}(z_1)-r, h^{l}_{i,j}(z_1)+r].
\end{array}
\right.\\[0.6cm]
\end{array}
\label{eq:fibappend}
\eequ
Here $r>0$ is the width of the micro-fibres and $C_{h^{l}_{i,j}}$ are constants that determine the maximum height of $\psi_{h^{l}_{i,j}}$ along the smooth paths $\{h^{l}_{i,j}\}_{i,j=1..5}$ in $\delta Y(x)$. Finally, $\psi_{\gamma}$ is the standard mollifier defined in Appendix \ref{standardmollifier}, with $\gamma_{_{0}}= h/16 $.\\

Finally, the initial spatial configuration of the pattern of macroscopic ECM fibre phase is selected according to a randomly generated matrix of labels $A=(a_{i,j})_{i,j=1..n}$ corresponding to the entire $n\times n$ grid discretising $Y$, in which the entries $a_{i,j}$ are allocated values randomly selected from the set of configuration labels $\{1,2,3,4,5\}$ that will dictate the choice of micro-fibres pattern among those described above that will be assigned to the micro-domains $\delta Y(j\Delta x, i\Delta y)$, for all $i,j=1..n$.

\section{Table for the parameter set $\Sigma_{1}$}    \label{paramSection}
Here we present a table for the parameter set $\Sigma_{1}$.
\begin{table}[htb!]
\centering
\begin{threeparttable}
\centering
\caption{The parameters in $\Sigma_{1}$}
\begin{tabular}{c c c c}
  \hline		
  Parameter & Value & Description & Reference \\
  \hline
  $D_1$ & $3.5 \times10^{-4}$ & diffusion coeff. for cell population $c_{1}$ & \cite{Domschke_et_al_2014}\\
  $D_2$ & $7\times10^{-4}$ & diffusion coeff. for cell population $c_{2}$ & \cite{Domschke_et_al_2014}\\
  $D_m$ & $10^{-3}$ & diffusion coeff. for MDEs & Estimated \\
  $\mu_1$ & $0.25$ & proliferation coeff. for cell population $c_{1}$ & \cite{Domschke_et_al_2014} \\
  $\mu_2$ & $0.25$ & proliferation coeff. for cell population $c_{2}$ & \cite{Domschke_et_al_2014} \\
  $\gamma_{1}$ & 2 & non-fibrous ECM degradation coeff. & \cite{Shutt_2018} \\
  $\gamma_{2}$ & 1.5 & macroscopic fibre degradation coeff. & \cite{Peng2016} \\
  $\alpha_{1}$ & $1$ & MDE secretion rate of $c_{1}$ & Estimated \\
  $\alpha_{2}$ & $1.5$ & MDE secretion rate of $c_{2}$ & Estimated \\
  $R$ & 0.15 & sensing radius & \cite{Shutt_2018} \\
  $r$ & 0.0016 & width of micro-fibres & \cite{Shutt_2018} \\
  $f_{\text{max}}$ & 0.6360 & max. micro-density of fibres & \cite{Shutt_2018} \\
  $p$ & 0.15-0.2 & percentage of non-fibrous ECM & Estimated \\
  $h$ & 0.03125 & macro-scale spatial discretisation size & \cite{Dumitru_et_al_2013}\\
  $\epsilon$ & 0.0625 & size of micro-domain $\epsilon Y$ & \cite{Dumitru_et_al_2013} \\
  $\delta$ & 0.03125 & size of micro-domain $\delta Y$ & \cite{Shutt_2018} \\
   \hline 
  \label{table:parameters} 
\end{tabular}
\end{threeparttable}
\end{table}

\newpage
\bibliography{ThesisReferences}

\end{document}